\documentclass[12pt]{article}
\pdfoutput=1

\usepackage{xcolor}
\definecolor{darkblue}{rgb}{0.1,0.1,.7}
\usepackage[colorlinks, linkcolor=darkblue, citecolor=darkblue, urlcolor=darkblue, linktocpage,pagebackref]{hyperref} 

\usepackage[english]{babel}
\usepackage{amsmath,amssymb,graphicx,enumerate,bbm}
\usepackage{booktabs}
\usepackage{geometry}
\usepackage[margin=10pt,font=small,labelfont=bf]{caption}
\usepackage{mcite}

\usepackage{titlesec}
\titlespacing\section{0pt}{12pt plus 4pt minus 2pt}{0pt plus 2pt minus 2pt}
\titlespacing\subsection{0pt}{12pt plus 4pt minus 2pt}{0pt plus 2pt minus 2pt}
\titlespacing\subsubsection{0pt}{12pt plus 4pt minus 2pt}{0pt plus 2pt minus 2pt}
\titleformat*{\section}{\large\bfseries}
\titleformat*{\subsection}{\normalsize\bfseries}
\titleformat*{\subsubsection}{\normalsize\it}
\titleformat*{\paragraph}{\normalsize\bfseries}
\titleformat*{\subparagraph}{\normalsize\bfseries}

\usepackage{cite}

\usepackage[utf8]{inputenc}

\usepackage{theorem}
\newtheorem{corollary}{Corollary}[section]
\newtheorem{lemma}{Lemma}[section]
{\theorembodyfont{\rmfamily}\newtheorem{remark}{Remark}[section]}
{\theorembodyfont{\rmfamily}\newtheorem{example}{Example}[section]}
\newtheorem{theorem}{Theorem}[section]

\newcommand{\backassign}{=:}
\newcommand{\cdummy}{\cdot}

\newcommand{\tmem}[1]{{\em #1\/}}
\newcommand{\tmmathbf}[1]{\ensuremath{\boldsymbol{#1}}}
\newcommand{\tmop}[1]{\ensuremath{\operatorname{#1}}}

\newcommand{\tmscript}[1]{\text{\scriptsize{$#1$}}}
\newcommand{\tmstrong}[1]{\textbf{#1}}
\newcommand{\tmtextbf}[1]{{\bfseries{#1}}}
\newcommand{\tmtextit}[1]{{\itshape{#1}}}
\newcommand{\tmtextrm}[1]{{\rmfamily{#1}}}
\newcommand{\tmtextup}[1]{{\upshape{#1}}}
\newenvironment{enumeratenumeric}{\begin{enumerate}[1.] }{\end{enumerate}}
\newenvironment{enumerateroman}{\begin{enumerate}[i.] }{\end{enumerate}}
\newenvironment{itemizedot}{\begin{itemize} }{\end{itemize}}

\newcommand{\reef}[1]{(\ref{1})}
\newcommand{\vareps}{\varepsilon}
\newcommand{\eps}{\varepsilon}
\newcommand{\del}{\partial}
\newcommand{\nn}{\nonumber}

\newcommand{\bR}{\mathbb{R}}

\newcommand{\calN}{\mathcal{N}}
\newcommand{\calM}{\mathcal{M}}

\newcommand{\calA}{\mathcal{A}}

\providecommand{\tfrac}[2]{\frac{#1}{#2}}
\newcommand{\half}{\frac{1}{2}}

\providecommand{\propto}{{\hspace{0.17em}}{\originalpropto}{\hspace{0.17em}}}

\newcommand{\Dpsi}{[\psi]}

\newcommand{\bx}{\mathbf{x}}

\newcommand{\bA}{\mathbf{A}}
\newcommand{\bB}{\mathbf{B}}
\newcommand{\barbB}{\overline{\mathbf{B}}}

\newcommand{\eff}{\rm{eff}}

\newcommand{\loc}{\tmop{loc}}

\newcommand{\Xterm}{\tmmathbf{\mathfrak{X}}}

\newcommand{\Resc}[2]{\Bigl[\hspace{-0.17em} \hspace{-0.17em} \begin{array}{c}
  #1\\
  #2
\end{array} \hspace{-0.17em} \hspace{-0.17em}\Bigr]}

\newcommand{\rest}{\text{R}}

\newcommand{\dbar}{{\kern}-.1em{\raise}.8ex-{\kern}-.6emd}

\newcommand{\comment}[1]{}

\providecommand{\qed}{{\hspace{fill}}{\raise}1pt{\vrule}height5pt width5pt
depth0pt}

\newcommand{\g}{\gamma}
\newcommand{\e}{\varepsilon}

\newcommand{\sign}{\mathrm{sign}}
\newcommand{\otto}{{\hspace{0.17em}}{\kern}-1.truept{\leftarrow}{\kern}-5.truept{\to}{\kern}-1.truept{\hspace{0.17em}}}

\providecommand{\qed}{{\hspace{fill}}{\raise}1pt{\vrule}height5pt width5pt
depth0pt}

\newcommand{\lis}[1]{\overline{#1}}

\newcommand{\pref}[1]{(\ref{1})}

\newcommand{\lb}[1]{\label{1}}

\newcommand{\betadiag}{I^{\resizebox{15pt}{!}{\includegraphics{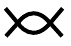}}}_2}
\newcommand{\cf}{- 8}

\renewcommand{\le}{\leqslant}

\renewcommand{\ge}{\geqslant}

\numberwithin{equation}{section}

\usepackage{tocloft}
\usepackage{chngcntr}
\counterwithin{figure}{section}

\usepackage{etoolbox}
\apptocmd{\thebibliography}{\setlength{\itemsep}{0em}}{}{}

\begin{document}
	
	\vspace*{-.6in} \thispagestyle{empty}
	\begin{flushright}
	\end{flushright}

	\vspace{.2in} {\Large
		\begin{center}
				\bf  Gentle introduction to rigorous Renormalization Group:\\ a worked
				fermionic example
		\end{center}
	}
	\vspace{.2in}
	\begin{center}
		{\bf 
			Alessandro Giuliani$^{a,b}$, Vieri Mastropietro$^{c}$, Slava Rychkov$^{d,e}$
		} 
		\\
		\vspace{.2in} 
		\it \small$^{a}$Dipartimento di Matematica e Fisica, Universit{\`a} degli
		Studi Roma Tre, \\L.go S. L. Murialdo 1, 00146 Roma - Italy\\[0.2em]
		$^{b}$Centro Linceo Interdisciplinare Beniamino Segre, Accademia Nazionale dei
		Lincei,\\ Via della Lungara 10, 00165 Roma - Italy\\[0.2em]
		$^{c}$Dipartimento di Matematica, Universit{\`a} degli Studi di Milano,\\ Via Cesare
		Saldini 50, 20133 Milano - Italy\\[0.2em]
		$^{d}$Institut des Hautes \'Etudes Scientifiques, Bures-sur-Yvette, France\\[0.2em]
		$^{e}$Laboratoire de Physique de l’Ecole normale sup\'erieure, ENS, \\
			Universit\'e PSL, CNRS, Sorbonne Universit\'e, Universit\'e de Paris, F-75005 Paris, France
	\end{center}
	
	\vspace{.2in}

\begin{abstract}
  Much of our understanding of critical phenomena is based on the notion of
  Renormalization Group (RG), but the actual determination of its fixed points
  is usually based on approximations and truncations, and predictions of
  physical quantities are often of limited accuracy. The RG fixed points can
  be however given a fully rigorous and non-perturbative characterization, and
  this is what is presented here in a model of symplectic fermions with a
  nonlocal (``long-range'') kinetic term depending on a parameter
  $\varepsilon$ and a quartic interaction. We identify the Banach space of
  interactions, which the fixed point belongs to, and we determine it via a
  convergent approximation scheme. The Banach space is not limited to relevant
  interactions, but it contains all possible irrelevant terms with
  short-ranged kernels, decaying like a stretched exponential at large
  distances. As the model shares a number of features in common with $\phi^4$
  or Ising models, the result can be used as a benchmark to test the validity
  of truncations and approximations in RG studies. The analysis is based on
  results coming from Constructive RG to which we provide a tutorial and
  self-contained introduction. In addition, we prove that the fixed point is
  analytic in $\varepsilon$, a somewhat surprising fact relying on the
  fermionic nature of the problem.
\end{abstract}
\vspace{.3in}
\hspace{0.7cm} 
{August 2020}
\newpage
\tableofcontents

\section{Introduction}\label{introduction}

Renormalization group (RG) is a pillar of theoretical physics, explaining how
long-distance collective behavior emerges from microscopic models. Critical
phenomena are thus understood in terms of RG fixed points (Wilson
{\cite{Wilson:1973jj,Wilson:1974mb,Wilson:1983dy}}) and universality is
explained in terms of basins of attractions. While this beautiful picture
qualitatively works very well, quantitative applications often lead to
practical difficulties. To compute critical exponents, one typically does perturbation theory in a
small parameter, like $\epsilon = 4 - d$ in the $\epsilon$-expansion
{\cite{Wilson:1971dc}}. The accuracy of this procedure is limited by the
proliferation of Feynman diagrams, and by the slow convergence of
Borel-resummed series (while without resummation it normally diverges). As a
consequence, predictions of critical exponents using
{{perturbative}} RG
\tmtextbf{}{\cite{Guida:1998bx,Kompaniets:2017yct}} are often less accurate
than lattice Monte Carlo simulations or the conformal bootstrap
{\cite{Poland:2018epd}}.

It should be stressed that Wilson did not consider RG limited to situations with a small coupling.
Two strongly coupled RG examples can be found in \cite{Wilson:1974mb}. One is his famous solution of the Kondo problem. The other is less known but no less impressive: an RG calculation for the 2D Ising model in a space of 217 couplings, concluding that ``one can do precise calculations using pure renormalization group methods with the only approximations being based on locality.''

Other developments in theoretical physics suggest, indirectly, that Wilson ideas are non-perturbatively correct. We can mention here exact results in two-dimensional field theory (see e.g.~\cite{Mussardo:2010mgq}), obtained by conformal field theory, integrable models, exact S-matrix bootstrap etc., which have provided many examples of exact non-perturbative RG flows, never finding any inconsistency with Wilsonian expectations. In higher dimensions, exact results in supersymmetric theories (see e.g.~\cite{Terning:2006bq}) as well as the gauge-gravity duality considerations (see e.g.~\cite{deBoer:1999tgo, Bianchi:2001de}) have always confirmed Wilson ideas.

It is therefore somewhat surprising that the most straightforward interpretation of Wilson's vision, as a non-perturbative machine which would allow non-perturbatively and with an essentially unlimited precision to compute the properties of any RG fixed point of interest, has not so far been achieved. It is fair to say that this was not for the lack of trying, see e.g.~\cite{burkhardt1982real} for the early attempts. 

Two notable, although not fully successful, attempts have been the Functional Renormalization Group (FRG) \cite{Berges:2000ew,Delamotte:2007pf} and Tensor Network Renormalization (TNR) \cite{Levin:2006jai,Evenbly,Gilt}. The FRG calculations include couplings with arbitrary powers of fluctuating field, but only up to some finite derivative order. Unfortunately, with more derivatives, FRG results tend to become more and more sensitive to the parameters specifying the regulating function
		{\cite{Balog:2019rrg,DePolsi:2020pjk}}. This has been traced to increasing
		violations of conformal invariance, except at some special parameter values
		satisfying a ``principle of minimal sensitivity'' {\cite{Balog:2020fyt}}. It
		remains to be understood why the convergence to the fixed point does not hold
		more robustly in FRG. As for the TNR, it works well for simple 2D lattice models such as the 2D Ising model but hasn't been yet as effective in higher dimensions.

So, in spite of these attempts, although it is generally believed that the non-perturbative Wilsonian RG fixed points do
exist, at present they often remain Platonic objects, confined to the world of
ideas and accessible to us only via approximations of rather limited accuracy.
Take e.g.~the RG fixed point for the 3D Ising model. If it exists, which
Banach space does it belong to? Can we access it via a provably convergent
approximation scheme? At the moment these questions are wide open.

Note that Wilson believed in the Ising RG fixed point very concretely: as a
fixed-point Hamiltonian invariant under a Kadanoff block-spin transformation.
As mentioned above, for 2D Ising, he found an approximate fixed-point Hamiltonian numerically,
truncating to a space of 217 lattice spin interactions \cite{Wilson:1974mb}.
But the convergence of his scheme has never been proven, nor has it been
implemented in 3D. Incidentally, Wilson did worry about rigorous convergence properties of RG maps; e.g. in \cite{Wilson:1970tp} a model RG transformation was shown to be convergent for a rescaling parameter larger than $4\times 10^6$.

Of course, as already mentioned, there are nowadays other methods to get precise values of
critical exponents, most notably the conformal bootstrap {\cite{Poland:2018epd}}.
However, this does not mean that the RG should be abandoned. First, RG is more general than the conformal bootstrap, since 
many RG fixed points important for physics do not have conformal invariance, such as any fixed point involving
time evolution or relaxation, and having a dynamical critical exponent $z \neq
1$. Second, it is quite possible that there exist much better RG implementations, and we just
haven't found them yet.

Since theoretical physicists have not been able to find a fully successful implementation of non-perturbative RG in spite of many attempts, can mathematical physics give any hint about what we have been doing wrong? In mathematical physics, the rigorous construction of
non-trivial RG fixed points has been achieved in different cases using
Constructive RG (CRG). It has been obtained in bosonic scalar field theories\footnote{See
	Section \ref{bosons} and Appendix \ref{literature} for more details about the rigorously constructed
	bosonic fixed points.}
{\cite{Brydges1998,Brydges:2002wq,Abdesselam:2006qg,Mitter-talk,Slade}} and interacting fermions {\cite{Gawedzki:1985jn}} with
long range interactions, in cases where the system has a scaling dimension
differing from marginality by an $\varepsilon$. It has also been achieved in
models with marginal interactions of strength $\lambda$ and asymptotically
vanishing beta function, such as 1D interacting fermions
{\cite{BGPS,BMXYZ,BMWICMP,BFMTh}} and 2D spin, vertex and dimer models
{\cite{MaCMP,GMCMP,GMPRL,BFMCMP,BFMPRL,Dimers,GMTJSM,GMTCMP}}. In all these
cases, the non-perturbative existence of a non-trivial RG fixed point, close
to the Gaussian or free Fermi one, has been proved for $\varepsilon$ or
$\lambda$ sufficiently small, and the critical exponents can be computed at an
arbitrary precision in terms of resummed perturbative expansions, with
rigorous bounds on the remainder. A feature of the CRG is that the fixed point
is found, without any approximation, in a Banach space of interactions where
all the irrelevant terms are nonlocal, even though fast decaying (e.g. like a
stretched exponential): this is in striking contrast with the FRG, where the
space of interactions is typically spanned by a sequence of {\tmem{local}} functions of
the fluctuation field and its derivatives.\footnote{For a fair comparison it
should be noted that FRG calculations are often performed in terms of the 1PI
effective action, not the Wilsonian effective action used here. Also, some FRG schemes do attempt to go beyond the local derivative expansion. See Section
\ref{FRG}.}

One lesson of all this body of rigorous work is that \emph{weakly coupled} non-perturbative RG is possible, both in the bosonic and fermionic case, although it is easier in the fermionic case because in this case convergent perturbation theory captures full non-perturbative physics. We emphasize that, in general, fermionic perturbation theory is expected to be convergent only in the running rather than the bare coupling, see Section \ref{sec:analyticnonpert}. On the other hand, \emph{strongly coupled} non-perturbative RG has so far been out of reach of mathematical physics research. For this reason we will, as a first step, focus in this paper on the weakly coupled fermionic case, well understood by mathematical physicists, and aim to transfer this knowledge into the theoretical physics realm. 

With this in mind, we will present here the rigorous construction of a non-Gaussian fixed
point for a fermionic model with weakly relevant quartic interaction. This is
possibly the simplest model of this kind where to test field-theoretical RG
methods, and a perfect example to provide an introduction to CRG accessible to a wider audience.
We will explain how these methods allow one to characterize a non-trivial fixed point without any ad-hoc
assumption or any uncontrolled approximation. The above mentioned crucial role played by the
space of {\tmem{mildly nonlocal}} interactions, as opposed to expanding all interactions in local functions of the field, will be evident from our presentation.

A complementary goal of our work will be to prove a new result, which is the {\tmem{analyticity}} of
the $\varepsilon$-expansion for our non-Gaussian fermionic fixed point. This
is in contrast with bosonic $\varepsilon$-expansions, which are, at best,
Borel summable. Although we will focus on $\varepsilon > 0$ in much of the
paper, eventually we will show analyticity in a complex disk around
$\varepsilon = 0$. Although the sign of $\varepsilon$ is correlated
with the sign of the fixed point quartic interaction, Dyson's argument against
analyticity does not apply for our model because fermions are allowed to have
quartic interaction of either sign. Moreover, analyticity in $\varepsilon$ of the fixed point is not 
	in contradiction with the divergence of perturbation theory in the bare couplings, see Section \ref{sec:analyticnonpert} for 
	further comments.
	
	Note that in this paper we only construct RG fixed points, and we do not discuss in detail the RG flow between the microscopic model and the constructed fixed points. In any case, the result about analyticity is only valid for the fixed point and does not extend to the full RG flow, whose very structure changes discontinuously with the sign of $\eps$. For positive $\eps$ we will have the gaussian model at short distances, perturbed by the relevant quadratic and quartic couplings and flowing at long distances to the nontrivial RG fixed point. For negative $\eps$ it will be the other way around: starting from the nontrivial RG fixed point we may flow to the gaussian model at long distances, with the quartic coupling then describing the leading irrelevant interaction at long distances.
The former situation is referred to as `IR fixed point', while the latter as `UV fixed point'. To avoid any ambiguity we stress that no reverse RG flow is implied: all RG flows are from short to long distances.

The convergence of our CRG scheme lets us hope that it will be used as a starting point for developing systematic and stable truncation
schemes for the {fermionic} FRG. Although we have been able to prove our theorems only for small $\eps$, it might be that the range of practical applicability of our scheme is order one and includes strongly coupled fixed points---in the future one should try to see if this is the case (see Section \ref{sec:increasing}).

Rigorous non-perturbative constructions of bosonic fixed points present several complications compared to the fermionic case discussed here (see Section \ref{bosons}). Extracting lessons from those constructions for practical RG calculations remains another important open problem for the future.

\subsection{The model}
The model we consider is schematically described by the following action:
\begin{equation}
  \tmop{aMFT} (\psi) + {\nu_0} \int d^d x\, \psi^2 + {\lambda_0} \int d^d x\, \psi^4,
  \label{aMFT}
\end{equation}
where aMFT is an `anticommuting Mean Field Theory' of the
fermionic\footnote{Fermionic$=$anticommuting$=$Grassmann in this paper. }
field $\psi$ with $N$ components, and the two additional terms are quadratic
and quartic interactions preserving $S p (N)$ global symmetry. Our fermions
will be scalars under rotation, rather than spinors. So the model is not
reflection-positive, but reflection positivity will play no role in the RG
analysis.

{Local} models of this kind were considered in
{\cite{Kausch:1995py}} {{under the name
`symplectic fermions'}}, and are relevant for the description of polymers and
loop-erased random walks \ {\cite{Saleur:1991hk,Wiese:2018dow}}. In 3D,
interacting symplectic fermion models have been recently considered in the
context of dS/CFT correspondence {\cite{Anninos:2011ui}}. See also
{\cite{Fei:2015kta,Stergiou:2015roa}} for theoretical studies of related
models, and {\cite{LeClair:2006kb,LeClair:2006mb,LeClair:2007iy}} for other
speculative appearances of \ symplectic fermions in physics. Our model is a
variant of those, with a nonlocal (``long-range'') kinetic term of the
schematic form $\psi \left( \del^2 \right)^{\#} \psi$ where $\left( \del^2
\right)^{\#}$ is a non-integer power of the Laplacian. Similar long-range
models in a large $N$ limit were recently considered in {\cite{Gross:2017vhb}}
in relation to the SYK model.

More precisely, our model is defined as follows: take an even number $N$ of
real Grassmann fields $\psi = (\psi^{}_a)_{a = 1}^N$ in $\mathbb{R}^d$,
\footnote{We will not make a distinction between the lower and upper $S p (N)$
and Euclidean indices whose position is determined only by typographic
convenience: $\psi^a \equiv \psi_a$ and $\partial_{\mu} \psi^{} \equiv
\partial^{\mu} \psi_{}$.} with $d = 1, 2, 3$. The reference Gaussian theory
(the aMFT mentioned above) is characterized by the following two-point
function:
\begin{equation}
  \langle \psi_a (x) \psi_b (y) \rangle = \Omega_{\mathit{a b}} P (x - y)
  \equiv G_{a b} (x, y), \label{eq:prop1.1}
\end{equation}
where $P (x) \propto \frac{1}{| x |^{d / 2 - \varepsilon}}$ at large distances
(see the next section for the explicit expression), with $\varepsilon$ small
and positive, and $\Omega_{ab}$ is the symplectic $N \times N$ matrix:
\begin{equation}
  \Omega_{ab} = {\small \left[ \begin{array}{ccccc}
    & 1 &  &  & \\
    - 1 &  &  &  & \\
    &  & \ddots &  & \\
    &  &  &  & 1\\
    &  &  & - 1 & 
  \end{array} \right]} 
\end{equation}
The quadratic and quartic monomials $\psi^2$ and $\psi^4$ in {\eqref{aMFT}} must be interpreted as
$\Omega_{a b} \psi_a \psi_b$ and $(\Omega_{a b} \psi_a \psi_b)^2$. Given the form of $P(x)$, the fields $\psi_a$ are assigned the 
scaling dimension $[\psi]=d / 4 - \varepsilon / 2$, so that the quadratic
and quartic terms in {\eqref{aMFT}} are both relevant, the quartic one being
barely so for $\varepsilon$ small and positive. The parameter $\varepsilon$
plays a role similar to the deviation of spatial dimension $d$ from 4,
$\epsilon = 4 - d$ in the Wilson-Fisher $\epsilon$-expansion, which, contrary
to ours, {is not at present suitable for a rigorous
non-perturbative RG analysis because the space of $4 - \epsilon$ dimensions
has not been rigorously defined so far.\footnote{See however
{\cite{El-Showk:2013nia}} for a non-perturbative analysis in non-integer $d$
using the conformal bootstrap.}} For $d = 1, 2, 3$, there are no other local
relevant or marginal terms in addition to those included in {\eqref{aMFT}}. In
perturbative RG, the lowest order RG equations for the fixed point are,
letting $\gamma$ be the scaling parameter:
\begin{equation}
  \nu = \gamma^{d / 2 + \varepsilon} (\nu + I_1 \lambda + \cdots), \qquad
  \lambda = \gamma^{2 \varepsilon} (\lambda + I_2 \lambda^2 + \cdots),
  \label{eq:fix1.4}
\end{equation}
where $I_1$ and $I_2$ are positive constants given by the
{{one-loop Feynman diagrams}}. Neglecting the
higher-order terms, we get a nontrivial fixed point $\lambda_{\ast} = (1 -
\gamma^{2 \varepsilon}) / I_2$, $\nu_{\ast} = I_1 \lambda_{\ast} / (1 -
\gamma^{d / 2 + \varepsilon})$, which is $O (\varepsilon)$, close to the
Gaussian one. This is just an approximation, and we want to be sure that the
existence of the fixed point is not spoiled by non-perturbative effects caused
by higher orders or irrelevant terms. Moreover, we want to define a scheme
whose truncations provide arbitrarily good approximations of the actual fixed
point, with apriori bounds on the error made.

\subsection{{Strategy and open questions}}

Our rigorous construction of the fixed point goes as follows.
First, we identity a space which is left invariant by the RG iteration. We
cannot restrict to the (finite) space of relevant couplings, since the RG
transformation generates the irrelevant interactions, whatever the input
action is. Similarly, we cannot restrict to the space of local irrelevant
interactions, because it too is not left invariant by the RG map. The right
choice turns out to be the span of all possible monomials in $\psi$ and
$\partial \psi$ with nonlocal, but sufficiently fast decaying, kernels: this
space is left invariant. We stress that this mild nonlocality is unrelated to
the long-range character of our reference Gaussian theory; it has to do with
the fact that the IR-cutoff propagator is not fully local although
short-range; it would be present also for the local kinetic term. Note that we
couldn't find an invariant space of nonlocal monomials involving $\psi$ only:
in our construction the presence of derivative fields $\partial \psi$ is
generated by what we call the {\tmem{trimming}} operation, which consists in
extracting from a nonlocal monomial of order $2$ or $4$ its local part, and in
re-expressing the nonlocal remainder in terms of irrelevant monomials of the
form $\partial \psi \partial \psi$ or $\psi^3 \partial \psi$. Note also that
our construction does not exclude the existence of other invariant spaces,
with different (non)locality properties of the kernels; in particular, it
remains to be seen whether there exists an invariant Banach space consisting
of local monomials in $\psi$ and its derivatives (of arbitrary order), but we
are not aware of any rigorous result in this sense. The construction of an
invariant Banach space of interactions comes, in particular, with a
non-perturbative definition of the RG map: this is achieved via combinatorial
cancellations due to the $\pm$ signs in the series expansion, ultimately due
to the fermionic nature of the fields. In order to take advantage of these
cancellations, we need to organize the perturbative expansion in the form of
series of determinants, rather than in the more standard form of series of
Feynman diagrams. This may be thought of as a smart
rearrangement and partial resummation of the perturbative series: while the
Feynman diagrams expansion is {\tmem{not}} absolutely convergent, the
determinant expansion is. Once the invariant space has been identified and the
RG map defined at a non-perturbative level, we prove that the RG map is
{\tmem{contractive}} in a suitable neighborhood of the approximate
lowest-order fixed point: this implies existence and uniqueness of the actual
fixed point in such a neighborhood. (More precisely,
the RG map is contractive near the fixed point along all directions but
$\psi^2$, but this complication is easily taken care of.) Remarkably, such
fixed point is analytic in $\varepsilon$. 

Therefore, the problem of obtaining the correct Banach space to which the
fixed point belongs, and of computing the fixed point via a provably
convergent approximation scheme, while still open for 3D Ising, is completely
solved in our fermionic case, at least when $\varepsilon$ is
sufficiently small. 

Our results have similarities with those of Gawedzki and Kupiainen (GK)
{\cite{Gawedzki:1985jn}}, with some differences. GK had fermions transforming
as spinors and the model (long-range Gross-Neveu) was reflection
positive. This is a minor difference and we could have considered their model,
the only complication being an extra spinorial index. Their quartic
interaction was weakly irrelevant rather than weakly relevant, and so they
have obtained an ultraviolet fixed point,\footnote{One of the purposes of
{\cite{Gawedzki:1985jn}} was to construct rigorously a healthy theory at short
distances from a non-renormalizable effective theory at long distances, hence
their title. This was made possible by the small parameter (weakly irrelevant
interaction). Unfortunately, their paper is often misunderstood as a license
to search for the UV theory in terms of IR degrees of freedom even when there
is no weak coupling in sight (as e.g. in the asymptotic safety program for
gravity).} while our fixed point for $\varepsilon > 0$ is in the infrared. Our
proof establishes estimates on the irrelevant fixed point interactions which
are of natural size suggested by perturbation theory. Finally, we establish
fixed point analyticity that, as far as we know, has not been previously
pointed out. Let us mention that the fixed point we construct can also be
obtained by using a different, rigorous, CRG scheme, based on a tree expansion
{\cite{Ga}}, which bypasses the use of the contraction mapping theorem, as
well as the apriori definition of an invariant Banach space of irrelevant
interactions (see Appendix \ref{App:trees}).

Open questions, to be addressed in future work, include: the computation of
critical exponents and their independence from the cutoff,
rigorous derivation of conformal invariance and
the operator product expansion (OPE) the connection between our mildly
nonlocal representation of the fixed point with the local operators used in
conformal field theory, the relation with analytic regularization and
Wilson-Fisher $\epsilon$-expansion, computer-assisted computation of the
radius of convergence, crossover to the local symplectic fermion fixed point
for $\varepsilon = \varepsilon_{\ast} = O (1)$ in $d = 3$ (Do critical
exponents coincide with the Wilson-Fisher $\epsilon$-expansion for the local
symplectic fermions in such a limiting case?), etc. See Section
\ref{conclusions} for a complete list of open problems (7 pages!).

\subsection{{Convergence, analyticity and non-perturbative nature of the fixed point}}\label{sec:analyticnonpert}

Some readers may feel that our result about the analyticity of the $\eps$-expansion contradicts quantum field theory lore, and here we wish to explain why this is not the case.
	
	There are two main reasons for the divergence of perturbation theory in quantum field theory: classical solutions (instantons) and renormalons. Since our theory is fermionic, it does not contain instantons. A related difference of fermions vs bosons is that bosons only make sense for positive quartic while fermions are defined for quartic of any sign, and indeed our fixed point coupling $\lambda_*$ will be positive or negative depending on the sign of $\eps$.
	
	As for the renormalons and associated divergences (see e.g.~reviews \cite{Altarelli:1995kz,Beneke:1998ui}), they exist both for fermions and bosons, but only if there is running over a long range of scales. Also in our model, the full RG flow from UV to IR would not be analytic, for reasons similar to renormalons in asymptotically free theories like QCD. However, in this paper we focus exclusively on the fixed point physics, so there is no running, and we are immune to renormalons.
	
	Let us illustrate this point by a short computation, considering for definiteness the weakly relevant quartic case (positive $\eps$). Note that the infrared fixed point can be constructed in two {\it equivalent} ways. The first, which is the one we use in the rest of this paper, is to construct it as the fixed point of the single step Wilsonian RG transformation. The second, which we briefly discuss here and in Appendix \ref{App:trees.sub}, is to construct it dynamically, as the infrared limit of the flow of the running couplings. We will not consider the full flow from the gaussian fixed point, but a ``half-flow'' which starts at an intermediate scale and flows to the IR fixed point. Even such a ``half-flow'' is already non-analytic, as we will see.

In our model, the beta function flow equation for $\lambda(t)$  (the running quartic coupling at scale $t$, where $t\le 0$ is the logarithm of the 
infrared cutoff scale) at lowest order has the following form: 
\begin{equation}\label{eq:toyflow}\frac{d\lambda(t)}{dt}=-2\varepsilon\lambda(t)+c_2\lambda^2(t),\end{equation}
for a suitable positive constant $c_2$. The solution to \eqref{eq:toyflow} with initial condition $\lambda(0)=\lambda_0$, which we assume to 
be positive and smaller than $2\varepsilon/c_2$, is: 
\begin{equation}\label{soltoyflow}\lambda(t)= \frac{\lambda_0}{e^{2\varepsilon t}+\tfrac{c_2\lambda_0}{2\varepsilon}(1-e^{2\varepsilon t})}.\end{equation}
The infrared fixed point is $\lambda_*=\lim_{t\to-\infty}\lambda(t)=2\varepsilon/c_2$, which is obviously analytic in $\varepsilon$. At any finite $t$, $\lambda(t)$ is analytic in $\lambda_0$, but {\it non-uniformly in $t$}, as $|t|$ grows. This effect is clearly due to the running and to the nontrivial structure of the RG flow: small positive $\lambda_0$ eventually flow to $\lambda_*$, while small negative $\lambda_0$ flow away to large negative values of the coupling. However, $\lambda(t)$ is Borel-summable in $\lambda_0>0$ uniformly in $t$. The complete flow is more complicated than the toy model \eqref{eq:toyflow}, but it retains 
the same qualitative features as the above illustration. In our fermionic setting, fixed point observables, such as critical exponents, are expected to be convergent power series in $\lambda_*$ and, therefore, analytic in $\varepsilon$. 
Observables (e.g.~correlation functions) at intermediate distance scales
can be expressed as convergent power series in
the {\it whole sequence} $\{\lambda(t)\}_{t\le 0}$ (see Appendix \ref{App:trees.sub} for further details on this point). Due to the non-analytic dependence of $\lambda(t)$ in the bare coupling $\lambda_0$, 
such observables are expected to be ``just'' Borel-summable in $\lambda_0$.

Finally, let us comment on the setup of massless perturbation theory, i.e.~when the gaussian fixed point is perturbed by only the (weakly relevant) quartic coupling, setting mass to zero and working directly in the continuum limit. Such a setup, under the name of ``conformal perturbation theory'' \cite{Z, Ludwig:1987gs, CL}, is often considered when perturbing non-gaussian fixed points (see e.g.~\cite{Komargodski:2016auf,Behan:2017emf} for recent applications), but it could be used in our problem as well. It is a form of perturbative expansion in the bare coupling. At a small but fixed $\eps$, the first $n$ terms of the resulting perturbative expansion will be finite, where $n\sim 1/\eps$, while subsequent terms have infinite coefficients (because the corresponding integrals diverge at long distances). Thus, the perturbative expansion itself is ill-defined beyond the first few terms in this framework. 

Some authors, e.g.~Ref.~\cite{CL}, argued that this pathology is a possible signal of the appearance of non-analytic terms in the infrared fixed point observables (although, as \cite{CL} admits, ``their actual presence is unclear at the moment''). It has to be emphasized that we are talking here about the situation when the RG flow leads to a fixed point, and only about the infrared fixed point observables, such as the critical exponents. We are not concerned with the situation when the flow leads to a massive phase, in which case the mass of the particles is indeed generically non-analytic in the bare couplings. While non-analytic terms may affect bosonic relatives of our model (see Section \ref{bosons}), in our fermionic model we rigorously exclude them, see Remark \ref{R2} and Appendix \ref{sec:non-pert}. Our analytic-in-$\varepsilon$ fixed point defines the infrared theory in a fully non-perturbative way. Thereby, results based on convergent perturbation theory lead in our case to a fully non-perturbative description. The key point allowing this to happen is that in finite volume Grassmann integrals are finite dimensional. Therefore, in presence of any finite-volume cutoff, there is no room for non-analytic terms. Furthermore, uniformity in the volume of our estimates, in combination with uniqueness theorems for the limit of uniformly convergent analytic functions, imply that the absence of non-analytic terms carries over to the infinite-volume limit, see Appendix \ref{sec:non-pert}
for details.

\subsection{{Summary}}

The paper is structured as follows. In Section \ref{sec:ModDef} we present
the model and we state informally our main results. In Section
\ref{sec:intuitive} we identify an approximate nontrivial fixed point by
truncating the RG map at lowest order (explicit perturbative computations are
in App. \ref{sec:I1I2}). The rest of the paper will be devoted to a
non-perturbative proof of its existence: In Section \ref{sec:Repr} we
introduce the Banach space of interactions consisting of monomials in the
fields with mildly nonlocal interactions, and we equip it with a suitable
norm, tailored for our smooth slicing cutoffs (whose properties are in App.
\ref{sec:gest}). In Section \ref{detailsRG} we show that the assumed form of
the interaction is left invariant by the RG map, a fact made apparent
rearranging the output via a trimming operation (more details are in Apps.
\ref{sec:Heff} and \ref{sec:Trim}). We show also there that the action of the
RG map can be expressed as a series which is absolutely convergent in norm;
this follows from a number of results described in App. \ref{sec:GH}, such as
determinant bounds for simple fermionic expectations and a suitable
representation of connected expectations. Absolute convergence allows to
rigorously estimate the action of the RG map, and this allow us in Section
\ref{sec:FP} to prove, see Theorem \ref{FPT}, the existence of the fixed
point, together with its independence of the slicing parameter and its
analyticity in $\varepsilon$. This result relies on the crucial Key Lemma \ref{Key} and
its variants, which ensures that the Banach space is invariant and the RG map
is contractive. The key lemma is in a sense optimal, as it predicts a
dependence on $\varepsilon$ of the effective interactions which is exactly the one
suggested by perturbation theory; this is obtained by a careful choice of
constants done in the proof, presented in Section \ref{sec:2k+2} and App.~\ref{Dkbounds}. 
Section \ref{conclusions} is devoted to conclusions and open
problems. The fact that our convergent analysis fully reconstruct the theory
and provides non-perturbative information is proved in App.~\ref{sec:non-pert}. 
In App.~\ref{sec:formal} we show that the fixed point can be obtained via a formal series expansion; 
perturbation theory is similar for boson or fermionic models but for fermions the series converges, 
a fact offering, see App.~\ref{App:trees}, a way to construct the RG fixed point
alternative to the path via Banach space and contraction method, using instead
the direct tree expansion technique. Finally in App.~\ref{literature} a review
and comparison with previous results in bosonic theories is presented.

\section{Definition of the model and formulation of the
problem}\label{sec:ModDef}

\begin{figure}[h]\centering
  \includegraphics{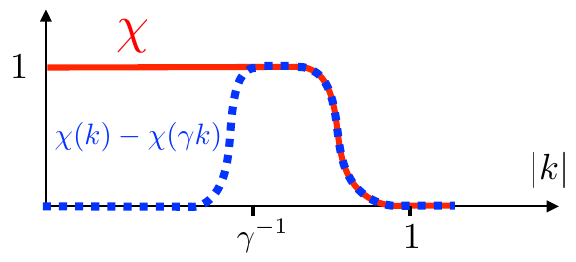}
  \caption{\label{fig:bump}The function $\chi (k)$ (red curve), and the
  resulting function $\chi (k) - \chi (\gamma k)$, Eq.~{\eqref{eq:split}}
  (blue dots).}
\end{figure}

Let us now discuss the model more in detail. The propagator
{\eqref{eq:prop1.1}} is defined in terms of $P (x)$, which is chosen in the
form
\begin{equation}
  P_{} (x) = \int \frac{d^d k}{(2 \pi)^d}  \hat{P} (k) e^{ikx}, \qquad \hat{P}
  (k) = \frac{\chi (k)}{|k|^{\frac{d}{2} + \e}} \hspace{0.17em}
  \hspace{0.17em} . \label{Pchi}
\end{equation}
The function $\chi (k)$ here is a ``UV cutoff'', \ a short-distance regulator
of the model. We will choose it satisfying the following conditions (see Fig.
\ref{fig:bump}):
\begin{equation}
  \chi \text{ is a radial $C^{\infty}$ function,} \quad 0 \leqslant \chi (k)
  \leqslant 1, \quad \chi (k) = \left\{ \begin{array}{ll}
    1, & (|k| \leqslant 1 / 2)\\
    0, & (|k| \geqslant 1),
  \end{array} \right. \label{chicond}
\end{equation}
In fact we will require something a bit stronger than $\chi \in C^{\infty}$,
namely:
\begin{equation}
  \chi \text{ belongs to the Gevrey class $G^s \tmop{for} \tmop{some}$ } s > 1.
  \label{chiG}
\end{equation}
This ``Gevrey condition'' will be defined in Section \ref{sec:Norms}, see Eq.
{\eqref{chiGdef}}, and is not used until then. As explained there, it is
needed so that the fluctuation propagator $g (x)$ (see Section \ref{RenMap})
decays at infinity as a stretched exponential. There are many cutoff functions
satisfying both conditions {\eqref{chicond}} and {\eqref{chiG}}; an explicit
example is given in Appendix \ref{exGevrey}.

As a consequence of {\eqref{Pchi}} and {\eqref{chicond}}, $P (x)$ is uniformly
bounded, and its large-$x$ asymptotics is proportional to $1 / |x|^{d / 2 -
\varepsilon}$, as stated after {\eqref{eq:prop1.1}}.

We denote by $d \mu_P (\psi)$ the Gaussian Grassmann integration with
propagator {\eqref{eq:prop1.1}}, which can be \ formally written as:
\begin{eqnarray}
  &  & d \mu_P (\psi) = D \psi e^{S_2 (\psi)}, \nonumber\\
  &  & S_2 (\psi) = \frac{1}{2} \int \frac{d^d k}{(2 \pi)^d} \hat{P} (k)^{-
  1} \Omega_{a b} \psi_a (k) \psi_b (- k) .  \label{S2}
\end{eqnarray}
[Since $\hat{P} (k)^{- 1}$ is non-analytic in $k^2$ near $k =0$, such an action is called ``long-range".] 

More precisely,
$d \mu_P (\psi)$ is characterized by the expectations of an even number $2 s$
of fields, via:
\begin{equation}
  \langle \psi_{a_1} (x_1) \ldots \psi_{a_{2 s}} (x_{2 s}) \rangle \equiv \int
  d \mu_P (\psi) \psi_{a_1} (x_1) \ldots \psi_{a_{2 s}} (x_{2 s}) = \sum_{\pi}
  (-)^{\pi} \prod_{i = 1}^s G_{\pi (a_{2 i - 1}) \pi (a_{2 i})} (x_{2 i - 1},
  x_{2 i}), \label{2kfields}
\end{equation}
where the sum is over all pairings of $2 s$ fields and $(-)^{\pi}$ is the sign
of the corresponding permutation.

The full model is defined by an ``interacting'' Grassmann measure
\begin{equation}
  Z^{- 1} d \mu_P (\psi) e^{H_{} (\psi)}, \label{ModDef}
\end{equation}
where the ``interaction'' $H_{} (\psi)$ is a bosonic function of Grassmann
fields, and $Z = \int d \mu_P (\psi) e^{H_{} (\psi)}$ is the partition
function. The simplest interaction includes only the local quadratic and
quartic terms:\footnote{Here and below we denote the local couplings by $\nu,\lambda$, rather than by $\nu_0,\lambda_0$, as in 
\eqref{aMFT}. The change of notation is meant to highlight the difference between the bare couplings $\nu_0,\lambda_0$, and the running ones, which will be their meaning from now on. In fact, in the following, 
we shall construct the interaction $H$ corresponding to the infrared fixed point, whose local quadratic and quartic couplings correspond to the fixed point values $\nu_*,\lambda_*$ 
computed at lowest order in Section \ref{sec:intuitive}. The notation $\nu,\lambda$ is used for generic values of the parameter entering the RG equations.}:
\begin{equation}
  \label{eq:V0} H_L (\psi) = \nu \int d^d x \hspace{0.17em} \Omega_{ab} \psi_a
  \psi_b + \lambda \int d^d x (\Omega_{ab} \psi_a \psi_b)^2 \hspace{0.17em} .
\end{equation}
This interaction has $Sp (N)$ global symmetry rotating the fermion indices, as
well as $O (d)$ spatial invariance. We will assume $N \geqslant 4$ so that the
quartic interaction does not vanish identically.\footnote{For $N = 2$ the
quartic interaction vanishes, while for $N = 4$ it is proportional to $\psi_1
\psi_2 \psi_3 \psi_4$.} We will furthermore assume
\begin{equation}
  d \in \{ 1, 2, 3 \}, \qquad 0 < \varepsilon < \min (2 - d / 2, d / 6) = d /
  6, \label{epsCond}
\end{equation}
where the second condition guarantees that the two terms in $H_L (\psi)$ are
the only $O (d) \times Sp (N)$-invariant interactions which are relevant, see
Section \ref{sec:Dil}.\footnote{\label{note:rel}The notion of relevance in our
setup involving mildly nonlocal kernels will be made precise below in Eq.
{\eqref{eq:normResc}}, and it will agree with the usual rule that the
interaction containing $l$ fields and $p$ derivatives is relevant if $l [\psi]
+ p < d$. }

RG transformation of the model will be acting in a more general space of
interactions
\begin{equation}
  H_{} (\psi) = H_L (\psi) + H_{\text{IRR}} (\psi), \label{HLHirr}
\end{equation}
where $H_{\text{IRR}} (\psi)$ stands for an infinite number of generally
nonlocal (although mildly so) terms corresponding to irrelevant interactions.
Like $H_L (\psi)$, interactions in $H_{\text{IRR}} (\psi)$ will respect $Sp
(N) \times O (d)$ invariance.\footnote{In the trimmed representation of
section \ref{sec:TrimRep}, $H_{\text{IRR}} (\psi)$ will consist of $H_{\text{2R}},
H_{\text{4R}}, H_{\text{6SL}}, H_{\text{6R}}$ and $H_{\ell}$ for $\ell \geqslant 8$.}

\begin{remark}
  \label{FinVol}Eq. {\eqref{ModDef}} as written is not immediately meaningful
  in infinite volume, because partition function is infinite: $Z = \infty$. To
  give it a rigorous meaning, we should e.g. put the model in finite volume
  and pass to the limit. To speed up this introductory part of the paper, let
  us work directly in infinite volume and consider the interacting measure
  {\eqref{ModDef}} in the sense of formal perturbative expansion in $H
  (\psi)$. In perturbation theory, the normalization factor $Z^{- 1}$ $^{}$in
  {\eqref{ModDef}} means that diagrams with disconnected interaction vertices
  should be excluded when computing expectations. In the main text we will
  show that infinite-volume perturbation theory is convergent (this is a
  general feature of fermionic models at weak coupling). The rigorous
  definition as a limit from finite volume is postponed to Appendix
  \ref{sec:non-pert}. Taking this limit will be easy once the infinite volume
  behavior is understood. See also Remark \ref{R2} below.
\end{remark}

\subsection{Renormalization map}\label{RenMap}

Let us fix a ``rescaling parameter''\footnote{Although at this point any
$\gamma > 1$ would do, we assume $\gamma \geqslant 2$ from the start, \ as
some estimates below, specifically Eq. {\eqref{TRfinal}}, will require that
$\gamma$ is separated from 1. The fixed point construction will require
raising $\gamma$ even further. } $\gamma \geqslant 2.$ We will define the
``renormalization map'' which maps $H (\psi)$ to another interaction $H'
(\psi)$. It will be a composition of integrating-out and dilatation.

Integrating-out consists in splitting the field $\psi$ as $\psi =
\psi_{\gamma} + \phi \hspace{0.17em}$ where $\psi_{\gamma}$ is the
``low-momentum component'' of $\psi$, and defining the effective interaction
$e^{H_{\eff} (\psi_{\gamma})}$ by eliminating $\phi$. Concretely, we split the
Grassmann propagator as (see Fig.\ref{fig:bump})
\begin{equation}
  P_{} (x) = P_{\gamma} (x) + g (x), \quad \widehat{P_{\gamma}} (k) =
  \frac{\chi (\gamma k)}{|k|^{\frac{d}{2} + \e}}, \quad \hat{g} (k) =
  \frac{\chi (k) - \chi (\gamma k)}{|k|^{\frac{d}{2} + \e}} . \label{eq:split}
\end{equation}

Note that $P_{\gamma}$ is just a rescaled version of $P$ (see Eq.
{\eqref{Pgamma}}), while $g (x)$ is called ``fluctuation propagator''. This
decomposition implies factorization of the integration measure $d \mu_{P_{}}
(\psi)$ as
\begin{equation}
  d \mu_{P_{}} (\psi) = d \mu_{P_{\gamma}} (\psi_{\gamma})  \hspace{0.17em} d
  \mu_g (\phi), \qquad \psi = \psi_{\gamma} + \phi \hspace{0.17em},\label{eq:decompos}
\end{equation}
where $\psi_{\gamma}$ and $\phi$ are two independent Grassmann fields with
propagators $P_{\gamma}$ and $g$. As mentioned, Eq.{\eqref{chiG}} will
guarantee that $g (x)$ decays at infinity as a stretched exponential.

Correlation functions of $\psi_{\gamma}$ with respect to the interacting
measure {\eqref{ModDef}} can equivalently be computed with respect to the
measure
\begin{equation}
  d \mu_{P_{\gamma}} (\psi_{\gamma})  \hspace{0.17em} e^{H_{\eff}
  (\psi_{\gamma})} \label{HeffMeas}
\end{equation}
(normalization understood) where $e^{H_{\eff} (\psi_{\gamma})}$ is defined by
``integrating out the fluctuation field'' $\phi$:\footnote{\label{infconst}We
will drop the $\psi_{\gamma}$-independent term in $H_{\eff} (\psi_{\gamma})$,
since this constant drops out when normalizing and does not affect the
expectations. As we will discuss in Appendix \ref{sec:non-pert}, this constant
is finite in finite volume although it becomes infinite in the infinite-volume
limit. }
\begin{equation}
  e^{H_{\eff} (\psi_{\gamma})} = \int d \mu_g (\phi) e^{H (\psi_{\gamma} +
  \phi)} \hspace{0.17em} . \label{eq:Heff0}
\end{equation}
Note that the propagator $P_{\gamma}$ is related to $P$ via
\begin{equation}
  P_{\gamma} (x) = \gamma^{- 2 [\psi]} P_{} (x / \gamma) \label{Pgamma}
\end{equation}
with $[\psi] = d / 4 - \varepsilon / 2$ as above. This motivates to consider
the dilatation transformation:
\begin{equation}
  \psi_{\gamma} (x) \mapsto \g^{- \Dpsi} \psi (x / \g), \label{Dsubs}
\end{equation}
which maps the measure {\eqref{HeffMeas}} to the measure $d \mu_P (\psi) e^{H'
(\psi)}$ with the same gaussian factor as in {\eqref{ModDef}} but with a
different interaction:
\begin{equation}
  H' (\psi) = H_{\tmop{eff}} \left[ \g^{- \Dpsi} \psi (\cdot / \g) \right] .
  \label{Hprime}
\end{equation}
This formula defines the renormalization map $R = R (\varepsilon, \gamma) : H
\mapsto H'$ (also called ``RG transformation''). Note that $R$ also depends on
$d, N, \chi$ but this dependence will be left implicit. As a function of
$\gamma$ for a fixed $\varepsilon$, the renormalization map satisfies the
semigroup property:
\begin{equation}
  R (\varepsilon, \gamma_1) R_{} (\varepsilon, \gamma_2) = R_{} (\varepsilon,
  \gamma_1 \gamma_2)_{} . \label{eq:Rgamma}
\end{equation}

Our main goal will be to construct the fixed point of the RG transformation. We would like to remind the reader that although our RG transformation is obtained by integrating out the degrees of freedom with momenta between $ \Lambda\sim 1$ and $\Lambda_{\rm IR}\sim \Lambda/\gamma$, one should not think of $\Lambda_{\rm IR}$ as some sort of mass which breaks criticality of our fixed point. The correct interpretation is that we have only one RG scale, $\Lambda$, while $\Lambda_{\rm IR}$ entered the game because we find it technically convenient to consider the discrete RG transformation rather than the continuous one, such as Polchinski's equation \cite{Polchinski:1983gv}. At an intuitive level discrete RG transformation can be obtained by integrating the continuous one, and they are expected to have the same fixed points (although to make rigorous sense of the continuous RG may be nontrivial, see Remark \ref{Polchinski} below). In particular, it would be wrong to think that some sort of `IR cutoff removal' has to be performed with our result to extract the fixed point physics. On the contrary, all of this physics is already contained in the fixed point $H_{\ast}$. E.g., the critical exponents can be obtained by linearizing the RG transformation (the same one which leads to the fixed point), near the fixed point, and computing the eigenvalues.

After this warning, the informal formulation of our main result goes as follows:\\[3pt]
{\tmem{\tmtextit{Fix $\chi, d \in \{ 1, 2, 3 \},$ and $N \geqslant 4, N \neq
8$. For $\gamma$ large enough and $\varepsilon > 0$ small enough, there exists
a nontrivial interaction $H_{\ast} (\varepsilon)$ which is a fixed point of $R
(\gamma, \varepsilon)_{_{}}$ for all $\gamma$:}
\begin{equation}
  R_{} (\varepsilon, \gamma) [H_{\ast} (\varepsilon)]_{} = H_{\ast}
  (\varepsilon)_{} . \label{RH=H}
\end{equation}
Moreover, $H_{\ast} (\varepsilon)$ can be extended to an analytic function of
$\varepsilon$ in a small neighborhood of the origin.}}

A precise statement is the content of Theorems \ref{FPT} and \ref{AFPT}, which
rely on Key Lemma \ref{Key} and Abstract Lemma \ref{AbstractLemma}. The
condition $N \neq 8$ comes from requiring a non-vanishing one-loop
beta-function. The condition that $\gamma_{}$ is sufficiently large arises for
the following technical reason: the fixed point $H_{\ast} (\varepsilon)$ will
live in a Banach space, and only for sufficiently large $\gamma$ will we be
able to show that $R_{} (\varepsilon, \gamma)$ is a bounded operator on this
Banach space, so that Eq. {\eqref{RH=H}} makes sense.

The definition of the Banach space requires a suitable representation of the
interactions, called {\tmem{trimmed representation}}, and discussed in Section
\ref{sec:Repr} below. In order to define it, we will distinguish, quite
naturally, the local (relevant) terms from the nonlocal (irrelevant) ones.
Moreover, we will rewrite the nonlocal quadratic or quartic interactions in
terms of derivative fields, via the trimming operation, defined in Section
\ref{sec:trim0} below: the usefulness of a representation in terms of
derivative fields is to make the irrelevance of these interactions apparent,
already at the level of the linearized RG map. An additional feature of the
trimmed representation is that it distinguishes a so-called ``semilocal''
sextic term from the fully nonlocal sextic interaction. This splitting may
look strange at first sight. In our setup with a smooth cutoff in momentum
space, it is needed to obtain the correct lowest-order approximation to the
fixed point, which is in turn important for defining a neighborhood in the
Banach space where the RG map (or, better, a suitable rewriting thereof) is
contractive.

In order to better motivate it, let us explain more explicitly the structure
of the splitting of the sextic term and the intuitive reason behind its
definition: we'll do it in the next section, before getting to the formal
definition of the trimmed representation.

\section{The fixed point equation at lowest order}\label{sec:intuitive}

Let us go back to the lowest order fixed point equation (FPE), whose structure
was anticipated in Eq.{\eqref{eq:fix1.4}}, and let us discuss its derivation
more carefully, in view of our choice of a smooth cutoff function. The most
naive approximation one can do is to compute the FPE by neglecting all
couplings but the relevant ones, $\nu$ and $\lambda$, and, assuming these
couplings to be of order $\varepsilon$, to retain only the dominant
contributions to the beta functions for $\nu$ and $\lambda$, which are of
order $\varepsilon$ and $\varepsilon^2$, respectively. While very natural, we
would like to convince the reader that such a naive approximation leads to a
wrong lowest order FPE, whose solution differs from the correct one by $O
(\varepsilon)$ rather than $O (\varepsilon^2)$: the important contribution
missed by this scheme is the $O (\varepsilon^2)$ contribution to the beta
function for the quartic coupling $\lambda$, due to the self contraction of
the ``semilocal'' sextic term (the tree graph contribution to the sextic
interaction, of order $O (\varepsilon^2)$), see below for details.

Let us start by describing the most naive approximation (the wrong one).
Consider a local interaction, $H (\psi) = H_L (\psi)_{}$, see {\eqref{eq:V0}},
and integrate the fluctuation field via {\eqref{eq:Heff0}}. After this
integrating-out step the local couplings are modified as follows:
\begin{equation}
  \label{eq:step1nl} \nu \to \nu + \Delta \nu \equiv \nu_{\tmop{eff}}, \quad
  \lambda \to \lambda + \Delta \lambda \equiv \lambda_{\tmop{eff}}
  \hspace{0.17em},
\end{equation}
where the leading contributions to $\Delta \nu$, $\Delta \lambda$ are given by
the diagrams
\begin{eqnarray}
  & \Delta \nu =
  \raisebox{-0.0119017178456031\height}{\includegraphics[width=1.52946674537584cm,height=0.705414534959989cm]{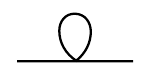}}
  + \cdots &  \nonumber\\
  & \Delta \lambda = \left(
  \raisebox{-7pt}{\includegraphics[width=1.49003017184835cm,height=0.705414534959989cm]{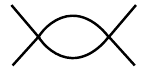}}
  \right)_{\loc} + \ldots &  \label{eq:lowest}
\end{eqnarray}
Here $(\cdot)_{\loc}$ stands for the local part of the nonlocal term generated
by the second diagram.\footnote{This operation is done in momentum space by
evaluating the diagram with all external momenta set to zero. Alternatively,
in position space one replaces the kernel of the nonlocal operator by its
integral.} Note that the diagram
$\raisebox{-7pt}{\includegraphics[width=0.616538764266037cm,height=0.705414534959989cm]{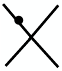}}
= O (\eps^2)$ does not contribute to $\Delta \lambda$ because its local part
vanishes. Indeed, the propagator $g$ vanishes in momentum space at $k = 0$, or
equivalently $\int \mathrm{} d^d x \hspace{0.17em} g (x) = 0$. For this
reason, $\nu$ insertions on external legs never give rise to local terms:
$\left(
\raisebox{-7pt}{\includegraphics[width=0.616538764266037cm,height=0.705414534959989cm]{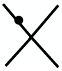}}
\right)_{\mathrm{loc}} = 0$. One can easily check by inspection that, starting
from $H = H_L$, there are no other contributions of $O (\varepsilon)$ to
$\Delta \nu$ and of $O (\varepsilon^2)$ to $\Delta \lambda$, beyond those
shown in {\eqref{eq:lowest}}.

We now rescale the fields as in {\eqref{Hprime}} and find $\nu' = \gamma^{d -
2 [\psi]} \nu_{\tmop{eff}}$ and $\lambda' = \gamma^{d - 4 [\psi]}
\lambda_{\tmop{eff}}$, that is, recalling {\eqref{eq:step1nl}},
\begin{gather}
  \nu' = \gamma^{\frac{d}{2} + \eps}  (\nu + I_1 \lambda + \ldots),
  \label{eq:modelRG1}\\
  \lambda' = \gamma^{2 \eps}  (\lambda + \betadiag \lambda^2 + \ldots)
  \hspace{0.17em}, \label{eq:modelRG2}
\end{gather}

where $I_1$ and $\betadiag$ are the one-loop diagrams in the two lines of
{\eqref{eq:lowest}}, respectively. Performing $\Omega$-tensor contractions,
one finds $\betadiag \propto N - 8 \neq 0$, since we are assuming $N \neq
8$.\footnote{See Appendix \ref{sec:I1I2} for the computations of these
coefficients, where we also comment that vanishing of $\betadiag$is an
accident which does not reproduce at higher loops.} It is now extremely
tempting to conclude that the fixed point equation for $\lambda$ is
\begin{equation}
  \lambda = \gamma^{2 \eps}  (\lambda + \betadiag \lambda^2 + \cdots),
  \label{FPEwrong}
\end{equation}
up to terms of $O (\varepsilon^3)$, so that the fixed point is $\lambda_{\ast}
= (1 - \gamma^{2 \varepsilon}) / \betadiag$ up to an error $O
(\varepsilon^2)$; plugging this into the fixed point equation for $\nu$, one
would find $\nu_{\ast} = I_1 \lambda_{\ast} / (1 - \gamma^{d / 2 +
\varepsilon})$ up to an error $O (\varepsilon^2)$. Even if extremely tempting,
{\tmstrong{this conclusion is wrong}}!

Where is the problem? The point is that neglecting the irrelevant terms, and
in particular the sextic one, leads to an error of $O (\varepsilon^2)$ in the
FPE for $\lambda$; such an error is comparable in size with the term
$\betadiag \lambda^2$ that we included above: therefore, dropping blindly the
irrelevant terms is not consistent even at the lowest order. To see this,
notice that, by starting with a local interaction, $H = H_L$, after having
integrated out the fluctuation field, we obtain an effective interaction
$H_{\tmop{eff}}$, whose sextic term contains the tree diagram
$\raisebox{-7pt}{\includegraphics[width=1.18982028072937cm,height=0.705414534959989cm]{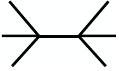}}
= O (\varepsilon^2)$. Therefore, in order to find an interaction $H$ solving
the fixed point equation $H' = H$ at $O (\varepsilon^2)$, we cannot avoid
assuming that $H$ contains a sextic irrelevant term with the same structure as
$\raisebox{-7pt}{\includegraphics[width=1.18982028072937cm,height=0.705414534959989cm]{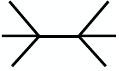}}$.
Let us then take $H = H_L + H_{\tmop{IRR}}$, with $H_{\tmop{IRR}}$ containing
the following sextic `$\Xterm$-term' interaction:
\begin{equation}
  \raisebox{-7pt}{\includegraphics[width=1.19960973370064cm,height=0.705414534959989cm]{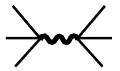}}
  = \Omega_{ab} \Omega_{a' b'} \Omega_{cc'}  \int d^d x d^d y (\psi_a \psi_b
  \psi_c) (x) \tmmathbf{\mathfrak{X}}  (x - y)  (\psi_{a'} \psi_{b'}
  \psi_{c'}) (y) \hspace{0.17em}, \label{eq:Xterm11}
\end{equation}
with the particular shown contraction of $\psi$ indices. This term might be
called `semilocal': there are two $\psi^3$ groups interacting via one nonlocal
kernel. The gothic $\Xterm$ is meant to remind about the shape of this
diagram. Upon integrating out, the unique new contribution to $\Xterm$ comes
from the tree-level diagram contracting two quartic vertices:
\begin{equation}
  \raisebox{-7pt}{\includegraphics[width=1.18982028072937cm,height=0.705414534959989cm]{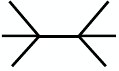}}
  \quad \Rightarrow \quad \Delta \Xterm (x) = \cf \lambda^2 g (x)
  \hspace{0.17em} .
\end{equation}
At the fixed point, we thus expect $\Xterm = O (\lambda_{\ast}^2) = O
(\eps^2)$. On the other hand, $\Xterm$ gives a direct contribution to $\Delta
\lambda$, which therefore has to be included explicitly: the equation for
$\Delta \lambda$ thus has to be corrected as follows:
\begin{equation}
  \Delta \lambda = \left(
  \raisebox{-7pt}{\includegraphics[width=1.49003017184835cm,height=0.705414534959989cm]{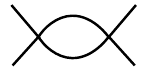}}
  +
  \raisebox{-7pt}{\includegraphics[width=1.02131706677161cm,height=0.705414534959989cm]{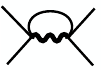}}
  \right)_{\loc} + \ldots \label{eq:DLnew}
\end{equation}
Assuming that $\Xterm$ is $O (\varepsilon^2)$ and that all the other
irrelevant interactions of order 6 or more in the fields are $O
(\varepsilon^3)$ or smaller (while the nonlocal, irrelevant, contributions of
order 2 or 4 are $O (\varepsilon^2)$), one can check by inspection that the
dots in {\eqref{eq:DLnew}} are $O (\varepsilon^3)$. At this point some readers
may be thrown out of balance: who has ever seen this second diagram? In fact
Wilson and Kogut discuss it, {\cite{Wilson:1973jj}}, Eq.(5.23) and below. They
do observe that it is $O (\eps^2)$ and thus would need to be included. Only if
one uses a sharp cutoff, then this diagram drops out because its local part
then vanishes (momenta along the wavy and curved lines do not overlap). Since
we use a smooth cutoff, we have to include both diagrams. The corrected
leading approximation to the FPE thus involves $\nu, \lambda$ and the
$\Xterm$-term parametrized by $\Xterm (x)$; it takes the form
\begin{eqnarray}
  \nu & = & \gamma^{\frac{d}{2} + \eps}  [\nu + I_1 \lambda + O (\eps^2)], \\
  \lambda & = & \gamma^{2 \eps}  [\lambda + \betadiag \lambda^2 + (N - 8) \int
  d^d x \hspace{0.17em} \Xterm (x) g (x) + O (\eps^3)], 
  \label{eq:RG-lambda}\\
  \Xterm (x) & = & \gamma^{2 d - 6 \Dpsi}  [\Xterm (x \gamma) \cf \lambda^2 g
  (x \gamma)],  \label{eq:RGX}
\end{eqnarray}
the factor $(N - 8)$ coming from $\Omega$-tensor contractions. This allows us
to compute the fixed point couplings $\nu_{\ast}$ and $\lambda_{\ast}$ at
order $\eps$, while $\Xterm_{\ast}$ will be order $\eps^2$. The approximation
is consistent: all the irrelevant terms not explicitly shown contribute to the
error terms only. Note that {\eqref{eq:RGX}} allows us to express the fixed
point $\Xterm_{\ast}$ in terms of $\lambda$ as a geometric series:
\begin{equation}
  \Xterm_{\ast} (x) = \cf \lambda^2  \sum_{n = 1}^{\infty} \gamma^{(2 d - 6
  \Dpsi) n} g (x \gamma^n) . \label{fixedX}
\end{equation}
Plugging $\Xterm = \Xterm_{\ast}$ in the right side of {\eqref{eq:RG-lambda}}
gives the FPE for the quartic coupling $\lambda = \gamma^{2 \varepsilon}
(\lambda + I_2 \lambda^2)$ up to an error $O (\varepsilon^3)$, with $I_2 =
\betadiag + I_2^{\Xterm}$ and $I_2^{\Xterm}$ the constant coming from the term
$(N - 8)  \int d^d x \hspace{0.17em} \Xterm (x) g (x)$. We are thus led to the
FPE {\eqref{eq:fix1.4}}, which we now expect to be correct at dominant order,
contrary to {\eqref{FPEwrong}}. Interestingly, in the $\eps \to 0$ limit the
sum of the two diagrams $I_2 = \betadiag + I_2^{\Xterm} \approx \bar{I}_2 \log
\gamma$, where $\bar{I}_2$ is independent of the choice of the cutoff function
(appendix \ref{sec:I1I2}). Therefore the fixed point coupling $\lambda_{\ast}$
is universal at order $\eps$. This is similar to the well-known scheme
independence of the first two beta-function coefficients.

In conclusion, the inclusion of the irrelevant sextic terms is crucial for
computing the correct coefficients in the lowest order FPE. One expects that
the inclusion of more and more irrelevant terms will produce better and better
approximation to the FPE, unless non-perturbative effects come into play.
Therefore, in order to compute the exact FPE, we'll assume that $H$ belongs to
a space of interactions including all possible irrelevant terms, of arbitrary
high order in the fields, as discussed in the next section.

\section{The Banach space of interactions}\label{sec:Repr}

\subsection{Representation of interactions by kernels}

In order to conveniently represent the interaction, we use the following
condensed notation for fields, their first derivatives, and products thereof:
\begin{equation}
  \Psi_A = \left\{ \begin{array}{ll}
    \psi_a, & A = a \hspace{0.17em},\\
    \del_{\mu} \psi_a, & A = (a, \mu) \hspace{0.17em}
  \end{array} \right., \qquad \Psi (\bA, \bx) = \prod_{i = 1}^l \Psi_{A_i}
  (x_i) \hspace{0.17em}, \label{products}
\end{equation}

where $\bA = (A_1, \ldots, A_l)$ and $\bx = (x_1, \ldots, x_l)$ are finite
sequences. $| \mathbf{A} |$ will denote the length of $\bA$, \ and $d (\bA)$
the number of derivative fields in $\Psi (\bA, \bx)$. See Section
\ref{sec:trimming} for why we allow fields with zero or one (but not more)
derivatives.

An interaction $H (\psi)$ is a sum of terms with some kernels $H (\mathbf{A},
\mathbf{x})$:
\begin{equation}
  H (\psi) = \sum_{\mathbf{A}} \int \mathrm{} d^d \bx \hspace{0.17em} H (\bA,
  \bx) \Psi (\bA, \bx), \label{Hpsi}
\end{equation}
where $d^d \bx = d^d x_1 \ldots d^d x_l$. The kernels satisfy various obvious
constraints following from the Grassmann nature of the fields and from the $Sp
(N) \times O (d)$ symmetry of the model. E.g., the kernels are assumed
antisymmetric.\footnote{\label{opA}This means $H (\pi \bA, \pi \bx) =
(-)^{\pi} H (\bA, \bx) \hspace{0.17em}$ for any permutation acting
simultaneously on $\bA$ and $\bx$. If not already antisymmetric, the
antisymmetrized kernels $H^{\calA} (\bA, \bx) = \frac{1}{l!}  \sum_{\pi \in
S_l} (-)^{\pi} H (\pi \bA, \pi \bx)$ define the same interaction. The kernel
dependence on $Sp (N)$ indices will be made out of $\Omega_{ab}$ tensors.
Their dependence on $x_i$ will be an $SO (d)$-invariant tensor built out of
$(x_i - x_j)_{\mu}$ where $\mu$ are spatial indices contained in $\bA$ (if
any), times a function of pairwise distances $|x_i - x_j |$. Finally, $O (d)$
also contains spatial parity $\psi_a (x) \to \psi_a  (- x)$, $\Psi (\bA, \bx)
\to (- 1)^{d (\bA)} \Psi (\bA, - \bx)$. Therefore, the kernels will satisfy $H
(\bA, - \bx) = (- 1)^{d (\bA)} H (\bA, \bx)$.} The individual interaction
terms being bosonic, ``the number of legs'' $l = | \mathbf{A} |$ must be even.

The local quadratic and quartic interactions in {\eqref{eq:V0}} correspond to
$\delta$-function kernels:
\begin{eqnarray}
  \nu \int d^d x \hspace{0.17em} \Omega_{ab} \psi_a \psi_b \hspace{0.17em} &
  \leftrightarrow & \nu \hspace{0.17em} \Omega_{ab} \delta^{}  (x_1 - x_2), 
  \label{loc2}\\
  \lambda \int d^d x (\Omega_{ab} \psi_a \psi_b)^2 & \leftrightarrow &
  \tfrac{1}{3} \lambda \hspace{0.17em} q_{abce} \delta^{}  (x_1 - x_2)
  \delta^{}  (x_1 - x_3) \delta (x_1 - x_4),  \label{loc4}
\end{eqnarray}
where $q_{abce} = \Omega_{ab} \Omega_{ce} - \Omega_{ac} \Omega_{be} +
\Omega_{ae} \Omega_{bc}$ is totally antisymmetric. We will represent
interactions of $H_{\tmop{IRR}}$ by nonlocal kernels rather than expanding
them in local interactions.

We divide the kernels into groups (``couplings'') $H_l$ according to their
number of legs $l$ ($l \geqslant 2$ even):
\begin{equation}
  H_l = \{H (\mathbf{A}, \mathbf{x})\}_{| \mathbf{A} | = l} .
\end{equation}
The interaction is thus represented by a coupling sequence $(H_l)_{l \geqslant
2}$. This is a \tmtextit{general} representation. It is useful to introduce a
notation also for the restriction of $H_l$ to the kernels with a specified
number of derivatives:
\begin{equation}
  H_{l, p} = \{H (\mathbf{A}, \mathbf{x})\}_{| \mathbf{A} | = l, d (\bA) = p}
  .
\end{equation}
We emphasize that $\bA$'s are \tmtextit{sequences}: the ordering is important
and terms with different orderings appear separately in {\eqref{Hpsi}}. This
convention leads to somewhat simpler combinatorics.

\subsubsection{Trimmed representation}\label{sec:TrimRep}

General representation has too much redundancy in the couplings with a small
number of legs. In view of the discussion of Section \ref{sec:intuitive}, it
is convenient to assume that the interaction has a more specific structure. In
particular, we want that: $H_2$ consists of a local term like {\eqref{loc2}}
plus an irrelevant term schematically of the form $(\partial \psi)^2$; $H_4$
consists of a local term like {\eqref{loc4}} plus an irrelevant term
schematically of the form $\psi^3 \partial \psi$; $H_6$ consists of a
semilocal term like {\eqref{eq:Xterm11}} plus higher order terms. More
precisely, we will assume that the interaction $H$, to be used as the input
for the RG map, is written in the \tmtextit{trimmed} representation, which
imposes the following extra requirements on kernels with $l \leqslant 6$:
\begin{enumeratenumeric}
  \item For $H_2$ we require:
  \begin{enumerateroman}
    \item $H_{2, 0}$ should be purely local, i.e. be the $\delta$-function
    kernel reproducing the local quadratic interaction $\nu \int \psi^2$ given
    in {\eqref{loc2}},
    
    \item $H_{2, 1} = 0$ (no kernels with one derivative).
  \end{enumerateroman}
  We will denote the nonzero parts of trimmed $H_2$ as
  \begin{equation}
    H_{\text{2L}} = H_{2, 0}, \quad H_{\text{2R}} = H_{2, 2}
  \end{equation}
  With some abuse of notation, we will identify $H_{\text{2L}}$ with the prefactor
  $\nu$ in front of the delta function, and similarly below for $H_{\text{4L}}$ with
  $\lambda$ and for $H_{\text{6SL}}$ with $\mathfrak{X}$.
  
  \item For $H_4$ we require that $H_{4, 0}$ should be purely local, i.e. be
  the $\delta$-function kernel reproducing the local quartic interaction
  $\lambda \int \psi^4$ given in {\eqref{loc4}}. We denote the parts of
  trimmed $H_4$ as
  \begin{equation}
    H_{\text{4L}} = H_{4, 0}, \quad H_{\text{4R}} = \{H_{4, p} \}_{p \geqslant 1} .
  \end{equation}
  \item For $H_6$ we demand that it comes split into two pieces:
  \begin{equation}
    H_6 = H_{\text{6SL}} + H_{\text{6R}},
  \end{equation}
  where $H_{\text{6SL}}$ contains only a `semi-local' interaction with no
  derivatives, of the form\footnote{\label{note:H6SLkernel}Its kernel is the
  antisymmetrization of $\Omega_{a_1 a_2} \Omega_{a_4 a_5} \Omega_{a_3 a_6}
  \delta (x_1 - x_2) \delta (x_1 - x_3) \tmmathbf{\mathfrak{X}}  (x_1 - x_4)
  \delta (x_4 - x_5) \delta (x_4 - x_6)$.} {\eqref{eq:Xterm11}}, parametrized
  by a function $\mathfrak{X} (x)$. Thus we have
  \begin{eqnarray}
    H_{6, 0} & = & H_{\text{6SL}} + H_{\text{6R}, 0},  \label{H60}\\
    H_{6, p} & = & H_{\text{6R}, p} \qquad (p \geqslant 1) . \nonumber
  \end{eqnarray}
  For the moment we do not make any specific requirement on the no-derivative
  part of $H_{\text{6R}}$.\footnote{Eventually, we shall impose a norm condition
  which will make the no-derivative part of $H_{\text{6R}}$ smaller by an extra
  $\eps$ factor, compared with $H_{\text{6SL}}$.} 
\end{enumeratenumeric}
Mnemonically, L stands for local, SL for semi-local, R for the rest. The
trimmed representation thus corresponds to a coupling sequence $(H_{\ell})$
where the index $\ell$ takes values from the `trimmed list'
\begin{equation}
  \ell \in \tmop{TL} = \{\text{2L}, \text{2R}, \text{4L}, \text{4R}, \text{6SL}, \text{6R}, 8, 10, 12 \ldots\} .
  \label{genind}
\end{equation}
The corresponding number of legs, an integer, will be denoted by $| \ell |$.
As for the general representation, we let $H_{\ell, p}$ be the restriction of
$H_{\ell}$ to the terms with $p$ derivatives. We will always use $\ell$ for
labels from the trimmed list {\eqref{genind}} and $l$ for integer labels: $l
\geqslant 2$ even. When $\ell$ and $l$ appear in the same equation, they are
related by $l = | \ell |$.

\begin{remark}
  \label{rationale}The reason for requirements 1,2 is as follows. The $H_{\text{2R}}$ and $H_{\text{4R}}$ interactions are irrelevant due to the presence of
  derivatives, while $H_{2, 0}, H_{2, 1}, H_{4, 0}$ would be relevant by the
  same criterion, see Section \ref{sec:Dil}. However, all these ``would-be
  relevant'' interactions with arbitrary kernels can be equivalently written
  as local quadratic and quartic couplings plus an irrelevant $H_{\text{2R}}$ and
  $H_{\text{4R}}$ (Section \ref{sec:trim0}). Therefore, requirements 1,2 make
  manifest the fact that our model has only two relevant couplings: $\nu$ and
  $\lambda$.
  
  Requirement 3 originates from the fact that, as discussed in Section
  \ref{sec:intuitive}, isolating the semilocal sextic term is important for
  deriving the correct lowest order FPE; this, in turn, will be crucial for
  defining the correct neighborhood on which the FPE (or, better, a suitable
  rewriting thereof) is contractive, see Section \ref{sec:FP}.
\end{remark}

\begin{remark}
  Even if the input interaction $H$ is in the trimmed representation, in
  general the interaction $H_{\tmop{eff}}$, obtained via the integrating-out
  step {\eqref{eq:Heff0}}, won't be trimmed. In order to put it in trimmed
  form, we will need to suitably manipulate the kernels of $H_{\tmop{eff}}$,
  via an operation called trimming, discussed in Section \ref{sec:trim0}
  below. This is one of the operations needed for proving that the space of
  interactions is left invariant by the action of the RG map.
\end{remark}

\subsection{Norms}\label{sec:Norms}

The interactions in the trimmed representation form a vector space. In order
to promote it to a Banach space, we need to equip it with a norm: for this
purpose, we will first specify the norm in the subspace associated with $\ell
\in$TL, see {\eqref{genind}}, and then the norm of a trimmed sequence.

\subsubsection{The norm of \texorpdfstring{$H_{\ell}$}{H l}}\label{Hlnorm}

We will be measuring the size of interaction kernels by means of the weighted
$L_1$ norm
\begin{equation}
  \|H (\bA)\|_w = \int_{x_1 = 0} \mathrm{} d^d \bx \hspace{0.17em} |H (\bA,
  \bx) | w (\bx) \hspace{0.17em}, \label{eq:normdef}
\end{equation}
where $w \left( \bx \right)$ is a translationally invariant weight function.
In view of translation invariance we perform the integral fixing one of the
$x_{}$ coordinates to zero. We also let
\begin{equation}
  \|H_l \|_w = \max_{| \bA | = l} \|H (\bA)\|_w \hspace{0.17em}, \label{eq:Hw}
\end{equation}
and similarly $\|H_{\ell} \|_w$ and $\|H_{\ell, p} \|_w$ are defined as the
maximum of weighted norms of all kernels belonging to the corresponding
trimmed coupling.\footnote{To avoid any misunderstanding, we stress that
$\|H_{\text{6SL}} \|_w$ and $\|H_{\text{6R}} \|_w$ are two independently defined
quantities.}

By choosing $w \left( \bx \right)$ growing at infinity appropriately, we will
incorporate the information about the decay of the kernels $H (\bA, \bx)$,
induced by the decay of the fluctuation propagator $g (x)$. Recall that we are
requiring the Gevrey condition {\eqref{chiG}}: $\chi \in G^s, s > 1$.
Concretely, this means that derivatives of $\chi$ of arbitrary order $\alpha$
are uniformly bounded by
\begin{equation}
  \max_{k \in \mathbb{R}^d} | \partial^{\alpha} \chi (k) |_{} \leqslant C^{|
  \alpha |} | \alpha |^{| \alpha | s} \label{chiGdef}
\end{equation}
for some constant $C = C (\chi) > 0$. The Gevrey condition is stronger than
$C^{\infty}$ but weaker than real analyticity. Importantly for us, Gevrey
class contains compactly supported functions. An explicit example of a cutoff
functions satisfying both condition {\eqref{chicond}} and {\eqref{chiG}} is
given in Appendix \ref{exGevrey}.

As usual, decay of $g (x)$ is related to the smoothness of its Fourier
transform $\hat{g} (k)$, i.e. to the smoothness of $\chi (k)$. It turns out
that the Gevrey condition {\eqref{chiGdef}} implies a stretched exponential
decay. Namely, we have the following bound for the fluctuation propagator, as
well as its first and second derivatives needed below:
\begin{equation}
  |g (x) |, | \del_{\mu} g (x) |, | \del_{\mu} \del_{\nu} g (x) | \leqslant M
  (x) \equiv C_{\chi 1} e^{- C_{\chi 2}  | x / \gamma |^{\sigma}} 
  \hspace{0.17em} \qquad (x \in \mathbb{R}^d), \label{gbound0}
\end{equation}
where $\sigma = 1 / s < 1$. The constants $C_{\chi 1}, C_{\chi 2}$ depend on
$\chi$ but are independent of $\gamma$. See Appendix \ref{stretched} for a
detailed proof, while here we only give two simple remarks. First, the decay
scale $x \sim \gamma$ in {\eqref{gbound0}} is as expected from the IR momentum
cutoff $\sim \gamma^{- 1}$. Second, stretched exponential is the best we could
hope for: exponential decay ($\sigma = 1$) would require analyticity of
$\chi$, incompatible with the compact support. \

\begin{figure}[h]\centering
  \begin{center}
    \resizebox{100pt}{!}{\includegraphics{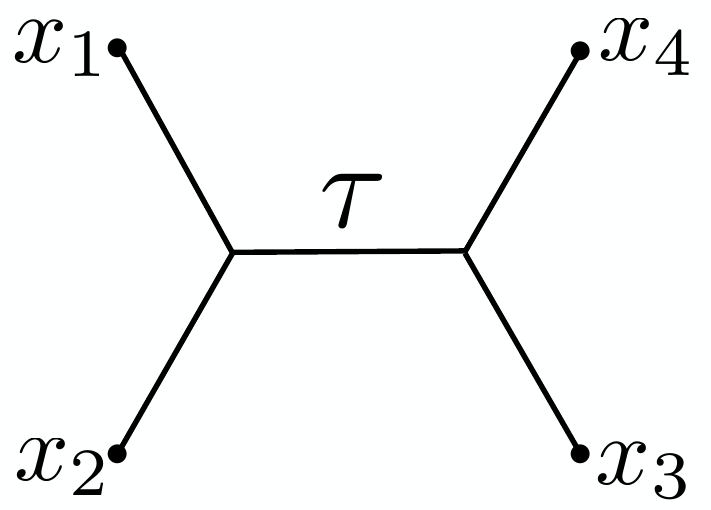}} 
  \end{center}
  \caption{\label{fig:steiner}The optimal Steiner tree $\tau$ for this
  configuration of 4 points contains two extra vertices.}
\end{figure}

Kernels $H (\bA, \bx)$ are expected to decay at large separation with the
same rate as {\eqref{gbound0}}. We choose $w (\bx)$ growing with a similar
rate. A convenient choice turns out to be
\begin{equation}
  w (\bx) = e^{\bar{C}_{}  (\tmop{St} (\bx) / \gamma)^{\sigma}}
  \hspace{0.17em}, \hspace{0.17em} \label{eq:ourW}
\end{equation}
where $\tmop{St} (\bx)$ is the Steiner diameter of the set $\bx$, defined
{\cite{Steiner}} as the length of the shortest tree $\tau$ connecting the
points in $\bx$ (the tree may contain extra vertices as in Fig.
\ref{fig:steiner}).\footnote{\label{note:St}To be precise, $\tmop{St} (\bx) =
\min_{\bx'} \min_{\tau (\bx \cup \bx')} L (\tau)$, the minimum taken over all
possible trees $\tau$ with vertices $\bx \cup \bx'$, with the tree length $L
(\tau)$ defined as the sum of the edge lengths. $\tmop{St} (\bx)$ coincides
with the usual diameter if all points lie on a line (e.g. for sets of 2
points). See Appendix \ref{sec:Snorm} for an explanation of why we use the
Steiner diameter.} We will fix $\bar{C}_{} = \frac{1}{2} C_{\chi 2}$ so that
$M (x)$ has a finite weighted norm:
\begin{equation}
  \| M \|_w = \int d^d x M (x) e^{\bar{C}_{}  (| x | / \gamma)^{\sigma}} =
  \tmop{Const} . \gamma^d < \infty . \label{Mw}
\end{equation}
Finally, it will be convenient to use definition {\eqref{eq:Hw}} also for the
$\delta$-function kernels of the local quadratic and quartic interactions.
Since $w = 1$ when all points coincide, we have (see {\eqref{loc2}},
{\eqref{loc4}})
\begin{equation}
  \| H_{\text{2L}} \|_w = | \nu |, \qquad \| H_{\text{4L}} \|_w = \frac{1}{3} | \lambda |
  . \label{H4Llambda}
\end{equation}

\subsubsection{The norm of a trimmed sequence}\label{sec:normtrimmedseq}

The norm of an interaction $H$ associated with the trimmed sequence
$(H_{\ell})$, $\ell \in$TL, will have the form $\|H\|= \sup_{\ell \in
\tmop{TL}} \|H_{\ell} \|_w / \delta_{\ell}$, for a sequence $\delta_{\ell}$ to
be fixed conveniently. In order to decide how to let $\delta_{\ell}$ scale
with $\ell$, let us first develop an intuition about the expected size of
$H_{\ell}$ at the fixed point. If we parametrize $H$ by $\left( \nu, \lambda,
\Xterm, u \right)$, with $u = (H_{\ell})_{\ell \neq \text{2L}, \text{4L}, \text{6SL}} \equiv
(u_{\ell})_{\ell \in \{\text{2R}, \text{4R}, \text{6R}, 8, 10, \ldots\}}$, we expect that at the
fixed point $\nu$ and $\lambda$ are of order $\varepsilon$, $\Xterm$ is equal
to the kernel $\Xterm_{\ast}$ in Eq.{\eqref{fixedX}}, which is of order
$\varepsilon^2$, and $u_{\ell}$ is of the order of the corresponding tree
graph (i.e., the leading Feynman diagram with vertices all of type $\lambda$
contributing to the interaction labelled $\ell$), namely: of order
$\varepsilon^2$ if $\ell = \text{2R}, \text{4R}$; of order $\varepsilon^3$ if $\ell = \text{6R}$; of order $\varepsilon^{l / 2 - 1}$ if $\ell = l \geqslant 8$.

In the following, in order to determine the fixed point, we will fix
$\Xterm_{} = \Xterm_{\ast}$, thought of as a function of $\lambda$, see
{\eqref{fixedX}}, and parametrize the fixed point interaction by the remaining
coordinates, $y = (\nu, \lambda, u)$. On this subspace, we will use the
following norm (depending on the parameters $\delta, A_0, A^{\text{R}}_{0},
A^{\text{R}}_{1}, A^{\text{R}}_{2}, A$):
\begin{equation}
  \|y\|_Y = \max \left\{ \frac{| \nu |}{A_0 \delta}, \frac{| \lambda |}{A_0
  \delta}, \frac{\|u_{\text{2R}} \|_w}{A^{\text{R}}_0 \delta^2}, \frac{\|u_{\text{4R}} \|_w}{A^{\text{R}}_1
  \delta^2}, \frac{\|u_{\text{6R}} \|_w}{A^{\text{R}}_2 \delta^3}_{}, \sup_{l \geqslant 8}
  \frac{\|u_l \|_w}{A \delta^{l / 2 - 1}} \right\}, \label{defBanach1st}
\end{equation}
where, motivated by the intuitive discussion above, the parameter $\delta$
will be chosen to be proportional to $\varepsilon$. Eq.{\eqref{defBanach1st}}
defines the Banach space of interest.\footnote{It is easy to see in particular
that the space is complete with respect to the introduced norm (because
weighted $L_1$ spaces are complete).} Eventually, the constants $A_0, A^{\text{R}}_{^{}
0}, A^{\text{R}}_{^{} 1}, A^{\text{R}}_{^{} 2}, A$ will be fixed in such a way that the action
of the RG map (or better, of a suitable equivalent rewriting thereof, called
$F$ in the following, see Section \ref{abstract}) on the sequence $y$ returns
a new sequence $y'$ in the same Banach space $Y$. Even more, we will show that
there is a neighborhood $Y_0$ in $Y$ on which the fixed point map $F$ is a
contraction and, therefore, $F$ admits a unique fixed point in $Y_0$. All this
will be proved in Sections \ref{sec:FP} and \ref{sec:2k+2} below. As a
preparation to these proofs, we need to specify how the RG map explicitly acts
on the space of trimmed sequences. This will be discussed in Section
\ref{detailsRG}. In particular, the result of the RG map on the trimmed
sequence $(H_{\ell})$ will be expressed in the form of a series, see
Eq.{\eqref{RHH'}} below, which is {\tmem{absolutely convergent}} in the norm
of interest, thanks to the bounds discussed in Section \ref{sec:NormBounds}
below.

\section{The renormalization map}\label{detailsRG}

In this section we detail the structure of the RG map, thought of as a map
from the vector space of trimmed sequences into itself. We proceed in steps:
we first describe the integrating-out map {\eqref{eq:Heff0}}, assuming the
input to be a trimmed sequence (Section \ref{sec:IntOut}). In general, the
output of the integrating-out map is not trimmed: therefore, we explain how to
make it so, via the trimming operation (Section \ref{sec:trim0}). Next, we
perform the rescaling {\eqref{Hprime}} (Section \ref{sec:Dil}). In Section
\ref{sec:RGtrimrep} we combine these three steps and derive the representation
Eq.{\eqref{RHH'}}, which expresses the image of the RG map as a series in the
multi-indices $(\ell_i)^n_{i = 1}$, $\ell_i \in$TL. Remarkably, this series
turns out to be {\tmem{absolutely convergent}} in the relevant norms, i.e.,
those introduced in Section \ref{sec:Norms} above, thanks to the norm bounds
discussed in Section \ref{sec:NormBounds}.

\subsection{Integrating-out map}\label{sec:IntOut}

Let $H$ be an interaction associated with the trimmed sequence
$(H_{\ell})_{}$, $\ell \in$TL, and consider the integrating-out map
{\eqref{eq:Heff0}}. For the effective interaction $H_{\eff}$ we have a
well-known perturbative formula in terms of connected expectations (see
Appendix \ref{sec:conn}):\footnote{In Eq.{\eqref{eq:Heff0}} the argument of
$H_{\eff}$ was $\psi_{\gamma}$, the low-momentum component of $\psi$. The
momentum-range restriction turns out unimportant for working out
{\eqref{eq:Heff2}}, so we replaced $\psi_{\gamma}$ by a generic $\psi$.}
\begin{equation}
  H_{\eff} (\psi) = \sum_{n = 1}^{\infty} \frac{1}{n!} \langle \underbrace{H
  (\psi + \phi) ; H (\psi + \phi) ; \ldots ; H (\psi + \phi)}_{n \hspace{1em}
  \mathrm{times}} \rangle_c . \label{eq:Heff2}
\end{equation}
We write the interaction in the trimmed representation using the same
Eq.{\eqref{Hpsi}} as in the general representation. Kernels $H (\mathbf{A},
\mathbf{x})$ now come from couplings $H_{\ell}$. The $H_{2, 1}$ kernels are
absent due to trimming requirements. The $H_{6, 0}$ kernels are understood as
a sums of $H_{\text{6SL}}$ and $H_{\text{6R}, 0}$ kernels, see {\eqref{H60}}. All
the other kernels $H (\mathbf{A}, \mathbf{x})$ are associated with a unique
coupling $H_{\ell}$.

Replacing $\psi \rightarrow \psi + \phi$ in {\eqref{Hpsi}}, each term gives
rise to `interaction vertices' with `external' $\psi$ legs and `internal'
$\phi$ legs. Parametrizing the external legs by a subsequence $\bB \subset
\bA$, \footnote{Sequences being ordered sets, a subsequence inherits ordering
from the parent sequence.} \ and the internal ones by $\barbB = \bA \setminus
\bB$, we write:
\begin{equation}
  H (\psi + \phi) = \sum_{\bA} \sum_{\bB \subset \bA} (-)^{\#} \int d^d \bx
  \hspace{0.17em} H (\bA, \bx) \Psi (\bB, \bx_{\bB}) \Phi (\barbB,
  \bx_{\barbB}) \hspace{0.17em}, \label{eq:HLtyp}
\end{equation}
where $(-)^{\#}$ is the sign, which we won't need to track, produced by
reordering the fields to put all $\psi$'s first. The $\bx_{\bB}$ and
$\bx_{\barbB}$ are the corresponding restrictions of the coordinate vector
$\bx$. Substituting {\eqref{eq:HLtyp}} into {\eqref{eq:Heff2}}, we obtain the
following formula for the kernels of the effective interaction (see Appendix
\ref{sec:Heff} for more details):
\begin{equation}
  H_{\eff} (\bB, \bx_{\bB}) =\mathcal{A} \sum_{n = 1}^{\infty} \frac{1}{n!} 
  \sum_{\tmscript{\begin{array}{c}
    \bB_1, \ldots, \bB_n\\
    \sum \bB_i = \bB
  \end{array}}} \sum_{\tmscript{\begin{array}{c}
    \bA_1, \ldots, \bA_n\\
    \bA_i \supset \bB_i
  \end{array}}} (-)^{\#} \int d^d \bx_{\barbB} \hspace{0.17em} \hspace{0.17em}
  \text{} \mathcal{C} \left( \bx_{\barbB} \right)  \prod_{i = 1}^n H (\bA_i,
  \bx_{\bA_i}), \label{eq:Hefflift}
\end{equation}
where the sum is over all ways to represent $\bB$ as a concatenation $\bB_1 +
\ldots + \bB_n$, and then over all ways to extend $\bB_i$'s to $\bA_i \supset
\bB_i$. The integration is over points $\bx_{\barbB}$, $\barbB = \barbB_1 +
\ldots + \barbB_n$, $\barbB_i = \bA_i \setminus \bB_i$, while the unintegrated
parts of the vectors $\bx_{\bA_i}$ form $\bx_{\bB} = \bx_{\bB_1} + \ldots +
\bx_{\bB_n}$, and the integration kernel $\mathcal{C} \left( \bx_{\barbB}
\right)$ is the connected expectation:
\begin{equation}
  \mathcal{C} \left( \bx_{\barbB} \right) = \left\langle \Phi (\barbB_1,
  \bx_{\barbB_1}) ; \ldots ; \Phi (\barbB_n, \bx_{\barbB_n}) \right\rangle_c .
  \label{CxB}
\end{equation}
Finally, $\mathcal{A}$ in Eq.{\eqref{eq:Hefflift}} denotes the
antisymmetrization operation, see footnote \ref{opA}. Note that the set of
kernels $H_{\eff} (\bB, \bx_{\bB})$ produced by this formula will not in
general satisfy the trimming requirements, even if the kernel $H (\bA_{},
\bx)$ did. This will be dealt with in the next section.

We write Eq.{\eqref{eq:Hefflift}} more compactly and abstractly as
\begin{equation}
  \label{Heffl} \left( H_{\eff} \right)_l = \sum_{(\ell_i)_1^n}
  \mathcal{S}_l^{\ell_1, \ldots, \ell_n} (H_{}), \qquad H = (H_{\ell})_{\ell
  \in \tmop{TL}} .
\end{equation}
Each term in {\eqref{Heffl}} is numbered by a sequence $(\ell_i)_1^n =
(\ell_1, \ldots, \ell_n)$, $n \geqslant 1$, $\ell_i \in \tmop{TL}$, and by an
even $l \geqslant 2$. The map $\mathcal{S}_l^{\ell_1, \ldots, \ell_n} (H_{})$
in {\eqref{Heffl}} is the sum of all terms in {\eqref{eq:Hefflift}} which have
$\left| \bB \right| = l$ and $H (\bA_i, \bx_{\bA_i}) \in H_{\ell_i},$while
$\bB_i$ can be arbitrary, subject to the requirements \ $\bB_i \subset \bA_i$,
$\sum \bB_i = \bB$.\footnote{As noted above, $H_{6, 0}$ kernels are a sum of
$H_{\text{6SL}}$ and $H_{\text{6R}, 0}$. Picking one or the other part of the sum
is understood when defining the map $\mathcal{S}_l^{\ell_1, \ldots, \ell_n}$
with $\ell_i = \text{6SL}$ or $\ell_i = \text{6R}$.}

Let $B_l$ and $B_{\ell}$ be the vector spaces of couplings $H_l$ and trimmed
couplings $H_{\ell}$, respectively, and let $B_{\tmop{trim}} =
\bigotimes_{\ell \in \tmop{TL}} B_{_{\ell}}$ be the vector space of trimmed
coupling sequences $H = (H_{\ell})_{\ell \in \tmop{TL}}$. The map
$\mathcal{S}_l^{\ell_1, \ldots, \ell_n}$ then acts from $B_{\tmop{trim}}$ to
$B_l$ and is homogeneous of degree $n$. Of course, this map only depends on
the couplings $H_{\ell_i}$ whose index $\ell_i$ occurs in the sequence
$(\ell_i)_1^n$. If index $\ell_i$ occurs $n_i$ times, this map has homogeneity
degree $n_i$ in $H_{\ell_i}$.

It will be also useful to define a closely related map $S_l^{\ell_1, \ldots,
\ell_n}$, obtained by replacing $\prod_{i = 1}^n H (\bA_i, \bx_{\bA_i})
\rightarrow \prod_{i = 1}^n h_i (\bA_i, \bx_{\bA_i})$ in {\eqref{eq:Hefflift}}
with independent $h_i \in B_{\ell_i}$. This gives a multilinear map:
\begin{equation}
  S_l^{\ell_1, \ldots, \ell_n} : B_{\ell_1} \times \ldots \times B_{\ell_n}
  \rightarrow B_l . \label{Slli}
\end{equation}
Note that this map is symmetric, i.e. invariant under the interchanges of
indices $\ell_i$ accompanied by the simultaneous interchange of arguments. By
identifying the arguments of $S_l^{\ell_1, \ldots, \ell_n}$, we get back the
map $\mathcal{S}_l^{\ell_1, \ldots, \ell_n}$:
\begin{equation}
  \mathcal{S}_l^{\ell_1, \ldots, \ell_n} (H_{}) = S_l^{\ell_1, \ldots, \ell_n}
  (h_1, \ldots, h_n), \qquad h_i = H_{\ell_i} \label{Sldef}
\end{equation}
It will be very important that the maps $S_l^{\ell_1, \ldots, \ell_n}$ and
$\mathcal{S}_l^{\ell_1, \ldots, \ell_n}$ vanish unless $\sum | \ell_i |_{}
\geqslant l + 2 (n - 1)$. Otherwise, the number of fields in the connected
expectation {\eqref{CxB}}, which is $\sum | \ell_i |_{} - l$, is not enough to
get a connected Wick contraction.

We will see that the defined maps are continuous with respect to the norms
from Section \ref{Hlnorm}, see Section \ref{sec:boundsSl} below.

While generally $S_l^{\ell_1, \ldots, \ell_n}$ acts into $B_l$, in two cases
it can be considered to act into $B_{\ell}$:
\begin{itemizedot}
  \item For $\ell \geqslant 8$ since then $B_{\ell} = B_l$.
  
  \item For $n = 1$ and $l = | \ell_1 |$, because in this case $S_l^{\ell_1}$
  is the identity map: $S^{\ell}_{l^{}} (H_{\ell}) = H_{\ell}$.
\end{itemizedot}
\begin{remark}
  \label{R2}Eq.{\eqref{eq:Hefflift}} is going to be the basis for all further
  considerations. Although derived so far by perturbation theory, in our model
  this equation will be non-perturbatively true. Let us discuss first why this
  may be expected on physical grounds. Non-perturbative validity of
  Eq.{\eqref{eq:Hefflift}} means that in our model the full effective action
  is captured by perturbation theory, with no extra contributions. Extra
  ``instantonic'' contributions are common in models involving bosonic fields,
  but these are absent in our model since we only have fermions. Perturbation theory may also break down if fermions form a bosonic bound state, but this
  typically requires a coupling that becomes large when iterating the RG, and
  in our model all couplings will stay weak. In the main text we will show in
  particular that the series in the r.h.s. is convergent provided that $H
  (\psi)$ is sufficiently small, and so $H_{\tmop{eff}} (\psi)$ is well
  defined.\footnote{\label{DiagMC}We wish to draw here a parallel with fermionic models of condensed-matter physics, which have convergent perturbation theory at finite temperatures (a notable exception being fermions at finite density in 3d continuous space, unstable with respect to collapse to a point for attractive interaction). One often studied example is the Fermi-Hubbard model, whose perturbative series in the onsite repulsion $U$ has a finite, $T$-dependent, radius of convergence. A simple extension of the GKL bound for connected expectations discussed in App. D.4 (see Sect.3 of \cite{BGPS}; see also \cite{PS08} and Sect.6 of \cite{Ale-review}) easily implies convergence of the series, but with a far-from-optimal temperature dependence ($U\propto T^{d+1}$, with $d$ the spatial dimension of the lattice). For realistic estimates on the convergence radius, see e.g. \cite{Benfatto2006,PhysRevB.79.201403,Graphene,2014JPhA47T5003M,Ma_jsp14,GMP19}. The perturbation series for Fermi-Hubbard can be evaluated to high order, and convergence checked, by Diagrammatic Monte Carlo (DiagMC) method \cite{VANHOUCKE201095,Prokofiev}. We thank Kris Van Houcke and Felix Werner for discussions about DiagMC. See also footnote \ref{Rossi}.} In Appendix \ref{sec:non-pert} we will give a rigorous
  justification of Eq.{\eqref{eq:Hefflift}}, by first deriving this equation
  in finite volume and then passing to the limit, in line with Remark
  \ref{FinVol}.
\end{remark}

\subsection{Trimming}\label{sec:trim0}

We wish to realize the renormalization map in the space of trimmed couplings.
Unfortunately, as mentioned, the kernels $(H_{\tmop{eff}})_l$ provided by
Eq.{\eqref{Heffl}} are not in general trimmed. To correct this, we need an
extra ``trimming'' step, which will find an equivalent trimmed representation
of the same interaction. This step corresponds, in different notation, to the
rewriting of $H_{\tmop{eff}}$ in the equivalent form
$\mathcal{L}H_{\tmop{eff}} +\mathcal{R}H_{\tmop{eff}}$ used in many papers in
CRG, see in particular {\cite{Ga}} and {\cite{BGbook}}. In the CRG literature,
$\mathcal{L}H_{\tmop{eff}}$ and $\mathcal{R}H_{\tmop{eff}}$ are usually called
the ``local'' and ``regularized'' parts of the effective interaction,
respectively. Before describing this step in detail, let us first discuss what
it means for two representations to be {\tmem{equivalent}}.

\subsubsection{Equivalent coupling sequences}\label{sec:interpol}

A coupling sequence $(N_l)$ is called null if the corresponding interaction
vanishes as a function of classical Grassmann fields $\psi^{}_a (x)$ (an
explicit example of what we mean is discussed below). Two coupling sequences
with a null difference are equivalent (represent the same interaction).
Interactions are thus identified with coupling sequences modulo this
equivalence relation.

The basic mechanism to produce equivalent couplings, referred to as
``interpolation'', starts with the Newton-Leibniz formula:
\begin{equation}
  \psi^{}_a (x) = \psi_a (y) + \int_0^1 ds \partial_s
  \psi_a  (y + s (x - y)) = \psi_a (y) + \int_0^1 ds
  (x - y)^{\mu} \partial_{\mu} \psi_a  (y + s (x - y))
  .
\end{equation}
We can also integrate the r.h.s. in $y$ against some function $f
(x, y)$ of unit total integral: $\int dyf (x, y) = 1$. We get a family of
``interpolation identities'' expressing $\psi^{a} (x)$
as a weighted linear combination of $\psi^{a} (y)$ and
$\partial_{\mu} \psi_a (y)$:
\begin{equation}
  \psi_a (x) = \int dy \hspace{0.17em} \left[ f (x, y)
  \psi_a (y) + \hspace{0.17em} f^{\mu} (x, y)
  \partial_{\mu} \psi_a (y) \right], \label{eqint}
\end{equation}
where $f^{\mu} (x, y)$ can be expressed in terms of
$f$.\footnote{Explicitly $f^{\mu} (x, y) = \int_0^1
\frac{ds}{(1 - s)^d}  [(x - z)^{\mu} f (x, z)]_{z = (y
- sx) / (1 - s)}$. This is finite if $f (x, y)$ decreases
sufficiently fast at large $y$.}

Now take a single interaction term $\int \mathrm{d} \bx \hspace{0.17em} H
(\bA, \bx) \Psi (\bA, \bx)$ corresponding to some $\bA$ with at least one
field not differentiated, e.g. the first one: $A_1 = a_1$. Pick a function $f
(x_, y)$ and replace $\psi^{}_{a_1} (x_1)$ inside this interaction term via
the identity {\eqref{eqint}}. This generates an equivalent representation of
the same interaction of the form
\begin{equation}
  \sum_{\mathbf{B}} \int d^d \bx \hspace{0.17em} \tilde{H} (\mathbf{B}, \bx)
  \Psi (\mathbf{B}, \bx), \label{Htilde00}
\end{equation}
where the sum has $d + 1$ terms: either $\mathbf{B}= \bA$ or it is obtained
from $\bA$ replacing $A_1 \rightarrow (a_1, \mu)$, $\mu = 1, \ldots, d$. The
corresponding kernels $\tilde{H}^{} (\mathbf{B}, \bx)$ are obtained
integrating $H (\bA, \bx)$ against $f$ and $f^{\mu}$. We can also
apply this procedure to multiple interaction terms in the original interaction
$H (\psi)$, summing up the new kernels $\tilde{H}^{} (\mathbf{B}, \bx)$ to the
kernels $H (\mathbf{B}, \bx)$ to which the transformation has not been
applied. The resulting total interaction, which we call $\tilde{H} (\psi)$, is
equivalent to $H (\psi)$. The difference of coupling sequences $(H_l)$ and
$(\tilde{H}_l)$ is null.

We stress that the Newton-Leibniz formula and the interpolation identities
will be applied only to those $\psi$'s in the interaction terms which do not
carry any derivatives. Then, all the produced terms contain $\psi$'s with at
most one derivative. This explains why in {\eqref{products}} we allowed fields
with zero or one (but not more) derivatives.

\subsubsection{Trimming map}\label{sec:trimming}

We now explain the trimming map, which maps the sequence $(H_{\tmop{eff}})_l$
in {\eqref{Heffl}} to an equivalent sequence of trimmed couplings. Of course,
the restriction of the sequence $(H_{\tmop{eff}})_l$ to $l \geqslant 8$ is
already trimmed (see the end of section \ref{sec:IntOut}), so we only need to
do something for $l \leqslant 6$.

For $l = 6$, we define
\begin{equation}
  (H_{\tmop{eff}})_{\text{6SL}} = H_{\text{6SL}} + S_6^{\text{4L}, \text{4L}} (H_{\text{4L}},
  H_{\text{4L}}), \label{Heff6SL}
\end{equation}
i.e. the two terms in the $(H_{\tmop{eff}})_6$ series which manifestly have
the form {\eqref{eq:Xterm11}}. We define $(H_{\tmop{eff}})_{\text{6R}}$ as the sum
of all the other terms in the $(H_{\tmop{eff}})_6$ series.

For $l = 2, 4$ trimming will involve ``localization'' and ``interpolation''.
``Localization'' extracts the local parts of $(H_{\tmop{eff}})_{2, 0}$ and
$(H_{\tmop{eff}})_{4, 0}$ (see the example below):
\begin{equation}
  (H_{\tmop{eff}})_{\text{2L}} = T^{2, 0}_{\text{2L}} (H_{\tmop{eff}})_{2, 0}, \qquad
  (H_{\tmop{eff}})_{\text{4L}} = T^{4, 0}_{\text{4L}} (H_{\tmop{eff}})_{4, 0} .
  \label{localization}
\end{equation}
``Interpolation'' rearranges the other components setting to zero the parts of
$(H_{\tmop{eff}})_{\text{2R}, p}$ and $(H_{\tmop{eff}})_{\text{4R}, p}$ in agreement with
the trimming requirements, and making sure that the resulting coupling
sequence is equivalent as in Section \ref{sec:interpol}. This operation will
have the following structure:
\begin{equation}
  (H_{\tmop{eff}})_{\text{4R}, p} = \left\{\begin{array}{ll}
    0 & \text{if } p = 0\\
    (H_{\tmop{eff}})_{4, 1} + T^{4, 0}_{\text{4R}} (H_{\tmop{eff}})_{4, 0} &
    \text{if } p = 1\\
    (H_{\tmop{eff}})_{4, p} & \text{if $p > 1$,}
  \end{array}\right. \label{def4R}
\end{equation}
\begin{equation}
  (H_{\tmop{eff}})_{\text{2R}, p} = \left\{\begin{array}{ll}
    0 & \text{if } p = 0, 1\\
    (H_{\tmop{eff}})_{2, 2} + T^{2, 1}_{\text{2R}} (H_{\tmop{eff}})_{2, 1} + T^{2,
    0}_{\text{2R}} (H_{\tmop{eff}})_{2, 0} & \text{if $p = 2$.}
  \end{array}\right. \label{def2L}
\end{equation}
This can be also written succinctly as
\begin{equation}
  (H_{\tmop{eff}})_{\text{4R}} = T^4_{\text{4R}} (H_{\tmop{eff}})_4, \qquad
  (H_{\tmop{eff}})_{\text{2R}} = T^2_{\text{2R}} (H_{\tmop{eff}})_2,
\end{equation}
where {\eqref{def4R}}, {\eqref{def2L}} define components of $T^4_{\text{4R}}, T_{\text{2R}}^2$ in subspaces with a definite number of derivatives.

Consider $l = 4$ as an example. The coupling $(H_{\tmop{eff}})_{4, 0}$
corresponds to an interaction the form
\begin{equation}
  \Omega_{a b} \Omega_{c e} \int d^d \mathbf{x}F (\mathbf{x}) \psi^{}_a (x_1)
  \psi_b (x_2) \psi_c (x_3) \psi_e (x_4)
\end{equation}
(plus two other terms with $\Omega_{a c} \Omega_{b e}$ and $\Omega_{a e}
\Omega_{b c}$). Substitute into this an interpolation identity
\begin{equation}
  \psi^{}_a (x_1) \psi_b (x_2) \psi_c (x_3) \psi_e (x_4) = (\psi_a \psi_b
  \psi_c \psi_e) (x_1) + \psi_a^{} (x_1) \int_0^1 d t \partial_t [\psi_b
  (x_2^t) \psi_c (x_3^t) \psi_e^{} (x_4^t)], \label{ExInt}
\end{equation}
where $x_i^t = x_1 + t (x_i - x_1)$. The first term gives a local quartic
interaction with
\begin{equation}
  \lambda = \int_{x_1 = 0} d^d \mathbf{x}F (\mathbf{x}), \label{lambdaF}
\end{equation}
which defines $T^{4, 0}_{\text{4L}}$ in
{\eqref{localization}}.\footnote{\label{note:mom}This equation can be
equivalently written in momentum space as \ $\lambda = \hat{F} (0, 0, 0, 0)$,
i.e. evaluating the kernel with all external momenta set to zero.} The second
term in {\eqref{ExInt}} gives a sum of interactions where one of $\psi_b,
\psi_c, \psi_e$ is differentiated: this defines $T^{4, 0}_{\text{4R}}
(H_{\tmop{eff}})_{4, 0}$.

The $l = 2$ maps $T^{2, 0}_{\text{2L}}, T^{2, 1}_{\text{2R}}, T^{2, 0}_{\text{2R}}$ are defined
analogously. For $T^{2, 0}_{\text{2R}}$ one needs to apply interpolation twice, to
get from a term with no derivatives to a term where both fields carry
derivatives. See Appendix \ref{sec:Trim} for the full construction of these
maps, and for the analysis of how they behave with respect to the norms
measuring the size of interaction kernels.

\begin{remark}
  Note that our trimming map $T$ is just one of infinitely many possible
  trimming maps, corresponding to different choices of interpolation
  identities. E.g. instead of {\eqref{ExInt}} we could have used
  \begin{multline}
    \psi_a (x_1) \psi_b (x_2) \psi_c (x_3) \psi_e (x_4) = \\\psi_a (x_1)
    \left[ \psi_b (x_1) + \int_0^1 d t \left. \frac{d \psi_b}{d t} (x_2^t)
    \right] \right. \left[ \psi_c (x_1) + \int_0^1 d s \frac{d \psi_c}{d s}
     (x_3^s)] \right] \left[ \psi_e (x_1) + \int_0^1 d u \left.
    \frac{d \psi_e}{d u} (x_4^u) \right] \right. . \label{ExIntAlt}
  \end{multline}
  Such alternative trimming maps $\tilde{T}$ differ from $T$ by a map which is
  null (gives a null sequence of couplings when applied to any interactions).
  In our construction we will use $T$, but any other trimming map satisfying
  the same norm bounds (see Appendix \ref{sec:Trim}) would work equally well.
\end{remark}

\subsection{Dilatation}\label{sec:Dil}

After having integrated out the fluctuation field and rearranged the result so
that it is equivalently rewritten in trimmed form, we rescale the fields, see
{\eqref{Dsubs}}-{\eqref{Hprime}}. We call this rescaling step
{\tmem{dilatation}}, and denote it by $D$. Note that $D$ preserves the trimmed
representation. The action on the kernels is:
\begin{equation}
  D : H_{\ell, p} (\bx) \mapsto \gamma^{- D_l - p} \gamma^{d (l - 1)} H_{\ell,
  p} (\gamma \bx) \hspace{0.17em}, \label{eq:Resclift}
\end{equation}
where we recall that $l = | \ell |$, $p$ denotes the number of derivatives in
the interaction term, and we denoted
\begin{equation}
  D_l = l [\psi] - d = l (d / 4 - \varepsilon / 2) - d . \label{Dl}
\end{equation}
For the special cases $\ell \in \{ \text{2L}, \text{4L}, 6 S L \}$
Eq.{\eqref{eq:Resclift}} becomes:
\begin{equation}
  \nu \mapsto \gamma^{- D_2} \nu = \gamma^{\frac{d}{2} + \eps} \nu, \qquad
  \lambda \mapsto \gamma^{- D_4} \lambda = \gamma^{2 \varepsilon} \lambda,
  \qquad \mathfrak{X} (x) \mapsto \gamma^{- D_6} \gamma^d \mathfrak{X} (\gamma
  x) .
\end{equation}
In terms of the norms of Section \ref{Hlnorm}, the irrelevance condition for
$H_{\ell, p}$ will be $D_l + p > 0$, see Eq.{\eqref{eq:normResc}} below. As
stated in Eq.{\eqref{epsCond}} we are assuming $d \in \{ 1, 2, 3 \}$ and $0 <
\varepsilon_{} < d / 6$. Under these conditions it's easy to check that $D_2,
D_4 < 0,$so that $\nu, \lambda$ are relevant, while
\begin{equation}
  \qquad D_2 + 2, D_4 + 1 > 0, \qquad D_l \geqslant D_6 > 0 \quad  (l
  \geqslant 6) . \label{Dbarpos}
\end{equation}
so that $H_{\text{2R}}, H_{\text{4R}}, H_{\text{6SL}}, H_{\text{6R}}$ and $H_{\ell}$ $(\ell
\geqslant 8)$ are irrelevant. These interactions comprise $H_{\text{IRR}}$ in
{\eqref{HLHirr}}.

\subsection{Renormalization map in the trimmed
representation}\label{sec:RGtrimrep}

The renormalization map $R$ is obtained composing the three operations:
integrating-out, then trimming, then dilatation. It is the map defined in
Section \ref{RenMap} but now written in a specific set of coordinates (the
trimmed representation). \ Summarizing Sections \ref{sec:IntOut},
\ref{sec:trimming}, \ref{sec:Dil}, we represent $R = R (\varepsilon, \gamma)$
follows: if $H \in B_{\tmop{trim}}$ (the vector space of sequences of trimmed
couplings), then $R : H \mapsto H' \in B_{\tmop{trim}}$, with
\begin{equation}
  H'_{\ell} = \sum_{(\ell_i)_1^n} \mathcal{R}_{\ell}^{\ell_1, \ldots, \ell_n}
  (H), \label{RHH'}
\end{equation}
where $\mathcal{R}_{\ell}^{\ell_1, \ldots, \ell_n}$ is a homogeneous map of
degree $n$ obtained by identifying the arguments in a multilinear map
$R_{\ell}^{\ell_1, \ldots, \ell_n}$:
\begin{equation}
  \mathcal{R}_{\ell}^{\ell_1, \ldots, \ell_n} (H_{}) = R_{\ell}^{\ell_1,
  \ldots, \ell_n} (h_1, \ldots, h_n), \qquad h_i = H_{\ell_i} .
\end{equation}
This multilinear map can be written explicitly as follows. For $(n ; (\ell_1,
\cdots, \ell_n)) = (1 ; \ell)$ we have
\begin{equation}
  R_{\ell}^{\ell_{}} = D, \label{Rlexpl0}
\end{equation}
since in this case $S_l^{\ell_{}} = \mathbbm{1}$ and trimming is not needed.
In all the other cases $(n ; (\ell_1, \cdots, \ell_n)) \neq (1 ; \ell)$,
recalling that $l = | \ell |$, we have
\begin{eqnarray}
  R_{\ell}^{\ell_1, \ldots, \ell_n} & = & D \left\{\begin{array}{ll}
    S_l^{\ell_1, \ldots, \ell_n} & \ell \geqslant 8\\
    T^l_{\ell} S_l^{\ell_1, \ldots, \ell_n} & \ell \in \{ \text{2L}, \text{2R}, \text{4L}, \text{4R}
    \}
  \end{array}\right. \nonumber\\
  R_{\text{6SL}}^{\ell_1, \ldots \ell_n} & = & D \left\{\begin{array}{ll}
    S_6^{\text{4L}, \text{4L}} & (\ell_i)_1^n = (\text{4L}, \text{4L})\\
    0 & \tmop{otherwise}
  \end{array}\right.  \label{Rlexpl}\\
  R_{\text{6R}}^{\ell_1, \ldots \ell_n} & = & D \left\{\begin{array}{ll}
    S_6^{\ell_1, \ldots, \ell_n} & (\ell_i)_1^n \neq (\text{6SL}), (\text{4L}, \text{4L})\\
    0 & \tmop{otherwise}
  \end{array}\right. \nonumber
\end{eqnarray}
where $T^l_{\ell} : B_l \rightarrow B_{\ell}$ is the trimming map whose
various components are defined by equations {\eqref{localization}},
{\eqref{def4R}}, {\eqref{def2L}}, and see Eq.{\eqref{Heff6SL}} for $\ell \in
\{ \text{6SL}, \text{6R} \}$.

Just as $S_{\ell}^{\ell_1, \ldots, \ell_n}$, the map $R_{\ell}^{\ell_1,
\ldots, \ell_n}$ is symmetric (invariant under the interchanges of indices
$\ell_i$ accompanied by the simultaneous interchange of arguments), and it
vanishes unless $\sum_i | \ell_i | \geqslant l + 2 (n - 1)$.

\subsection{Fixed point equation}\label{sec:beta}

The fixed point equation (FPE) that we will study is
\begin{equation}
  (H_{\ell}') = (H_{\ell}) . \label{FPErestr}
\end{equation}
with $(H_{\ell}')$ given by {\eqref{RHH'}}. If we distinguish the components
$\ell = \text{2L}, \text{4L}, 6$SL from the other couplings, denoted $u = (H_{\ell})_{\ell
\neq \text{2L}, \text{4L}, \text{6SL}} \equiv (u_{\ell})_{\ell \in \{\text{2R}, \text{4R}, \text{6R}, 8, 10,
\ldots\}}$, it reads:
\begin{eqnarray}
  \nu & = & \gamma^{\frac{d}{2} + \eps} \nu + R_{\text{2L}^{}}^{\text{4L}} (\lambda) +
  \sum_{(\ell_i)_1^n \neq (\text{2L}), (\text{4L})}^{} R_{\text{2L}}^{\ell_1, \ldots, \ell_n}
  (H_{\ell_1}, \ldots, H_{\ell_n}), \nonumber\\
  \lambda & = & \gamma^{2 \eps} \lambda + R_{\text{4L}^{}}^{\text{4L}, \text{4L}} (\lambda,
  \lambda) + R_{\text{4L}}^{6 S L} (\mathfrak{X}) + \sum_{(\ell_i)_1^n \neq (\text{4L}),
  (\text{4L}, \text{4L}), (\text{6SL})} R_{\text{4L}}^{\ell_1, \ldots, \ell_n} (H_{\ell_1},
  \ldots, H_{\ell_n}),  \label{betainiz}\\
  \mathfrak{X} (x) & = & R_{\text{6SL}}^{\text{6SL}} (\mathfrak{X}) + R_{6
  \tmop{SL}}^{\text{4L}, \text{4L}} (\lambda, \lambda) = \gamma^{2 d - 6 \Dpsi} 
  [\mathfrak{X}(x \gamma) - 8 \lambda^2 g (x \gamma)], \nonumber\\
  u_{\ell} & = & \sum_{(\ell_i)_1^n}^{} R_{\ell}^{\ell_1, \ldots, \ell_n}
  (H_{\ell_1}, \ldots, H_{\ell_n}), \quad \text{if} \quad \ell \neq \text{2L}, \text{4L},
  \text{6SL}. \nonumber
\end{eqnarray}
We already observed that, given $\lambda$, the FPE for $\mathfrak{X}$ is
solved exactly by $\mathfrak{X}=\mathfrak{X}_{\ast}$, with
$\mathfrak{X}_{\ast}$ as in {\eqref{fixedX}}. Substituting $\mathfrak{X}_{}
(x) =\mathfrak{X}_{\ast} (x)$ in the remaining equations, the variable
$\mathfrak{X}$ is eliminated.\footnote{Note that although in this paper we
take advantage of this possibility, in principle we could have treated
$\mathfrak{X}$ on par with all the other irrelevant couplings. The RG map
would end up contractive also in the $\mathfrak{X}$ direction, and RG
iterations would converge to the same solution
$\mathfrak{X}=\mathfrak{X}_{\ast}$.} Denoting $y = (\nu, \lambda, u)$, we are
left with the fixed point equation $y = R (y)$, or in components:
\begin{eqnarray}
  &  & \nu = \gamma^{\frac{d}{2} + \eps}  (\nu + I_1 \lambda) + e^{(0)}_{\nu}
  (y), \nonumber\\
  &  & \lambda = \gamma^{2 \eps}  (\lambda + I_2 \lambda^2) +
  e^{(0)}_{\lambda} (y),  \label{FPeqs}\\
  &  & u = e_u (y), \nonumber
\end{eqnarray}
with $e^{(0)}_{\nu}, e^{(0)}_{\lambda}, e_u$ defined via the infinite sums in
the right sides of {\eqref{betainiz}}, and $I_1$, $I_2$ defined by
\begin{equation}
  R_{\text{2L}^{}}^{\text{4L}} (\lambda) = \gamma^{\frac{d}{2} + \eps} I_1 \lambda, \qquad
  R_{\text{4L}^{}}^{\text{4L}, \text{4L}} (\lambda, \lambda) + R_{\text{4L}}^{6 S L}
  (\mathfrak{X}_{\ast}) = \gamma^{2 \eps} I_2 \lambda^2 . \label{I1I2def}
\end{equation}
These coefficients $I_1, I_2$ are the same as in Section \ref{sec:intuitive}.
They are given by one-loop Feynman integrals evaluated in Appendix
\ref{sec:I1I2}.

Moving the l.h.s. into the r.h.s. and rescaling, we rewrite the system
{\eqref{FPeqs}} as
\begin{equation}
  f (y) = 0, \qquad f (y) : = \left( \begin{array}{c}
    \nu + a \lambda + e_{\nu} (y)\\
    \varepsilon \lambda + b \lambda^2 + e_{\lambda} (y)\\
    u - e_u (y)
  \end{array} \right) \label{fy=0},
\end{equation}
where
\begin{eqnarray}
  &  & (a, e_{\nu}) = \frac{1}{1 - \gamma^{- d / 2 - \varepsilon}} (I_1,
  \gamma^{- \frac{d}{2} - \eps} e^{(0)}_{\nu}), \nonumber\\
  &  & (b, e_{\lambda}) = \frac{\varepsilon}{1 - \gamma^{- 2 \varepsilon}}
  (I_2, \gamma^{- 2 \varepsilon} e^{(0)}_{\lambda}) .  \label{eresc}
\end{eqnarray}
Eq.{\eqref{FPeqs}} or its equivalent Eq.{\eqref{fy=0}} are the main equations
that we will be solving. Of course, part of the problem is to show that these
equations make sense: that is, we need to prove that the infinite sums
entering the definitions of $e^{(0)}_{\nu}, e^{(0)}_{\lambda}, e_u$ are
convergent. We will actually show that, if $y = (\nu, \lambda, u)$ has bounded
norm {\eqref{defBanach1st}}, say $\|y\|_Y \leqslant 1$, with $\delta$
sufficiently small, then the sums defining $e^{(0)}_{\nu}, e^{(0)}_{\lambda},
e_u$ are {\tmem{absolutely}} convergent and $e_u$ is contractive. This will be
proved in Sections \ref{sec:FP} and \ref{sec:2k+2} below. In preparation to
this, in the next subsection we state the norm bounds satisfied by the
multilinear operators $R_{\ell}^{\ell_1, \ldots, \ell_n}$, which will be
central for our proof of convergence.

\begin{remark}
  \label{generalFPE}From a more general viewpoint, a fixed point is a sequence
  of couplings $(H_{\ell})$ such that $(H_{\ell}') = R (\varepsilon, \gamma)
  [(H_{\ell})]$ given by {\eqref{RHH'}} describes the same
  {\tmem{interaction}} as $(H_{\ell})$. This will be the case if $(H_{\ell}')
  = (H_{\ell})$, as stated in {\eqref{FPErestr}}, or, more generally, if the
  two sequences differ by a null sequence of couplings (see Section
  \ref{sec:interpol}). In this sense, the FPE {\eqref{FPErestr}} discussed
  above is not the most general we could (and should) consider: the general
  FPE to be considered reads $(H_{\ell}') = (H_{\ell}) + (N_{\ell})$, with
  $(N_{\ell})$ a null sequence. In this paper, for simplicity, we focus only
  on the restricted FPE {\eqref{FPErestr}}, and we will show that it has a
  non-trivial, non-null, solution\footnote{The fixed points we will construct
  will have nonzero $\lambda$ and $\nu$, and will therefore be nontrivial.
  \tmtextbf{Lemma.} \tmtextit{Any trimmed coupling sequence with nonzero $\nu$
  and/or $\lambda$ is not null.} Proof is left as an exercise.}, which is
  unique in some neighborhood. The same methods of proof would allow us to
  show that, for each sufficiently small $(N_{\ell})$, the general FPE has a
  unique solution, which differs from the one with $N_{\ell} \equiv 0$ by a
  null sequence $(N_{\ell}')$. In this sense, we expect that there is a unique
  interaction (equivalent class of couplings) solving the general FPE. This
  remains to be shown in full detail, but we prefer not to present this
  additional proof here, in order not to overwhelm the presentation.
\end{remark}

\begin{remark}
  \label{Polchinski}Recall that we are considering renormalization maps $R =
  R_{} (\varepsilon, \gamma)$ with rescaling factor $\gamma \geqslant 2$, in
  particular $\gamma$ is separated from 1. Such RG transformations are called
  ``finite'' or ``discrete''. Eq.{\eqref{betainiz}} thus sets to zero the
  ``beta-functions'' expressing the change of the interaction under a finite
  RG transformation. In theoretical physics, it is more common to take the
  limit $\gamma \rightarrow 1$ and define an ``infinitesimal'' or
  ``continuous'' RG transformation formally given by the derivative $(d / d
  \gamma) R_{\gamma}$ at $\gamma = 1$. At a formal level the continuous RG
  equation (Polchinski's equation {\cite{Polchinski:1983gv}}) has fewer terms
  and looks much simpler than the discrete RG. However, so far it has not been
  possible to take advantage of this formal simplicity in rigorous
  constructions of RG fixed points. The problem is to show that solutions to
  Polchinski's equation have sufficiently good boundedness properties in a
  Banach space of interactions, and it is not known how to do this without
  dealing with the finite RG at the intermediate steps of the argument, which
  brings back the complexity. This problem is open even in fermionic
  theories.\footnote{See {\cite{Brydges1987}} for some global solvability
  results for Polchinski's equation in bosonic theories with bounded
  interactions. Ref. {\cite{Brydges1988}} attempted to prove local solvability
  for fermionic theories but their argument has a gap, see
  {\cite{Wright1999}}. See also an interesting discussion in the conclusions
  of {\cite{Salmhofer2000}}. Ref. {\cite{Disertori:1998qe}} considered
  continuous RG in a fermionic theory, although that construction was not fully
  based on continuous RG: they define the effective action via a
  convergent tree expansion (morally equivalent to using a finite
  RG), then verify that the continuous RG equations hold when applied to this
  effective action. 
}
\end{remark}

\subsection{Norm bounds}\label{sec:NormBounds}

As anticipated in the previous subsection, we now state the norm bounds
satisfied by the multilinear operators $R_{\ell}^{\ell_1, \ldots, \ell_n}$
entering the definitions of $e^{(0)}_{\nu}, e^{(0)}_{\lambda}, e_u$. In the
case $(n ; (\ell_1, \cdots, \ell_n)) = (1 ; \ell)$, in which $R^{\ell}_{\ell}$
is defined as in {\eqref{Rlexpl0}}, we have
\begin{equation}
  \|R_{\ell}^{\ell} (H_{\ell})\|_w \leqslant \left\{\begin{array}{ll}
    \gamma^{- D_2 - 2} \|H_{\text{2R}} \|_w & \text{if } \ell = \text{2R},\\
    \gamma^{- D_4 - 1} \|H_{\text{4R}} \|_w & \text{if $\ell = \text{4R}$},\\
    \gamma^{- D_l} \|H_{\ell} \|_w & \text{if } l = | \ell | \geqslant 6,
  \end{array}\right. \label{Rll}
\end{equation}
while $| R^{\text{2L}}_{\text{2L}} (\nu) | = \gamma^{- D_2} | \nu |$ and $| R^{\text{4L}}_{\text{4L}}
(\lambda) | = \gamma^{- D_4} | \lambda |$. In all the other cases $(n ;
(\ell_1, \cdots, \ell_n)) \neq (1 ; \ell)$, in which $R^{\ell_1, \ldots,
\ell_n}_{\ell}$ is defined as in {\eqref{Rlexpl}}, we have
\begin{equation}
  \|R_{\ell}^{\ell_1, \ldots, \ell_n} (h_1, \ldots, h_n)\|_w \leqslant
  \gamma^{- D_l} \rho_l (h_1, \ldots, h_n), \qquad h_i \in B_{\ell_i}
  \label{Rest},
\end{equation}
\begin{equation}
  \rho_l (h_1, \ldots, h_n) : = \left\{\begin{array}{ll}
    C^{n - 1}_{\gamma} \prod_{i = 1}^n C_0^{| \ell_i |} \|h_i \|_w & \tmop{if}
    \quad \sum_i | \ell_i | \geqslant l + 2 (n - 1)\\
    0 & \tmop{otherwise}
  \end{array}\right. \label{rholbound},
\end{equation}
where, as usual, $l = | \ell |$, and, in {\eqref{rholbound}}, $C_{\gamma}$,
$C_0$ are constants independent of $l, n, \ell_i .$ In addition, $C_0$ does
not depend on $\gamma$, while $C_{\gamma}$ does.

The proof of {\eqref{Rll}} readily follows from the definition of
$R^{\ell}_{\ell}$, see {\eqref{Rlexpl0}}, and from the fact that, using the
definition of $D$, see {\eqref{eq:Resclift}}, and of weighted norm, see
Section \ref{Hlnorm}, we have:
\begin{equation}
  \|DH_{\ell, p} \|_w = \gamma^{- D_l - p} \|H_{\ell, p} \|_{w (\cdummy /
  \gamma)} \leqslant \gamma^{- D_l - p} \|H_{\ell, p} \|_w \hspace{0.17em} .
  \label{eq:normResc}
\end{equation}
Besides proving {\eqref{Rll}}, this justifies the rule stated in Section
\ref{sec:Dil} that the terms with $D_l + p > 0$ are irrelevant. Since $D_l + p
= l [\psi] + p$, this rule turns out the same as for the local interactions
(see footnote \ref{note:rel}).

The proof of {\eqref{Rest}} is more subtle, see the next two subsections,
\ref{sec:boundsSl} and \ref{sec:boundsRl}.

\subsubsection{Bounds for \texorpdfstring{$S_l^{\ell_1, \ldots, \ell_n}$}{S l l1,...,ln}}\label{sec:boundsSl}

Recall that, if $(n ; (\ell_1, \cdots, \ell_n)) \neq (1 ; \ell)$, then
$R^{\ell_1, \ldots, \ell_n}_{\ell}$ is defined in terms of $S^{\ell_1, \ldots,
\ell_n}_l$ via {\eqref{Rlexpl}}. Therefore, in order to prove {\eqref{Rest}},
we first need a bound on $S^{\ell_1, \ldots, \ell_n}_l$. This is similar to
{\eqref{Rlexpl}}, with the important difference that there is no scaling
factor $\gamma^{- D_l}$ in the right side:
\begin{equation}
  \|S_l^{\ell_1, \ldots, \ell_n} (h_1, \ldots, h_n)\|_w \leqslant \rho_l (h_1,
  \ldots, h_n), \qquad h_i \in B_{\ell_i} \label{Sest},
\end{equation}
with the same $\rho_l$ as in {\eqref{rholbound}} (with, possibly, a different
constant $C_0$). For the full proof of {\eqref{Sest}} see Appendix
\ref{sec:Snorm}. Here are the main ideas: from its definition, the map
$S_l^{\ell_1, \ldots, \ell_n}$ is an integral operator whose kernel is the
connected expectation $\text{} \mathcal{C} \left( \bx_{\barbB} \right)$ (more
precisely, it is a sum of $O \left( \tmop{const}^{\sum l_i} \right)$ integral
operators corresponding to different choices of $\bB_i$ and $\bA_i$). The
fermionic connected expectation $\text{} \mathcal{C} \left( \bx_{\barbB}
\right)$ satisfies a crucial bound due to Gawedzki-Kupiainen-Lesniewski
(Appendix \ref{sec:GKL}):
\begin{equation}
  \left| \mathcal{C} \left( \bx_{\barbB} \right) \right| = \left| \left\langle
  \Phi (\barbB_1, \bx_{\barbB_1}) ; \ldots ; \Phi (\barbB_n, \bx_{\barbB_n})
  \right\rangle_c \right| \leqslant C_{\mathrm{}}^s  \sum_{\mathcal{T}_{}}
  \prod_{(x x') \in \text{} \mathcal{T}_{}} M (x_{} - x'), \label{curGKL}
\end{equation}
where the sum is over all \ ``anchored trees $\mathcal{T}$ on $n$ groups of
points $\bx_{\barbB_i}$''. These are graphs which become connected trees when
each group of points $\bx_{\barbB_i}$ is collapsed to a point. There is at
least one anchored tree within each connected Wick contraction, and bounding
each propagator along the anchored tree by {\eqref{gbound0}} we get the
product in {\eqref{curGKL}}. The contribution of remaining $s = \half  (\sum |
\barbB_i | - 2 (n - 1))$ propagators is bounded by $C^s$. This explains the
general structure of \ {\eqref{curGKL}}, but the full proof is rather more
subtle. The sum of connected graphs defining the connected expectation has to
be rewritten as a sum over anchored trees without double counting. For each
anchored tree, we then have to sum over the remaining propagator choices, and
this whole sum with factorially many terms has to be bounded by $C^s$. This
turns out possible due to fermionic cancelations.

The number of anchored trees is $\leqslant n! 4^{\sum \left| \barbB_i
\right|}$ (Appendix \ref{sec:anchored-trees}), which by the way is much
smaller than the total number of connected graphs. This $n!$ cancels with $1 /
n!$ in {\eqref{eq:Hefflift}}, leaving only exponential factors. When
evaluating the weighted norm, the product of $M$'s in {\eqref{curGKL}} gives
the factor $C_{\gamma}^{n - 1}$ with \ $C_{\gamma} = \| M \|_w = O (\gamma^d)$
by {\eqref{Mw}}. This finishes our brief exposition of {\eqref{Sest}}; see
Appendix \ref{sec:Snorm} for the details.

By {\eqref{Sest}}, the multilinear map $S_l^{\ell_1, \ldots, \ell_n}$ is
continuous. The homogeneous map $\mathcal{S}_l^{\ell_1, \ldots, \ell_n}$
related to \ $S_l^{\ell_1, \ldots, \ell_n}$ by identifying some arguments, Eq.
{\eqref{Sldef}}, is also continuous.

We will also need Frechet derivatives of these maps.\footnote{Recall that
Frechet derivative is a generalization of ordinary derivative to Banach
spaces. In general, for a map $f (x)$ from a Banach space $Z$ to another space
$Z'$, its Frechet derivative at a point $x$ is defined as a linear operator
$\nabla f (x) \in \mathcal{L} (Z, Z')$ having the property that
\[ \lim_{\| \delta x\|_Z \rightarrow 0}  \frac{\|f (x + \delta x) - f (x) -
   \nabla f (x) \delta x\|_{Z'}}{\| \delta x\|_Z} = 0. \]
In some of our cases of interest, one of the two spaces $Z$ or $Z'$ may be
$\mathbb{R}$. When $Z =\mathbb{R}$ we have $\nabla f (x) \in Z'$, and when $Z'
=\mathbb{R}$ we have $\nabla f (x) \in \mathcal{L} (Z, \mathbb{R})$, i.e. a
linear functional on $Z$.} Since $S_l^{\ell_1, \ldots, \ell_n}$ is a
multilinear map, it is Frechet-differentiable and its Frechet derivative in
each argument coincides with the map itself. The homogeneous map
$\mathcal{S}_l^{\ell_1, \ldots, \ell_n}$ is also Frechet-differentiable. The
derivative $\nabla_{} \mathcal{S}_l^{\ell_1, \ldots, \ell_n} (H)$ is, for a
fixed $H$, a linear operator from $B_{\tmop{trim}}$ to $B_l$. Using Eq.
{\eqref{Sldef}}, the value of this operator on $\delta H_{} \in
B_{\tmop{trim}}$ is:
\begin{equation}
  [\nabla_{H_{}} \mathcal{S}_l^{\ell_1, \ldots, \ell_n} (H)] \delta H_{} =
  \sum_{i = 1}^n S_l^{\ell_1, \ldots, \ell_n} (H_{\ell_1}, \ldots, \delta
  H_{\ell_i}, \ldots, H_{\ell_n}) .
\end{equation}
Estimating each term in the r.h.s. via {\eqref{Sest}} we get a bound:
\begin{equation}
  \| [\nabla_{H_{}} \mathcal{S}_l^{\ell_1, \ldots, \ell_n} (H)] \delta H_{}
  \|_w \leqslant \sum_{i = 1}^n \rho_l (H_{\ell_1}, \ldots, \delta H_{\ell_i},
  \ldots, H_{\ell_n}) . \label{Sderbound}
\end{equation}

\subsubsection{Bounds for \texorpdfstring{$R_{\ell}^{\ell_1, \ldots,\ell_n}$}{R l l1,...,ln}}\label{sec:boundsRl}

From the definition {\eqref{Rlexpl}}, we have that $R_{\ell}^{\ell_1, \ldots,
\ell_n}$ is related to $S^{\ell_1, \ldots, \ell_n}_l$ via the dilatation
operator, which we already bounded in {\eqref{eq:normResc}}, and via the
trimming operator $T$ introduced in section \ref{sec:trimming}, whose
components we still need to bound. The easiest components to bound are the
localization maps $T^{2, 0}_{\text{2L}}$ and $T^{4, 0}_{\text{4L}}$, which map kernels to
local kernels and do not increase the norm (see {\eqref{TLfinal}}):
\begin{equation}
  \|(H_{\tmop{eff}})_{\text{2L}} \|_w \leqslant \|(H_{\tmop{eff}})_{2, 0} \|_w,
  \qquad \|(H_{\tmop{eff}})_{\text{4L}} \|_w \leqslant \|(H_{\tmop{eff}})_{4, 0}
  \|_w . \label{TLmain}
\end{equation}
In turn, the interpolation maps satisfy the bounds (see Appendix
\ref{sec:Trim}):
\begin{eqnarray}
  &  & \|T^{4, 0}_{\text{4R}} (H_{\tmop{eff}})_{4, 0} \|_{w (\cdummy /
  \gamma)} \leqslant C_{\text{R}} \gamma \|(H_{\tmop{eff}})_{4, 0} \|_w,
  \nonumber\\
  &  & \|T^{2, 1}_{\text{2R}} (H_{\tmop{eff}})_{2, 1} \|_{w (\cdummy / \gamma)}
  \leqslant C_{\text{R}} \gamma \|(H_{\tmop{eff}})_{2, 1} \|_w,  \label{TRfinal}\\
  &  & \|T^{2, 0}_{\text{2R}} (H_{\tmop{eff}})_{2, 0} \|_{w (\cdummy / \gamma)}
  \leqslant C_{\text{R}} \gamma^2 \|(H_{\tmop{eff}})_{2, 0} \|_w, \nonumber
\end{eqnarray}
where $C_{\text{R}}$ depends on $C_w$ and $\sigma$ in {\eqref{eq:ourW}} but not on
$\gamma$, and we use the fact that $\gamma \geqslant 2$.

Putting together {\eqref{eq:normResc}}, {\eqref{Sest}}, {\eqref{TLmain}} and
{\eqref{TRfinal}}, we readily obtain {\eqref{Rest}}. In fact, for $\ell
\geqslant 8$ and $\ell \in \left\{ \text{6R}, \text{6SL} \right\}$,
{\eqref{Rest}} is a consequence of {\eqref{Sest}}, and {\eqref{eq:normResc}}
with the worst possible $p = 0$. For $\ell \in \{ \text{2L}, \text{4L} \}$ we also need to
use {\eqref{TLmain}}. Finally, for $\ell \in \{ \text{2R}, \text{4R} \}$ we additionally
have to use {\eqref{TRfinal}} and rely on the first equality in
{\eqref{eq:normResc}}. A power of $\gamma$ that we lose in the r.h.s. of
{\eqref{TRfinal}} is compensated during dilatation, due to the presence of
derivatives in the couplings $H_{\text{2R}}$, $H_{\text{4R}} .$ Because of the
sums in the r.h.s. of {\eqref{def4R}} and {\eqref{def2L}}, we get Eq.
{\eqref{Rest}} with an extra factor of $\gamma^{- 2} + \gamma^{- 1} C_{\text{R}} + C_{\text{R}}
\leqslant 1 + 2 C_{\text{R}}$ for $\ell = 2 \text{R}$ and $\gamma^{- 1} + C_{\text{R}} \leqslant
1 + C_{\text{R}}$ for $\ell = 4 \text{R}$. We absorb this factor by increasing the
constant $C_0$ in the definition of the function $\rho_l$ in the right side of
{\eqref{Rest}}\footnote{\label{CRabsorb}Since $\sum | \ell_i | \geqslant 4$ in
any of these cases, it's enough to increase $C_0 \rightarrow C_0 (1 + 2
C_{\text{R}})^{1 / 4}$.}.

We will also need Frechet derivatives of the homogeneous map
$\mathcal{R}_{\ell}^{\ell_1, \ldots, \ell_n}$. Since
$\mathcal{R}_{\ell}^{\ell_{}} = R_{\ell}^{\ell}$ is a linear map, its Frechet
derivative coincides with it and satisfies the same bound {\eqref{Rll}}. In
all the other cases $(n ; (\ell_1, \cdots, \ell_n)) \neq (1 ; \ell)$ we have
the bound
\begin{equation}
  \| [\nabla_{H_{}} \mathcal{R}_{\ell}^{\ell_1, \ldots, \ell_n} (H)] \delta
  H_{} \|_w \leqslant \gamma^{- D_l} \sum_{i = 1}^n \rho_l (H_{\ell_1},
  \ldots, \delta H_{\ell_i}, \ldots, H_{\ell_n}), \label{Rderbound}
\end{equation}
which follows from {\eqref{Sderbound}} just as {\eqref{Rest}} followed from
{\eqref{Sest}}.

\section{Construction of the fixed point}\label{sec:FP}

In this and the following section, we finally construct a solution of the FPE
$f (y) = 0$, see {\eqref{fy=0}}, and discuss its uniqueness and regularity
properties. The presentation is organized as follows: in Section
\ref{sec:key}, we state the main bound on the components of $f (y)$, whose
proof (which is one of the main technical contributions of this paper, and
uses in a crucial way the bounds stated in Section \ref{sec:NormBounds}) is
postponed to Section \ref{sec:2k+2}. Given the bounds of Section
\ref{sec:key}, existence and uniqueness at fixed $\gamma$ of the fixed point
follow by a rather general and straightforward argument, discussed in Sections
\ref{abstract} and \ref{FPth}. The independence of the fixed point from
$\gamma$ and its analyticity in $\varepsilon$ are simple but remarkable
corollaries of our construction, discussed in Sections \ref{semigroup} and
\ref{analyticity}, respectively.

\subsection{Key lemma}\label{sec:key}

In this subsection we formulate, as promised, the estimates for the functions
$e_{\nu}^{(0)}, e_{\lambda}^{(0)}, e_u$ entering the definition of $f (y)$,
see {\eqref{fy=0}}-{\eqref{eresc}}. We will assume that $\gamma$ is large
enough and that the norm of $y$, see {\eqref{defBanach1st}}, is bounded, say
smaller than $1$; the constants $A_0, A_0^R, A_1^R, A_2^R, A$ in
{\eqref{defBanach1st}} will be fixed in a suitable, $\gamma$-dependent, way,
and the parameter $\delta$ will be chosen sufficiently small (in a
$\gamma$-dependent way). The smallness of $\delta$ is conceptually independent
from any stringent requirement on the physical parameter $\varepsilon$: the
only needed condition on $\varepsilon$ will be that all $u_{\ell}$ directions
are irrelevant, as guaranteed by Eqs.{\eqref{epsCond}}, {\eqref{Dbarpos}}. The
conditions that $\lambda$ is weakly relevant ($\varepsilon$ small), and that
its one-loop beta-function does not vanish ($N \neq 8$) won't be used here. To
emphasize that for the moment the smallness of $\varepsilon$ is not used, here
we assume that $\delta$ is independent of $\varepsilon$. Eventually, the
smallness of $\varepsilon$ will come back into play in the full contraction
argument involving all couplings $\nu, \lambda, u_{\ell}$ (Sections
\ref{abstract} and \ref{FPth}): there, $\delta$ will be identified with
$\varepsilon$ up to a constant factor, but here it is logically convenient to
keep them separate.

Given $\gamma$-dependent constants $A_0 = A_0 (\gamma), A^{\text{R}}_0 = A^{\text{R}}_0
(\gamma), A^{\text{R}}_1= A^{\text{R}}_1 (\gamma), A^{\text{R}}_2 = A^{\text{R}}_2 (\gamma), A = A (\gamma)$, we
denote by $\|u\|_{B (\gamma, \delta)}$ the following norm of a vector $u =
(H_{\ell})_{\ell \neq \text{2L}, \text{4L}, \text{6SL}} \equiv (u_{\ell})_{\ell \in \{\text{2R}, \text{4R},
\text{6R}, 8, 10, \ldots\}}$ of irrelevant components (6SL excluded):
\begin{equation}
  \|u\|_{B (\gamma, \delta)} = \max \left\{ \frac{\|u_{\text{2R}} \|_w}{A_0^{\rest}
  (\gamma) \delta^2}, \frac{\|u_{\text{4R}} \|_w}{A_1^{\rest} (\gamma) \delta^2},
  \frac{\|u_{\text{6R}} \|_w}{A_2^{\rest}  (\gamma) \delta^3}, \sup_{\ell \ge 8} 
  \frac{\|u_{\ell} \|_w}{A (\gamma) \delta^{k (\ell)}} \right\},
  \label{defnormB}
\end{equation}
where $k (\ell) = \frac{| \ell |}{2} - 1$, in terms of which the norm
{\eqref{defBanach1st}} of $y = (\nu, \lambda, u)$ can be rewritten
\begin{equation}
  \|y\|_{Y (\gamma, \delta)} = \max \left\{ \frac{| \nu |}{A_0 (\gamma)
  \delta}, \frac{| \lambda |}{A_0 (\gamma) \delta}, \|u\|_{B (\gamma, \delta)}
  \right\} . \label{normYB}
\end{equation}
Note that, compared with {\eqref{defBanach1st}}, the symbol $Y$ in
{\eqref{normYB}} has an explicit dependence upon $\gamma$ and $\delta$; the
dependence on $\delta$ is obvious, the one on $\gamma$ is meant to emphasize
the $\gamma$-dependence of the constants $A, A_0$, etc. and of the weight $w$
(see {\eqref{eq:ourW}}). We are now ready to state the main result of this
section.

\begin{lemma}[Key lemma]
  \label{Key}Choose $d \in \{ 1, 2, 3 \}$, cutoff $\chi$, $N \geqslant 4$, and
  an $\varepsilon$ satisfying {\eqref{epsCond}}. There exists
  $\gamma_{\tmop{key}} \geqslant 2$ and 
  \begin{equation}
    \delta_0 (\gamma), A_0 (\gamma), \{A^{\rm{R}}_k (\gamma) \}_{k = 0, 1, 2}, A
    (\gamma), E_0 (\gamma), E_1 (\gamma), \label{eq:tofix}
  \end{equation}
  positive continuous functions on $\gamma \geqslant \gamma_{\tmop{key}}$
  [whose dependence on $\gamma$ is omitted in
  Eqs.{\eqref{eq:31}}-{\eqref{eq:36}} below], with the following property.
  Take any $\gamma \geqslant \gamma_{\tmop{key}}$, any $0 < \delta \leqslant
  \delta_0 (\gamma)$, and any sequence $y = (\nu, \lambda, u)$ satisfying
  \begin{equation}
    \|y\|_{Y (\gamma, \delta)} \leqslant 1. \label{keyass}
  \end{equation}
  Then the infinite sums defining the functions
$e_{\nu}^{(0)}, e_{\lambda}^{(0)}, e_u$ in the right
  side of {\eqref{FPeqs}}, see {\eqref{betainiz}} and following lines, are
  absolutely convergent, and their sums satisfy:
  \begin{equation}
    |e_{\nu}^{(0)} (y) | \leqslant E_0 \delta^2, \qquad |e_{\lambda}^{(0)} (y)
    | \leqslant E_1 \delta^3, \qquad \|e_u (y) \|_{B (\gamma, \delta)}
    \leqslant \gamma^{- \bar{D}}, \label{eq:31}
  \end{equation}
  where $\bar{D} = \frac{1}{2} \min \{D_2 + 2, D_4 + 1, D_6 \}$. In addition,
  \begin{eqnarray}
    &  & | \partial_i e_{\nu}^{(0)} (y) | \leqslant E_0 \delta^2 / (A_0
    \delta), \quad | \partial_i e_{\lambda}^{(0)} (y) | \leqslant E_1 \delta^3
    / (A_0 \delta), \quad \| \partial_i e_u (y) \|_B \leqslant \gamma^{-
    \bar{D}} / (A_0 \delta) \quad (i = \nu, \lambda), \nonumber\\
    &  & \| \partial_u e_{\nu}^{(0)} (y) \|_{\mathcal{L} (B,
    \mathbb{R})} \leqslant E_0 \delta^2, \qquad \| \partial_u
    e_{\lambda}^{(0)} (y) \|_{\mathcal{L} (B, \mathbb{R})}
    \leqslant E_1 \delta^3, \qquad \| \partial_u e_u (y) \|_{\mathcal{L} (B,
    B)} \leqslant \gamma^{- \bar{D}},  \label{eq:36}
  \end{eqnarray}
  where $B = B (\gamma, \delta)$, $\mathcal{L} (B, \mathbb{R})$ is the space
  of linear operators from $B$ to $\mathbb{R}$, and similarly for $\mathcal{L}
  (B, B) .$
\end{lemma}

We wrote {\eqref{eq:36}} in the form which makes apparent that the
$u$-derivatives satisfy the same bounds as the functions themselves, while the
bounds for $\nu, \lambda$-derivatives are worse by $1 / (A_0 \delta)$ factor.
This pattern is natural in view of the assumptions $| \nu | \leqslant A_0
\delta$, $| \lambda | \leqslant A_0 \delta$ and $\|u\|_B \leqslant 1$ (which
are the same as $\|y\|_Y \leqslant 1$). Before presenting the proof of the Key
lemma, which is postponed to Section \ref{sec:2k+2}, we will show that its
bounds straightforwardly imply that the FPE $f (y) = 0$ has a unique solution
in a suitable neighborhood of the Banach space $Y$; see the next two
subsections, \ref{abstract} and \ref{FPth}.

\begin{remark}
  The third inequality in {\eqref{eq:31}} means that the RG map restricted to
  the irrelevant directions $\ell = \text{2R}, \text{4R}, \text{6R}, 8, 10, \ldots_{}$ is
  contractive, as it is natural to expect. Contractivity along the directions
  with $| \ell | \geqslant 6$ is ``easy'' to establish: it follows
  straightforwardly from the bounds {\eqref{Rll}} and {\eqref{Rest}}; note, in
  fact, that $\gamma^{- D_l} \leqslant \gamma^{- \bar{D}} < 1$ for $l = | \ell
  | \geqslant 6$. See Sections \ref{sec:l8} and \ref{sec:l=6R} for the full
  proof. On the other hand, contractivity along the directions $\text{2R}$ and $\text{4R}$
  is more subtle to prove, due to the factor $\gamma^{- D_l}$, which is larger
  than 1 for $l = 2, 4$, in the right side of {\eqref{Rest}}. In these cases,
  we take advantage of the fact that the linearization of the RG map,
  {\eqref{Rll}}, has the good factor $\gamma^{- D_2 - 2}$ and $\gamma^{- D_4 -
  1}$ in the directions $\ell = \text{2R}$ and $\text{4R}$, respectively. On the other
  hand, the nonlinear contributions bounded in {\eqref{Rest}} are small
  because they are of higher order: loosely speaking, the higher order can be
  used to compensate the additional bad factor $\gamma^2$ or $\gamma$, which
  ultimately originates from the bounds {\eqref{TRfinal}}. More technically,
  here is where we use the freedom in the choice of the constants $A, A_0,
  A^{\text{R}}_{^{} 0}$ etc., entering the definition the norm {\eqref{defBanach1st}}
  defining the Banach space: by carefully playing with these
  $\gamma$-dependent constants, we can reabsorb the bad factors $\gamma^2$ or
  $\gamma$ into their definitions, see Sections \ref{sec:l=4R}, \ref{sec:l=2R}
  and \ref{sec:choices} for the technical details.
\end{remark}

\begin{remark}
  In connection with the end of previous remark, we note that the use of a
  norm involving several constants $A, A_0, A^{\text{R}}_{^{} 0}$ etc, rather than a
  single one, is one original aspect of our proof, and it is the key
  ingredient allowing us to choose an optimal powers of $\delta$ in
  {\eqref{defBanach1st}} (recall that eventually $\delta$ will be chosen
  proportional to $\varepsilon$, and that the $\delta$-exponents $2, 2, 3, k
  (\ell) = | \ell | / 2 - 1$ in the right side of {\eqref{defBanach1st}} are
  dictated by the lowest order contribtions to $u_{\ell}$ in perturbation
  theory and cannot be improved; see the discussion at the beginning of
  Section \ref{sec:normtrimmedseq}). \ If we tried to repeat the proof of Key
  lemma with a simplified norm with $A = A_0 = A^{\text{R}}_0 = \cdots$ we would not
  succeed in proving the analogues of {\eqref{eq:31}} and {\eqref{eq:36}}. One
  can however use a simplified norm, and a simplified proof of Key lemma, if
  one changes the optimal powers of $\delta$ to sub-optimal ones, strictly
  smaller than $2, 2, 3, k (\ell)$. This was the strategy followed in
  {\cite{Gawedzki:1985jn}} (see the non-optimal powers in their Eq.(2.17)).
  Naively, this strategy leads to an estimate on the fixed-point couplings
  (analogue of Corollary \ref{FPaction} below) with sub-optimal powers.
  However, armed with our analyticity argument from Section \ref{analyticity},
  this limitation can be overcome. Namely, once the fixed point existence is
  proven by working in the sub-optimal Banach space, the argument from Section
  \ref{analyticity} still works and shows that it is analytic in a disk around
  $\varepsilon = 0$. From analyticity, we could then recover the optimal
  estimates on the fixed-point couplings. Although such a mixed real/complex
  strategy is possible, here we prefer to keep these two lines of development
  independent. So, we work with the optimal powers from the start and obtain
  the optimal estimates with purely real methods (even though it leads to some
  mild complications in the proof of Key Lemma).
\end{remark}

\subsection{Abstract analysis}\label{abstract}

Recall that we are solving $f (y) = 0$ with $y = (\nu, \lambda, u)$ and $f$
given in {\eqref{fy=0}}. In this subsection we consider $u$ as a vector living
in an abstract Banach space $B$ endowed with some norm $\| u \|_B$. This norm
will be used to state conditions on the maps $e_j$ guaranteeing the existence
and uniqueness of a solution in some neighborhood of $y_0$, see Eq.
{\eqref{y0}}. In the next subsection these conditions will be verified with
the help of Key Lemma, identifying the norm $\| u \|_B$ with
{\eqref{defnormB}}.

Concerning the rescalings {\eqref{eresc}}, note that (the $O$ symbols here and
in {\eqref{ab}} have $\gamma$- and $\varepsilon$-independent constants):
\begin{equation}
  \frac{\varepsilon}{1 - \gamma^{- 2 \varepsilon}} = (2 \log \gamma)^{- 1} (1
  + O (\varepsilon \log \gamma)) . \label{logg}
\end{equation}
By the small $\varepsilon$ asymptotics of $I_1, I_2$ from Lemma \ref{lemI1I2}
in Appendix \ref{sec:I1I2}, we have
\begin{eqnarray}
  a & = & 2 (N - 2) \left[ \int \frac{d^d k}{(2 \pi)^d} \frac{\chi (k)}{|k|^{d
  / 2}} +  O (\varepsilon \log \gamma ) \right],
  \nonumber\\
  b & = & - 2 (N - 8)  \left[ \frac{S_d}{(2 \pi)^d} + O (\varepsilon \log
  \gamma) \right] .  \label{ab}
\end{eqnarray}
As mentioned in the introduction, we are assuming $N \neq 8$ so that $b \neq
0$. We will also assume $\varepsilon \leqslant c / \log \gamma$ where $c$ is a
small $\gamma$-independent constant. Under these conditions $a, b, b^{- 1} = O
(1)$. In particular $b \neq 0$.

Setting $e_j$ ($j = \nu, \lambda, u$) to zero in {\eqref{fy=0}}, we get an
``approximate equation''
\begin{equation}
  f_0 (y) = 0, \quad f_0 (y) = \left( \begin{array}{c}
    \nu + a \lambda\\
    \varepsilon \lambda + b \lambda^2\\
    u
  \end{array} \right), \label{f0}
\end{equation}
which has a nontrivial solution
\begin{equation}
  y_0 = (\nu_0, \lambda_0, u_0) = \left( \frac{a}{b} \varepsilon, -
  \frac{1}{b} \varepsilon, 0 \right) \label{y0} .
\end{equation}
Our goal will be to show that the full equation $f (y) = 0$ has a solution of
the form $y_0 + O (\varepsilon^2)$. Aiming to apply a contraction argument, we
rewrite equation $f (y) = 0$ one last time as a fixed point equation for a map
$F (y)$. We choose the following rewrite:
\begin{equation}
  f (y) = 0 \quad \Longleftrightarrow \quad y = F (y), \qquad F (y) = y - G^{-
  1} f (y), \label{rewrite}
\end{equation}
with $G$ an arbitrary invertible linear operator. We would like to choose $G$
so that $F (y)$ is a contraction in a small neighborhood of $y_0$. Recall that
Newton's method for solving nonlinear equations would correspond to $G =
\nabla f (y)$. We do not want to deal with the full gradient of the
complicated map $f (y)$, and we will instead choose $G = \nabla f_0 (y_0)$,
cf. {\eqref{f0}}. This ``approximated gradient'' will be sufficient to make $F
(y)$ a contraction. We have
\begin{equation}
  G = \left( \begin{array}{ccc}
    1 & a & 0\\
    0 & - \varepsilon & 0\\
    0 & 0 & \mathbbm{1}
  \end{array} \right), \qquad G^{- 1} = \left( \begin{array}{ccc}
    1 & a \varepsilon^{- 1} & 0\\
    0 & - \varepsilon^{- 1} & 0\\
    0 & 0 & \mathbbm{1}
  \end{array} \right) .
\end{equation}
With this choice, the map $F (y)$ takes the form:
\begin{equation}
  F (y) \equiv \left( \begin{array}{c}
    F^{\nu} (y)\\
    F^{\lambda} (y)\\
    F^u (y)
  \end{array} \right) = \left( \begin{array}{c}
    - 2 a \lambda - ab \frac{\lambda^2}{\varepsilon} - e_{\nu} - a
    \frac{e_{\lambda}}{\varepsilon}\\
    2 \lambda + b \frac{\lambda^2}{\varepsilon} +
    \frac{e_{\lambda}}{\varepsilon}\\
    e_u
  \end{array} \right) . \label{Fy}
\end{equation}
\begin{remark}
  The reader may be puzzled: why introduce the new map $F$ rather than use in
  its place the renormalization map itself, given that Eq. {\eqref{FPeqs}}
  already has the fixed point form $y = R (y)$? The reason is that contraction
  argument cannot be applied directly to $R$, since it is not fully
  contracting: it is contracting near the fixed point along all directions
  except $\nu$. Note as well that $R$ is only ``barely contracting'' in
  direction $\lambda$: its linearization around $y_0$ has the corresponding
  eigenvalue equal to $2 - \gamma^{2 \varepsilon} = 1 - 2 \varepsilon \log
  \gamma + \cdots$, smaller than 1 but only by $O (\varepsilon)$. This \
  ``barely contracting'' direction is the reason why we apply the contraction
  argument in a neighborhood of size $\varepsilon^2$ of $y_0$ (outside of
  which even $F$ would not be contracting).
\end{remark}

We will aim to apply a contraction argument to $F (y)$ in a neighborhood $Y_0$
of $y_0$ defined as
\begin{equation}
  Y_0 = \{ y : | \nu - \nu_0 | \leqslant M_0 \varepsilon^2, | \lambda -
  \lambda_0 | \leqslant M_0 \varepsilon^2, \| u \|_B \leqslant 1 \},
  \label{Y0}
\end{equation}
whose size depends on $\varepsilon$ and on an additional parameter $M_0$.
First of all let us arrange that $F$ maps $Y_0$ to itself. Writing $\lambda =
\lambda_0 + \delta \lambda$, we express $F (y)$ as
\begin{equation}
  F (y) = \left( \begin{array}{c}
    \nu_0 - \frac{a b}{\varepsilon} (\delta \lambda)^2 - e_{\nu} - a
    \frac{e_{\lambda}}{\varepsilon}\\
    \lambda_0 + \frac{b}{\varepsilon} (\delta \lambda)^2 +
    \frac{e_{\lambda}}{\varepsilon}\\
    e_u
  \end{array} \right) . \label{Fdefalt}
\end{equation}
We see that $F (Y_0) \subset Y_0$ provided that for any $y \in Y_0$
\begin{equation}
  \text{} K_1 \max (M_0^2 \varepsilon^3, | e_{\nu} (y_{}) |, \varepsilon^{- 1}
  | e_{\lambda} (y_{}) |) \leqslant M_0 \varepsilon^2, \qquad \| e_u (y_{})
  \|_B \leqslant 1, \label{someconds}
\end{equation}
where
\begin{equation}
  K_1 = K_1 (a, b) = \max (1 + | a |^{} + | a b |, 1 + | b |^{}) .
  \label{Kdef}
\end{equation}
Then {\eqref{someconds}} are satisfied as long as $K_1 M_0 \varepsilon
\leqslant 1$ and provided that
\begin{equation}
  | e_{\nu} (y) | \leqslant \frac{M_0}{K_1} \varepsilon^2, \qquad |
  e_{\lambda} (y) | \leqslant \frac{M_0}{K_1} \varepsilon^3, \qquad \| e_u (y)
  \|_B \leqslant 1 \qquad (y \in Y_0) . \label{contained}
\end{equation}
We next proceed to arrange that $F$ is a contraction in $Y_0$. For this we
need to specify a Banach space norm on $y = (\nu, \lambda, u)$. We will use
the norm
\begin{equation}
  \|y\|_Y = \max \{ \tilde{\varepsilon}^{- 1} | \nu |, \tilde{\varepsilon}^{-
  1} | \lambda |, \|u\|_B \}, \label{yY}
\end{equation}
depending on a parameter $\tilde{\varepsilon}$. Since in our application $\nu$
and $\lambda$ are $O (\varepsilon)$, while $\|u\|_B$ will be $O (1)$ when
identified with {\eqref{defnormB}}, the natural value for
$\tilde{\varepsilon}$ is order $\varepsilon$ so that all terms in {\eqref{yY}}
have the same order. Eventually in Section \ref{FPth} we will fix
$\tilde{\varepsilon} = A_0 \delta$ so that this norm will coincide with
{\eqref{normYB}}. However in this section let us keep the ratio
$\tilde{\varepsilon} / \varepsilon$ as a free parameter.

We will next study the gradient $\nabla_{} F$ and arrange that its operator
norm is less than 1. Here the gradient $\nabla F$ is the Frechet derivative
which was already discussed in Section \ref{sec:key}.

From {\eqref{Fy}}, we compute the gradient $\nabla F$ in components:
\begin{equation}
  \frac{\partial_{} (F^{\nu}, F^{\lambda}, F^u)}{\partial (\nu, \lambda, u)} =
  \left( \begin{array}{ccc}
    - \partial_{\nu} e_{\nu} - a \frac{\partial_{\nu}
    e_{\lambda}}{\varepsilon} & - \frac{2 a b}{\varepsilon}  (\lambda -
    \lambda_0) - \partial_{\lambda} e_{\nu} - a \frac{\partial_{\lambda}
    e_{\lambda}}{\varepsilon} & - \partial_u e_{\nu} - a \frac{\partial_u
    e_{\lambda}}{\varepsilon}\\
    \frac{\partial_{\nu} e_{\lambda}}{\varepsilon} & \frac{2 b}{\varepsilon} 
    (\lambda - \lambda_0) + \frac{\partial_{\lambda} e_{\lambda}}{\varepsilon}
    & \frac{\partial_u e_{\lambda}}{\varepsilon}\\
    \partial_{\nu} e_u & \partial_{\lambda} e_u & \partial_u e_u
  \end{array} \right) . \label{gradF1}
\end{equation}
Various partial derivatives of maps $e_j (\nu, \lambda, u)$ are understood as
Frechet derivatives, sometimes with $Z$ or $Z'$ being equal to $\mathbb{R}$.
E.g. $\partial_{\nu} e_u$ is, just like $e_u$, a $B$-valued function on $Y$.
On the other hand $\partial_u e_{\lambda} \in \mathcal{L} (B, \mathbb{R})$, a
linear functional on $B$.

We next proceed to study the norm of $\nabla F (y) \in \mathcal{L} (Y, Y)$
where $y \in Y_0 .$ Let $\delta y \in Y$, $\| \delta y \|_Y \leqslant 1$ which
means
\begin{equation}
  | \delta \nu | \leqslant \tilde{\varepsilon}^{}, \quad | \delta \lambda |
  \leqslant \tilde{\varepsilon}^{}, \quad \| \delta u \|_B \leqslant 1 .
\end{equation}
We have
\begin{equation}
  \nabla F (y) \delta y = \left(\begin{array}{c}
    \partial_{\nu} F^{\nu} \delta \nu + \partial_{\lambda} F^{\nu} \delta
    \lambda + \partial_u F^{\nu} \delta u\\
    \partial_{\nu} F^{\lambda} \delta \nu + \partial_{\lambda} F^{\lambda}
    \delta \lambda + \partial_u F^{\lambda} \delta u\\
    \partial_{\nu} F^u \delta \nu + \partial_{\lambda} F^u \delta \lambda +
    \partial_u F^u \delta u
  \end{array}\right), \label{gradF2}
\end{equation}
where all partial derivatives in the r.h.s. are evaluated at $y$. This implies
\begin{align}
  \| \nabla F (y) \|_{\mathcal{L} (Y, Y)} = \sup_{\| \delta y\|_Y \leqslant 1}
  \| \nabla F (y) (\delta y) \|_Y & = \sup_{\| \delta y\|_Y \leqslant 1}
  \max \left(\begin{array}{c}
    \tilde{\varepsilon}^{- 1} | \partial_{\nu} F^{\nu} \delta \nu +
    \partial_{\lambda} F^{\nu} \delta \lambda + \partial_u F^{\nu} \delta u
    |\\
    \tilde{\varepsilon}^{- 1} | \partial_{\nu} F^{\lambda} \delta \nu +
    \partial_{\lambda} F^{\lambda} \delta \lambda + \partial_u F^{\lambda}
    \delta u |\\
    \| \partial_{\nu} F^u \delta \nu + \partial_{\lambda} F^u \delta \lambda +
    \partial_u F^u \delta u \|_B
  \end{array}\right) \nonumber\\
  & \leqslant \max \left(\begin{array}{c}
    | \partial_{\nu} F^{\nu} | + | \partial_{\lambda} F^{\nu} | +
    \tilde{\varepsilon}^{- 1} \| \partial_u F^{\nu} \|_{\mathcal{L} (B,
    \mathbb{R})}\\
    | \partial_{\nu} F^{\lambda} | + | \partial_{\lambda} F^{\lambda} | +
    \tilde{\varepsilon}^{- 1} \| \partial_u F^{\lambda} \|_{\mathcal{L} (B,
    \mathbb{R})}\\
    \tilde{\varepsilon} \| \partial_{\nu} F^u \|_B + \tilde{\varepsilon} \|
    \partial_{\lambda} F^u \|_B + \| \partial_u F^u_{} \|_{\mathcal{L} (B, B)}
  \end{array}\right) 
\end{align}
Finally using the explicit form of $\nabla F$ components we get that for $y
\in Y_0$
\begin{align}
  \| \nabla F (y) \|_{\mathcal{L} (Y, Y)} \leqslant K_2 &\max \Bigl\{ 
  | \partial_{\nu} e_{\nu} |, \frac{| \partial_{\nu} e_{\lambda}
  |}{\varepsilon}, M_0 \varepsilon, | \partial_{\lambda} e_{\nu} |, \frac{|
  \partial_{\lambda} e_{\lambda} |}{\varepsilon}, \frac{\| \partial_u e_{\nu}
  \|_{\mathcal{L} (B, \mathbb{R})^{^{}}}}{\tilde{\varepsilon}}, \frac{\|
  \partial_u e_{\lambda} \|_{\mathcal{L} (B,
  \mathbb{R})}}{\varepsilon^{} \tilde{\varepsilon}}, \nonumber\\
  &\hspace{1.2cm}  \tilde{\varepsilon} \| \partial_{\nu} e_u \|_B, \tilde{\varepsilon} \|
  \partial_{\lambda} e_u \|_B, \| \partial_u e_u \|_{\mathcal{L} (B, B)} \Bigr\}
  ,  \label{gradFbound}
\end{align}
where we used that $| \lambda - \lambda_0 | \leqslant M_0 \varepsilon^2$ in
$Y_0 $ and defined a constant
\begin{equation}
  K_2 (a, b) = \max (3 + 3 | a | + 2 | a b |, 3 + 2 | b |) \label{Cgrad} .
\end{equation}
We will demand that the following conditions hold uniformly for $y \in Y_0$:
\begin{eqnarray}
  &  & | \partial_i e_{\nu} | \leqslant M_0 \varepsilon, \qquad | \partial_i
  e_{\lambda} | \leqslant M_0 \varepsilon^2, \qquad \| \partial_i e_u \|_B
  \leqslant \alpha \tilde{\varepsilon}^{- 1} \quad (i = \nu, \lambda),
  \nonumber\\
  &  & \| \partial_u e_{\nu} \|_{\mathcal{L} (B, \mathbb{R})^{^{}}} \leqslant
  M_0 \varepsilon \tilde{\varepsilon}^{}, \qquad \| \partial_u e_{\lambda}
  \|_{\mathcal{L} (B, \mathbb{R})^{^{}}} \leqslant M_0 \varepsilon^2
  \tilde{\varepsilon}, \qquad \| \partial_u e_u \|_{\mathcal{L} (B, B)}
  \leqslant \alpha,  \label{grad}
\end{eqnarray}
where $\alpha$ is yet another parameter. Under these conditions Eq.
{\eqref{gradFbound}} implies:
\begin{equation}
  \| \nabla F (y) \|_{\mathcal{L} (Y, Y)} \leqslant \max (K_2 M_0 \varepsilon,
  K_2 \alpha) \qquad (y \in Y_0) . \label{gradFget}
\end{equation}
We restate the conclusions of the above discussion as

\begin{lemma}[Abstract Lemma]
  \label{AbstractLemma}Suppose that, for a given $\varepsilon,$the constants
  $M_0, \tilde{\varepsilon}, \alpha$ are such that maps $e_j$ satisfy bounds
  {\eqref{contained}} and {\eqref{grad}} everywhere in $Y_0$ defined by
  {\eqref{Y0}}. Suppose in addition that (see {\eqref{Kdef}}, {\eqref{Cgrad}}
  for the definition of $K_1$ and $K_2$)
  \begin{equation}
    K_1 M_0 \varepsilon \leqslant 1, \qquad K_2 M_0 \varepsilon \leqslant 1 /
    2, \qquad K_2 \alpha \leqslant 1 / 2 . \label{contr1}
  \end{equation}
  Then $F (Y_0) \subset Y_0$ and $\| \nabla F (y) \|_{\mathcal{L} (Y, Y)}
  \leqslant 1 / 2$ in $Y_0$, so that $F$ is a contraction in $Y_0$ and has a
  unique fixed point there.
\end{lemma}

\subsubsection{Complex version of the Abstract Lemma}

By a few minor modifications of the proof of the Abstract Lemma we can get a
complex-$\varepsilon$ version thereof. This is needed in the proof of fixed
point analyticity (Section \ref{analyticity}) and is not used anywhere else.
We let $\varepsilon \in \mathbb{C}$, $y$ be an element of the {\tmem{complex}}
Banach space $\mathbb{Y}$ with the norm {\eqref{yY}}, and $\mathbb{Y}_0$ (the
complex analogue of $Y_0$, see {\eqref{Y0}}) be defined as:
\begin{equation}
  \mathbb{Y}_0 = \{ y : | \nu - \nu_0 | \leqslant M_0 | \varepsilon |^2, |
  \lambda - \lambda_0 | \leqslant M_0 | \varepsilon |^2, \| u \|_B \leqslant 1
  \} . \label{Y1}
\end{equation}
Then the following generalization of Lemma \ref{AbstractLemma} holds.

\begin{lemma}
  {\tmstrong{\tmtextup{(Complex Abstract
  Lemma)}}}\label{AbstractLemma_analytic} Suppose that, for a given
  $\varepsilon \in \mathbb{C},$the constants $M_0, \tilde{\varepsilon},
  \alpha$ are such that maps $e_j$ satisfy bounds the complex analogues of
  {\eqref{contained}} and {\eqref{grad}}, i.e.,
  \begin{equation}
    | e_{\nu} (y) | \leqslant \frac{M_0}{K_1}  | \varepsilon |^2, \qquad |
    e_{\lambda} (y) | \leqslant \frac{M_0}{K_1}  | \varepsilon |^3, \qquad \|
    e_u (y) \|_B \leqslant 1 \qquad (y \in \mathbb{Y}_0), \label{containedbis}
  \end{equation}
  and
  \begin{eqnarray}
    &  & | \partial_i e_{\nu} | \leqslant M_0 | \varepsilon |, \qquad |
    \partial_i e_{\lambda} | \leqslant M_0 | \varepsilon |^2, \qquad \|
    \partial_i e_u \|_B \leqslant \alpha \tilde{\varepsilon}^{- 1} \quad (i =
    \nu, \lambda), \nonumber\\
    &  & \| \partial_u e_{\nu} \|_{\mathcal{L} (B, \mathbb{R})^{^{}}}
    \leqslant M_0 | \varepsilon | \tilde{\varepsilon}^{}, \qquad \| \partial_u
    e_{\lambda} \|_{\mathcal{L} (B, \mathbb{R})^{^{}}} \leqslant M_0 |
    \varepsilon |^2 \tilde{\varepsilon}, \qquad \| \partial_u e_u
    \|_{\mathcal{L} (B, B)} \leqslant \alpha,  \label{gradbis}
  \end{eqnarray}
  everywhere in $\mathbb{Y}_0$ defined by {\eqref{Y1}}. Suppose in addition
  that (see {\eqref{Kdef}}, {\eqref{Cgrad}} for the definition of $K_1$ and
  $K_2$)
  \begin{equation}
    K_1 M_0 | \varepsilon | \leqslant 1, \qquad K_2 M_0 | \varepsilon |
    \leqslant 1 / 2, \qquad K_2 \alpha \leqslant 1 / 2 . \label{contr1bis}
  \end{equation}
  Then $F (\mathbb{Y}_0) \subset \mathbb{Y}_0$ and $\| \nabla F (y)
  \|_{\mathcal{L} (Y, Y)} \leqslant 1 / 2$ in $\mathbb{Y}_0$, so that $F$ is a
  contraction in $\mathbb{Y}_0$ and has a unique fixed point there.
\end{lemma}

{\tmem{Proof.}} The proof of this lemma is a straightforward repetition of the
one of Lemma \ref{AbstractLemma}, modulo the replacement of $\varepsilon$ by
$| \varepsilon |$ in a few inequalities. More precisely, a simple critical
rereading of the proof shows that, if we leave the definitions of $F$, see
{\eqref{Fdefalt}}, and of $\nabla F$, see {\eqref{gradF1}} and
{\eqref{gradF2}}, as they are, and we replace $Y_0$ by $\mathbb{Y}_0$ and
$\varepsilon$ by $| \varepsilon |$ everywhere in the rest of the proof (in
particular in the following places: 1 line after {\eqref{Kdef}}; in
Eq.{\eqref{contained}}; in Eq.{\eqref{gradFbound}}; in Eq.{\eqref{grad}}; and
in Eq.{\eqref{gradFget}}), then we readily obtain the desired claim.

\subsection{Fixed point theorem}\label{FPth}

In this section we will put Key Lemma and Abstract Lemma together and will
finally show that the FPE {\eqref{fy=0}} has a solution. Namely, we will prove
the following result:

\begin{theorem}
  \label{FPT}There exists a $\gamma_0 \geqslant 2$ and a positive continuous
  function $\varepsilon_0 (\gamma)$ defined for $\gamma \geqslant \gamma_0$
  such that for each $\gamma \geqslant \gamma_0$ and $0 < \varepsilon
  \leqslant \varepsilon_0 (\gamma)$ the fixed point equation {\eqref{fy=0}}
  has a nontrivial solution. \ 
\end{theorem}

\tmtextit{Proof. }We will show, with the help of Key Lemma \ref{Key}, that for
$\gamma \geqslant \gamma_0$ and for $0 < \varepsilon \leqslant \varepsilon_0
(\gamma)$ conditions of Abstract Lemma \ref{AbstractLemma} can be satisfied.

We thus identify the abstract Banach space $B$ in Section \ref{abstract} with
the space $B (\gamma, \delta)$ in {\eqref{defnormB}}. We also put
\begin{equation}
  \tilde{\varepsilon} = A_0 \delta, \label{etd}
\end{equation}
and identify the space $Y$ from {\eqref{yY}} with $Y (\gamma, \delta)$ in
{\eqref{normYB}}. The parameter $\delta$ in the Key Lemma will be chosen
proportional to $\varepsilon$:
\begin{equation}
  \delta = h \varepsilon, \label{dep}
\end{equation}
with $h$ to be fixed momentarily.

Abstract Lemma requires us to examine the neighborhood $Y_0$ defined in
{\eqref{Y0}}. By $K_1 M_0 \varepsilon \leqslant 1$, the first of conditions
{\eqref{contr1}} (we will make sure to satisfy all of these conditions below),
the points of $Y_0$ will satisfy
\begin{equation}
  | \nu |, | \lambda | \leqslant K_3 \varepsilon, \qquad K_3 = K_3 (a, b) =
  \max \left( \left| \frac{a}{b} \right| + \frac{1}{K_1}, \frac{1}{| b |} +
  \frac{1}{K_1} \right) . \label{C1}
\end{equation}
Let us choose
\begin{equation}
  h = K_3 / A_0 . \label{defdeltah}
\end{equation}
By {\eqref{C1}}, we have
\begin{equation}
  Y_0 \subset \{ y : \| y \|_Y \leqslant 1 \} .
\end{equation}
Thus, the basic assumption {\eqref{keyass}} holds in $Y_0$, and we can use Key
Lemma to estimate $e_j$ and their derivatives in $Y_0$.

We will also fix (see Key Lemma for the definition of $\bar{D}$)
\begin{equation}
  \alpha = \gamma^{- \bar{D}} .
\end{equation}
With this identification and {\eqref{etd}}, the bounds on the derivatives of
$\partial_i e_u, \partial_u e_u$ requested in {\eqref{grad}} coincide with the
bounds for the same derivatives in {\eqref{eq:36}} of the Key Lemma. The
request $\| e_u \|_B \leqslant 1$ in $Y_0$ (Eq. {\eqref{contained}}) is also
satisfied by the bound on $\| e_u \|_B$ in {\eqref{eq:31}}.

Furthermore, we choose $\gamma_0$ as
\begin{equation}
  \gamma_0 = \max (\gamma_{\tmop{key}}, (2 K_2)^{1 / \bar{D}}) . \label{6.39}
\end{equation}
Then for $\gamma \geqslant \gamma_0$ we have $\gamma \geqslant
\gamma_{\tmop{key}}$ so that we can use Key Lemma, and in addition we satisfy
the third condition in {\eqref{contr1}}.

Let us now arrange for the conditions in {\eqref{contained}} and \
{\eqref{grad}} concerning $e_{\nu}, e_{\lambda}$, and their derivatives. By
Eq.{\eqref{eresc}}, $e_{\nu}, e_{\lambda}$ equal $e^{(0)}_{\nu},
e^{(0)}_{\lambda}$ times factors bounded by a $\gamma$-dependent constant
$f_{\gamma}$. Key Lemma gives estimates for $e^{(0)}_{\nu}, e^{(0)}_{\lambda}$
and their derivatives with constants $E_0, E_1$ in the r.h.s., and $e_{\nu},
e_{\lambda}$ will satisfy the same estimates with $E_i \rightarrow E_i' =
f_{\gamma} E_i$. Using the proportionality {\eqref{dep}} between $\delta$ and
$\varepsilon$, these estimates take the form
\begin{align}
&|e_{\nu}^{} | \leqslant E'_0 h^2 \varepsilon^2, 
 && |e_{\lambda}^{} | \leqslant E'_1 h^3 \varepsilon^3,  \nonumber\\
 &| \partial_i e_{\nu}^{} | \leqslant (E'_0 / A_0) h \varepsilon, 
  &&| \partial_i e_{\lambda}^{} | \leqslant (E'_1 / A_0) h^2 \varepsilon^2
  \qquad (i = \nu, \lambda), \qquad \nonumber\\
& \| \partial_u e_{\nu}^{} \|_{\mathcal{L} (B, \mathbb{R})}
  \leqslant E'_0 h^2 \varepsilon^2, &&\| \partial_u e_{\lambda}^{} (\nu,
  \lambda, u)\|_{\mathcal{L} (B, \mathbb{R})} \leqslant E'_1 h^3
  \varepsilon^3 . 
\end{align}
These have the same scaling in $\varepsilon$ as the corresponding estimates in
{\eqref{contained}}, {\eqref{grad}} (recall that $\tilde{\varepsilon} /
\varepsilon = A_0 h$). So, to satisfy {\eqref{contained}}, {\eqref{grad}}, we
simply choose $M_0$ sufficiently large, namely:
\begin{equation}
  M_0 = \max (K_1 E'_0 h^2, K_1 E'_1 h^3, (E'_0 / A_0) h, (E'_1 / A_0) h^2) .
\end{equation}
We still have to satisfy the first two conditions in {\eqref{contr1}}, as well
as to make sure that $\delta = h \varepsilon \leqslant \delta_0$. We achieve
this by choosing
\begin{equation}
  \varepsilon_0 (\gamma) = \min \left( \frac{\delta_0}{h}, \frac{1}{K_1 M_0},
  \frac{1}{2 K_2 M_0} \right) . \label{e0g}
\end{equation}
For any $0 < \varepsilon \leqslant \varepsilon_0 (\gamma)$, conditions of
Abstract Lemma are satisfied, and hence a fixed point exist.

\begin{corollary}
  \label{FPaction}The fixed point whose existence we proved belongs to the
  neighborhood
  \begin{eqnarray}
    &  & | \nu - \nu_0 |, | \lambda - \lambda_0 | \leqslant M_0
    \varepsilon^2, \nonumber\\
    &  & \| H_{\text{2R}} \|_w \leqslant A_0^R h^2 \varepsilon^2, \| H_{\text{4R}} \|_w
    \leqslant A_1^R h^2 \varepsilon^2, \| H_{\text{6R}} \|_w \leqslant A_2^R h^3
    \varepsilon^3, \| H_l \|_w \leqslant A h^{l / 2 - 1} \varepsilon^{l / 2 -
    1},  \label{FPactbnds}
  \end{eqnarray}
  where $C_0, A, A_k^R, h$ are some $\gamma$-dependent quantities. Moreover in
  this neighborhood this is a unique solution of the fixed point equation.
\end{corollary}

This follows from writing in full the condition $\| u \|_B \leqslant 1$.

\subsection{Semigroup property and \texorpdfstring{$\gamma$}{gamma}-independence}\label{semigroup}

Theorem \ref{FPT} shows that the renormalization map $R (\varepsilon, \gamma)$
has a fixed point provided that $\gamma \geqslant \gamma_0$ is sufficiently
large and $\varepsilon \leqslant \varepsilon_0 (\gamma)$ is sufficiently
small. We would now like to study how this fixed point depends on various
parameters. In this section we will { discuss $\gamma$-independence}, while in the next one we will show that it depends on
$\varepsilon$ analytically.

The $\gamma$-independence at $O (\varepsilon)$ is visible in Eq.{\eqref{y0}},
since both $a$ and $b$ become $\gamma$-independent as $\varepsilon \rightarrow
0$. That it should hold in general can be suspected from the semigroup
property {\eqref{eq:Rgamma}}. Indeed, if a certain interaction $H_{\ast}$ is a
fixed point of $R (\varepsilon, \gamma)$, then by the semigroup property it is
also a fixed point of $R (\varepsilon, \gamma^n)$ for any $n \geqslant 2$, as
long as $R (\varepsilon, \gamma^n)$ is defined on $H_{\ast}$ as a continuous
map acting on a neighborhood of a Banach space to which $H_{\ast}$ belongs.
Combining this simple argument with continuity in $\gamma$, {one should be able to} prove that,
in fact, the fixed point is unique and completely independent of
$\gamma$, at least on a suitable interval of values of $\gamma$, {such as the one} denoted by
$J$ below. {A full proof of this fact requires a critical re-reading of the proofs of the Key Lemma, of the Abstract Lemma and
of the Fixed point theorem, as well as a generalization thereof, providing existence and uniqueness of the general FPE in the space of equivalency classes of couplings modulo 
null sequences, see Remark \ref{generalFPE} and Remark \ref{rem:below} below. We won't belabor all the required details here, 
but we will provide all the elements sufficient for a willing reader to sit down and check the various claims, most of which are just 
straightforward corollaries of the previous discussion.}

Fix an interval $I = [\gamma_{\tmop{key}}, \bar{\gamma}]$, with
$\gamma_{\tmop{key}}$ the constant of the Key Lemma. From the proof the Key
Lemma, see in particular Section \ref{sec:choices}, we see that, with no loss
of generality, all the functions in {\eqref{eq:tofix}} can be chosen to be
decreasing in $\gamma$, so that their smallest values in $I$ are those at
$\gamma = \bar{\gamma}_{}$, which we denote by $\bar{\delta}_0, \bar{A}_0, \{
\bar{A}_k^R \}_{k = 0, 1, 2}, \bar{A}, \bar{E}_0, \bar{E}_1$. It is easy to
check that the Key Lemma \ref{Key} admits the following ``uniform'' version on
$I$: take $\gamma \in I$, $0 \leqslant \delta \leqslant \bar{\delta}_0$, and
any sequence $(\nu, \lambda, u)$ satisfying the inequalities {\eqref{keyass}}
with $A_0$ replaced by $\bar{A}_0$ and $\|u\|_B$ replaced by $\|u\|_{\bar{B}}
=\|u\|_{B (\bar{\gamma}, \delta)}$; then the conclusions of the lemma,
{\eqref{eq:31}} and {\eqref{eq:36}} hold, with $\delta_0, A_0, \{A_k^{R_{}}
\}_{k = 0, 1, 2}, A, E_0, E_1$ replaced by $\bar{\delta}_0, \bar{A}_0, \{
\bar{A}_k^R \}_{k = 0, 1, 2}, \bar{A}, \bar{E}_0, \bar{E}_1$, and $B$ replaced
by $\bar{B}$.

Similarly, we can easily obtain a uniform version of the Abstract Lemma and
Fixed point theorem for $0 < \varepsilon \leqslant \bar{\varepsilon}_0$, with
$\bar{\varepsilon}_0 = \min_{\gamma \in I} \varepsilon_0 (\gamma)$, and
$\gamma \in I_0 \equiv [\bar{\gamma}_0, \bar{\gamma}]$, with $\bar{\gamma}_0$
defined by the analogue of {\eqref{6.39}} with $K_2$ replaced by $\bar{K}_2 =
\min_{0 < \varepsilon \leqslant \bar{\varepsilon}_0, \gamma \in I_{}} K_2$
(note that $I_0$ is non empty for $\bar{\gamma}$ large enough). We denote by
$F_{\gamma}$ the function $F$ of Section \ref{abstract}, in order to emphasize
its dependence upon $\gamma$. We define $\bar{\nu}_0 = \frac{a_0}{b_0}
\varepsilon$ and $\bar{\lambda}_0 = - \frac{1}{b_0} \varepsilon$, with $a_0 =
a |_{\varepsilon = 0} $ and $b_0 = b |_{\varepsilon = 0}
$, see {\eqref{ab}}, and let
\begin{equation}
  \overline{Y_0} = \{ y : | \nu - \bar{\nu}_0 | \leqslant \bar{M}_0
  \varepsilon^2, | \lambda - \bar{\lambda}_0 | \leqslant \bar{M}_0
  \varepsilon^2, \| u \|_{\bar{B}} \leqslant 1 \} .
\end{equation}
A critical re-reading of the Abstract Lemma and of the Fixed point theorem
shows that there exist constants $\bar{h}$ and $\bar{M}_0$ such that, fixing
$\delta = \bar{h} \varepsilon$ and using the uniform version of the Key Lemma,
then, for any $\gamma \in I_0$, $F_{\gamma} (\bar{Y}_0) \subset \bar{Y}_0$ and
$F_{\gamma}$ is continuous for $\gamma \in I_0$. Moreover, letting $\bar{Y}$
be the Banach space with norm
\begin{equation}
  \|y\|_{\bar{Y}} = \max \{ (\bar{A}_0 \delta)^{- 1} | \nu |, (\bar{A}_0
  \delta)^{- 1} | \lambda |, \|u\|_{\bar{B}} \},
\end{equation}
we have $\| \nabla F_{\gamma} (y) \|_{\mathcal{L} (\bar{Y}, \bar{Y})}
\leqslant 1 / 2$ in $\bar{Y}_0$, so that, for any $\gamma \in I_0$, \
$F_{\gamma}$ is a contraction in $\bar{Y}_0$, uniformly in $\gamma$, and has a
unique fixed point there, denoted $y_{\ast} (\gamma)$. Of course, $y_{\ast}
(\gamma) = \lim_{n \rightarrow \infty} F^n_{\gamma^{}} (\bar{y}_0)$, with
$\bar{y}_0 = (\bar{\nu}_0, \bar{\lambda}_0, 0)$. Recalling that $F_{\gamma}$
is continuous in $\gamma$ for $\gamma \in I_0$ and is uniformly contractive
there, we find that $y_{\ast} (\gamma)$ is continuous in $\gamma$ for $\gamma
\in I_0$, being the uniform limit of a sequence of uniformly continuous
functions.

{\begin{remark}\label{rem:below} The previous discussion, as well as the one of the previous sections, 
shows that $y_{\ast}(\gamma)$ is the unique solution of the {\it restricted} FPE \eqref{FPErestr}, in the sense 
of Remark \ref{generalFPE}. As discussed there, we expect that a generalization of the methods 
of this paper will allow us to prove the uniqueness of the solution of the general FPE 
$(H_{\ell}') = (H_{\ell}) + (N_{\ell})$ modulo null couplings, provided the null sequence $(N_{\ell})$
is sufficiently small in norm. We will denote by the symbol 
$\mathfrak{h}_{\ast}(\gamma)$ such a (presumed) unique solution in the space of equivalency classes of couplings.
Of course, continuity of $y_{\ast}(\gamma)$ implies the continuity of $\mathfrak{h}_{\ast}(\gamma)$ in the appropriate topology.
\end{remark}}

By construction, $y_{\ast} (\gamma), \gamma \in I_0,$ is the unique solution
in $\bar{Y}_0$ to the fixed point equation $y = R_{\gamma} (y)$, where
$R_{\gamma}$ is the original form of the RG map (before the manipulations
{\eqref{rewrite}}), given by the right side of {\eqref{FPeqs}}. By its very
definition, $R_{\gamma}$ satisfies the semigroup property $R_{\gamma} \circ
R_{\gamma'} = R_{\gamma \cdot \gamma'}{+}${null}, so that, if ${\mathfrak{h}_{\ast}}(\gamma) =
{\mathfrak{h}_{\ast}}(\gamma')$, then ${\mathfrak{h}_{\ast}}(\gamma) = {\mathfrak{h}_{\ast}}(\gamma') = {\mathfrak{h}_{\ast}}(\gamma \cdot \gamma')$. From this, it follows that ${\mathfrak{h}_{\ast}}(\gamma) \equiv {\mathfrak{h}_{\ast}}(\bar{\gamma})$ for all the values $\gamma$ in the subset $X$ of
$I_0$ characterized by the following properties: {\tmem{(i)}}
$\overline{\gamma_{}} \in X$; {\tmem{(ii)}} if $\gamma \in X$, then $\gamma^{1
/ n} \in X$, for all natural $n$ such that $\gamma^{1 / n} \in I_0$;
{\tmem{(iii) }}if $\gamma, \gamma' \in X$, then $\gamma \cdot \gamma' \in X$,
as long as $\gamma \cdot \gamma' \in I_0$. Of course, by the continuity of
${\mathfrak{h}_{\ast}}(\gamma)$, the fixed point is constant and equal to ${\mathfrak{h}_{\ast}}(\bar{\gamma})$ on the closure of $X$, as well. For $\bar{\gamma} \geqslant
(\bar{\gamma}_0)^3$, the closure of $X$ contains the sub-interval $J =
[\bar{\gamma}^{2 / 3}, \bar{\gamma}]$.\footnote{For $\bar{\gamma} \geqslant
(\bar{\gamma}_0)^3$, we have $\bar{\gamma}^{1 / 3} \in X$ \ by $(i i)$, and
then $\bar{\gamma}^{2 / 3} \in X$ by $(i i i)$. So both endpoints of $J$ are
in $X$. Also, if $\gamma_1, \gamma_2 \in J$, then both $\gamma^{1 / 2}_1,
\gamma_2^{1 / 2} \in X$ by $(i i)$ and hence the geometric mean $(\gamma_1
\gamma_2)^{1 / 2} \in X$ by $(i i i)$. Applying this last statement
recursively starting from the endpoints of $J$, we obtain that $X$ is dense in
$J$.} This proves the independence of the fixed point from $\gamma$, for any
$\gamma \in J$.

\begin{remark}
  \label{chiindep}Another parameter which entered into the renormalization map
  is the cutoff function $\chi$. The fixed point coupling $\nu_{\ast}$ depends
  on $\chi$ already at $O (\varepsilon)$, as seen from Eq. {\eqref{y0}},
  because $a$ depends on $\chi$. That $\lambda_{\ast}$ is $\chi$-independent
  at $O (\varepsilon)$ is in agreement with the usual lore that the
  beta-functions for near-marginal couplings and the corresponding fixed-point
  coupling should not depend on the UV regularization scheme at the first
  nontrivial order. In higher orders in $\varepsilon$ we expect that all
  couplings will acquire $\chi$ dependence. So, in contrast with the
  $\gamma$-independence, the fixed point does depend on $\chi$. In spite of
  this, we expect on physical grounds that the critical exponents (i.e.
  eigenvalues of the renormalization map linearized near the fixed point)
  should be $\chi$-independent. Showing this rigorously is one of the open
  problems for the future (see Section \ref{conclusions}).
\end{remark}

\subsection{Analyticity}\label{analyticity}

In view of the Complex Abstract Lemma \ref{AbstractLemma_analytic}, and of the
complex version of the Key Lemma, stated and proved in Section \ref{sec:2k+2},
see Lemma \ref{KeyLemmaC}, it is easy to show that the fixed point of Theorem
\ref{FPT} can be extended to an analytic function of $\varepsilon$ in a small
neighborhood of the origin. More precisely, we get the following:

\begin{theorem}
  \tmtextbf{\tmtextup{(Analytic Fixed Point Theorem)}} \label{AFPT}There
  exists a $\gamma_0 \geqslant 2$ and a positive continuous function
  $\varepsilon_0 (\gamma)$ defined for $\gamma \geqslant \gamma_0$ such that
  for each $\gamma \geqslant \gamma_0$ and $\varepsilon \in \{ z \in
  \mathbb{C}: | z | \leqslant \varepsilon_0 (\gamma) \} \equiv \mathbb{E}_0$
  the fixed point equation {\eqref{FPeqs}} has a solution, analytic in
  $\varepsilon$, extending the one of Theorem \ref{FPT}. For any $\varepsilon
  \in \mathbb{E}_0$, such a solution is the unique solution of the fixed point
  equation in the complex neighborhood defined by the analogue of
  {\eqref{FPactbnds}} with $| \varepsilon |$ replacing $\varepsilon$.
\end{theorem}

{\tmem{Proof.}} We let $\delta = h | \varepsilon |$, with $h$ the same as in
{\eqref{defdeltah}}. By proceeding as in the proof of Theorem \ref{FPT}, with
$\bar{D}$ defined as in {\eqref{eq:8.3}}, we find that $F$ is a contraction on
$\mathbb{Y}_0$ for each $\varepsilon \in \mathbb{E}_0 \setminus \{ 0 \}$.
Moreover, by Lemma \ref{KeyLemmaC}, $F$ is analytic in $\varepsilon$ on the
punctured disk $\overline{\mathbb{Y}}_0 (\gamma) \equiv \cup_{0 < |
\varepsilon | \leqslant \varepsilon_0 (\gamma) } \mathbb{Y}_0$, and
so is $y_0$. Therefore, $y_n \equiv F^n (y_0)$ is analytic in $\varepsilon$ on
$\overline{\mathbb{Y}}_0 (\gamma)$. Since $F$ is a contraction, $y_n$
converges to a fixed point, call it $y_{\ast}  (\varepsilon),$as $n
\rightarrow \infty$, for any $\varepsilon \in \mathbb{E}_0 \setminus \{ 0 \}$;
for any such $\varepsilon$, $y_{\ast}  (\varepsilon)$ is the unique solution
of the fixed point equation in $\mathbb{Y}_0$. By Vitali's theorem on the
convergence of sequences of analytic functions, $y_{\ast}  (\varepsilon)$ is
holomorphic in $\varepsilon$ on $\overline{\mathbb{Y}}_0 (\gamma)$ (monodromy
follows from the uniqueness of the solution to the fixed point equation in
$\mathbb{Y}_0$). Note that $\lim_{\varepsilon \rightarrow 0} \varepsilon
y_{\ast}  (\varepsilon) = 0$; therefore, by Riemann's theorem on removable
singularities, $y_{\ast}  (\varepsilon)$ can be extended to an analytic
function of $\varepsilon$ on the complex disk of radius $\varepsilon_0
(\gamma)$ by letting $y_{\ast} (0) = 0$.

This result has various consequences. One clear consequence is that since the
fixed point is analytic around $\varepsilon = 0$, it has a convergent power
series expansion around this point. This is just the perturbative
$\varepsilon$-expansion discussed at the level of formal power series in
Appendix \ref{sec:formal} which is therefore convergent. Another consequence
is that the fixed points with real $\varepsilon > 0$ analytically continue to
the fixed points with $\varepsilon < 0$. For negative real $\varepsilon$, the
quartic interaction is an irrelevant perturbation of the gaussian fixed point
(at the linearized level). Thus, the $\varepsilon < 0$ fixed points should be
interpreted as UV fixed points: one can RG-flow from them to the gaussian
theory, not the other way around. We expect analyticity to be valid also in
the long-range Gross-Neveu model of {\cite{Gawedzki:1985jn}} (see the
introduction), and in other similar models. See also Appendix \ref{App:trees}
for an alternative proof of fixed point analyticity via the tree expansion.

\section{Proof of Key lemma}\label{sec:2k+2}

Here we finally prove the Key lemma that, as seen above, is the crucial
ingredient for showing the existence and uniqueness of the nontrivial RG fixed
point. Rather than proving the Key lemma in the formulation of Section
\ref{sec:key}, here we state and prove a generalization of the lemma with
complex $\varepsilon$, which is the version used in Section \ref{analyticity}
in the discussion on the analyticity of the fixed point. This does not create
any additional complications in the proof. \

Let us start by observing that both the fluctuation propagator
{\eqref{eq:split}} and the rescaling factor $\gamma^{- [\psi]}$ in
{\eqref{Hprime}} depend analytically on $\varepsilon$. So each individual term
$R_{\ell}^{\ell_1, \ldots, \ell_n}$ is analytic in $\varepsilon$, and the sum
{\eqref{RHH'}} will be analytic when convergent. Let $T$ be a compact subset
of the half-plane (see Eq.{\eqref{epsCond}})
\begin{equation}
  T \subset \{ \varepsilon \in \mathbb{C}: \tmop{Re} \varepsilon < d / 6 \} .
  \label{Tsubs}
\end{equation}
By Lemma \ref{CGHbound}, the constant $C_{\tmop{GH}}$ is uniformly bounded for
$\varepsilon \in T$. As a result the multilinear maps $S_l^{\ell_1, \ldots,
\ell_n}$ will satisfy estimates {\eqref{Sest}} with uniform ($T$-dependent)
constants for $\varepsilon \in T$. The action of dilatation for complex
$\varepsilon$ is still given by {\eqref{eq:Resclift}}, where $D_l = D_l
(\varepsilon)$ are complex. We have to replace $D_l \rightarrow \tmop{Re} D_l$
in the norm bounds {\eqref{eq:normResc}} for dilatation, which become
\begin{equation}
  \|DH_{\ell, p} \|_w = \gamma^{- \tmop{Re} D_l - p} \|H_{\ell, p} \|_{w
  (\cdummy / \gamma)} \leqslant \gamma^{- \tmop{Re} D_l - p} \|H_{\ell, p}
  \|_w \hspace{2.17em} (\varepsilon \in \mathbb{C}) \label{eq:normRescC}
\end{equation}
The criterion for irrelevance becomes $\tmop{Re} D_l - p > 0$. The same
replacement has to be done in the right-hand-sides of the estimates for
multilinear maps $R_{\ell}^{\ell_1, \ldots, \ell_n}$ in Section
\ref{sec:NormBounds}, see Eqs.{\eqref{Rll}} and {\eqref{Rest}}.

The parameter $\bar{D}$ from Section \ref{sec:key} is redefined for complex
$\varepsilon \in T$ as
\begin{equation}
  \bar{D} = \bar{D} (T) = \frac{1}{2} \min_{\varepsilon \in T}\{\tmop{Re}
  D_2 (\varepsilon) + 2, \tmop{Re} D_4 (\varepsilon) + 1, \tmop{Re} D_6
  (\varepsilon) \} . \label{eq:8.3}
\end{equation}
Note that $\bar{D}_{} > 0$ by assumptions on $T$. We can now state the
generalization of Lemma \ref{Key} to $\varepsilon\in\mathbb{C}$.

\begin{lemma}[Complex Key Lemma]
  \label{KeyLemmaC}Choose $d \in \{ 1, 2, 3 \}$, cutoff $\chi$, $N \geqslant
  4$, and a compact set $T \subset \mathbb{C}$ satisfying {\eqref{Tsubs}}.
  There exists $\gamma_{\tmop{key}} \geqslant 2$ and positive continuous
  functions {\eqref{eq:tofix}} on $\gamma \geqslant \gamma_{\tmop{key}}$, with
  the following property. Take any $\gamma \geqslant \gamma_{\tmop{key}}$, any
  $0 < \delta \leqslant \delta_0 (\gamma)$ and any sequence $y = (\nu,
  \lambda, u)$ satisfying $\|y\|_{Y (\gamma, \delta)} \leqslant 1$, and apply
  to it the renormalization map $R (\varepsilon, \gamma)$ with any
  $\varepsilon \in T$. Then the functions $e_{\nu}^{(0)}, e_{\lambda}^{(0)},
  e_u$ in Eq.{\eqref{FPeqs}} and their derivatives satisfy the bounds
  {\eqref{eq:31}} and {\eqref{eq:36}} uniformly in $\varepsilon \in T$. These
  functions are analytic in $\varepsilon$, being given by convergent series
  consisting of analytic terms.
\end{lemma}

The proof of the Complex Key Lemma is presented in the next subsections,
distinguishing various subcases. For instance, in order to prove that $\|e_u
\|_{B (\gamma, \delta)} \leqslant \gamma^{- \bar{D}}$, recalling the
definition {\eqref{defnormB}} of the norm, we will seperately prove that $\|
(e_u)_{\ell} \|_w \leqslant A^{}_{} (\gamma) \delta^{k (\ell)}$ for all $\ell
\geqslant 8$, that $\| (e_u)_{\text{6R}} \|_w \leqslant A^{\text{R}}_2 (\gamma) \delta^3$,
$\| (e_u)_{\text{4R}} \|_w \leqslant A^{\text{R}}_1 (\gamma) \delta^2$, and $\| (e_u)_{\text{2R}}
\|_w \leqslant A^{\text{R}}_0 (\gamma) \delta^2$. For ease of notation, we will drop
the dependence on $\gamma$ from the constants $A (\gamma), A_0 (\gamma)$, etc,
and simply denote them by $A, A_0$, etc. Similarly for $Y (\gamma, \delta)$
and $B (\gamma, \delta)$, to be denoted by $Y$ and $B$, respectively.

\subsection{Case \texorpdfstring{$\ell \ge 8$}{l gtreq 8}}\label{sec:l8}

We start from the bound on $\|(e_u (y))_{\ell} \|_w$
with $\ell \ge 8$. From the definitions, see {\eqref{betainiz}} and
{\eqref{FPeqs}}, we have
\begin{equation}
  (e_u (y))_{\ell} = \sum_{(\ell_i)_1^n}
  R_{\ell}^{\ell_1, \ldots, \ell_n} (H_{\ell_1}, \ldots, H_{\ell_n}) .
\end{equation}
Using bounds {\eqref{Rll}} and {\eqref{Rest}} on $R_{\ell}^{\ell_1, \ldots,
\ell_n}$ collected we find that
\begin{equation}
  \|(e_u (y))_{\ell} \|_w \le \gamma^{- \tmop{Re} D_{\ell}} \|u_{\ell} \|_w
  + \gamma^{- \tmop{Re} D_{\ell}}  \sum_{(\ell_i)_1^n \neq (\ell)} \rho_l
  [(\ell_i)_1^n], \label{eu8}
\end{equation}
where we denoted
\begin{equation}
  \rho_l [(\ell_i)_1^n] = \rho_l (H_{\ell_1}, \ldots, H_{\ell_n}),
\end{equation}
and $\rho_l (H_{\ell_1}, \ldots, H_{\ell_n})$ is given in Eq.
{\eqref{rholbound}}. Here $H_{\ell_i}$ should be interpreted as equal to:
$\nu$, if $\ell_i = \text{2L}$; $\lambda$, if $\ell_i = \text{4L}$; $\mathfrak{X}_{\ast}$,
if $\ell_i = 6$SL; $u_{\ell_i}$, otherwise. Recall that $\rho_l = 0$ unless
$\sum_i | \ell_i | \ge l + 2 (n - 1)$.

By using the assumption $\|y\|_{Y (\gamma, \delta)} \leqslant 1$ of Key lemma,
writing in full the meaning of this condition (recall the definition of
$\|y\|_{Y (\gamma, \delta)}$, Eq.{\eqref{defBanach1st}}), we find:
\begin{eqnarray}
  &  & \|H_{\text{2L}} \|_w + \|H_{\text{2R}} \|_w \le A_0 \delta + A_0^{\rest} \delta^2
  \backassign b_0, \nonumber\\
  &  & \|H_{\text{4L}} \|_w + \|H_{\text{4R}} \|_w \le A_0 \delta + A_1^{\rest} \delta^2
  \backassign b_1, \nonumber\\
  &  & \|H_{\text{6SL}} \|_w + \|H_{\text{6R}} \|_w \le C_{\gamma 3} A_{0}^2
  \delta^2 + A_2^R \delta^3 = : b_2, \nonumber\\
  &  & \|H_{\ell} \|_w \le A \delta^{k (\ell)} \backassign b_{k (\ell)},
  \qquad \text{if} \qquad \ell \ge 8. 
\end{eqnarray}
It will be convenient to arrange so that
\begin{equation}
  b_k \leqslant A \delta^{\max \{k, 1\}}, \quad k \geqslant 0.
\end{equation}
For $k \geqslant 3$ this is true as an equality by the definition of $b_k$. To
have this for $k = 0, 1, 2$ as well, we will assume (we will see later how to
satisfy simultaneously all $\spadesuit$-constraints):
\begin{equation}
  (\spadesuit)  \quad 2 \max (A_0, A_0^R \delta_0, A_1^R \delta_0, C_{\gamma
  3} A_{0}^2 + A_2^R \delta_0) \leqslant A. \label{eq:heart1}
\end{equation}
Using these bounds in {\eqref{eu8}} we find that:
\begin{equation}
  \|(e_u (y))_{\ell} \|_w \le \gamma^{- \tmop{Re} D_{\ell}}  [A \delta^{k
  (\ell)} + \Delta_{k (\ell)}^{(1)} + \Delta_{k (\ell)}^{(2)}], \label{eu8.1}
\end{equation}
where we defined [here $C = C_2^2$]:
\begin{eqnarray}
  &  & \Delta^{(1)}_k = \sum_{k' = k + 1}^{\infty} C_{}^{k' + 1} b_{k'} 
  \nonumber,  \label{Deltak1}\\
  &  & \Delta^{(2)}_k = \sum_{(k_i)_{i = 1}^n, n \geqslant 2} F_k
  [(k_i)_1^n],  \label{Deltak2}\\
  &  & F_k [(k_i)_1^n] = \left\{ \begin{array}{ll}
    C_{\gamma}^{n - 1}  \prod_{i = 1}^n C_{}^{k_i + 1} b_{k_i} & \tmop{if}
    \quad \sum_i k_i \ge k,\\
    0 & \tmop{otherwise} .
  \end{array} \right.  \label{Fkdef}
\end{eqnarray}
We will estimate these sums with the help of the following lemma, imposing
assumptions {\eqref{eq:constr2}} which we will arrange in the end by choosing
$\delta_0$ and $A$ appropriately. For the proof see Appendix \ref{Dkbounds}.

\begin{lemma}
  \label{Dkest}Suppose the nonnegative constants $C_{\gamma}, C, \delta, A$
  satisfy
  \begin{equation}
    (\spadesuit) \quad C_{} \delta_{} \leqslant 1 / 4, \qquad C_{\gamma}
    C_{} A \delta_{} \leqslant 1 / 2, \qquad C_{\gamma}
    C_{} A \leqslant 1 / 2, \label{eq:constr2}
  \end{equation}
  and that $0 \leqslant b_k \leqslant A \delta^{\max \{k, 1\}}$ for all $k
  \geqslant 0$. Then $\Delta^{(1)}_k$ and $\Delta^{(2)}_k$ defined in terms of
  $C, C_{\gamma}, b_k$ by {\eqref{Deltak1}}, {\eqref{Deltak2}} satisfy
  \begin{eqnarray}
    \Delta^{(1)}_k & \leqslant & A \delta^{k + 1} (2 C_{}^{k +
    2}),  \label{Dk1}\\
    \Delta^{(2)}_k & \leqslant & A \delta^{\max \{k, 2\}} \cdot \left\{
    \begin{array}{ll}
      C_0 = 4 C_{} + 8 C^2_{} + 16 C^3 & \text{if $k = 0$,} 1\\
      2 (2 C_{})^{k + 1} & \text{if $k \ge 2$,}
    \end{array} \right.  \label{Dk2}
  \end{eqnarray}
\end{lemma}

Using {\eqref{Dk1}}, {\eqref{Dk2}} in {\eqref{eu8.1}}, and recalling that we
are assuming $\ell \geqslant 8$ (so that $k (\ell) \geqslant 3$), we find
\begin{equation}
  \|(e_u (y))_{\ell} \|_w \leqslant \gamma^{- \tmop{Re} D_{\ell}} A \delta^{k
  (\ell)}  [1 + 2 C_{}^{k (\ell) + 2} \delta + 2 (2 C_{})^{k
  (\ell) + 1}], \label{eu8.2}
\end{equation}
It follows that
\begin{equation}
  \|(e_u (y))_{\ell} \|_w \leqslant \gamma^{- \bar{D}_{}} A \delta^{k (\ell)}
  \qquad (\ell \ge 8), \label{eubndobtained}
\end{equation}
as long as we impose
\begin{equation}
  1 + C_{}^{k (\ell) + 1} + 2 (2 C_{})^{k (\ell) + 1} \leqslant
  \gamma^{\tmop{Re} D_{\ell} - \bar{D}} .
\end{equation}
Given the form of this inequality, it is sufficient to check that it holds for
$\ell = 8$, and that the l.h.s. grows slower than r.h.s. as $\ell \to \ell +
2$, which amounts to two requirements:
\begin{equation}
  (\spadesuit) \quad 1 + C_{}^4 + 2 (2 C_{})^4 \leqslant
  \gamma^{\tmop{Re} D_8 - \bar{D}}, \qquad \max (1, C_{}, 2 C_{}) \leqslant
  \gamma^{d / 2 - \tmop{Re} \eps} . \label{eq:dust2}
\end{equation}
Next let us estimate derivatives. Consider a vector $\delta y = (\delta \nu,
\delta \lambda, \delta u)$ satisfying $\| \delta y\|_Y
\leqslant 1$. Consider also a trimmed coupling sequence $\delta H_{\ell}$
which contains the couplings in $\delta y$ and, in addition, the coupling
$\delta H_{\text{6SL}}$ corresponding to the variation of
$\mathfrak{X}_{\ast}$. We have
\begin{equation}
  \nabla_y (e_u (y))_{\ell} \delta y = \sum_{(\ell_i)_1^n} \sum_{i = 1}^n
  R^{\ell_1, \ldots, \ell_n}_{\ell} (H_{\ell_1}, \ldots, \delta H_{\ell_i},
  \ldots, H_{\ell_n}),
\end{equation}
and thus
\begin{equation}
  \| \nabla_y (e_u (y))_{\ell} \delta y \|_w \leqslant \gamma^{- \tmop{Re}
  D_{\ell}} \| \delta H_{\ell} \|_w + \gamma^{- \tmop{Re} D_{\ell}}
  \sum_{(\ell_i)_1^n \neq (\ell)} \sum_{i = 1}^n \rho_l (H_{\ell_1}, \ldots,
  \delta H_{\ell_i}, \ldots, H_{\ell_n}) \label{gradeuest}
\end{equation}
Note that $\| \delta H_{\text{6SL}} \|_w \leqslant 2 C_{3 \gamma}
A_{0}^2 \delta^2$. We will increase $C_{3 \gamma}$ by factor 2.
Then all couplings $\delta H_{\ell}$ satisfies the same bounds as the bounds
on couplings $H_{\ell}$ used to estimate $\|(e_u (y))_{\ell} \|_w$. It follows
that the functions $\rho_l$ in the r.h.s. of {\eqref{gradeuest}} can be
estimated in exactly the same way. This gives an estimate of the same form as
{\eqref{eu8.1}}, namely
\begin{equation}
  \| \nabla_y (e_u (y))_{\ell} \delta y \|_w \leqslant \gamma^{- \tmop{Re}
  D_{\ell}}  [A \delta^{k (\ell)} + \Delta_{k (\ell)}^{(1)} +
  \tilde{\Delta}_{k (\ell)}^{(2)}],
\end{equation}
where $\tilde{\Delta}^{(2)}_k$ differs from $\Delta^{(2)}_k$ in that $F_k
[(k_i)]$ is replaced by
\begin{equation}
  \tilde{F}_k [(k_i)_1^n] = n F_k [(k_i)_1^n],
\end{equation}
where the factor $n$ accounts for the sum $\sum_{i = 1}^n$ in
{\eqref{gradeuest}}. We will increase $C_{\gamma}$ in {\eqref{Fkdef}} by 2 to
absorb this factor (note $n \leqslant 2^{n - 1}$), so that both $\tilde{F}_k$
and $F_k$ can be considered to satisfy the same bound {\eqref{Fkdef}}.

Then, under the same assumptions that {\eqref{eubndobtained}} was obtained, we
will have
\begin{equation}
  \| \nabla_y (e_u (y))_{\ell} \delta y \|_w \leqslant \gamma^{- \bar{D}} A
  \delta^{k (\ell)} \qquad (\ell \ge 8) . \label{euderbndobtained}
\end{equation}
Taking into account the assumed bounds on couplings $\delta y$, this
inequality is precisely what is asserted in the last line of {\eqref{eq:36}}
concerning the part of $e_u$ with $\ell \geqslant 8$.

Incidentally, convergence of the series {\eqref{gradeuest}} also proves that
the functions $e_u (y)$ are in fact Frechet differentiable.

The shown method of bounding derivatives is general and will apply to all the
other functions that we still have to consider, i.e. $(e_u)_{\text{2R}}, (e_u)_{4
R}, (e_u)_{\text{6R}}, e_{\nu}^{(0)}, e_{\lambda}^{(0)}$. They are all given by sums
of multilinear operators applied to the sequence $H_{\ell}$, and will be
estimated using the basic bound {\eqref{rholbound}}. Whenever we manage to
bound such a function by an $X$, the shown method will naturally bound its
$u$-derivative by the same $X$, while its $\nu, \lambda$ derivatives by $X /
(A_0 \delta)^{}$. Note that all bounds {\eqref{eq:36}} are of precisely such a
form. So we no longer need to discuss derivative bounds, but can focus on
estimating the functions themselves.

\subsection{Case \texorpdfstring{$\ell = \text{6R}$}{l eq 6R}}\label{sec:l=6R}

From the definitions (see {\eqref{betainiz}} and {\eqref{FPeqs}} and the third
of {\eqref{Rlexpl}}) and the bounds on $R_{\text{6R}}^{\ell_1, \ldots, \ell_n}$ we
find that
\begin{equation}
  \|(e_u (y))_{\text{6R}} \|_w \le \gamma^{- \tmop{Re} D_6} \|u_{\text{6R}} \|_w +
  \gamma^{- \tmop{Re} D_6}  \sum_{(\ell_i)_1^n \neq (\text{6SL}), (\text{6R}), (\text{4L}, 4
  L)} \rho_6 [(\ell_i)_1^n] . \label{eu6}
\end{equation}
By repeating a discussion analogous to that of Section \ref{sec:l8}, we get
the analogue of {\eqref{eu8.1}}, namely
\begin{equation}
  \|(e_u (y))_{\text{6R}} \|_w \le \gamma^{- \tmop{Re} D_6}  \left[ A_2^{\rest}
  \delta^3 + \Delta_2^{(1)} + \Delta_{2 ; \text{6R}}^{(2)} \right], \label{eu6.1}
\end{equation}
where $\Delta_{2 ; \text{6R}}^{(2)}$ is defined analogously to $\Delta_2^{(2)}$,
modulo the fact that the contribution from the sequence $(k_i)_{i = 1}^n = (1,
1)$ is now proportional to $b_1 b_1^{\rest}$, with $b_1^{\rest} = A_1^{\rest}
\delta^2$, rather than to $b_1^2$ (this comes from the constraint $(\ell_i)_{i
= 1}^n \neq (\text{4L}, \text{4L})$ in {\eqref{eu6}}):
\begin{equation}
  \Delta^{(2)}_{2 ; \text{6R}} = 2 C_{\gamma} C_{}^4 b_1 b_1^{\rest} +
  \sum_{(k_i)_1^n \neq (1, 1)}^{n \geqslant 2} F_2 [(k_i)_1^n] . \label{D26R}
\end{equation}
It is convenient to define, for any sequence $\varkappa = (k_i)_1^n$,
\begin{equation}
  F_{\tmop{ext}} [\varkappa] = \sum_{\text{$\varkappa' :$extends $\varkappa$
  by $\geqslant 0$ zeros}} F [\varkappa'] . \label{Fextdef0}
\end{equation}
Using this definition, we split the second term in the r.h.s. of
{\eqref{D26R}} into (a) the contributions of sequences $(1, 1, 0)$, $(2, 0)$,
their permutations and extensions by zero and (b) sequences with $\sum k_i
\geqslant 3$ which form $\Delta_3^{(2)}$. We get
\begin{equation}
  \Delta^{(2)}_{2 ; \text{6R}} = 2 C_{\gamma} C_{}^4 b_1 b_1^{\rest} + 2
  F_{\tmop{ext}} [(2, 0)] + 3 F_{\tmop{ext}} [(1, 1, 0)] + \Delta_3^{(2)} .
\end{equation}
It is shown in Appendix \ref{Dkbounds}, see Eq. {\eqref{eq:Fextbd2}}, that, in
the assumptions of Lemma \ref{Dkest},
\begin{equation}
  F_{\tmop{ext}} [(k_i)_1^n] \leqslant 4 C_{}^{k + 1} A \delta^{k + m}
  \label{eq:Fextbd20},
\end{equation}
where $k = \sum k_i$ and $m$ is the number of zeros in the sequence
$(k_i)_1^n$.

Using {\eqref{Dk2}} for $\Delta_3^{(2)}$, the basic estimates $b_1^R \leqslant
A_1^R \delta^2$, $C_{\gamma} C A \leqslant 1 / 2$, and {\eqref{eq:Fextbd20}}
we get
\begin{equation}
  \Delta^{(2)}_{2 ; \text{6R}} \leqslant (C_{}^3 A_1^R + [8 C_{}^3 + 12 C^3 + 2 (2
  C_{})^4] A) \delta^3, \label{eq:D22bound}
\end{equation}
so that
\begin{equation}
  \|(e_u (y))_{\text{6R}} \|_w \le \gamma^{- \tmop{Re} D_6} \delta^3  \left\{
  A_2^{\rest} + 2 C_{}^4 A + C_{}^3 A_1^R + [20 C_{}^3 + 2 (2
  C_{})^4] A \right\}, \label{eu6.2}
\end{equation}
which is smaller than $\gamma^{- \bar{D}} A_2^{\rest} \delta^3$, provided that
\begin{equation}
  (\spadesuit) \qquad A_2^{\rest} + 2 C_{}^4 A + C_{}^3 A_1^R + [20 C_{}^3 + 2
  (2 C_{})^4] A \le \gamma^{\tmop{Re} D_6 - \bar{D}}
  A_2^{\rest} . \label{eq:H6Rreq}
\end{equation}
\subsection{Case \texorpdfstring{$\ell = \text{4R}$}{l eq 4R}}\label{sec:l=4R}

From the definitions and the bounds on $R_{\text{4R}}^{\ell_1,
\ldots, \ell_n}$ we find that
\begin{equation}
  \|(e_u (y))_{\text{4R}} \|_w \le \gamma^{- \tmop{Re} D_4 - 1} \|u_{\text{4R}} \|_w +
  \gamma^{- \tmop{Re} D_4}  \sum_{(\ell_i)_1^n \neq (\text{4L}), (\text{4R})} \rho_4
  [(\ell_i)_1^n], \label{eu4}
\end{equation}
(the condition $(\ell_i)_1^n \neq (\text{4L})$ comes from the fact that $R_{\text{4R}}^{4
L}$ is identically zero, see the first of {\eqref{Rlexpl}} and the definition
of $T_{\text{4R}}^4$ in Section \ref{sec:trimming}; note in particular that, by
construction, $T_{\text{4R}}^{4, 0}$ annihilates the local quartic kernel associated
with $H_{\text{4L}}$) so that
\begin{equation}
  \|(e_u (y))_{\text{4R}} \|_w \le \gamma^{- \tmop{Re} D_4 - 1}  \left[
  A_1^{\rest} \delta^2 + \gamma \Delta_1^{(1)} + \gamma \Delta_1^{(2)}
  \right], \label{eu4.1}
\end{equation}
which gives
\begin{equation}
  \|(e_u (y))_{\text{4R}} \|_w \le \gamma^{- \tmop{Re} D_4 - 1} \delta^2  \left[
  A_1^{\rest} + \gamma A (2 C_{}^3) + \gamma C_0 A \right] . \label{eu4.2}
\end{equation}
This is smaller than $\gamma^{- \bar{D}} A_1^{\rest} \delta^2$, provided that
\begin{equation}
  (\spadesuit)  \qquad A_1^{\rest} + \gamma A (2 C_{}^3 + C_0) \le
  \gamma^{\tmop{Re} D_4 + 1 - \bar{D}_{}} A_1^{\rest} . \label{eq:H4Rreq}
\end{equation}
\subsection{Case \texorpdfstring{$\ell = \text{2R}$}{l eq 2R}}\label{sec:l=2R}

From the definitions and the bounds on $R_{\text{2R}}^{\ell_1, \ldots, \ell_n}$ we
find that
\begin{equation}
  \|(e_u (y))_{\text{2R}} \|_w \le \gamma^{- \tmop{Re} D_2 - 2} \|u_{\text{2R}} \|_w +
  \gamma^{- \tmop{Re} D_2}  \sum_{(\ell_i)_{ 1}^n \neq (\text{2L}), (2
  R), (\text{4L})} \rho_2 [(\ell_i)_1^n], \label{eu2}
\end{equation}
(the conditions $(\ell_i)_1^n \neq (\text{2L}), (\text{4L})$ come from the fact that $R_{2
R}^{\text{2L}}$ and $R_{\text{2R}}^{\text{4L}}$ are identically zero, see the first of
{\eqref{Rlexpl}} and the definition of $T_{\text{2R}}^2$ in Section
\ref{sec:trimming}; note in particular that, by construction, $T_{\text{2R}}^{2, 0}$
annihilates the local quadratic kernels associated with $H_{\text{2L}}$ and with
$S^{\text{4L}}_2 (H_{\text{4L}})$) so that
\begin{equation}
  \|(e_u (y))_{\text{2R}} \|_w \le \gamma^{- \tmop{Re} D_2 - 2}  \left[
  A_0^{\rest} \delta^2 + \gamma^2 \Delta_{0 ; \text{2R}}^{(1)} + \gamma^2
  \Delta_0^{(2)} \right], \label{eu2.1}
\end{equation}
where
\begin{equation}
  \Delta^{(1)}_{0 ; \text{2R}} = C_{}^2 b_1^{\rest} + \Delta^{(1)}_1 \le C_{}^2
  A_1^{\rest} \delta^2 + A \delta^2  (2 C_3^3) .
\end{equation}
Therefore,
\begin{equation}
  \|(e_u (y))_{\text{2R}} \|_w \le \gamma^{- \tmop{Re} D_2 - 2} \delta^2  \left[
  A_0^{\rest} + \gamma^2 (C_{}^2 A_1^{\rest} + 2 C_{}^3 A + C_0 A) \right] .
  \label{eu2.2}
\end{equation}
This is smaller than $\gamma^{- \bar{D}_{}} A_0^{\rest} \delta^2$, provided
that
\begin{equation}
  (\spadesuit) \qquad A_0^{\rest} + \gamma^2  (C_{}^2 A_1^{\rest} + 2 C_{}^3 A
  + C_0 A) \le \gamma^{\tmop{Re} D_2 + 2 - \bar{D}_{}} A_0^{\rest} .
  \label{eq:H2Rreq}
\end{equation}

\subsection{\texorpdfstring{$e_{\nu}^{(0)}$}{e nu 0}}

From the definitions and the bounds on $R_{\text{2L}}^{\ell_1, \ldots, \ell_n}$ we
find that
\begin{eqnarray}
  &  & |e_{\nu}^{(0)} (y) | \le \gamma^{- \tmop{Re} D_2} 
  \sum_{(\ell_i)_{ 1}^n \neq (\text{2L}), (\text{2R}), (\text{4L})} \rho_2
  [(\ell_i)_1^n], 
\end{eqnarray}
(the conditions $(\ell_i)_1^n \neq (\text{2L}), (\text{4L})$ come directly from the
definition of $e_{\nu}^{(0)}$, see {\eqref{betainiz}} and {\eqref{FPeqs}},
while $(\ell_i)_1^n \neq (\text{2R})$ comes from the fact that $R_{\text{2L}}^{\text{2R}}$ is
identically zero, see the first of {\eqref{Rlexpl}} and the definition of
$T_{\text{2L}}^2$ in Section \ref{sec:trimming}; note in particular that, by
construction, $T_{\text{2L}}^{2, 0}$ annihilates the nonlocal quadratic kernel
associated with $H_{\text{2R}}$) so that
\begin{equation}
  |e_{\nu}^{(0)} (y) | \le \gamma^{- \tmop{Re} D_2}  [\Delta_{0 ; 2
  R}^{(1)} + \Delta_0^{(2)}], \label{enu.1}
\end{equation}
which gives
\begin{equation}
  |e_{\nu}^{(0)} (y) | \le \gamma^{- \tmop{Re} D_2} \delta^2  \left[ C_{}^2
  A_1^{\rest} + 2 C_{}^3 A + C_0 A \right] . \label{enu.2}
\end{equation}
We thus get the first of {\eqref{eq:31}}, with
\begin{equation}
  E_0 = \gamma^{- \tmop{Re} D_2}  \left[ C_{}^2 A_1^{\rest} + 2 C_{}^3 A +
  C_0 A \right] .
\end{equation}

\subsection{\texorpdfstring{$e_{\lambda}^{(0)}$}{e lambda 0}}

From the definitions and the bounds on $R_{\text{4L}}^{\ell_1, \ldots, \ell_n}$ we
find that
\begin{eqnarray}
  &  & |e_{\lambda}^{(0)} (y) | \le \gamma^{- \tmop{Re} D_4} 
  \sum_{\tmscript{\begin{array}{c}
    (\ell_i)_{ 1}^n \neq (\text{4L}), (\text{4R}), (\text{6SL}),\\
    (\text{4L}, \text{4L}), (\text{4L}, \text{2L}), (\text{4R}, \text{2L})
  \end{array}}} \rho_4 [(\ell_i)_1^n], 
\end{eqnarray}
(the conditions $(\ell_i)_1^n \neq (\text{4L}), (\text{6SL}), (\text{4L}, \text{4L})$ come
directly from the definition of $e_{\lambda}^{(0)}$, see {\eqref{betainiz}}
and {\eqref{FPeqs}}, while $(\ell_i)_1^n \neq (\text{4R}), (\text{4L}, \text{2L}), (\text{4R}, \text{2L})$
come from the fact that $R_{\text{4L}}^{\text{4R}}$, $R^{\text{4L}, \text{2L}}_{\text{4L}}$ and $R^{\text{4R}, 2
L}_{\text{4L}}$ are identically zero, see the first of {\eqref{Rlexpl}} and the
definition of $T_{\text{4L}}^4$ in Section \ref{sec:trimming}; note in particular
that, by construction, $T_{\text{4L}}^{4, 0}$ annihilates the nonlocal quartic
kernels associated with $H_{\text{4R}}$, $S^{\text{4L}, \text{2L}}_4 (H_{\text{4L}}, H_{\text{2L}})$ and
$S^{\text{4R}, \text{2L}}_4 (H_{\text{4R}}, H_{\text{2L}})$) so that
\begin{equation}
  |e_{\lambda}^{(0)} (y) | \le \gamma^{- \tmop{Re} D_4}  [\Delta_{1 ;
  \lambda}^{(1)} + \Delta_{1 ; \lambda}^{(2)}], \label{elambda.1}
\end{equation}
where
\begin{eqnarray}
  &  & \Delta^{(1)}_{1 ; \lambda} = C_{}^3 b_2^{\rest} + \Delta^{(1)}_2 \le
  C_{}^3 A_2^{\rest} \delta^3 + A \delta^3  (2 C_{}^4), \\
  &  & \Delta^{(2)}_{1 ; \lambda} = 2 C_{\gamma} C_{}^4 b_1 b_1^{\rest} +
  \sum_{(k_i)_1^n \neq (1, 1)}^{n \geqslant 2} F_2 [(k_i)_1^n] . 
\end{eqnarray}
The sequences with $\sum k_i = 1$ such as $(1, 0)$, $(1, 0, 0)$, etc are
excluded from the second term because $b_0$ insertions then happen on the
external legs of a quartic interaction, and they give rise to a vertex with a
vanishing local part. We see that $\Delta^{(2)}_{1 ; \lambda}$ is identical to
{\eqref{D26R}} and therefore satisfies the same bound {\eqref{eq:D22bound}}
\begin{equation}
  \Delta^{(2)}_{1 ; \lambda} \leqslant (C_{}^3 A_1^R + [8 C_{}^3 + 12 C^3 + 2
  (2 C_{})^4] A) \delta^3,
\end{equation}
Therefore we get
\begin{equation}
  |e_{\lambda}^{(0)} (y) | \le \gamma^{- \tmop{Re} D_4} \delta^3  \left[
  C_{}^3 A_2^{\rest} + 2 C_{}^4 A + C_{}^3 A_1^R + [20 C_{}^3 + 2 (2
  C_{})^4] A \right] . \label{elambda.2}
\end{equation}
We thus get the second equation of {\eqref{eq:31}}, with
\begin{equation}
  E_1 = \gamma^{- \tmop{Re} D_4}  \left[ C_{}^3 A_2^{\rest} + 2 C_{}^4 A +
  C_{}^3 A_1^R + [20 C_{}^3 + 2 (2 C_{})^4] A \right] .
\end{equation}
\subsection{Possibility of all choices}\label{sec:choices}

Finally, we need to show that all the $\spadesuit$-constraints above can be
satisfied consistently: Eqs.{\eqref{eq:heart1}}, {\eqref{eq:constr2}},
{\eqref{eq:dust2}}, {\eqref{eq:H6Rreq}}, {\eqref{eq:H4Rreq}},
{\eqref{eq:H2Rreq}}. To write them in a more manageable form, let us replace
al $\gamma$-independent constants in the l.h.s. of the
$\spadesuit$-constraints by their maximum $\bar{C}$ (Recall that $C_0$ was
fixed in terms of $C$ in {\eqref{Dk2}}). Also let $\bar{C}_{\gamma} = \max
(C_{\gamma}, C_{\gamma 3})$. Finally let $Z$ be the minimal of the exponents
of $\gamma$ in the r.h.s. of {\eqref{eq:dust2}}, {\eqref{eq:H6Rreq}},
{\eqref{eq:H4Rreq}}, {\eqref{eq:H2Rreq}} over $\varepsilon \in T$:
\begin{eqnarray}
  Z & = & \min_{\varepsilon \in T} \left\{ \tmop{Re} D_8 - \bar{D}, d / 2 -
  \tmop{Re} \eps, \tmop{Re} D_6 - \bar{D}, \tmop{Re} D_4 + 1 -
  \bar{D}, \tmop{Re} D_2 + 2 - \bar{D} \right\} . 
\end{eqnarray}
Crucially $Z > 0$ by the assumption on $T$ and the definition of $\bar{D}$. We
then get the following list of constraints which, if satisfied, imply the
$\spadesuit$-constraints for any $\varepsilon \in T$:
\begin{eqnarray}
  &  & \bar{C} \le \gamma^Z,  \label{5051}\\
  &  & \bar{C} \delta_0 \le 1, \quad \bar{C}  \bar{C}_{\gamma} A \le 1, 
  \label{4546}\\
  &  & \max (A_0, A_0^R \delta_0, A_1^R \delta_0, \bar{C}_{\gamma}
  A_{0}^2 + A_2^R \delta_0) \leqslant A / 2,  \label{from38}\\
  &  & A_2^R + \bar{C}  (A_1^R + A) \le \gamma^Z A_2^R, \qquad A_1^R +
  \bar{C} \gamma A \leqslant \gamma^{Z_{}} A_1^R, \qquad A_0^R + \bar{C}
  \gamma^2  (A_1^R + A) \le \gamma^{Z_{}} A_0^R .  \label{last3}
\end{eqnarray}
The only remaining varying parameter is $\gamma$. We should now choose
$\gamma_{\tmop{key}}$ and $\delta_0, A_0, \{A_k^R \}_{k = 0, 1, 2}, A, E_0,
E_1,$ which are $\gamma$-dependent and positive, so that all these constraints
hold for $\gamma \geqslant \gamma_{\tmop{key}}$.

We can satisfy the first two lines taking $\gamma$ large, then $A$ and
$\delta_0$ small (in this order, because $\bar{C}_{\gamma} $depends on
$\gamma$). The remaining constraints are a bit more subtle because $A$ and
$A_k^R$ occur both in the l.h.s. and in the r.h.s. To satisfy {\eqref{last3}}
we will require:
\begin{eqnarray}
  &  & A_1^R, A \leqslant A_2^R, \qquad \gamma A \leqslant A_1^R, \qquad
  \gamma^2 A, \gamma^2 A_1^R \leqslant A_0^R,  \label{eq:line1}\\
  &  & 1 + 2 \bar{C} \le \gamma^{Z_{}}, \qquad 1 + 2 \bar{C} \le
  \gamma^{Z_{}}, \quad 1 + \bar{C} \leqslant \gamma^{Z_{}} .  \label{eq:line2}
\end{eqnarray}
The last three constraints on $\gamma$ are of the same type as {\eqref{5051}}.
Joining inequalities in {\eqref{eq:line1}} to {\eqref{from38}}, the resulting
set of constraints reduces to:
\begin{eqnarray}
  &  & A_0 \leqslant 0.5 A, \quad \bar{C}_{\gamma} A_{0}^2
  \leqslant 0.25 A,  \label{eq:A0A}\\
  &  & A_2^R \in [A, 0.25 \delta_0^{- 1} A], \qquad A_1^R \in [\gamma A, 0.5
  \delta_0^{- 1} A], \qquad A_0^R \in [\gamma^2 A, 0.5 \delta_0^{- 1} A], 
  \label{eq:A2RA}\\
  &  & A_1^R \leqslant A_2^R, \quad \gamma^2 A_1^R \leqslant A_0^R .  
  \label{eq:A1RA2R}
\end{eqnarray}
Here's then the final order in which all choices have to be made:
$\gamma_{\tmop{key}}$ is chosen as the minimal $\gamma \geqslant 2$ satisfying
{\eqref{5051}} and {\eqref{eq:line2}}. We then pick any $\gamma \geqslant
\gamma_{\tmop{key}}$ and compute the constant \ $\bar{C}_{\gamma}$. We then
satisfy {\eqref{4546}} by choosing:
\begin{equation}
  A = (\bar{C}  \bar{C}_{\gamma})^{- 1} .
\end{equation}
We then choose $A_0$ sufficiently small to satisfy {\eqref{eq:A0A}}. Finally,
we choose
\begin{equation}
  \delta_0 = \min (\bar{C}^{- 1}, 1 / (2 \gamma^3)), \label{d0g}
\end{equation}
which satisfies {\eqref{4546}} and at the same time, thanks to $\delta_0
\leqslant 1 / (2 \gamma^3)$, allows us to choose $A_k^R$ as follows:
\begin{equation}
  A_2^R = 0.25 \delta^{- 1}_0 A, \qquad A_1^R = \gamma A, \qquad
  A_0^R = 0.5 \delta_0^{- 1} A.
\end{equation}
Then {\eqref{eq:A2RA}} is satisfied, and {\eqref{eq:A1RA2R}} holds as well.
Key lemma is proved.

\section{Discussion and open problems}\label{conclusions}

In this paper, we discussed what is perhaps the simplest theoretical model to
study field-theoretic non-Gaussian fixed points, which is amenable to rigorous
analysis: \tmtextbf{symplectic fermions with a long-range kinetic term and
local quartic interaction}. Our model is translation and rotation invariant,
and the structure of the RG equations is quite similar to models with local
kinetic term. This makes our model more realistic than, for example, models
with hierarchical interactions (see \cite{Bauerschmidt} for an
introduction). \

Our model depends on 3 physical parameters: the number of dimensions $d$, the
number of fermion species $N$ (assuming $\tmop{Sp} (N)$ invariance), and a
parameter $\varepsilon$ in the long-range fermion propagator, which controls
the relevance of the quartic fermion interaction. For $0 < \varepsilon \ll 1$
this interaction is weakly relevant, and the beta-function equation for the
quartic coupling $\lambda$ takes the forms $\beta_{\lambda} = - \varepsilon
\lambda + \tmop{const} \cdot (N - 8) \lambda^2 + \cdots$. One thus has the right
to expect that, for $N \neq 8$, there exists an RG fixed point with $\lambda =
O (\varepsilon)$. Our main result (Theorem \ref{FPT}) establishes the
existence of this fixed point rigorously and non-perturbatively.

Although the path towards this rigorous result was somewhat long, most of the
ingredients are rather natural. We introduce an infinite-dimensional Banach
space of interactions, whose kernels are essentially local (have to decay very
fast at point separation). We work with a smooth momentum space cutoff, so
that the UV and IR-cutoff fermion propagator decays very fast in position
space, and the almost-locality of the interaction is preserved by an RG step.

An essential feature of our model is that a single RG step leads to a
convergent effective action (for weak coupling). Intuitively, this property of
fermionic models is due to the Pauli principle, or, equivalently, to fermionic
signs leading to cancellations between Feynman diagrams. The formal derivation
is somewhat delicate, and we review it pedagogically in Appendix \ref{sec:GH}.
This is standard in the constructive field theory community, but may appear
unexpected to the others. A related detail is that exhibiting these fermionic
cancellations requires considering a finite RG step with a rescaling parameter
$\gamma > 1$. That's what we do in this paper, as opposed to performing
continuous RG {\`a} la Polchinski's equation (see Remark \ref{Polchinski}).

With these ingredients, we show that the Wilsonian RG map is a well-defined
nonlinear operator in the Banach space of interactions, and is a contraction
(has derivative whose operator norm is less than 1) along the irrelevant
directions. The behavior along the mass direction $\nu$ and the quartic
$\lambda$ has to be analyzed separately. These directions are both relevant at
the linearized level, with $\lambda$ becoming irrelevant near the approximate
one-loop fixed point. Rigorous bounds on error terms show that these
statements remain true at the nonlinear level, at weak coupling. The proofs of
these results rely just on some elementary combinatorics, geometric series
convergence, and chasing $\gamma^{- D_l}$ factors suppressing the irrelevant
interactions. Given one relevant and infinitely many irrelevant directions,
the fixed point equation can then be rigorously solved (for $\varepsilon$
sufficiently small) via a variant of Newton's method, appealing to the Banach
fixed point theorem.

Although our main interest is in $\varepsilon$ real and positive, in which
case the fixed point can be thought of as the IR fixed point of an RG flow
originating at the gaussian theory, our methods apply as long as $\varepsilon$
is small and nonzero. E.g. we can also consider $\varepsilon < 0$. In this
case the quartic interaction is irrelevant around the gaussian theory, but
relevant around the fixed point whose existence we can prove (which in this
case is classified as a UV fixed point). We can also consider complex nonzero
$\varepsilon$. Although perhaps lacking clear physical meaning, we can use
this as a formal device to show that the fixed point is analytic in
$\varepsilon$ in a punctured disk around the origin, and thus in the whole
disk by Riemann's removable singularity theorem (Section \ref{analyticity}).
This is a dramatic conclusion, which implies that our fixed points can be
obtained via the convergent perturbative $\varepsilon$-expansion around
$\varepsilon = 0$.

\subsection{Open questions}

We will now list many open questions raised by our work. Some of them are
theoretical, while others have potential practical applications to numerical
calculations of critical exponents.

\subsubsection{Extensions to other nonlocal models}\label{extensions}

It should be relatively easy to extend our results to many other similar
models:
\begin{itemizedot}
  \item Models with a symmetry group $G \subset \tmop{Sp} (N)$, which have
  several mass terms and quartic couplings consistent with this symmetry. One
  should be able to find a non-perturbative fixed point in a neighborhood of
  any isolated solution to the one-loop beta-function equations (as long as
  all quartic directions have eigenvalues $O (\varepsilon)$, the condition
  which generalizes non-vanishing one-loop beta-function used in this work).
  
  \item Models where different species of fermions have different propagator
  scaling (different $\varepsilon$). This may include models where some
  fermions $\psi$ have local kinetic terms (and thus a fixed scaling dimension
  for a given $d$), while others $\psi'$ are long-range with tunable
  dimensions, so that the interaction $\psi^2 (\psi')^2$ can be made
  near-marginal.
  
  \item Models with a vanishing one-loop beta function, like our model with
  $N = 8$. As discussed in Appendix \ref{sec:I1I2}, the two-loop beta-function
  term $\lambda^3$ has a nonzero coefficient {\cite{Benedetti:2020rrq}},
  giving a perturbative fixed point with $\lambda = O \left(
  \sqrt{\varepsilon} \right)$. The non-perturbative existence of such a fixed
  point and its analyticity properties in $\varepsilon$ can be understood
  almost immediately using the tree expansion method described in Appendix
  \ref{App:trees}, and a contraction argument should also be possible.
  
  \item Our model in $d = 4$. Compared to $d \in \{ 1, 2, 3 \}$ treated here,
  the local term $(\partial \psi)^2$ would be (weakly) relevant for
  $\varepsilon > 0$. One thus has to treat it on equal footing with the local
  $\psi^2$ and $(\psi^2)^2$ terms. One should be able to construct a
  non-perturbative fixed point for small $\varepsilon$, helped by the fact
  that the new coupling is quadratic in $\psi$. This would be the fermionic
  analogue of the bosonic problem considered in {\cite{Brydges1998}}.
  
  \item Models where the sextic or higher power $(\psi^2)^p$ is near-marginal,
  i.e. $[\psi] \approx d / (2 p)$, \ $p \geqslant 3$. 
\end{itemizedot}
More ambitiously, time may be ripe for a ``general theory of fermionic
fixed points with scale-invariant kinetic terms (local or long-range) and
near-marginal local polynomial interactions''. One should be able to prove
that any such fixed point showing up in perturbative analysis exists
non-perturbatively, rather than writing a new paper for each particular model.
The main challenge is to choose an efficient notation, and to cleanly separate
the algebraic and analytic aspects of the problem.\footnote{One may be
inspired by how somewhat similar difficulties have been solved for nonlinear
stochastic partial differential equations, another problem which involves
renormalization {\cite{Hairer2014,Bruned2019}}.} This future general theory
should cover all the above examples, as well as fermionic fields transforming
in other rotation representations (e.g. spinors {\cite{Gawedzki:1985jn}}), and
even non-rotationally invariant (Lifshitz-type) fixed points having
anisotropic scaling.

\subsubsection{Further properties of the RG fixed points}

Here we proved that the RG fixed points exist, and established a few of there
basic properties such as $\gamma$-independence (modulo some loose ends), and analyticity in $\varepsilon$. Future
work should investigate several other interesting properties, such as:
\begin{itemizedot}
  \item Uniqueness of the fixed point as an equivalence class of interactions
  (i.e. uniqueness of solutions of the general fixed point equation; see
  Remark \ref{generalFPE});
  
  \item Dependence of the fixed point on the UV cutoff function $\chi$ (see
  Remark \ref{chiindep}). In spite of this dependence, the critical exponents
  are expected to be $\chi$-independent. It is instructive to compare the
  family of long-range models discussed here with one-parameter families of
  short-range fixed points, such as the Ashkin-Teller model, 6- and 8-vertex
  models, and interacting dimer models (see e.g. {\cite{BFMCMP,GMTCMP}}). In
  the latter case, the deformation parameter is an exactly marginal coupling,
  which can renormalize along the RG flow, and so the critical exponents
  depend on the microscopic details, although if one critical exponent is
  known, others can be expressed via it (the so called weak universality). In
  our case, $\varepsilon$ is not a coupling but a parameter controlling the
  nonlocal part of the action, so it does not renormalize. Therefore, the
  situation is similar to the usual universality, and all critical exponents
  should be universal functions of $\varepsilon$ independent of microscopic
  details such as the UV cutoff $\chi$.\footnote{Note in this respect that the
  IR scaling dimension of $\psi$ is exactly known and equal to its UV
  dimension $d / 4 - \varepsilon / 2$. Therefore the exponent $\eta$ is
  trivially known as a function of $\varepsilon$. Even for weak universality,
  all exponents can be found if one exponent is known, making the conclusion
  that in our situation all exponents are universal functions of $\varepsilon$
  less surprising.} It would be interesting to establish this rigorously. See
  {\cite{DombGreenVol6}} for a classic intuitive discussion of these issues,
  in the context of local models.
  
  \item Critical exponents. These can be defined, most generally, as
  eigenvalues of the RG transformation linearized around the fixed point
  (removing the eigenvalues corresponding to the ``redundant operators''
  {\cite{DombGreenVol6}}).\footnote{Sometimes this is equivalently expressed
  by introducing perturbing ``source terms'' and studying their
  beta-functions.} From the densities of the corresponding eigenvectors, one
  should be able to define the ``scaling operators'', whose correlation
  functions with respect to the fixed point interaction have exact scale
  invariance. One can also study correlation functions of simple operators
  such as $\psi^2$. While not exactly scale invariant, they should become so
  at asymptotically long distances.
  
  \item Full RG trajectory. By this we mean the theory which interpolates
  between the gaussian fixed point at short distances and the non-gaussian
  fixed point at long distances (for $\varepsilon > 0$, while for $\varepsilon
  < 0$ it is the other way around).
\end{itemizedot}
\subsubsection{Conformal invariance}\label{conf}

The RG fixed points constructed here are expected to be conformally invariant,
based on the same intuitive arguments as for the long-range bosonic models
{\cite{Paulos:2015jfa}}. Conformal invariance means the invariance of
correlation functions of scaling operators (see above) under the
finite-dimensional conformal group $\tmop{SO} (d + 1, 1)$. For $d = 1$ these
are M{\"o}bius transformations, and for $d = 2$ the product of holomorphic and
antiholomorphic M{\"o}bius transformations.\footnote{Because of the nonlocal
kinetic term, there will be no invariance under more general holomorphic
transformations, unlike in the case of fixed points of fully local models.}
This invariance also implies correspondence between correlation functions in
infinite volume as studied here, and correlation functions on a sphere of
finite radius (which for $d = 1$ is just a circle with periodic boundary
conditions), putting the two manifolds in correspondence via the stereographic
projection. Such properties are expected to be generally true based on
intuitive physics arguments, and it would be very interesting to see how they
emerge rigorously in an explicit model such as ours. In particular, this would
provide the first rigorous non-gaussian conformal theory in $d =
3$.\footnote{In the last 20 years, starting with Smirnov {\cite{SMIRNOV}},
there was significant progress in showing rigorously conformal invariance of
various critical observables of specific 2d lattice models (see
{\cite{Smirnov-review}} for review). Many of these models are exactly solvable
in infinite volume, and the main challenge is to show conformal invariance of
correlators defined in an arbitrary planar region (see e.g. {\cite{Chelkak}}
for the 2d Ising model). A key method used in these works is discrete
holomorphicity, which is limited to 2d and to specific models, while RG does
not play much of a role. The proof of conformal invariance of our RG fixed
points will require very different methods, which should work for any $d$.}

Conserved stress tensor operator plays key role in intuitive discussions of
conformal invariance of local theories. Our model being nonlocal (long-range),
it does not possess a local stress tensor in $d$ dimensions. One way around
this difficulty is to represent the nonlocal kinetic term as arising from a
local quadratic action in the $(d + 1)$-dimensional Anti-de-Sitter (AdS)
space, of which the $d$-dimensional space is the boundary, where the
quadratic, quartic, and all the irrelevant interactions are localized. This
construction is useful for intuitive understanding of conformal invariance (as
discussed for bosonic models in {\cite{Paulos:2015jfa}}), and perhaps also for
proving it rigorously.\footnote{On the contrary it is probably hard to make
rigorous sense of the Caffarelli-Silvestre construction from
{\cite{Paulos:2015jfa}}, where the higher-dimensional ambient space is flat,
but it has non-integer dimension.}

A key property of local conformally invariant theories is the convergent
Operator Product Expansion (OPE).\footnote{See {\cite{Rychkov:2016iqz}} for an
introduction for physicists, and {\cite{Rychkov:2020rcd}} for a more
mathematical one.} Though nonlocal (long-range), our model also should have
this property due to the local AdS representation.\footnote{Bosonic cousins of
our model have been studied via the numerical conformal bootstrap in
{\cite{Behan:2018hfx}}.} It would be very interesting to establish this
rigorously. This appears somewhat nontrivial due to the fact that the scaling
operators, introduced as densities of linearized RG eigenvectors (see above),
will not be exactly local but ``mildly nonlocal'', with kernels of \
stretched-exponential decay. It is puzzling why this mild nonlocality does not
invalidate the usual intuitive arguments for the OPE, which treat scaling
operators as living at a point. Somehow, the mild nonlocalities of the scaling
operators and of the fixed-point interaction should conspire to produce a
fully local OPE. Note that this issue is not specific to our model with a
long-range kinetic term, as mild nonlocality of scaling operators would be
present also for fully local models such as the 3D Ising model.

\subsubsection{Relations with analytic regularization}

Analytic dependence of our fixed point on $\varepsilon$ implies that the
critical exponents should also be analytic in $\varepsilon$. Analyticity of
the correlation-length and correction-to-scaling exponents $\nu$ and $\omega$
follows easily from the tree expansion construction (Appendix
\ref{App:trees}), as they can be computed by linearizing the analytic right
sides of {\eqref{nulambdaTree}} near the fixed point.\footnote{These series
have been computed, via another regulator, for the bosonic $O (N_b)$
long-range models in {\cite{Benedetti:2020rrq}} up to three loops. We thanks Dario Benedetti for sharing a Mathematica
notebook. Fermionic series should be obtainable by setting $N_b = - N$. These
series are not sufficiently long to test our claim that they are convergent in
the fermionic case.} The exponent $\eta$ is trivial in our model due to the
absence of wavefunction renormalization. Higher exponents may have a subtle
analytic structure because of degeneracies of linearized RG eigenvalues at
$\varepsilon = 0$. Since our model is non-unitary, some higher critical
exponents may become complex even for real $\varepsilon$, forming
complex-conjugate pairs.\footnote{This is similar to how some higher
Wilson-Fisher critical exponents become complex in $4 - \epsilon$ dimensions
{\cite{Hogervorst:2015akt}}.}

As already mentioned, it would be interesting to show that the critical
exponents are $\chi$-independent. Another problem is to prove rigorously that
our critical exponents agree with perturbative techniques by which these
exponents are computed in theoretical physics. This is especially interesting
given that, as we have shown, perturbation theory converges in the problem at
hand.

In theoretical physics, higher-order perturbative computations of critical
exponents are usually done by working with a bare action containing only the
relevant and nearly-marginal couplings. This uses the fact that, due to the
renormalizability of the theory at short distances, one can always find an RG
trajectory leading to the fixed point from such a UV theory where all
irrelevant couplings are set to zero. Furthermore, theoretical physics
calculations are greatly simplified by choosing a ``mass-independent
regularization scheme'', which allows to simply set the mass terms to zero.
Examples of such schemes are dimensional regularization and analytic
regularization with minimal subtraction, which amount to analytically continue
Feynman diagrams in $\varepsilon$, dropping the poles. It is universally
believed that any scheme, and in particular a mass independent one, should
give the same power series in $\varepsilon$ for the critical exponents, but to
our knowledge this has never been discussed in full rigor.

\subsubsection{Increasing the range of \texorpdfstring{$\varepsilon$}{varepsilon}}
\label{sec:increasing}

Existence proofs of fixed points in this paper work for $| \varepsilon |
\leqslant \varepsilon_0 (\gamma)$. We have not attempted to evaluate
$\varepsilon_0  (\gamma)$ explicitly, although it would be straightforward to
do this, following step-by-step our arguments. This may be a good exercise for
someone wishing to understand our methods in depth. Both of our methods
(contraction and the tree expansion) can be obtimized to enlarge the range of
$\varepsilon$ where the fixed point is under control. One simple strategy is
to increase the number of terms which are computed explicitly, or estimated
more carefully than what is currently done. For $\varepsilon$ of order 1, one
might have to resort to computer-assisted methods.\footnote{Inspired by
Lanford's construction of the Feigenbaum fixed point {\cite{lanford1982}}.}

An interesting feature which might be revealed by such exploration is the
cross-over to the short-range universality class. Namely consider the local
symplectic fermion model with the bare action (cf {\eqref{aMFT}})
\begin{equation}
  \int d^d x (\Omega^{}_{a b} \partial_{\mu} \psi^a \partial^{\mu} \psi^b +
  \nu \psi^2 + \lambda \psi^4) . \label{SRS}
\end{equation}
Some literature concerning such models was cited in the Introduction. This
model is expected to flow to a non-gaussian fixed point for $d = 3$ (although,
by the usual arguments, not for $d \leqslant 2$). This fixed point is strongly
coupled, and we cannot access it using the techniques of this
paper.\footnote{We could still prove a result like Key Lemma \ref{Key}, but
we would not be able to derive the Fixed Point Theorem \ref{FPT}, for lack of
a small parameter analogous to $\varepsilon$. Perhaps a computer-assisted
method could help.} Physicists study such fixed points by the usual
Wilson-Fisher $\epsilon$-expansion working formally in $d = 4 - \epsilon$ and
then resumming the series at $\epsilon = 1$. [We will use $\epsilon$ to denote
$4 - d$ as opposed to the long-range parameter $\varepsilon$.] As mentioned in
Appendix \ref{sec:I1I2}, $\epsilon$-expansion for these models is
perturbatively equivalent to the $\epsilon$-expansion of bosonic $O (N_b)$
models with $N_b = - N$.

So for $d = 3$ we have a family of long-range fixed points studied here whose
critical exponents depend on $\varepsilon$, and the fixed point of
{\eqref{SRS}} which we will call ``short-range''. The scaling dimension of
$\psi$ is $[\psi]_{\tmop{LR}} (\varepsilon) = d / 4 - \varepsilon / 2$ in our
models, while it is $[\psi_{}]_{\tmop{SR}} = d / 2 - 1 + \eta_{\tmop{SR}} / 2$
at the fixed point of {\eqref{SRS}}. The short-range $\eta_{\tmop{SR}}$ is
given by $\eta_{\tmop{SR}} = \epsilon^2 \frac{2 + N_b}{2 (8 + N_b)^2} + O
(\epsilon^3)$, $N_b = - N$, with the series which needs to be Borel-resummed
at $\epsilon = 1$.\footnote{This series is known up to 7 loops \cite{Schnetz:2016fhy,Schnetz-maple}, see
{\cite{Kompaniets:2017yct}} for the earlier 6 loop results. \label{spec}It is tempting to
speculate, by analogy with the long-range case, that the $\epsilon$-expansion
series have a finite radius of convergence for negative $N_b$, while they are
known to be only Borel-summable for positive $N_b$. Numerically, the 6-loop
series for $\nu_{\tmop{SR}}, \eta_{\tmop{SR}}, \omega_{\tmop{SR}}$ for $N_b =
- 4$ seem to be remarkably well behaved. E.g. $\eta_{\tmop{SR}} (N_b = - 4) =
- 0.25 \epsilon^3 - 0.25 \epsilon^2 - 0.535957 \epsilon^4 - 1.25122 \epsilon^5
- 3.14893 \epsilon^6$. We are grateful to Kay Wiese for communicating this to
us.}

The subsequent discussion applies for any $N$ for which
$[\psi_{}]_{\tmop{SR}} < d / 4$, as appears to be the case at least for $N =
4$ (see footnote \ref{spec}). For such $N$, we will have $[\psi]_{\tmop{LR}}
(\varepsilon) = [\psi_{}]_{\tmop{SR}}$ for $\varepsilon = \varepsilon_{\ast} =
2 (d / 4 - [\psi_{}]_{\tmop{SR}}) > 0$. It can then be conjectured that, for
$\varepsilon = \varepsilon_{\ast}$, the long-range to short-range crossover
will take place. Namely, the long-range fixed point at $\varepsilon =
\varepsilon_{\ast}$ should become identical to the short-range fixed-point
plus a non-interacting gaussian theory of an $\tmop{Sp} (N)$ symplectic
fermion $\zeta$ of scaling dimension $d - [\psi]_{\tmop{LR}}
(\varepsilon_{\ast})$. {This would be analogous to the bosonic
long-range models, for which such a crossover has been studied since a long
time theoretically (starting in {\cite{Fisher:1972zz,Sak}}, reviewed in
{\cite{Cardy}}, section 4.3) and is supported by Monte Carlo simulations
{\cite{Blote,Angelini}}. The extra gaussian field $\zeta$ is expected by the
same arguments as in {\cite{Behan:2017emf,Behan:2017dwr}} for the bosonic
case.\footnote{We will give three reasons: (1) Since the LR (long-range)
theory is nonlocal, the theory to which it crosses over cannot be fully local.
(2) (Counting of degrees of freedom) The leading spin 2 operator is not
conserved in the LR theory. At the crossover its dimension goes to $d$, which
is the stress-tensor dimension of the SR (short-range) fixed point. However
it's still not conserved by continuity, so its divergence represents extra
states not present in SR. (3) At the coalescence of LR and SR fixed points
there must be a marginal operator, on general grounds and because logarithmic
corrections are seen in Monte Carlo simulations. The SR fixed point by itself
does not have a marginal operator; it can however be constructed as $\psi
\zeta$. More arguments are given in {\cite{Behan:2017emf,Behan:2017dwr}} where
this picture was proposed and thoroughly tested for consistency.}} Note that
the operator $\psi \zeta$ is marginal for $\varepsilon = \varepsilon_{\ast}$
(it should be marginally irrelevant for the conjecture to hold). Furthermore,
$\varepsilon = \varepsilon_{\ast}$ marks the boundary of the region of
analyticity of the long-range fixed point, and for $\varepsilon >
\varepsilon_{\ast}$ the long-range fixed point with real couplings seizes to
exist. It would be extremely interesting to provide rigorous evidence for
these phenomena.

\subsubsection{Extension to non-integer \texorpdfstring{$N$}{N}? To non-integer \texorpdfstring{$d$}{d}?}

In quantum field theory, one often likes to continue the number of fields from
a positive integer, as it nominally should be, to an arbitrary real or even
complex value. \ For lattice models, such continuation often have geometric
meaning, as for the $O (N)$ and Potts models when it can be interpreted
respectively in terms of loops and Fortuin-Kastelein clusters. Symmetry
meaning of such continuations in terms of Deligne categories was recently
discussed in {\cite{Binder:2019zqc}}. We wish to discuss how such a
continuation can be rigorously performed for the model studied here. First one
has to factor out explicitly the products of $\Omega$-tensors out of the
couplings, i.e. write
\[ H_{2 k} (\bA, \bx) = \Omega_{a_1 a_2} \Omega_{a_3 a_4} \cdots \Omega_{a_{2
   k - 1} a_{2 k}} \tilde{H}_{2 k} (\widetilde{\bA}, \bx) \pm
   \tmop{permutations}, \]
where the kernels $\tilde{H}_{2 k}$ are ``$\Omega$-free'', i.e. no longer
depend on the $a$ indices, the sequence $\widetilde{\bA}$ containing only
$\mu$ indices. The RG equations can be rewritten in terms of such
$\Omega$-free kernels, with contractions of $\Omega$-tensors giving rise in
each order to some factors depending polynomially on $N$. One should then
study such $\Omega$-free fixed point equations. It is tempting to conjecture
that one can prove fixed point existence for any $N$ and its analytic
dependence on $N$.\footnote{If this is achieved, the coefficient $N-8$ of the one-loop beta function becomes a new small parameter for $N$ near 8. One could then work for $\varepsilon=0$ where the quartic is marginal, and construct a fixed point with $\lambda=O(N-8)$ balancing the one-loop term against the two-loop term which has an $O(1)$ coefficient. We are grateful to Dario Benedetti for this comment. This could then be done even in $d=4$, for a theory of local symplectic fermions \eqref{SRS}. This would be a rigorous version of the Banks-Zaks fixed points in 4d gauge theories \cite{Caswell:1974gg,Belavin:1974gu,Banks:1981nn}.}

More speculatively, one could also try to perform analytic continuation in the
space dimension $d$. One would have to use rotation invariance to come up with
a parametrization of the kernel in terms of scalar functions depending on
distances between points, times polynomials in point differences. Expressing
the RG equations in terms of scalar functions only, dimension $d$ becomes just
a parameter which can in principle be continued to non-integer values.
Controlling this continuation will likely require major changes in our
arguments (much more so than the continuation in $N$), since we relied on the
existence of the physical position space carrying a positive integration
measure in several crucial points. But the stakes are high: if one could prove
non-perturbative analyticity in $d$, it would be the first rigorous result of
this kind in almost 50 years since the space of 3.99 dimensions was ushered in by Wilson and Fisher {\cite{Wilson:1971dc}}.

\subsubsection{Connections to Functional Renormalization Group}\label{FRG}

FRG represents the most systematic attempt to implement Wilsonian RG in
absence of small parameter; see references in the introduction. We are not
aware of any FRG studies of specifically symplectic fermions (local or
long-range), although other fermionic models (such as Gross-Neveu, Thirring,
or Nambu-Jona-Lasinio), or mixed fermion-boson models with Yukawa
interactions have been studied via FRG-like techniques; see e.g.
{\cite{Jakovac:2014lqa,Gehring:2015vja,Gies:2017tod,Dabelow:2019sty}}.

Let us compare our results to FRG calculations. In our theorems, all
irrelevant couplings were included, and an RG fixed point was rigorously
located in a Banach space of interactions, with a provably convergent way to
approach it (for a wide range of cutoff procedures). Any FRG calculation
always truncates the space of interactions, so that only a subset of
irrelevant couplings is included (even if an infinite one). To our knowledge,
there are no rigorous results about the best way to exhaust the space of
interactions. What is done instead often looks like a matter of prejudice or
of convenience. E.g., for bosonic models with the field $\varphi$ one typically allows
an arbitrary potential $V (\varphi)$ but only a handful of derivatives of
$\varphi$, because the former is believed (although unproven) to be more
important, but also because an arbitrary potential is easy (the so called
local potential approximation), while derivatives are hard. This state of
affairs is both an invitation to mathematical physicists to weigh in and
provide rigorous criteria, and to FRG practitioners to explore different
exhaustion schemes. 

FRG experts may find instructive that our construction used nonlocal
interactions terms parametrized by kernels having fast decay at infinity.\footnote{See also \cite{Blaizot:2005wd,Blaizot:2006vr,Benitez:2011xx,Hasselmann:2012xw,Rose:2018tpn} for FRG setups allowing nonlocal momentum dependence in the vertices.} In
principle, our interactions could be expanded in the basis of local monomials
with fields carrying an arbitrarily high number of derivatives (the expansion
coefficients would be all finite because of the stretched exponential decay of
the kernels). However, we have not found such an expansion necessary. It is an
open question if rigorous RG analysis can be carried out with interactions
expanded in local monomials, and what would be the appropriate Banach space.

Another difference between our result and FRG is that we work with the full
Wilsonian effective action, while most FRG calculations are nowadays performed
in terms of the one-particle irreducible (1PI) effective action,
which flows according to the Wetterich equation {\cite{WETTERICH}}, as opposed
to Polchinski's equation.\footnote{See also a non-partisan review in
{\cite{Gurau:2014vwa}}, Chapter 5.} Empirically, this seems to give better
results. The 1PI effective action may be expected to be a somewhat more local
object than the Wilsonian effective action, but it too cannot be fully local.
We are not aware of any rigorous fixed point results in terms of the 1PI
effective action.

\subsubsection{Bosonic fixed points}\label{bosons}

While this paper deals with fermionic fixed point, most fixed points of
interest to physics do contain bosonic fields. A few available rigorous
bosonic fixed points are listed in Appendix \ref{literature}. Notably, they
include the bosonic analogues of the models that we studied here, i.e.
long-range bosonic $O (N)$ field theories with weakly relevant quartic bare
interactions. Unfortunately, these rigorous constructions remain rather
daunting, in spite of serious pedagogical work which went into trying to
simplify them (e.g. \cite{Bauerschmidt}). Further simplification is
desirable, however unlikely. A very accessible review can be found in
{\cite{Mitter:2005zx}}.

A major complication in the bosonic case, compared to the fermionic one, is
that, in defining the RG map $H \mapsto H'$, the terms involving fluctuation
fields $\varphi$ that are (on a local scale) large compared with their
standard deviation must be treated in a distinguished way: rather than dealing
with them via resummations of perturbation theory, they are controlled via
apriori bounds on the probability of such ``large fields'' configurations in
combination with a ``polymer expansion''\footnote{Loosely speaking, one
proceeds as follows: the volume is paved into boxes of typical length
comparable with the correlation length of the fluctuation field; each box is
called ``good'' or ``bad'' depending on whether the typical size of $\varphi$
in the box is smaller or larger than a large multiple of the standard
deviation, respectively; the probability of a bad box is bounded apriori and
proved to be very small in the perturbative parameter: therefore, bad boxes
are typically far apart from each other; in other words, they form a rarefied
gas, whose partition function can be computed via an analogue of the low
fugacity expansion for the pressure of lattice gases (this is the polymer
expansion which we referred to in the main text).}. These additional
small/large fields decomposition and polymer expansion add a whole new
combinatorial level to the construction, which inevitably leads to several
technical complications. To date, essentially all the rigorous works on the
construction of bosonic fixed points use a parameterization of the full
probability distribution of the form
\begin{equation}
  e^{H_{}} +\mathcal{P}_{}, \label{8.2}
\end{equation}
rather than of the more standard Gibbs form $e^H$: in {\eqref{8.2}}, $H_{}$
includes the relevant and marginal parts of the interaction, which are
exponentiated, while $\mathcal{P}$ includes the ``non-perturbative'' large
field contributions, which are kept non-exponentiated; this mixed form turns
out to be optimal for proving that the RG map preserves the space of
interactions. Assuming {\eqref{8.2}}, RG fixed point equation becomes $H'_{} =
H_{}, \mathcal{P}_{} =\mathcal{P}'$, whose form is quite different from (and
quite more involved than) the standard ``Exact RG equations'', such as
Polchinski's or Wetterich's, which typically neglect the contribution from the
polymer expansion of the large fields contributions (the
\ensuremath{\mathcal{P}}-term).

\section*{Acknowledgements}

We warmly thank Giuseppe Benfatto for his generous contributions to this work
in its preliminary phases. We thank David Brydges and Manfred Salmhofer for
communications about the rigorous status of Polchinski's equation, and Marco Serone for discussions about renormalons. We thank Malek Abdesselam, Dario Benedetti, David Brydges,  Krzysztof Gawedzki, Antti Kupiainen, Pronob Mitter, Gordon Slade, Kay Wiese for comments on the draft. 

SR is grateful to Accademia Nazionale dei Lincei for organizing the conference
``Advances in Mathematics and Mathematical Physics'' (Rome, September 2017)
where the first seeds of this project were planted, and to the organizers of
the program ``Renormalisation in quantum field theory'' (Isaac Newton
Institute, Cambridge, Fall 2018) where part of the work was done. AG is
grateful to IHES for hospitality in January-February 2020, where part of this
work was completed.

The work of AG and VM has been supported by: the European Research Council
(ERC) under the European Union's Horizon 2020 research and innovation
programme (ERC CoG UniCoSM, grant agreement n.724939), and by MIUR, PRIN 2017
project MaQuMA cod.~2017ASFLJR, which are gratefully acknowledged. SR is
supported by the Simons Foundation grant 488655 (Simons Collaboration on the
Nonperturbative Bootstrap), and by Mitsubishi Heavy Industries as an ENS-MHI
Chair holder.

\appendix\section{Gevrey classes and fluctuation propagator
bounds}\label{sec:gest}

In this appendix we give an explicit example of a compactly supported cutoff
function satisfying the Gevrey condition, and prove the stretched exponential
bound {\eqref{gbound0}} for the fluctuation propagator.

\subsection{Cutoff function of Gevrey class}\label{exGevrey}

Here we will explain that the set of cutoff functions satisfying conditions
{\eqref{chicond}} and {\eqref{chiG}} is not empty. Bump $C^{\infty}$ functions
being standard, we will explain how to satisfy in addition to
{\eqref{chicond}} the condition \
{\eqref{chiG}}$\Longleftrightarrow${\eqref{chiGdef}}, which we copy here:
\begin{equation}
  \sup_{k \in \mathbb{R}^d} | \del^{\alpha} \chi (k) | \leqslant C^n n^{ns},
  \qquad n = | \alpha | = 0, 1, 2 \ldots \label{eq:condX}
\end{equation}
We will not assume any knowledge about Gevrey classes; see e.g.
{\cite{rodino1993,GevreyAMS}} and {\cite{Dimers}}, App.~C.

Recall the following classic result for analytic function. Let $F (k)$ be a
function which allows an analytic continuation from real $k \in \mathbb{R}^d$
to a polydisk $D_R$, i.e. the region of complex $z \in \mathbb{C}^d$ such that
$|z_i - (k_0)_i | \leqslant R$ ($i = 1, \ldots, d$). Then, by the Cauchy
integral representation, the derivatives of $F (k)$ at the midpoint of the
polydisk are bounded by ($n = | \alpha |$)
\begin{equation}
  | \del^{\alpha} F (k_0) | \leqslant n!R^{- n} A, \qquad A = \max_{D_R} |F
  (z) | \hspace{0.17em} . \label{eq:Cauchy}
\end{equation}
By {\eqref{eq:Cauchy}}, an analytic $\chi (k)$ would satisfy
{\eqref{eq:condX}} with $s = 1$. However, by {\eqref{chicond}} our $\chi (k)$
is compactly supported, hence cannot be analytic. The best we can hope for is
{\eqref{eq:condX}} with $s > 1$.

\begin{figure}[h]\centering
  (a)\raisebox{-0.00323146640411002\height}{\includegraphics[width=5cm]{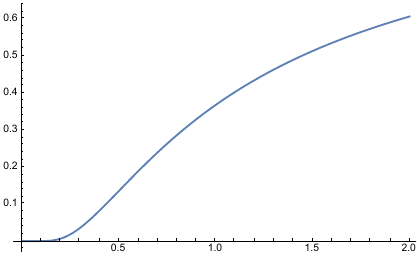}}\quad 
    (b)\raisebox{-0.0032286543069744\height}{\includegraphics[width=5cm]{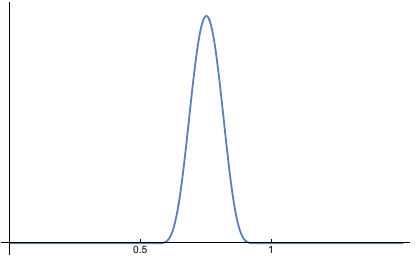}}\quad 
  (c)\raisebox{-0.00328102070503496\height}{\includegraphics[width=5cm]{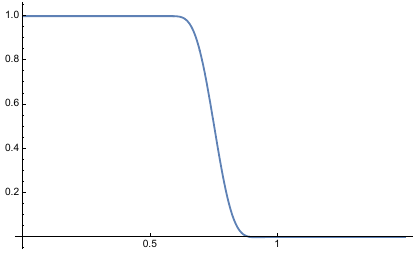}}
  \caption{\label{fig:X} (a) Function $X_0 (t)$ for $r = 1$; (b,c) The
  corresponding functions $X_1 (t)$ and $X (t)$. }
\end{figure}

Let us first construct a $d = 1$ example of a compactly supported Gevrey-class
function. Fix $r > 0$ and consider a $C^{\infty}$ function (see
Fig.~\ref{fig:X})
\begin{equation}
  X_0 (t) = \left\{\begin{array}{ll}
    0 & t \leqslant 0\\
    e^{- 1 / t^r}, & t > 0 \hspace{0.17em} .
  \end{array}\right. \label{X0}
\end{equation}
This function is not compactly supported, but this will be corrected below.
For now let us check that it is Gevrey class, namely that it satisfies the $d
= 1$ analogue of {\eqref{eq:condX}}:
\begin{equation}
  | \del^n_{} X_0 (t) | \le C^n n^{ns}, \qquad n = 0, 1, 2 \ldots
  \label{eq:condX0}
\end{equation}
with $s = 1 + 1 / r$. ($C$ stands for various positive constants which can
change from one line to the next.) Indeed, consider the function
\begin{equation}
  X_0  (t + z) = e^{- 1 / (t + z)^r} . \label{eq:X1}
\end{equation}
In the disk of complex $|z| < \kappa t$, where $\kappa > 0$ is sufficiently
small, this function is analytic and bounded in absolute value by $e^{- C /
t^r}$.\footnote{$\kappa$ here depends on $r$. We can choose it so that
\tmtextrm{Re}$[(1 + \zeta)^{- r}] > 1 / 2$ for $| \zeta | < \kappa$.} By the
$d = 1$ case of the Cauchy estimate {\eqref{eq:Cauchy}}, we have:
\begin{equation}
  | \del^n X_0 (t) | \le n! (\kappa t)^{- n} e^{- C / t^r},
\end{equation}
from where {\eqref{eq:condX0}} follows via elementary maximization of the
r.h.s. over $t$.

From $X_0 (t)$ which vanishes at $t \leqslant 0$, we pass to a function of
compact support $[1 / 2, 1]$:
\begin{equation}
  X_1 (t) = X_0  (t - 1 / 2) X_0  (1 - t) .
\end{equation}
By the Leibniz rule, it's easy to verify that $X_1 (t)$ also satisfies
{\eqref{eq:condX0}}. Finally, we put
\begin{equation}
  X (t) = \int_{| t |}^{\infty} X_1 (t') dt' \qquad
\end{equation}
which is constant for $| t | \leqslant 1 / 2$, vanishes for $| t | \geqslant
1$, and still satisfies {\eqref{eq:condX0}}. We rescale it so that $X (0) =
1$. See Fig.~\ref{fig:X}.

The function $X (t)$ is an explicit example of a cutoff function satisfying
conditions {\eqref{chicond}} and {\eqref{chiG}} in $d = 1.$ Infinitely many
examples of this sort can be given multiplying $X_0 (t)$ in {\eqref{X0}} by an
analytic function and repeating the construction.

The function $\chi (k)$ in $d$ dimensions will be given in terms of the 1d
function by
\begin{equation}
  \chi (k) = X (|k|) .
\end{equation}
While Eq. {\eqref{eq:condX}} can be verified using the chain rule, we will
instead give a more robust argument via analytic continuation and the Cauchy
estimate. The function $f (k) = |k|$ is real analytic for $| k | \in [1 / 2,
1]$ where derivatives of $X (t)$ are nonzero. Generally, a composition $X (f
(k))$ of a Gevrey class function $X (t)$ and a real analytic function $f (k)$
remains in the Gevrey class. For the proof, let $f (z)$ be analytic
continuation into a polydisk $|z_i - k_i | \leqslant R$ (we can choose $R = 1
/ 4$ for $f (k) = |k|$, $1 / 2 \leqslant |k| \leqslant 1$). Writing $X (f (z))
= X (f (k) + [f (z) - f (k)])$ and Taylor-expanding we have
\begin{equation}
  X (f (z)) = \sum_{i = 0}^{\infty} \frac{1}{i!}  \del^i X (f (k))  [f (z) - f
  (k)]^i \hspace{0.17em} . \label{eq:Taylor0}
\end{equation}
Suppose we want to compute $\del^{\alpha} [X (f (k))]$, $| \alpha | = n$. We
can compute this derivative by differentiating the Taylor series
{\eqref{eq:Taylor0}} truncated to $i \leqslant n$, since all terms with $i >
n$ are anyway higher order:
\begin{equation}
  \del_k^{\alpha} [X (f (k))] = \del_z^{\alpha} \Phi (z) |_{z = k}, \qquad
  \Phi (z) = \sum_{i = 0}^n \frac{1}{i!}  \del^i X (f (k)) \hspace{0.17em}  [f
  (z) - f (k)]^i .
\end{equation}
The function $\Phi (z)$ is analytic. It can be bounded in the polydisk by
\begin{equation}
  | \Phi (z) | \le C^n n^{n (s - 1)} \hspace{0.17em},
\end{equation}
using {\eqref{eq:condX0}} for $X (t)$, and that $f (z)$ is bounded in the
polydisk. From here using {\eqref{eq:Cauchy}} we get {\eqref{eq:condX}}.

\subsection{Fluctuation propagator bounds}\label{stretched}

\subsubsection{\texorpdfstring{$k$}{k}-space}

Recall that the Fourier transform of $g (x)$ is given by Eq.
{\eqref{eq:split}} which we copy here:
\begin{equation}
  \hat{g} (k) = [\chi (k) - \chi (\gamma k)] / |k|^{\frac{d}{2} + \e} .
\end{equation}
In this subsection we will show, using {\eqref{eq:condX}}, that, for any $k
\in \mathbb{R}^d$ and any $n = | \alpha | \geqslant 0$,
\begin{equation}
  | \del^{\alpha}  \hat{g} (k) | \leqslant C \frac{(C \gamma)^n n^{ns}}{|k|^{d
  / 2 + \eps}} . \label{eq:ghatbound}
\end{equation}
($C$ will denote $\gamma$- and $n$-independent constants which may change from
one equation to the next.)

We first estimate the derivatives of $1 / |k|^{d / 2 + \eps}$. Consider the
analytic continuation of $F (k) = 1 / |k|^{d / 2 + \eps}$ into the polydisk
centered at $k \neq 0$ of radius $R = \frac{1}{2} \max_i |k_i |$. The maximum
of $|F (z) |$ in this polydisk is bounded by $C F (k)$. We conclude by
{\eqref{eq:Cauchy}} that\footnote{For $d = 1$ bound {\eqref{eq:step1}} would
be easy to get by repeated differentiation deriving an explicit formula for
the l.h.s. The analytic continuation argument is more robust to show that the
same estimate is true for any $d$.}
\begin{equation}
  \left| \del^{\alpha} \frac{1}{|k|^{d / 2 + \eps}} \right| \leqslant n!R^{-
  n} C F (k) \leqslant \frac{C^n n!}{|k|^{d / 2 + \eps + n}} .
\end{equation}
Finally since $| k | \geqslant 1 / (2 \gamma)$ on $\tmop{supp} [\chi (k) -
\chi (\gamma k)]$ we conclude
\begin{equation}
  \left| \del^{\alpha} \frac{1}{|k|^{d / 2 + \eps}} \right| \leqslant \frac{(C
  \gamma)^n n!}{|k|^{d / 2 + \eps}}  \qquad \tmop{on} \tmop{supp} [\chi (k) -
  \chi (\gamma k)] \hspace{0.17em} . \label{eq:step1}
\end{equation}
Now let us prove {\eqref{eq:ghatbound}}. By the Leibniz rule we have
\begin{equation}
  \del^{\alpha}  \hat{g} (k) = \sum_{\beta \le \alpha} Q_{\beta} 
  \hspace{0.17em} \del^{\beta} \frac{1}{|k|^{d / 2 + \eps}} \times
  \del^{\alpha - \beta}  [\chi (k) - \chi (\gamma k)], \label{eq:ghat0}
\end{equation}
where $Q_{\beta} = \prod_{i = 1}^d \binom{\alpha_i}{\beta_i}$. We estimate the
$\del^{\beta}$ derivative on the support of $\chi (k) - \chi (\gamma k)$ via
{\eqref{eq:step1}}, while the $\del^{\alpha - \beta}$ factor by
{\eqref{eq:condX}} with $C \to C \gamma$. Combining these two estimates via
($n = | \alpha |$)
\begin{eqnarray}
  &  & (C \gamma)^{| \beta |}  (C \gamma)^{| \alpha - \beta |} = (C \gamma)^n
  \hspace{0.17em}, \nonumber\\
  &  & | \beta |^{| \beta |}  | \alpha - \beta |^{s | \alpha - \beta |}
  \leqslant n^{s | \beta |} n^{s | \alpha - \beta |} = n^{sn} \hspace{0.17em},  
\end{eqnarray}
and using that $\sum_{\beta} Q_{\beta} = 2^n$, we get {\eqref{eq:ghatbound}}.

\subsubsection{\texorpdfstring{$x$}{x}-space}\label{gxbound}

Finally we show {\eqref{gbound0}}. Consider first the bound for $g (x)$. We
use the standard trick that the Fourier transform of $(- ix)^{\alpha} g (x)$
is $\del^{\alpha}  \hat{g} (k)$, hence
\begin{equation}
  \sup_x |x^{\alpha} g (x) | \leqslant (2 \pi)^{- d} \| \partial^{\alpha}
  \hat{g} \|_{L^1} .
\end{equation}
Bound~{\eqref{eq:ghatbound}} then implies (note that $\hat{g} (k)$ has compact
support and that $\int_{|k| \le 1} \frac{d^d k}{|k|^{d / 2 + \eps}} < \infty$)
\begin{equation}
  \sup_x |x^{\alpha} g (x) | \leqslant C (C \gamma)^n n^{ns} \hspace{0.17em},
  \label{eq:xag}
\end{equation}
which can be rewritten as
\begin{equation}
  |g (x) | \leqslant C u^{- n} n^{n s} \equiv C e^{n s \log \frac{n}{u^{1 /
  s}}}, \qquad u = C | x | / \gamma . \label{eq:estgunif}
\end{equation}
From here {\eqref{gbound0}} for $g (x)$ follows by choosing $n$ optimally as
$n = \lfloor u^{1 / s} / e \rfloor$.

The Fourier transforms of the first and second derivatives of $g (x)$ are $g_1
(k) = k_{\mu}  \hat{g} (k)$ and $g_2 (k) = k_{\mu} k_{\nu}  \hat{g} (k)$.
Using {\eqref{eq:ghatbound}} for $\hat{g} (k)$, it's easy to see that $g_1
(k), g_2 (k)$ satisfy the same type of bounds. Thus the bounds for the first
and second derivatives of $g (x)$ follow by the same argument. In fact
derivative of any order will have the same kind of decay, only the constants
will degrade.

\begin{figure}[h]\centering
  \raisebox{-0.00192205900570987\height}{\includegraphics[width=6.98701298701299cm,height=4.36804735668372cm]{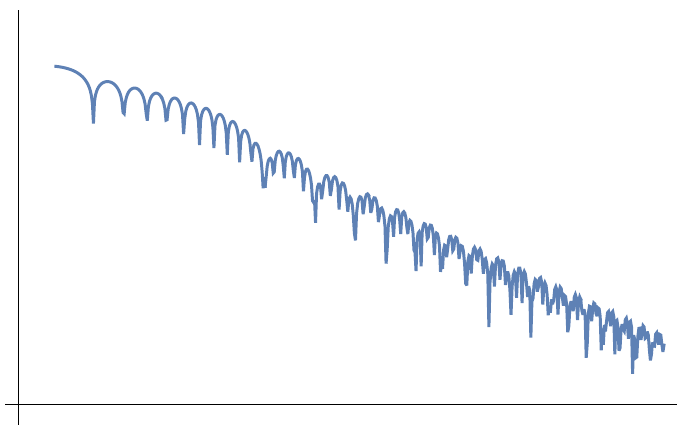}}
  \caption{\label{fig:FTX}The Fourier transform of the function $X (t)$ from
  Fig. \ref{fig:X}(c), plotted in log scale against $|x|^{1 / s}$ where $s = 1
  + 1 / r = 2$. The expected $\exp (- C |x|^{1 / s}$) decay is visible.}
\end{figure}

As an illustration, we plot in Fig.~\ref{fig:FTX} the numerically computed
Fourier transform of the function $X (t)$ from Fig.~\ref{fig:X}(c). The plot
shows the expected $\exp (- C |x|^{1 / 2}$) decay.

\section{Details about \texorpdfstring{$H_{\tmop{eff}}$}{Heff}}\label{sec:Heff}

In this appendix we give more details about the derivation of Eq.
{\eqref{eq:Hefflift}}. Plugging {\eqref{eq:HLtyp}} into {\eqref{eq:Heff2}}, we
represent $H_{\tmop{eff}} (\psi)$ as
\begin{eqnarray}
  H_{\eff} (\psi) & = & \sum_{n = 1}^{\infty} \frac{1}{n!}  \sum_{\bA_1,
  \ldots, \bA_n} \sum_{\tmscript{\begin{array}{c}
    \bB_1, \ldots, \bB_n\\
    \bB_i \subset \bA_i
  \end{array}}} (-)^{\#}  \label{eq:Heffwriteup}\\
  &  & \times \int d^d \mathrm{} \bx \hspace{0.17em} \Psi (\bB_1,
  \bx_{\bB_1}) \ldots \Psi (\bB_n, \bx_{\bB_n})  \prod_{i = 1}^n H (\bA_i,
  \bx_{\bA_i}) \left\langle \Phi (\barbB_1, \bx_{\barbB_1}) ; \ldots ; \Phi
  (\barbB_n, \bx_{\barbB_n}) \right\rangle_c . \nonumber
\end{eqnarray}
Let us explain this in words. We sum over even-length sequences $\bA_i$
indexing terms in $H$. We introduce a coordinate sequence $\bx$ of length $|
\bA_1 | + \ldots + | \bA_n |$ to be integrated over. We further sum over
subsequences $\bB_i \subset \bA_i$ selecting which fields inside the $H
(\bA_i, \bx_{\bA_i})$ term are external. The fields from the complements
$\barbB_i = \bA_i \setminus \bB_i$ are internal, to be contracted in the
connected expectation. We use $\bx_{\bA_i}$, $\bx_{\bB_i}$, $\bx_{\barbB_i}$
to denote the part of the vector $\bx$ for the corresponding subsequence. The
$(-)^{\#}$ is the sign, which we don't need to track, of the permutation
reordering sequence $\bA_1 + \ldots + \bA_n$ to $\bB_1 + \ldots + \bB_n +
\barbB_1 + \ldots + \barbB_n$.

\begin{figure}[h]\centering
  \raisebox{-0.00450791526528025\height}{\includegraphics[width=3.27010363374cm,height=1.86242293060475cm]{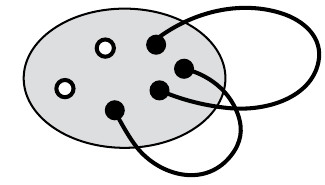}}
  \caption{\label{fig:n1} This figure represents a term in $H_{\eff}$
  corresponding to $n = 1$ in {\eqref{eq:Heffwriteup}}: just one vertex with
  $| \bA | = 6$ (gray oval). Empty circles are the external fields and filled
  circles are the internal ones. The generated $H_{\eff}$ term has $l = | \bB
  | = 2$. The 4 internal fields are contracted (just one possible contraction
  is shown).}
\end{figure}

\begin{figure}[h]\centering
  \raisebox{-0.00307232566652065\height}{\includegraphics[width=10.4721238357602cm,height=2.73266758494031cm]{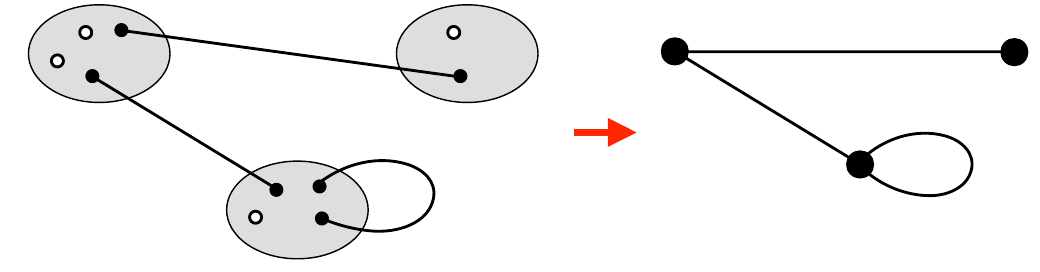}}
  \caption{\label{fig:n2} {\tmem{Left:}} A contraction of three vertices, $n =
  3$ in {\eqref{eq:Heffwriteup}}. The generated $H_{\eff}$ term has $l = 4$
  (the total number of empty circles). \tmtextit{Right:} The graph of
  contractions obtained when every gray oval is shrunk to a point (denoted by
  a fat dot). This graph is connected, as it should be because we are
  considering connected expectations.}
\end{figure}

There are several distinguished groups of terms in {\eqref{eq:Heffwriteup}}:
\begin{itemizedot}
  \item Terms with $n = 1$ and $\bB_1 = \bA_1$. They involve no contractions
  and their sum gives back $H$.
  
  \item Terms with $n = 1$ and $\bB_1 \ne \bA_1$. These involve a single $H
  (\bA, \bx)$ vertex with several fields identified as internal and contracted
  among themselves, while the rest remaining external (see
  Fig.~\ref{fig:n1}){\hspace{0.17em}}.
  
  \item Terms with $n \geqslant 2$, which therefore correspond to contractions
  of several vertices. Since {\eqref{eq:Heffwriteup}} involves connected
  expectation, we have to sum over contractions such that the graph becomes
  connected when every interaction vertex is shrunk to a point (see
  Fig.~\ref{fig:n2}).
  
  \item Terms for which all $\bB_k$ are empty, meaning that all fields are
  contracted. These sum up to a $\psi$-independent constant (infinite when
  working in infinite volume as we are). As mentioned in Section \ref{RenMap},
  footnote \ref{infconst}, this constant will be dropped.
\end{itemizedot}

Finally, Eq. {\eqref{eq:Hefflift}} follows by rewriting
{\eqref{eq:Heffwriteup}} in the form {\eqref{Hpsi}}.

\section{Trimming details}\label{sec:Trim}

This appendix deals with the trimming map introduced in Section
\ref{sec:trimming}, and with how it behaves with respect to the weighted
norms. This map takes the interaction $H_{\tmop{eff}}$ in a general
representation and returns an equivalent trimmed representation. Consider the
parts of $\mathcal{H}= H_{\tmop{eff}}$ which need to be set to zero:
$\mathcal{H}_{4, 0}$, $\mathcal{H}_{2, 0}$, $\mathcal{H}_{2, 1}$. Recall that
$\mathcal{H}_{l, p}$ corresponds to $l$-leg interactions with $p$ derivatives:
\begin{equation}
  \mathcal{H}_{l, p} \leftrightarrow \sum_{| \mathbf{A} | = l, d (\mathbf{A})
  = p} \int d^d \mathbf{x}\mathcal{H}_{} (\mathbf{A}, \mathbf{x}) \Psi_{}
  (\mathbf{A}, \mathbf{x}) .
\end{equation}
The dependence on $S p (N)$ indices should be given by possible invariant
tensors:
\begin{eqnarray}
  \mathcal{H}_{4, 0} & \leftrightarrow & \mathbf{A}= (a, b, c, e), \quad
  \mathcal{H}_{} (\mathbf{A}, \mathbf{x}) = \Omega_{a b} \Omega_{c e} F_1
  (\mathbf{x}) - \Omega_{a c} \Omega_{b e} F_2 (\mathbf{x}) + \Omega_{a e}
  \Omega_{b c} F_3 (\mathbf{x}) \\
  \mathcal{H}_{2, 0} & \leftrightarrow & \mathbf{A}= (a, b), \hspace{3em}
  \mathcal{H}_{} (\mathbf{A}, \mathbf{x}) = \Omega_{a b} G_{} (\mathbf{x}) \\
  \mathcal{H}_{2, 1} & \leftrightarrow & \mathbf{A}= ((a, \mu), b), \quad
  \mathcal{H}_{} (\mathbf{A}, \mathbf{x}) = \Omega_{a b} K_1^{\mu}
  (\mathbf{x}), \qquad  \label{HH21}\\
  &  & \mathbf{A}= (a, (b, \mu)), \quad \mathcal{H}_{} (\mathbf{A},
  \mathbf{x}) = \Omega_{a b} K_2^{\mu} (\mathbf{x}) . 
\end{eqnarray}
Recall that kernels $\mathcal{H}_{} (\mathbf{A}, \mathbf{x})$ are
antisymmetric, $\mathcal{H}_{} (\pi \mathbf{A}, \pi \mathbf{x}) = (-)^{\pi}
\mathcal{H}_{} (\mathbf{A}, \mathbf{x})$. This implies various symmetry
relations for the functions $F, G, K$. E.g. $G$ is symmetric, $F_1, F_2, F_3$
are all related by permutations of their arguments, and finally $K_1^{\mu}
(x_{1,} x_2) = K_2^{\mu} (x_2, x_1)$.

By spatial parity (see footnote \ref{opA}) we have $K_1^{\mu} (- x_{1,} - x_2)
= - K_1^{\mu} (x_{1,} x_2)$. Combined with translational and rotational
invariance this implies that $K_1^{\mu} (x_{1,} x_2) = (x_1 - x_2)^{\mu}
\bar{K} (| x_1 - x_2 |)$.

\textbf{\boldmath The map $T_{\text{2R}}^{2, 1}$} takes $\mathcal{H}_{2, 1}$ and returns an
equivalent interaction of $H_{2, 2}$ type. Consider the part of
$\mathcal{H}_{2, 1}$ with $\mathbf{A}= ((a, \mu), b)$. Using the interpolation
identity
\begin{equation}
  \psi_b (x_2) = \psi^{}_b (x_1) + \int_0^1 d t \partial_t [\psi_b (x_1 + t
  (x_2 - x_1))], \label{intid2}
\end{equation}
this term is mapped onto the sum of two terms. The first one is
\begin{equation}
  \Omega_{a b} \int d^d x_1 \left[ \int d^d x_2^{} K_1^{\mu} (x_{1,} x_2)
  \right] (\partial_{\mu} \psi_a \psi_b) (x_1),
\end{equation}
which vanishes because as mentioned above $K_1^{\mu} (x_{1,} x_2)$ is odd in
$x_1 - x_2$. So no local terms are generated in the case at hand. The second
term is
\begin{equation}
  \Omega_{a b} \int d^d x_1 d^d x_2 K_1^{\mu} (x_{1,} x_2) \partial_{\mu}
  \psi_a (x_1)  (x_2 - x_1)_{\nu} \int_0^1 d t \partial_{\nu} \psi_b (x_1 + t
  (x_2 - x_1)) . \label{T2Rdef}
\end{equation}
Changing integration variables from $d^d x_2$ to $d^d y$ with $y = x_1 + t
(x_2 - x_1)$ and doing the integral over $t$ we have an identity
\begin{equation}
  \int d^d x_2 K_1^{\mu} (x_{1,} x_2)  (x_2 - x_1)_{\nu} \int_0^1 d t
  \partial_{\nu} \psi_b (x_1 + t (x_2 - x_1)) = \int d^d y L_{}^{\mu \nu}
  (x_1, y) \partial_{\nu} \psi_b (y), \label{KL}
\end{equation}
with the help of which we rewrite {\eqref{T2Rdef}} as
\begin{equation}
  \Omega_{a b} \int d^d x_1 d^d y L_{}^{\mu \nu} (x_1, y) \partial_{\mu}
  \psi^{}_a (x_1) \partial_{\nu} \psi^{}_b (y) . \label{Lmn}
\end{equation}
which is an interaction of type $H_{2, 2}$ as promised, and we associate it
with $T_{\text{2R}}^{2, 1} (\mathcal{H}_{2, 1})$. The action on the part of \
$\mathcal{H}_{2, 1}$ with $\mathbf{A}= (a, (\mu, b))$ is analogous and we have
to add it to the previous result.

\textbf{\boldmath The maps $T_{\text{2L}}^{2, 0}$ and $T_{\text{2R}}^{2, 0}$} take
$\mathcal{H}_{2, 0}$ and return an equivalent interaction which is a sum of
$H_{\text{2L}}$ and $H_{2, 2}$ type interactions. Using the interpolation identity
{\eqref{intid2}}, an $\mathcal{H}_{2, 0}$ interaction is mapped to a sum of
two terms. The first term is the local quadratic interaction and we associate
it with $T_{\text{2L}}^{2, 0}$:
\begin{equation}
  T_{\text{2L}}^{2, 0} (\mathcal{H}_{2, 0}) = \nu \Omega_{a b} \int d^d x_1 (\psi_a
  \psi_b) (x_1), \qquad \nu = \int d^d x_2^{} G (x_{1,} x_2) .
\end{equation}
(By translational invariance the integral $\int d^d x_2^{} G (x_{1,} x_2)$ is
$x_1$-independent). The second term is
\begin{equation}
  \Omega_{a b} \int d^d \mathbf{x}\, G (x_{1,} x_2) \psi^{}_a (x_1) (x_2 -
  x_1)_{\nu} \int_0^1 d t \partial_{\nu} \psi^{}_b (x_1 + t (x_2 - x_1)),
\end{equation}
and similarly to {\eqref{Lmn}} we can write it after a change of variable and
$t$-integration as
\begin{equation}
  \Omega_{a b} \int d^d x_1 d^d y\, G_{}^{\nu} (x_1, y) \psi_a (x_1)
  \partial_{\nu} \psi_b (y) .
\end{equation}
This is of type $H_{2, 1}$ which we already considered. Acting on it with the
map $T_{\text{2R}}^{2, 1}$ we will get an equivalent interaction of type $H_{2, 2}$.
This final result is $T_{\text{2R}}^{2, 0} (\mathcal{H}_{2, 0})$.

\textbf{\boldmath The maps $T_{\text{4L}}^{4, 0}$ and $T_{\text{4R}}^{4, 0}$} are constructed
with the help of the interpolation identity {\eqref{ExInt}}, which maps
$\mathcal{H}_{4, 0}$ to an equivalent sum of two interactions, the first of
which defines $T_{\text{4L}}^{4, 0}$ as it is a local quartic interaction with the
coupling
\begin{equation}
  \lambda = \int_{x_1 = 0} d^d \mathbf{x} [F_1 (\mathbf{x}) + F_2 (\mathbf{x})
  + F_3 (\mathbf{x})] = 3 \int_{x_1 = 0} d^d \mathbf{x}F_1 (\mathbf{x}),
  \label{T4L}
\end{equation}
while the second term is an interaction of $H_{4, 1}$ type which is associated
with $T_{\text{4R}}^{4, 0} .$

We now consider \tmtextbf{weighted norm estimates} for the introduced maps.
Since $w \geqslant 1$, the localization maps have simply norm one (factor 3 in
{\eqref{T4L}} cancels with $1 / 3$ in {\eqref{H4Llambda}}):
\begin{equation}
  \| T_{\text{2L}}^{2, 0} (\mathcal{H}_{2, 0}) \|_w \leqslant \| \mathcal{H}_{2, 0}
  \|_w, \qquad \| T_{\text{4L}}^{4, 0} (\mathcal{H}_{4, 0}) \|_w \leqslant \|
  \mathcal{H}_{4, 0} \|_w . \label{TLfinal}
\end{equation}
On the other hand, due to the factors like $(x_2 - x_1)$ in {\eqref{T2Rdef}},
the interpolation maps will not preserve the norm $\| \cdot \|_w$. Let us aim
instead for an inequality of the type $\| T (\mathcal{H}) \|_{w'_{}} \leqslant
\tmop{Const} . \| \mathcal{H} \|_{w''_{}}$ where $w'$ is a slightly weaker
weight than $w''_{}$ (i.e. growing slower than $w''_{}$ at infinity).
Eventually we will choose $w'_{} = w (\cdot / \gamma)$, $w''_{} = w$.

\textbf{\boldmath For $T_{\text{2R}}^{2, 1}$}, we need to estimate the $w'_{}$-norm of
$L^{\mu \nu}$ in {\eqref{Lmn}} in terms of the $w''$-norm of $K_1^{\mu}$. The
relation between $L^{\mu \nu}$ and $K_1^{\mu}$ is encoded by the identity
{\eqref{KL}} which by translational invariance and renaming $\partial_{\nu}
\psi^b$ by $f$ can be written equivalently as
\begin{equation}
  \int d^d y L_{}^{\mu \nu} (0, y) f (y) = \int d^d x_{} K_1^{\mu} (0_, x_{})
  x_{}^{\nu} \int_0^1 d t f (t x) \label{KL0},
\end{equation}
where $f (y)$ is an arbitrary function. The actual expression for $L^{\mu
\nu}$ in terms of $K_1^{\mu}$ can be written by e.g. choosing $f (y) = \delta
(y - y_0)$ but we don't need it. We write the norm of $L^{\mu \nu}_{}$ as
\begin{eqnarray}
  \|L^{\mu \nu}_{} \|_{w'_{}} & = & \int \mathrm{} d^d y \hspace{0.17em}
  |L^{\mu \nu}_{} (0, y) | w'_{} (0, y) = \int \mathrm{} d^d y L_{}^{\mu \nu}
  (0, y) \Sigma (y) w'_{} (0, y) \nonumber\\
  & = & \int d^d x_{} K_1^{\mu} (0, x_{}) x^{\nu} \int_0^1 dt [\Sigma (y) w'
  (0, y)]_{y = tx_{}} \nonumber\\
  & \leqslant & \int d^d x_{} | K_1^{\mu} (0, x_{}) |  | x | w' (0, x), 
  \label{LKnorm}\\
  & \leqslant & C_1 \| K_1^{\mu} \|_{w''}, \nonumber
\end{eqnarray}
where in the first line we defined $\Sigma (y) = \sign L_{}^{\mu \nu} (0, y)$,
in the second line we used identity {\eqref{KL0}} with $f (y) = \Sigma (y) w'
(0, y)$, in the third line we used that $| \Sigma | = 1$ and assumed that the
weight $w'$ is monotonically increasing. Finally, in the last line we assumed
the inequality:
\begin{equation}
  | x | w' (0, x) \leqslant C_1 w'' (0, x) \qquad (x \in \mathbb{R}^d) .
  \label{C1tr}
\end{equation}
Multiplying the bound {\eqref{LKnorm}} by 2 to account for the contribution of
$K_2^{\mu}$, we conclude
\begin{equation}
  \| T_{\text{2R}}^{2, 1} (\mathcal{H}_{2, 1}) \|_{w'} \leqslant 2 C_1 \|
  \mathcal{H}_{2, 1} \|_{w''}
\end{equation}
\textbf{\boldmath For $T_{\text{2R}}^{2, 0}$}, very similar considerations will apply.
Recall that we have to apply the interpolation identity twice, and each time
we will pay a factor of $| x |$ in the weight function. So we get a bound
\begin{equation}
  \| T_{\text{2R}}^{2, 0} (\mathcal{H}_{2, 0}) \|_{w'} \leqslant C_2 \|
  \mathcal{H}_{2, 0} \|_{w''}
\end{equation}
under the condition
\begin{equation}
  | x |^2 w' (0, x) \leqslant C_2 w'' (0, x) \qquad (x \in \mathbb{R}^d) .
  \label{C2tr}
\end{equation}
\textbf{\boldmath For $T_{\text{4R}}^{4, 0}$}, the interpolation identity {\eqref{ExInt}}
will give rise to an extra factor $|x_2 | + |x_3 | + |x_4 |$ in the weight
function. We will therefore obtain:
\begin{equation}
  \| T_{\text{4R}}^{4, 0} (\mathcal{H}_{4, 0}) \|_{w'} \leqslant C_3 \|
  \mathcal{H}_{4, 0} \|_{w''},
\end{equation}
assuming
\begin{equation}
  (|x_2 | + |x_3 | + |x_4 |) w' (0, x_2, x_3, x_4) \leqslant C_3 w'' (0, x_2,
  x_3, x_4)  \qquad (x_2, x_3, x_4 \in \mathbb{R}^d) . \label{C3tr}
\end{equation}
Finally, we specialize to the case of interest for us: $w'' = w$, $w' = w
(\cdot / \gamma)$ where $w$ is our weight {\eqref{eq:ourW}}. We leave it as an
elementary exercise to show that, for $\gamma \geqslant 2$, inequalities
{\eqref{C1tr}}, {\eqref{C2tr}}, {\eqref{C3tr}} hold with $C_1, C_3 = O
(\gamma)$, $C_2 = O (\gamma^2)$ and the constants in the $O$ bounds depend
only on $C_w$ and $\sigma$ in \ {\eqref{eq:ourW}}. [Here $\gamma \geqslant 2$
is useful as $C_i$ would blow up in the limit $\gamma \rightarrow 1$, $w'
\rightarrow w''$.] Bounds {\eqref{TRfinal}} follow.

\section{Determinant bounds for fermionic expectations}\label{sec:GH}

This appendix discusses the determinant bounds (Gram-Hadamard and \
Gawedzki-Kupiainen-Lesniewski) for the simple and connected\footnote{Connected
expectations are referred to as `truncated' in mathematical physics.}
fermionic expectations. These are standard in mathematical physics, but will
be unfamiliar to most theoretical physicists, usually concerned with
computing, not bounding. These bounds are closely related to the Pauli
principle and, at a formal level, to the fermionic expectation being a
determinant (hence the name). We will also review the Brydges-Battle-Federbush
(BBF) formula, a clever integral representation for connected expectations,
useful to derive bounds (and perhaps for other things).

Classic sources (citing previous literature) are {\cite{Lesniewski:1987ha}}
for the bounds and {\cite{Brydges:1984vu}} for the BBF formula. Other
presentations are in {\cite{Gentile:2001gb,Giuliani_LesHouches,Vieri-book}}.

\subsection{Simple expectations}\label{sec:simple}

We are interested in the expectations of the fluctuation field $\phi_a^{} (x)$
which is a gaussian Grassmann field with the propagator $\langle \phi_a^{} (x)
\phi_b (y) \rangle = \Omega_{ab} g (x - y)$. We use the notation $\Phi \left(
\bA, \bx) \right.$ for field products as in {\eqref{products}}. The simple
fermionic expectations are given by, see {\eqref{2kfields}},
\begin{equation}
  \label{eq:usexp} \langle \Phi (\bA, \bx) \rangle \equiv \langle \Phi_{A_1}
  (x_1) \ldots \Phi_{A_{2 s}} (x_r) \rangle = \sum (-)^p \times \text{Wick
  contractions},
\end{equation}
where a Wick contraction is a product of $s$ propagators like
\begin{equation}
  \langle \Phi_{A_1} (x_1) \Phi_{A_2} (x_2) \rangle \ldots \langle \Phi_{A_{2
  s - 1}} (x_{r - 1}) \Phi_{A_{2 s}} (x_r) \rangle,
\end{equation}
or any other pairing where fields are ordered as $p (1) \ldots p (r)$ and
$(-)^p = \pm 1$ in {\eqref{eq:usexp}} is the sign of the corresponding
permutation $p$. \

Eq.~{\eqref{eq:usexp}} contains factorially many terms, but there are
cancellations because of the signs. To see this, we rewrite {\eqref{eq:usexp}}
as a determinant {\cite{Caianiello:1956pm}}. As a model, take gaussian
Grassmann fields $\xi$ and $\bar{\xi}$ with propagator
\begin{equation}
  \langle \xi (x) \bar{\xi} (y) \rangle = g (x - y) .
\end{equation}
Then,
\begin{equation}
  \langle \xi (x_1) \ldots \xi (x_s) \bar{\xi} (y_1) \ldots \bar{\xi} (y_s)
  \rangle = \pm \det M, \quad M_{ij} = g (x_i - y_j) .
\end{equation}
In the general case {\eqref{eq:usexp}}, fields $\phi_a^{}$ carry indices $a =
1 \ldots N$, and propagator is $\Omega_{ab} g (x - y)$. Renaming odd-$a$
fields as $\xi_{\alpha}$, even-$a$ as $\bar{\xi}_{\alpha}$, the
$\xi_{\alpha}$-$\bar{\xi}_{\alpha}$ pairs ($\alpha = 1 \ldots N / 2$) are
decoupled, with propagator $\propto \delta_{\alpha \alpha'}$. The number of
$\xi$ and $\bar{\xi}$ fields in the non-vanishing expectation must be the
same, let $x_i$ and $y_j$ be their coordinates. Then the expectation
{\eqref{eq:usexp}} is, up to a sign, the determinant of the $s \times s$
matrix:
\begin{equation}
  \det \calM, \quad \calM_{ij} = \delta_{\alpha_i \alpha_j} \Gamma_{ij}  (x_i
  - y_j) \hspace{0.17em} . \label{eq:ourM}
\end{equation}
Here $\Gamma_{ij}$ is either $g (x - y)$ or its derivative if some fields
carry derivatives. We will estimate it with the help of

\begin{lemma}[Gram-Hadamard inequality]
  For $(f_i)$, $(h_i)$ ($i = 1 \ldots s$) two lists of vectors in a Hilbert
  space, let $M_{ij}$ be the $s \times s$ matrix of their inner products:
  $M_{ij} = (f_i, h_j)$. Then $D (f, h) = \det M$ satisfies an upper bound:
  \begin{equation}
    |D (f, h) | \leqslant \prod_{i = 1}^s \|f_i \| \|h_i \| .
    \label{eq:GHbound}
  \end{equation}
\end{lemma}

{\tmem{Proof.}} For $h_i$ orthonormal, this holds interpreting the determinant
as the volume of parallelepiped formed by $f_i$ (this case is known as
Hadamard's inequality). By rescaling, the inequality remains true for $h_i$
orthogonal of arbitrary length. We will next reduce the general case to this
special case.

Out of general $h_i$, we build $\tilde{h}_i$ by Gram-Schmidt: $\tilde{h}_1 =
h_1$, $\tilde{h}_2 = h_2 - \alpha h_1 \perp \tilde{h}_1$ (projection of $h_2$
on the subspace orthogonal to $\tilde{h}_1$), $\tilde{h}_3$ projection of
$h_3$ on the subspace orthogonal to $\tilde{h}_1, \tilde{h}_2$ etc. By
properties of determinants:
\begin{equation}
  D (f, h) = D (f, \tilde{h}) .
\end{equation}
From the special case, the r.h.s. is bounded by $\prod \|f_i \| \| \tilde{h}_i
\|$, and $\| \tilde{h}_i \| \leqslant \|h_i \|$ since it's a projection.
Q.E.D.

To apply this result, we have to write the matrix elements {\eqref{eq:ourM}}
as products of vectors in a Hilbert space. Without indices and derivatives we
have
\begin{equation}
  g (x_i - y_j) = (f_i, h_j), \label{E8}
\end{equation}
introducing two families of $L^2$ functions in momentum space (a trick due to
{\cite{Feldman1986}}):\footnote{$\hat{g} (k)$ will be non-negative if $\chi
(k)$ is non-negative and monotonic, but we state this part of the argument for
a general complex $\hat{g} (k)$.}
\begin{equation}
  f_i (k) = e^{- ikx_i} | \hat{g} (k) |^{1 / 2} \hat{g} (k) / | \hat{g} (k) |,
  \quad h_j (k) = e^{- iky_j} | \hat{g} (k) |^{1 / 2}, \label{eq:fh}
\end{equation}
When some fields carry derivatives, we just include a factor $ik_{\mu}$ into
the corresponding function. Finally with indices, we view $f_i$ and $h_j$ as
vector functions, multiplying {\eqref{eq:fh}} by the unit vectors in the
directions $\alpha_i$, $\alpha_j$, whose inner product reproduces the
Kronecker $\delta_{\alpha_i \alpha_j}$. We just proved

\begin{lemma}[Gram-Hadamard bound]
  We have the bound
  \begin{eqnarray}
    &  & | \langle \Phi_{A_1} (x_1) \ldots \Phi_{A_{2 s}} (x_{2 s}) \rangle |
    \leqslant (C_{\mathrm{GH}})^s \quad \text{with} \nonumber\\
    &  & C_{\mathrm{GH}} = \max \left( \int \frac{d^d k}{(2 \pi)^d} | \hat{g}
    (k) |, \int \frac{d^d k}{(2 \pi)^d} (k_1)^2 | \hat{g} (k) | \right) . 
    \label{eq:GHcond}
  \end{eqnarray}
\end{lemma}

This bound is related to the Pauli principle, as can be seen from the
following alternative proof. We can represent $\Phi_A (x)$ as operators acting
on a Hilbert space (fermionic Fock space). Fermionic occupation numbers being
either zero or one, operators $\Phi_A (x)$ turn out to have a finite norm $\|
\Phi_A \|$, and expectation then grow at most as a power $\| \Phi_A \|^{2 s}$.
This should be contrasted with the bosonic case, when the operator norm would
have been infinite (even for a simple harmonic oscillator).

\begin{lemma}
  \label{CGHbound}Let $\hat{g} (k)$ be as in {\eqref{eq:split}}, with $\chi
  (k)$ satisfying {\eqref{chicond}}. Then $C_{\tmop{GH}}$ is uniformly bounded
  over $\gamma \geqslant 2$ and $\varepsilon \in T$ where $T \subset
  \mathbb{C}$ is any compact subset of the complex half-plane {\eqref{Tsubs}}.
\end{lemma}

\tmtextit{Proof}. We have $C_{\tmop{GH}} \leqslant \int_{| k | \leqslant 1}
d^d k (2 \pi | k |)^{- \left( d / 2 + \tmop{Re} \varepsilon \right)}$,
uniformly bounded since $\max_T \tmop{Re} \varepsilon < d / 6$ by
{\eqref{Tsubs}}.

\subsection{Connected expectations}\label{sec:conn}

Dividing the points $\bx$ into $n$ groups $\bx_1, \ldots, \bx_n$, connected
(also called `truncated') expectations are given by
\begin{equation}
  \label{eq:connexp} \langle \Phi (\bA_1, \bx_1) ; \ldots ; \Phi (\bA_n,
  \bx_n) \rangle_c = \sum (-)^p \times \text{connected Wick contractions} .
\end{equation}
Connected Wick contractions form a subset of terms from~{\eqref{eq:usexp}},
those for which the graph of propagators becomes connected when each group of
points $\bx_i$ is collapsed into one point. The signs $(-)^p$ are the same as
in {\eqref{eq:usexp}}.

Because $g (x)$ decays at infinity, connected expectations are small when any
two groups $\bx_i$ and $\bx_j$ get far apart. We need a bound incorporating
both this fact and the cancellations due to signs. This will be done via a
clever generalization of the determinant representation to connected
expectations.

To begin with, via {\eqref{eq:usexp}} and {\eqref{eq:connexp}}, the simple and
connected expectations satisfy two relations ($\Phi_i \equiv \Phi (\bA_i,
\bx_i)$). First, they coincide for a single group of points:
\begin{equation}
  \langle \Phi_i \rangle_c = \langle \Phi_i \rangle \label{eq:0step} .
\end{equation}
Second, simple expectation can be computed by partitioning $n$ group of fields
into subsets, taking products of connected expectations within each subset,
and summing over all ways of partitioning:\footnote{\label{Rossi}This formula is more rapid that  \eqref{eq:connexp} to compute connected expectations: one recursively expresses them via the usual expectations, which in turn are evaluated by the determinant formula \eqref{eq:ourM} \cite{Rossi}. This observation speeds up the Diagrammatic Monte Carlo algorithm from footnote \ref{DiagMC} \cite{Rossi} achieving polynomial complexity \cite{Polynomial}. One wonders if the BBF formula \eqref{eq:BBFfinal} below could give an alternative practical way to evaluate connected expectations.}
\begin{equation}
  \langle \prod_{i = 1}^n \Phi_k \rangle = \sum_{\Pi \in \text{partitions of }
  \{1 \ldots n\}} (-)^{\pi} \prod_{Y \in \Pi} \langle \Phi_{Y_1} ; \Phi_{Y_2}
  ; \ldots \rangle_c \hspace{0.17em} . \label{eq:altconn}
\end{equation}
The $(-)^{\pi}$ is the parity of the permutation bringing fields in the
r.h.s.~into the original order in the l.h.s. (it's not the same permutation as
in {\eqref{eq:connexp}}.

Reading {\eqref{eq:altconn}} from right to left, one can recursively computes
connected expectations from simple ones. E.g. for $n = 2, 3$ we have
\begin{eqnarray}
  \langle \Phi_1 \Phi_2 \rangle & = & \langle \Phi_1 ; \Phi_2 \rangle_c +
  \langle \Phi_1 \rangle \langle \Phi_2 \rangle,  \label{eq:k=2}\\
  \langle \Phi_1 \Phi_2 \Phi_3 \rangle & = & \langle \Phi_1 ; \Phi_2 ; \Phi_3
  \rangle_c + \langle \Phi_1 ; \Phi_2 \rangle_c \langle \Phi_3 \rangle +
  \langle \Phi_1 \rangle \langle \Phi_2 ; \Phi_3 \rangle_c \nonumber\\
  &  & + (-)^{N_2 N_3} \langle \Phi_1 ; \Phi_3 \rangle_c \langle \Phi_2
  \rangle_{} + \langle \Phi_1 \rangle \langle \Phi_2 \rangle \langle \Phi_3
  \rangle .  \label{eq:k=3}
\end{eqnarray}
In the r.h.s. we replaced $\langle \Phi_k \rangle_c = \langle \Phi_k \rangle$
by {\eqref{eq:0step}}. $N_k$ is the number of fields in $\Phi_k$. From
{\eqref{eq:k=2}} we find $\langle \Phi_1 ; \Phi_2 \rangle_c = \langle \Phi_1
\Phi_2 \rangle - \langle \Phi_1 \rangle \langle \Phi_2 \rangle$; substituting
this into {\eqref{eq:k=3}} we find $\langle \Phi_1 ; \Phi_2 ; \Phi_3
\rangle_c$; etc. So {\eqref{eq:altconn}} provides an alternative definition of
connected expectations, a useful starting point for what follows.

Let $m = \sum_{k = 1}^n | \mathbf{x}_k |$ be the total number of points. In
Section \ref{sec:simple} we wrote the simple expectation as a determinant of
the matrix $\calM$ defined in {\eqref{eq:ourM}}. Introducing auxiliary
Grassmann variables $\eta_i$ and $\bar{\eta}_j$ ($m / 2$ of each type), we
write it then as a Grassmann integral
\begin{equation}
  \langle \prod_{k = 1}^n \Phi (\bA_k, \bx_k) \rangle = \int \prod d \eta_i d
  \bar{\eta}_j  \hspace{0.17em} e^V \hspace{0.17em}, \label{eq:grrep}
\end{equation}
where we defined the potential function
\begin{equation}
  \label{eq:potVM} V = \sum_{i, j} \calM_{ij} \eta_i  \bar{\eta}_j
  \hspace{0.17em} .
\end{equation}
Here $i, j$ index the fields classified in Section \ref{sec:simple} as
$\xi_{\alpha_i} (x_i)$ and $\bar{\xi}_{\alpha_j} (y_j)$. Depending in which
group $\bx_k$ their positions $x_i$ and $y_j$ belong, we subdivide $V$ as
\begin{eqnarray}
  &  & V = \frac{1}{2}  \sum_{k, l = 1}^n V_{kl},  \label{eq:doublesum}\\
  &  & V_{kl} = \sum_{i, j : x_i \in \bx_k, y_j \in \bx_l} \calM_{ij} \eta_i 
  \bar{\eta}_j + (k \leftrightarrow l) \hspace{0.17em} .  \label{eq:Vkl}
\end{eqnarray}
Define $V (X)$ and $\psi (X)$ on any finite subset $X \subset \{1 \ldots n\}$
by
\begin{equation}
  V (X) = \frac{1}{2}  \sum_{k, l \in X} V_{kl}, \quad \psi (X) = e^{V (X)}
  \hspace{0.17em} .
\end{equation}
We can think of $V (X)$ as the total potential energy for a group of points
with pairwise interactions. Define connected part $\psi_c (X)$ recursively by
the following equations:
\begin{eqnarray}
  &  & \psi_c (X) = \psi (X) \quad \text{if } |X| = 1 \hspace{0.17em}, 
  \label{eq:conn11}\\
  &  & \psi (X) = \sum_{\Pi \in \text{partitions of } X} \prod_{Y \in \Pi}
  \psi_c (Y) .  \label{eq:conn12}
\end{eqnarray}
Crucially, the form these equations is such that integrating them in $\eta_i$,
$\bar{\eta}_j$, we land precisely on {\eqref{eq:0step}}, {\eqref{eq:altconn}}
(including the ($-)^{\pi}$ sign), provided we identify
\begin{equation}
  \langle \Phi_{Y_1} ; \Phi_{Y_2} ; \ldots \rangle_c = \int \prod d \eta_i d
  \bar{\eta}_j \psi_c (Y), \label{eq:connviapsi}
\end{equation}
where the integral is over the subset of Grassmann variables belonging to $Y$
(which means $x_i \in \bx_k, y_j \in \bx_l$, where $k, l \in Y$). Computing
connected expectations is thus reduced to finding $\psi_c (Y)$ in terms of
$V$. We will consider this problem in general, for an arbitrary symmetric
$V_{kl}$. That our $V_{kl}$ is given by Eq. {\eqref{eq:Vkl}} will become
important again only in Section \ref{sec:GKL}.

There is a standard formula for $\psi_c$:
\begin{equation}
  \label{eq:psistandard} \psi_c (X) = \sum_{G \in \text{connected graphs on
  $X$, } |G| = |X|} \prod_{kl \in G} (e^{V_{kl}} - 1)  \prod_{k \in X}
  e^{\frac{1}{2} V_{kk}} .
\end{equation}
But this is not very useful for our purposes: plugging it into
{\eqref{eq:connviapsi}} just gives back Eq.~{\eqref{eq:connexp}} (perhaps not
surprisingly as both {\eqref{eq:connexp}} and {\eqref{eq:psistandard}} involve
connected graphs; we leave the proof as an exercise.) The number of terms in
{\eqref{eq:psistandard}} is asymptotically $2^{\binom{n}{2}}$, since for large
$n$ almost all graphs on $n$ points are connected (e.g.~{\cite{Flajolet}},
Example II.15).

\subsection{Brydges-Battle-Federbush (BBF) formula}\label{sec:BBF}

We will now derive a remarkable formula for $\psi_c (X)$ with much fewer
terms. First two simplifying remarks: 1) The diagonal interactions $V_{kk}$
enter into $\psi_c (X)$ as a trivial common factor $\exp (\tfrac{1}{2} \sum_{k
\in X} V_{kk})$. So we will set them to zero and reinstate in the final
result. We consider a complete graph on $n$ points with pairs $kl$ as the
graph edges $e$, and we write
\begin{equation}
  V (X) = \sum_e V (e) \hspace{0.17em} .
\end{equation}
2) It is enough to aim for the equation
\begin{equation}
  e^{V (X)} = \sum_{Y \ni 1} e^{V (X \setminus Y)} \psi_c (Y),
  \label{eq:toaim}
\end{equation}
from which {\eqref{eq:conn12}} follows by iterating. Every term in the r.h.s.
has $Y$ and $X \setminus Y$ ``decoupled'' in the sense that no edges linking
them are involved. Let us describe a general decoupling mechanism.

Let $F (X)$ be any sum of pairwise interactions $F (e)$, and $Z \subset X$ a
subset we wish to decouple. We say that an edge $e$ ``exits $Z$'' (written $e
\dashv Z$) if it is of the form $e = kl$ where $k \in Z$, $l \in X \setminus
Z$. In other words, an edge exits $Z$ if it connects $Z$ to $X \setminus Z$.
Introduce a variable $s \in [0, 1]$ and a new pairwise interaction with
exiting edges rescaled by $s$, others left intact:
\begin{equation}
  F \Resc{s}{Z} (e) = \left\{\begin{array}{ll}
    sF (e) \hspace{0.17em}, & \text{if } e \dashv Z \hspace{0.17em},\\
    F (e) & \text{otherwise} .
  \end{array}\right.
\end{equation}
We also define $F \Resc{s}{Z} (X)$ summing this new interaction:
\begin{equation}
  F \Resc{s}{Z} (X) = \sum_{e \subset X} F \Resc{s}{Z} (e),
\end{equation}
Then $F \Resc{1}{Z} (X) = F (X)$ and $F \Resc{0}{Z} (X) = F (X \setminus Z) +
F (Z)$ is decoupled. Therefore we have
\begin{equation}
  e^{F (X)} = e^{F \Resc{0}{Z} (X)} + \int_0^1 ds \hspace{0.17em} \partial_s
  \exp F \Resc{s}{Z} (X) = e^{F (X \setminus Z)} e^{F (Z)} + \sum_{e \dashv Z}
  \int_0^1 ds \hspace{0.17em} F (e) \exp F \Resc{s}{Z} (X) . \label{eq:decmec}
\end{equation}
The first term is decoupled. In the second one we have a sum over exiting
edges. Below, each of these summands will be decoupled further with respect to
the sets $Z \sqcup e$ obtained by joining to $Z$ the outside vertex of the
edge $e$. ($\sqcup$ denotes ``vertex union'').

Think of $\Resc{s}{Z}$ as an operation, which can be applied repeatedly. E.g.
$F \Resc{s_1}{Z_1} \Resc{s_2}{Z_2}$ means first rescale by $s_1$ all edges
$e_1 \dashv Z_1$, then by $s_2$ all edges $e_2 \dashv Z_2$ by $s_2$. In
general these operations don't commute.

To derive {\eqref{eq:toaim}}, we first apply this general mechanism with $Z =
\{1\}$ consisting of one point, and $F (X) = V (X)$. We obtain
\begin{equation}
  e^{V (X)} = e^{V (X \setminus \{1\})} + \sum_{e_1 \dashv \{1\}} \int_0^1
  ds_1  \hspace{0.17em} V (e_1) \exp V \Resc{s_1}{\{1\}} (X) .
  \label{eq:apply0}
\end{equation}
The first term gives the first term in the r.h.s. of {\eqref{eq:toaim}}, with
$Y = \{1\}$ and $\psi_c (Y) = 1$. For each term in the sum over $e_1$, we
define the set $Z_{e_1} = \{1\} \sqcup e_1$ and decouple with respect to
$Z_{e_1}$, i.e.~apply {\eqref{eq:decmec}} for $F = V \Resc{s_1}{\{1\}}$ and
the rescaled interaction $F \Resc{s_2}{Z_{e_1}} = V \Resc{s_1}{\{1\}}
\Resc{s_2}{Z_{e_1}}$:
\begin{equation}
  e^{V \Resc{s_1}{\{1\}} (X)} = e^{V (X \setminus Z_{e_1})} e^{V
  \Resc{s_1}{\{1\}} (Z_{e_1})} + \sum_{e_2 \dashv Z_{e_1}} \int_0^1 ds_2 
  \hspace{0.17em} V \Resc{s_1}{\{1\}} (e_2)  \hspace{0.17em} \exp V
  \Resc{s_1}{\{1\}} \Resc{s_2}{Z_{e_1}} (X) . \label{eq:apply1}
\end{equation}
In the first term we used $F (X \setminus Z_{e_1}) = V (X \setminus Z_{e_1})$,
since $F = V$ outside $Z_{e_1}$. Plugging {\eqref{eq:apply1}} into
{\eqref{eq:apply0}} we get:
\begin{eqnarray}
  e^{V (X)} & = & e^{V (X \setminus \{1\})} \nonumber\\
  &  & + \sum_{e_1 \dashv \{1\}} e^{V (X \setminus Z_{e_1})}  \int_0^1 ds_1 
  \hspace{0.17em} V (e_1) \exp V \Resc{s_1}{\{1\}} (Z_{e_1}) \nn \nonumber\\
  &  & + \sum_{e_1 \dashv \{1\}} \sum_{e_2 \dashv Z_{e_1}} \int_0^1 ds_1 
  \hspace{0.17em} ds_2  \hspace{0.17em} V (e_1) V \Resc{s_1}{\{1\}} (e_2) 
  \hspace{0.17em} \exp V \Resc{s_1}{\{1\}} \Resc{s_2}{Z_{e_1}} (X) . 
  \label{eq:applytot}
\end{eqnarray}
To compare with {\eqref{eq:toaim}}, we rewrite the second line as
\begin{equation}
  \sum_{e_1 \dashv \{1\}} e^{V (X \setminus Z_{e_1})} \psi_c (Z_{e_1}), \qquad
  \psi_c (Z_{e_1}) = \int_0^1 ds_1  \hspace{0.17em} V (e_1) \exp V
  \Resc{s_1}{\{1\}} (Z_{e_1}) .
\end{equation}
This equation defines $\psi_c$ for sets of two points.

We then continue iterating. For each term in the third line of
{\eqref{eq:applytot}}, we define $Z_{e_1 e_2} = Z_{e_1} \sqcup e_2$ and
decouple with respect to this set. This allows us to rewrite the third line
as:
\begin{equation}
  \sum_{e_1 \dashv \{1\}} \sum_{e_2 \dashv Z_{e_1}} e^{V (X \setminus Z_{e_1
  e_2})}  \int_0^1 ds_1  \hspace{0.17em} ds_2  \hspace{0.17em} V (e_1) V
  \Resc{s_1}{\{1\}} (e_2)  \hspace{0.17em} \exp V \Resc{s_1}{\{1\}}
  \Resc{s_2}{Z_{e_1}} (Z_{e_1 e_2}),
\end{equation}
plus a triple integral term which we don't write. To compare this with
{\eqref{eq:toaim}}, we write it as
\begin{equation}
  \sum_{Z \ni \{1\}, |Z| = 3} e^{V (X \setminus Z)} \psi_c (Z) .
\end{equation}
We should define $\psi_c (Z)$ summing over all possible orders of adding edges
so that the final set $Z_{e_1 e_2} = Z$, which means the two added edges
should be picked from $Z$. We thus have
\begin{equation}
  \psi_c (Z) = \sum_{e_i \subset Z, e_1 \dashv \{1\}, e_2 \dashv Z_{e_1}}
  \int_0^1 ds_1  \hspace{0.17em} ds_2  \hspace{0.17em} V (e_1) V
  \Resc{s_1}{\{1\}} (e_2)  \hspace{0.17em} \exp V \Resc{s_1}{\{1\}}
  \Resc{s_2}{Z_{e_1}} (Z) . \label{eq:BBF3}
\end{equation}
Continuing to iterate, we obtain the following general formula for $|Z| = n$,
$Z \ni \{ 1 \}$, which involves $n - 1$ added edges and integrations:
\begin{eqnarray}
  \psi_c (Z) & = & \sum_{e_i \subset Z, e_1 \dashv \{1\}, e_2 \dashv Z_{e_1},
  e_3 \dashv Z_{e_1 e_2}, \ldots}  \label{eq:BBF}\\
  &  & \times \int_0^1 \prod_{k = 1}^{n - 1} ds_k  \left\{ V (e_1) V
  \Resc{s_1}{\{1\}} (e_2) V \Resc{s_1}{\{1\}} \Resc{s_2}{Z_{e_1}} (e_3) \cdots
  \right\} \exp V \Resc{s_1}{\{1\}} \Resc{s_2}{Z_{e_1}} \cdots \Resc{s_{n -
  1}}{Z_{e_1 e_2 \ldots e_{n - 2}}} (Z) . \nonumber
\end{eqnarray}
This is valid under the simplifying assumption $V_{kk} = 0$ made above, while
for general $V$ we must multiply by $\exp (\tfrac{1}{2} \sum_{k \in Z}
V_{kk})$. Another remark: definition {\eqref{eq:conn11}}, {\eqref{eq:conn12}}
shows that $\psi_c (Z)$, like $\psi (Z)$, must be symmetric with respect
to permutations. Eq.~{\eqref{eq:BBF}} is not manifestly symmetric as it
selects $\{1\} \in Z$, although of course it produces symmetric results, see
an example below. This lack of manifest symmetry will not be a problem in our
applications.

\begin{example} It's instructive to check this formula for three points:
  $Z = \{1, 2, 3\}$. We thus have three edges $e = 12, 13, 23$. The rescaled
  interactions are ($s_1 = s, s_2 = t$):
  \begin{equation}
    V \Resc{s}{\{1\}} :
    \raisebox{-0.435643056993226\height}{\includegraphics[width=2.10786435786436cm,height=1.59997704315886cm]{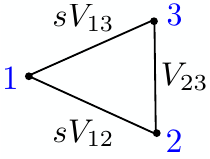}}
    \qquad V \Resc{s}{\{1\}} \Resc{t}{\{12\}} :
    \raisebox{-0.395218860088639\height}{\includegraphics[width=2.28210678210678cm,height=1.58725239407058cm]{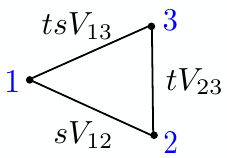}}
  \end{equation}
  Thus the term in {\eqref{eq:BBF3}} corresponding to $e_1 = 12$ reads:
  \begin{equation}
    \int_0^1 ds \hspace{0.17em} dt \hspace{0.17em} V_{12}  (sV_{13} + V_{23})
    e^{sV_{12} + tsV_{13} + tV_{23}} \hspace{0.17em} .
  \end{equation}
  For $e_1 = 13$ we get the same integral with $2 \leftrightarrow 3$. Doing
  the two integrals and summing, we obtain the same expression as by expanding
  out {\eqref{eq:psistandard}}:
  \begin{equation}
    \psi_c (123) = 2 - e^{V_{12}} - e^{V_{13}} - e^{V_{23}} + e^{V_{12} +
    V_{13} + V_{23}} \hspace{0.17em} .
  \end{equation}
\end{example}

Continuing with the general case, it is not hard to write out all parts of
{\eqref{eq:BBF}} explicitly. Define a function $g : \{1, \ldots, n\} \to \{1,
\ldots, n\}$ which tells us the order of adding vertices when adding edges
$e_l$: $g (1) = 1$ and $g (l)$ is the endpoint of the edge $e_{l - 1}$. So
$e_{l - 1} = g (k) g (l)$ with some $k < l$. In the exponential of
{\eqref{eq:BBF}} the contribution $V_e$ of this edge appears rescaled by
\begin{equation}
  r_e = s_k s_{k + 1} \cdots s_{l - 1}, \label{eq:pref}
\end{equation}
while in the prefactor $\{\ldots\}$ $V_{e_{l - 1}}$ appears rescaled by
\begin{equation}
  s_k s_{k + 1} \cdots s_{l - 2}
\end{equation}
(or 1 if $k = l - 1$). This expression is consistent with applying $\del_{s_{l
- 1}}$ to {\eqref{eq:pref}}, as it must be by {\eqref{eq:decmec}}.

It follows from this discussion that every summand in {\eqref{eq:BBF}} can be
written as
\begin{equation}
  \prod_{k = 1}^{n - 1} V_{e_k} \times \int_0^1 \prod_{k = 1}^{n - 1} ds_i f
  (\mathbf{s}) \exp \sum_e r_e V_e \label{eq:BBFrewr},
\end{equation}
where $r_e$ are given by {\eqref{eq:pref}} and $f (\mathbf{s})$ is some
product of $s$'s. Its dependence on the choice of added edges can be made
completely explicit, but for us it will suffice to know that $f \ge 0$.
Furthermore, we would like to view the integral in {\eqref{eq:BBFrewr}} as
performed over $r_e$'s, not over $s$'s, writing it as
\begin{equation}
  \prod_{k = 1}^{n - 1} V_{e_k} \times \int d \mu (\mathbf{r}) \exp \sum_e r_e
  V_e \label{eq:BBFrewr1},
\end{equation}
where $d \mu (\mathbf{r})$ is some nonnegative measure, push-forward of the
measure $\prod_{k = 1}^{n - 1} ds_k f (\mathbf{s})$ to $r_e$'s. This singular,
delta-function-like, measure is concentrated on $r_e$'s which can be
represented as {\eqref{eq:pref}}.

For the final repackaging, consider the graph which is the union of all added
edges:
\begin{equation}
  T = e_1 \cup e_2 \cup \ldots \cup e_{n - 1} .
\end{equation}
By construction, this is a tree with $n$ vertices $\{1 \ldots n\}$. Note that
the same tree $T$ may appear from different sequences of edges, all satisfying
the constraints $e_1 \dashv \{1\}, e_2 \dashv Z_{e_1}$, etc in
{\eqref{eq:BBF}}. E.g.~the following tree may arise from $\{e_1, e_2, e_3 \} =
\{13, 12, 24\}$ or $\{12, 13, 14\}$ or $\{12, 24, 13\}$:
\begin{equation}
  \raisebox{-0.00613982491905504\height}{\includegraphics[width=2.80488324806507cm,height=1.36740784468057cm]{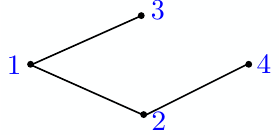}}
\end{equation}
Terms corresponding to the same tree $T$ have the same prefactor $\prod_{k =
1}^{n - 1} V_{e_k}$ in {\eqref{eq:BBFrewr1}}, although the measures $d \mu
(\mathbf{r})$ will be different. Let us group all such terms into one term per
tree $T$, summing their measures into some total measure. Reintroducing as
well the trivial $\exp (\tfrac{1}{2} \sum V_{kk})$ factor, we obtain the
remarkable BBF formula {\cite{Battle:1982pv,Brydges:1978tn}}; our presentation
followed {\cite{Brydges:1984vu}}.

\begin{lemma}[BBF formula]
  Let $T$ run through all trees with $n$ vertices $\{1 \ldots n\}$. There
  exist non-negative measures $d \mu_T (\mathbf{r})$ so that, for any pairwise
  interaction $V$:
  \begin{equation}
    \psi_c  (1 \ldots n) = \sum_T \prod_{e \in T} V_e  \int d \mu_T
    (\mathbf{r}) \exp (\sum_e r_e V_e + \tfrac{1}{2} \sum_{k = 1}^n V_{kk}) .
    \label{eq:BBFfinal}
  \end{equation}
  In addition, these measures have the following two properties:
  \begin{itemize}
    \item For each $\mathbf{r}$ in the support of $d \mu_T (\mathbf{r})$,
    there exists a bijection $g : \{1, \ldots, n\} \to \{1, \ldots, n\}$ and a
    set of $n - 1$ numbers $s_k \in [0, 1]$ such that for all $k < l$
    \begin{equation}
      r_{g (k) g (l)} = s_k s_{k + 1} \cdots s_{l - 1} . \label{eq:pref1}
    \end{equation}
    \item The measures $d \mu_T (\mathbf{r})$ are probability measures, i.e.
    they have total weight 1: $\int d \mu_T (\mathbf{r}) =
    1${\hspace{0.17em}}.
  \end{itemize}
\end{lemma}

The first property is clear, since $d \mu_T (\mathbf{r})$ were obtained as
sums of such measures. The second property can be checked by a trick: apply
the general formula to a particular conveniently chosen potential $V$. Pick a
tree $T$ and consider $V$ such that $V_e = \eps$ for $e \in T$ and $V = 0$
otherwise. That $\int d \mu_T (\mathbf{r}) = 1$ follows by comparing, for
small $\eps$, the equations
\begin{equation}
  \psi_c  (1 \ldots n) = \left\{\begin{array}{ll}
    \eps^n  \int d \mu_T (\mathbf{r}) + \text{higher order} \hspace{0.17em} &
    \text{from {\eqref{eq:BBFfinal}}} \hspace{0.17em},\\
    \eps^n + \text{higher order} & \text{from {\eqref{eq:psistandard}}} .
  \end{array}\right.
\end{equation}
\begin{remark}
  The measure $d \mu_T$ may be thought as having several components,
  corresponding to different bijections $g$, whose number equals the number of
  ways to grow the tree $T$ by adding edges. Each component can be pulled back
  to an integral over $s_k$'s, and the total weight of all components is 1.
  This can also be verified by an explicit computation (Lemma A.4 in
  {\cite{Gentile:2001gb}} or Lemma 2.3 in {\cite{Vieri-book}}).
\end{remark}

The number of terms in the BBF formula is much smaller than in the standard
formula {\eqref{eq:psistandard}}: it grows as the number of trees on $n$
points, which is $n^{n - 2}$ (Cayley).

We we will use the BBF formula to prove bounds on the connected fermionic
expectations. One wonders if this formula can also be useful to
{\tmem{evaluate}} connected expectations, e.g.~performing the integral
numerically, rather than just prove bounds. We are not aware of such
applications (see also footnote \ref{Rossi}).

\subsection{Gawedzki-Kupiainen-Lesniewski (GKL) bound}\label{sec:GKL}

We will now present a bound on fermionic connected expectations due to
Gawedzki and Kupiainen {\cite{Gawedzki:1985ez}}. Its physical origin, like for
the Gram-Hadamard bound {\eqref{eq:GHcond}}, is the Pauli principle. Our
exposition follows Lesniewski {\cite{Lesniewski:1987ha}}, who gave an elegant
proof based on the BBF formula.\footnote{Original proofs
{\cite{Gawedzki:1985ez}}, {\cite{Feldman1986}} were based on an improved
Gram-Hadamard inequality for the simple expectations, transferred to connected
expectations via cluster expansion techniques. See also {\cite{Salmhofer2000}}
for an alternative approach.}

We start with Eq.~{\eqref{eq:connviapsi}} copied here for $Y = \{1, \ldots,
n\}$:
\begin{equation}
  \langle \Phi (\bA_1, \bx_1) ; \ldots ; \Phi (\bA_n, \bx_n) \rangle_c = \int
  \prod d \eta_i d \bar{\eta}_j \psi_c  (1 \ldots n) . \label{eq:connviapsi1}
\end{equation}
Recall that we introduced the potential function {\eqref{eq:potVM}} (see
{\eqref{eq:ourM}} for matrix $\calM$) subdivided in symmetric pairwise
interactions $V_{kl}$, $1 \leqslant k, l \leqslant n$, see
{\eqref{eq:doublesum}}, {\eqref{eq:Vkl}}. The BBF formula
{\eqref{eq:BBFfinal}} gives a general expression for $\psi_c  (1 \ldots n)$ in
terms of $V_{kl}$.

What happens with every term in the BBF formula when we plug in expressions
{\eqref{eq:Vkl}} for $V_{kl}$ and do the Grassmann integral? First of all let
us look at the prefactor $\prod_{e \in T} V_e$. By {\eqref{eq:Vkl}}, every
$V_e = V_{kl}$ is a sum of terms $\calM_{ij} \eta_i  \bar{\eta}_j$ where $x_i
\in \bx_k$, $y_j \in \bx_l$ or vice versa and $\calM_{ij}$ is a propagator,
see {\eqref{eq:ourM}}. We choose in each $V_e$, $e \in T$, one of such
possible propagator terms. The graph with all chosen propagators as edges is
called an ``anchored tree $\mathcal{T}_{}$ on the $n$ groups of points
$\bx_i$''. When each group of points $\bx_i$ is contracted to one points, the
anchored tree $\mathcal{T}_{}$ becomes a tree (the tree $T$ in the case at
hand). We will say that ``$\mathcal{T}$ comes from $T$''.

Let us look at a term corresponding to a fixed anchored tree $\mathcal{T}$
coming from $T$. The variables $\eta$ and $\bar{\eta}$ of the vertices along
the anchored tree are saturated by the prefactor. Doing the Grassmann integral
over these ``tree-saturated'' variables gives a product of $n - 1$ propagators
along $\mathcal{T}$:
\begin{equation}
  \prod_{\text{along } \mathcal{T}_{}} \Gamma_{ij}  (x_i - y_j) .
  \label{Gammapref}
\end{equation}
We are left with the integral of the exponent over the remaining $m - 2 (n -
1)$ variables.\footnote{Recall that $m = \sum_{k = 1}^n | \mathbf{x}_k |$ is
the total number of points, and we have one Grassmann variable per point.} (By
the rules of Grassmann integration, the tree-saturated variables can now be
set to zero in the exponent.) The potential function in the exponent of
{\eqref{eq:BBFfinal}} will have the form
\begin{equation}
  \sum_{ij} \calN_{ij} \eta_i  \bar{\eta}_j, \qquad \calN_{ij} = r_{k (i) k
  (j)} \calM_{ij}, \label{NM}
\end{equation}
summing over the remaining variables, so $\calN = \calN (\mathbf{r})$ is an $s
\times s$ matrix with $s = \half  (m - 2 (n - 1))$. If variables $i$, $j$
belong to two different groups of points $\bx_{k (i)}$, $\bx_{k (j)}$, then
$r_{k (i) k (j)} = r_e$ in {\eqref{NM}}, where $e = k (i) k (j)$ is the edge
of the tree $T$, progenitor of the anchored tree, and $r_e$ is the rescaling
factor in the BBF formula. Terms from the same group, coming from $\frac{1}{2}
\sum V_{kk}$ in {\eqref{eq:BBFfinal}}, should not be rescaled: so we set
$r_{kk} = 1$. The Grassmann integral over the remaining variables is then the
determinant of the so defined matrix $\calN$. To summarize, the connected
expectation {\eqref{eq:connviapsi1}} can be represented as
\begin{equation}
  \sum_{\mathcal{T}_{}} \prod_{\text{along } \mathcal{T}_{}} \Gamma_{ij}  (x_i
  - y_j) \int d \mu_T (\mathbf{r}) \det \calN . \label{eq:detNrep}
\end{equation}
Aiming to bound \tmtextrm{det}$\hspace{0.17em} \calN$ by the Gram-Hadamard
inequality {\eqref{eq:GHbound}}, we wish to represent $\calN_{ij}$ as a
product of vectors in a Hilbert space. For $\calM_{ij}$ such a representation
was given in Section \ref{sec:simple}. To deal with the extra factor $r_{k (i)
k (j)}$ we will use

\begin{lemma}
  Let $s_k \in [0, 1], k = 1 \ldots n - 1$. There exist $n$ unit vectors $u_k
  = u_k (\mathbf{s}) \in \mathbb{R}^n$ such that $(u_k, u_l) = s_k \ldots s_{l
  - 1}$ for all $1 \leqslant k < l \leqslant n$.
\end{lemma}

\tmtextit{Proof.} Let $v_k$ be the standard orthonormal basis in
$\mathbb{R}^n$. We put $u_1 = v_1$. Take $u_2$ the unit-length linear
combination of $u_1$ and $v_2$ which has projection $s_1 u_1$ on $V_1 =
\mathrm{span} (v_1)$, explicitly $u_2 = s_1 u_1 + (1 - s_1^2)^{1 / 2} v_2$.
Take $u_3$ the unit-length linear combination of $u_2$ and $v_3$ which has
projection $s_2 u_2$ on $V_2 = \mathrm{span} (v_1, v_2)$, explicitly $u_3 =
s_2 u_2 + (1 - s_2^2)^{1 / 2} v_3$. Continuing in this fashion, we end up with
a sequence of unit vectors $u_k \in V_k$ whose orthogonal projections on the
previous $V_{k - 1}$ are
\begin{equation}
  P_{V_{k - 1}} (u_k) = s_{k - 1} u_{k - 1} .
\end{equation}
Computing $(u_k, u_l)$, $k < l$, via orthogonal projections $u_l \to V_{l - 1}
\to V_{l - 2} \ldots$ gives precisely $s_k \cdots s_{l - 1}$. Q.E.D.

Now for any component of $d \mu_T$ measure, i.e.~one particular bijection $g$
in the BBF formula, we satisfy {\eqref{eq:pref1}} via
\begin{equation}
  r_{kl} = (u_{g^{- 1} (k)}, u_{g^{- 1} (l)}),
\end{equation}
which is also symmetric in $k, l$ and $r_{kk} = 1$. Finally the rescaling
factor in the $\calN$ matrix:
\begin{equation}
  r_{k (i) k (j)} = (u_{g^{- 1} (k (i))}, u_{g^{- 1} (k (j))}) .
\end{equation}
In Section \ref{sec:simple} we showed that $\calM_{ij} = (f_i, h_j)$ where $f,
h$ are vectors in a Hilbert space. Considering the tensor product of those
vectors with $u_{g^{- 1} (k (i))} \in \bR^n$, we obtain an inner product
representation for $\calN_{ij}$ elements. Since $u$'s have unit length, by the
same argument which led to {\eqref{eq:GHcond}} we obtain a bound with the same
constant $C_{\mathrm{GH}}$:
\begin{equation}
  | \det \calN | \leqslant (C_{\mathrm{GH}})^s, \quad s = \half  (m - 2 (n -
  1)) \hspace{0.17em} .
\end{equation}
This is true for any $r_e$ lying in the support of the measure $d \mu_T$. We
can also integrate this bound since $d \mu_T$ has weight 1. We conclude that
the connected expectation is bounded by
\begin{equation}
  (C_{\mathrm{GH}})^s \sum_T \sum_{\text{$\mathcal{T}$ comes from $T$}} \left|
  \text{Eq. {\eqref{Gammapref}}} \right| = (C_{\mathrm{GH}})^s_{}
  \sum_{\text{$\mathcal{T}$}} \left| \text{Eq. {\eqref{Gammapref}}} \right| .
\end{equation}
where we used that every anchored tree comes from one and only one tree. We
finally get:

\begin{lemma}[GKL bound]
  Fermionic connected expectations are bounded by
  \begin{equation}
    | \langle \Phi (\bA_1, \bx_1) ; \ldots ; \Phi (\bA_n, \bx_n) \rangle_c |
    \leqslant (C_{\mathrm{GH}})^s  \sum_{\mathcal{T}_{}} \prod_{\text{along }
    \mathcal{T}_{}} | \Gamma_{ij} (x_i - y_j) |, \label{eq:GKL}
  \end{equation}
  where $s = \half  \sum_{i = 1}^n | \bx_i | - (n - 1)$, $C_{\mathrm{GH}}$ is
  from {\eqref{eq:GHcond}}, the sum is over all anchored trees $\mathcal{T}$
  on $n$ groups of points $\bx_k$, and the product of propagators is along
  $\mathcal{T}_{}$. 
\end{lemma}

\begin{figure}[h]\centering
  \raisebox{-0.00502482972500834\height}{\includegraphics[width=10.4721238357602cm,height=1.67083169355897cm]{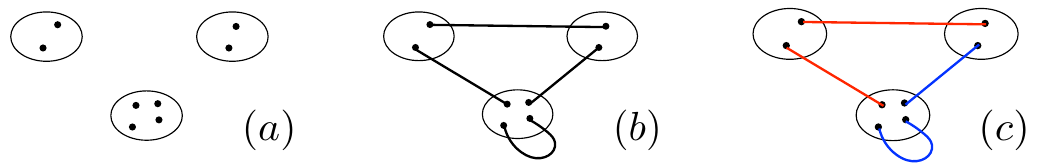}}
  \caption{\label{fig:trees1}This illustrates the $n = 3$ case of the
  connected expectation. (a) Three groups of points. (b) A particular
  connected Wick contraction. (c) Red: an anchored tree consisting of $n - 1$
  propagators. Blue: remaining $s$ propagators. }
\end{figure}

The $\Gamma_{ij}$ here are either propagators, or their first derivatives with
respect to $x_i$ and/or $y_j$. Since we are assuming {\eqref{gbound0}}, we can
replace $| \Gamma_{ij} (x_i - y_j) |$ by $M (x_i - y_j)$ in the r.h.s. of the
bound. See also Fig. \ref{fig:trees1} for an illustration. For $n = 1$ the GKL
bound reduces to the Gram-Hadamard bound {\eqref{eq:GHcond}}.

\subsection{Bound on the number of anchored trees}\label{sec:anchored-trees}

We will prove that the number of anchored trees on $n$ groups of points
$\bx_i$ is bounded by
\begin{equation}
  N_{\mathcal{T}} \leqslant n! 4^{\sum_{i = 1}^n | \bx_i |} . \label{NTbound}
\end{equation}
Recall that an anchored tree is a graph which becomes a tree when each group
$\bx_i$ is collapsed to a point. The graphs are labeled (i.e. the vertices are
distinguishable).

By Cayley's formula, the number of labeled trees $T$ on $n$ points is $n^{n -
2}$. Its proof via Pr{\"u}fer sequences {\cite{Prufer}} simultaneously gives a
finer result: the number of labeled trees specifying degrees $d_k$ of each
vertex $k$ is
\begin{equation}
  \frac{(n - 1) !}{(d_1 - 1) ! \cdots (d_n - 1) !} \hspace{0.17em} .
  \label{eq:prufer}
\end{equation}
To get an anchored tree $\mathcal{T}$ out of $T$, we choose a propagator for
each edge $e = kl$. There are at most $m_k m_l$ choices per edge, with $m_k =
| \bx_k |$ the number of points in the group $k$, so at most $\prod_{k = 1}^n
(m_k)^{d_k}$ choices in total. Multiplying by {\eqref{eq:prufer}} and summing
over all possible degrees, gives an upper bound on the number of anchored
trees:
\begin{equation}
  N_{\mathcal{T}} \leqslant (n - 1) ! \sum_{(d_k)_{k = 1}^n}  \prod_{k = 1}^n
  \frac{(m_k)^{d_k}}{(d_k - 1) !} \hspace{0.17em} . \label{eq:prufer1}
\end{equation}
Summing over each $d_k$ independently from $1$ to $\infty$ (which is an
overestimate since e.g. degrees are limited from above by $n - 1$), results in
a further bound
\begin{equation}
  N_{\mathcal{T}} \leqslant (n - 1) ! \prod_{k = 1}^n F (m_k), \qquad F (x) =
  \sum_{p = 1}^{\infty} \frac{x^p}{(p - 1)^{} !} = xe^x \hspace{0.17em} .
  \label{eq:prufer2}
\end{equation}
We have $F (x) \leqslant B^x$ with $B = e^{1 + 1 / e} \approx 3.93$, and so
{\eqref{NTbound}} follows. For a proof not relying on {\eqref{eq:prufer}} see
{\cite{Vieri-book}}, Lemma 2.4.

\section{Proof of \texorpdfstring{$S_l^{\ell_1, \ldots, \ell_n}$}{S l l1,...,ln} norm bound}\label{sec:Snorm}

Our goal here is to prove the bound {\eqref{Sest}}. Let $h_i \in H_{\ell_i}$
and denote $\widetilde{H_l} = S_l^{\ell_1, \ldots, \ell_n} (h_1, \ldots, h_n)
\in B_l$. Unpacking the definition in Section \ref{sec:IntOut}, the kernels of
$\widetilde{H_l}$ are
\begin{eqnarray}
  &  & \widetilde{H_l} (\bB, \bx_{\bB}) =\mathcal{A} \frac{1}{n!} 
  \sum_{\tmscript{\begin{array}{c}
    \bB_1, \ldots, \bB_n\\
    \sum \bB_i = \bB
  \end{array}}} \sum_{\tmscript{\begin{array}{c}
    \bA_1, \ldots, \bA_n\\
    \bA_i \supset \bB_i
  \end{array}}} (-)^{\#} K_{\left( \bB_i, \bA_i \right)_1^n} \left( \bx_{\bB}
  \right),  \label{Htildel}\\
  &  & K_{\left( \bB_i, \bA_i \right)_1^n} \left( \bx_{\bB} \right) = \int
  d^d \mathrm{} \bx_{\barbB} \hspace{0.17em} \hspace{0.17em} \text{}
  \mathcal{C} \left( \bx_{\barbB} \right)  \prod_{i = 1}^n h_i (\bA_i,
  \bx_{\bA_i}) . \nonumber
\end{eqnarray}
Let us count the terms in the sum {\eqref{Htildel}} corresponding to a fixed
$\bB$ and fixed lengths $| \bA_i | = l_i$. It is easy to see that these terms
are in one-to-one correspondence with sequences $\mathbf{R}$ of total length
$l_1 + l_2 + \cdots + l_n$ extending the sequence $\bB$. [Every such sequence
$\mathbf{R}$ can be cut into sequences $\bA_1, \ldots, \bA_n$, uniquely since
the lengths $| \bA_i |$ are kept fixed, and then $\bB_i$ can be extracted,
uniquely, as the part of $\bB$ falling into $\bA_i$.] It follows that the
number of terms in {\eqref{Htildel}} is bounded by:
\begin{equation}
  \text{Number of terms in {\eqref{Htildel}} } = \binom{\sum l_i}{l} (Nd +
  N)^{\sum l_i - l} \leqslant 2^{\sum l_i} \times (N d + N)^{\sum l_i},
  \label{terms213}
\end{equation}
where in the first equality $N d + N$ is the maximal number of choices for
every element of the sequence $\mathbf{R} \backslash \bB$, see
{\eqref{products}}, assuming that they are assigned independently. [This
counting does not take into account that some of these terms would vanish by
constraints imposed by the rotation and $\tmop{Sp} (N)$ invariances.]

We next pick some $\bB_i, \bA_i$ and consider the corresponding term $K_{}
\left( \bx_{\bB} \right) = K_{\left( \bB_i, \bA_i \right)_1^n} \left(
\bx_{\bB} \right)$ in {\eqref{Htildel}}. Its integration kernel $\text{}
\mathcal{C} \left( \bx_{\barbB} \right)$, Eq. {\eqref{CxB}}, is bounded by the
GKL bound {\eqref{eq:GKL}} with $\bx_i \equiv \bx_{\barbB_i}$:
\begin{equation}
  \left| \mathcal{C} \left( \bx_{\barbB} \right) \right| = \left| \left\langle
  \Phi (\barbB_1, \bx_{\barbB_1}) ; \ldots ; \Phi (\barbB_n, \bx_{\barbB_n})
  \right\rangle_c \right| \leqslant (C_{\mathrm{GH}})^s  \sum_{\mathcal{T}_{}}
  \prod_{(x x') \in \text{} \mathcal{T}_{}} M (x_{} - x'), \label{CxBbound}
\end{equation}
where $s = \frac{1}{2} \sum \left| \barbB_i \right| - (n - 1) \leqslant
\frac{1}{2} \sum l_i$. We wish to bound the norm of $K_{} \left( \bx_{\bB}
\right)$:
\begin{equation}
  \| K_{} \|_w = \int_{x_1 = 0} \mathrm{} d^d \bx_{\bB} \hspace{0.17em} |K_{}
  \left( \bx_{\bB} \right) | w \left( \bx_{\bB} \right)  \label{Kneed}
\end{equation}
with the weight function {\eqref{eq:ourW}}. Let $\mathcal{T}_{}$ be any
anchored tree in {\eqref{CxBbound}}. If $\tau_i$ are trees connecting points
of $\bx_{\bA_i}$, then $\mathcal{T}_{} \cup \tau_1 \cup \ldots \cup \tau_n$
connects points in $\bx_{\bB}$. Therefore, we have a bound:
\begin{equation}
  \tmop{St} \left( \bx_{\bB} \right) \leqslant \sum_{i = 1}^n \tmop{St} \left(
  \bx_{\bA_i} \right) + \prod_{(x x') \in \text{} \mathcal{T}_{}} | x_{} - x'
  | . \label{StSt}
\end{equation}
Raising this to the power $\sigma$ and using the elementary inequality
\begin{equation}
  (\sum p_i)^{\sigma} \leqslant \sum p_i^{\sigma} \qquad (p_i \geqslant 0, 0 <
  \sigma \leqslant 1), \label{elemineq}
\end{equation}
we conclude
\begin{equation}
  w (\bx_{\bB}) \leqslant \prod_{i = 1}^n w (\bx_{\bA_i})  \prod_{(x x') \in
  \text{} \mathcal{T}} w (\{ x, x' \}) \hspace{0.17em} . \label{eq:wbound}
\end{equation}

Using in {\eqref{Kneed}} this bound, the definition of $K_{} \left( \bx_{\bB}
\right)$, and the bound {\eqref{CxBbound}}, we get
\begin{eqnarray}
  \| K_{} \|_w & \leqslant & (C_{\mathrm{GH}})^s  \sum_{\mathcal{T}} \int_{x_1
  = 0} \mathrm{} d^d \bx_{\bB \cup \barbB} \prod_{(x x') \in \text{}
  \mathcal{T}_{}} M (x_{} - x') w (\{ x, x' \}) \prod_{i = 1}^n h_i (\bA_i,
  \bx_{\bA_i}) w (\bx_{\bA_i}) \nonumber\\
  & = & (C_{\mathrm{GH}})^s N_{\mathcal{T}} \| M \|_w^{n - 1} \prod_{i = 1}^n
  \| h_i \|_w .  \label{Kbound}
\end{eqnarray}
The latter equality is shown via an ``amputating tree leaves'' argument. Given
an anchored tree $\mathcal{T}_{}$, we can find a leaf: a group of points
$\mathbf{x}_{\barbB_k} \subset \bx_{A_k}$ connected to the rest by just one
edge of $\mathcal{T}_{}$, call it $(zz')$ where $z \in \bx_{A_k}$. Amputating
the leaf consists of two steps. First, keep $z$ fixed and integrate over all
the other leaf points, which gives a factor $\|h_k \|_w$. Second, integrate
over $z$ (keeping $z'$ fixed), which gives a factor $\|M\|_w$. Then find the
next leaf and continue the amputation. We get the same result for each
$\mathcal{T}$. By Appendix \ref{sec:anchored-trees}, the number of anchored
trees
\[ N_{\mathcal{T}} \leqslant n! 4^{\sum | \bx_{\barbB_i} |} = n! 4^{\sum l_i -
   l} \leqslant n! 4^{\sum l_i} . \label{NTbound1} \]
To summarize, the number of terms in {\eqref{Htildel}} is bounded by
{\eqref{terms213}}, the norm of each individual term by {\eqref{Kbound}}, $s$
by $\half \sum l_i$ and $N_{\mathcal{T}}$ by $n! 4^{\sum l_i}$. The latter
$n!$ cancels with $\frac{1}{n!}$ in {\eqref{Htildel}}. A further useful fact
is that the antisymmetrization operator $\mathcal{A}$ (footnote \ref{opA})
does not increase the norm (it averages over all permutations with signs, and
the norm is defined as the maximum over all permutations). Taking all of these
into account, we get a bound
\begin{equation}
  \| \tilde{H}_l \|_w \leqslant C_{\gamma}^{n - 1} \prod_{i = 1}^n C_0^{l_i}
  \| h_i \|_w
\end{equation}
with $C_{\gamma} = \| M \|_w^{}$, $C_0 = 8 (N d + N)
\sqrt{C_{\mathrm{GH}}}^{}$. This is the bound {\eqref{Sest}} in the case $\sum
l_i \geqslant l + 2 (n - 1)$. In the opposite case $S_l^{\ell_1, \ldots,
\ell_n}$ vanishes (Section \ref{sec:IntOut}), so there is nothing to prove.

\begin{remark}
  We can see how some of the above steps justify choices made in the main
  text. The Steiner diameter is tailor-made for {\eqref{StSt}}. Stretched
  exponential weight is handy because of {\eqref{elemineq}}. The $L_1$ norm
  (as opposed to any other $L_p$) works great when recursively amputating tree
  leaves. 
\end{remark}

\section{Estimates of \texorpdfstring{$\Delta_k^{(1)} , \Delta_k^{(2)}$}{Delta k(1),Delta k(2)}}\label{Dkbounds}

In this appendix we prove Lemma \ref{Dkest}. To estimate $\Delta_k^{(1)}$, we
use $b_k \leqslant A \delta^k $for $k \geqslant 1$, sum the geometric
progression, and use $C \delta \leqslant 1 / 2$, which gives what we need:
\begin{equation}
  \Delta^{(1)}_k \leqslant C_{}^{k + 2} A \delta^{k + 1}_{} \frac{1}{1 - C_{}
  \delta} \leqslant 2 C_{}^{k + 2} A \delta^{k + 1}_{} \qquad (k \geqslant 0)
  . \label{eq:Dk1est}
\end{equation}
The estimate of $\Delta_k^{(2)}$ is more subtle and will require several
steps. First we introduce the tool of \tmtextbf{extending by zeros}. For any
sequence $\varkappa_{} = (k_i)_1^n$, we say that a sequence $\varkappa'$
``extends $\varkappa_{}$ by $m$ zeros'' if it is obtained from $\varkappa$
inserting $m$ zeros in arbitrary places. This definition is also valid if
$\varkappa$ itself already contains some zeros (this will be useful). For $m =
0$ we have $\varkappa' = \varkappa$. E.g. (0,2,0) extends (2) by 2 zeros, and
(0,1,0,1) extends $(0, 1, 1)$ by 1 zero.

We define $F_{\tmop{ext}} [\varkappa]$ as the sum of $F [\varkappa']$ over all
$\varkappa'$ extending $\varkappa$ by an arbitrary number of zeros $m
\geqslant 0$:
\begin{equation}
  F_{\tmop{ext}} [\varkappa] = \sum_{m = 0}^{\infty} \sum_{\text{$\varkappa'
  :$extends $\varkappa$ by $m$ zeros}} F [\varkappa'] . \label{Fextdef}
\end{equation}
For a fixed $m$, the number of sequences $\varkappa'$ is $\leqslant \binom{m +
n_{}}{m}$ where $n$ is the length of $\varkappa$. [$= \binom{m + n_{}}{m}$ if
the original sequence $\varkappa$ does not have zeros.] Since they all have $F
[\varkappa'] = F [\varkappa] (C_{\gamma} C_{} b_0)^m$, we obtain
\[ F_{\tmop{ext}} [\varkappa] \leqslant F [\varkappa] \sum_{m = 0}^{\infty}
   \binom{m + n}{m} (C_{\gamma} C_{} b_0)^m = \frac{1}{(1 - C_{\gamma} C_{}
   b_0)^{n + 1}} F [\varkappa] . \]
Recall $b_0 \leqslant A \delta$. Thus using $C_{\gamma} C^{}_{} A \delta_{}
\leqslant 1 / 2$ we have
\begin{equation}
  F_{\tmop{ext}} [\varkappa] \leqslant 2^{n (\varkappa) + 1} F [\varkappa] .
  \label{eq:Fextbd}
\end{equation}
Next we will \tmtextbf{sum over sequences with a fixed $\sum k_i$}. Namely we
define
\begin{equation}
  \Phi_k = \sum_{n = 2}^{\infty} \sum_{(k_i)_1^n, \sum_{}^{} k_i = k} F
  [(k_i)_1^n] \label{Phikdef}  \quad (k \geqslant 0)
\end{equation}
We first estimate $\Phi_k$'s and then convert into the estimate for
$\Delta^{(2)}_k = \sum_{k' = k}^{\infty} \Phi_{k'}$.

Consider first $k \geqslant 2$ (see below for the simpler $k = 0, 1$). We can
obtain all sequences in $\Phi_k$ extending by zeros sequences which satisfy
$\sum_{}^{} k_i = k$, $k_i \geqslant 1$ (and whose length is therefore at most
$k$). Thus
\begin{equation}
  \Phi_k \leqslant \sum_{n = 1}^k \sum_{(k_i)_1^n, \sum_{}^{} k_i = k, k_i
  \geqslant 1} F_{\tmop{ext}} [(k_i)_1^n] \leqslant \sum_{n = 1}^k 2^{n + 1}
  \sum_{(k_i)_1^n, \sum_{}^{} k_i = k, k_i \geqslant 1} F_{} [(k_i)_1^n],
  \label{eq:Fk1}
\end{equation}
where we used {\eqref{eq:Fextbd}}. Note that although $n \geqslant 2$ in
{\eqref{Phikdef}}, we need $n \geqslant 1$ in {\eqref{eq:Fk1}} to include the
one-term sequence $(k)$ whose two-term extensions $(k, 0)$ and $(0, k)$ appear
in {\eqref{Phikdef}}.\footnote{Their contribution is suppressed by $b_0$
compared to $(k)$, but taking this suppression into account does not lead to a
better estimate because of other terms present in {\eqref{eq:Fk1}}.}

Using $b_{k_i} \leqslant A \delta^{k_i}$ and $\sum k_i = k$ we have, for
sequences with all $k_i \geqslant 1$,
\begin{equation}
  F [(k_i)_1^n] \leqslant (C_{\gamma})^{n - 1} \prod_{i = 1}^n C_{}^{k_i + 1}
  A \delta_{}^{k_i} = (C_{\gamma})^{n - 1} C_{}^{k + n} A^n \delta^k =
  (C_{\gamma} C_{} A)^{n - 1} C_{}^{k + 1} A \delta^k . \label{eq:Fest}
\end{equation}
Finally, by elementary combinatorics the number of $n$-term sequences
$(k_i)_1^n$ satisfying $\sum_{}^{} k_i = k, k_i \geqslant 1$ is $\binom{k -
1}{n - 1}$. Plugging all this information into {\eqref{eq:Fk1}}, we have
\begin{equation}
  \Phi_k \leqslant 4 C_{}^{k + 1} A \delta^k  \sum_{n = 1}^k \binom{k - 1}{n -
  1}  (2 C_{\gamma} C_{} A)^{n - 1} = 4 C_{}^{k + 1} A \delta^k (1 + 2
  C_{\gamma} C_{} A)^{k - 1},
\end{equation}
which implies
\begin{equation}
  \Phi_k \leqslant 2 C (2 C \delta_{})_{}^k A \qquad (k \geqslant 2) .
  \label{Phikest}
\end{equation}
once we use
\begin{equation}
  C_{\gamma} C_{} A \leqslant 1 / 2 . \label{eq:heart4}
\end{equation}
For future use, let us derive a bound on $F [(k_i)]$, $F_{\tmop{ext}} [(k_i)]$
under the condition {\eqref{eq:heart4}}, this time allowing for $k_i \geqslant
0$. Using $b_k \leqslant A \delta^{\max (k, 1)}$ we then have the same bound
{\eqref{eq:Fest}} but with an extra factor $\delta^m$ where $m$ is the number
of zeros in the sequence. Using {\eqref{eq:heart4}}, {\eqref{eq:Fextbd}} we
have
\begin{equation}
  F [(k_i)_{i = 1}^n] \leqslant 2^{^{1 - n}} C_{}^{k + 1} A \delta^{k + m}
  \qquad \left( \sum k_i = k, k_i \geqslant 0, m =\# \{ k_i = 0 \} \right),
  \label{eq:Fkbd2}
\end{equation}
\begin{equation}
  F_{\tmop{ext}} [(k_i)_1^n] \leqslant 4 C_{}^{k + 1} A \delta^{k + m} .
  \label{eq:Fextbd2}
\end{equation}
Finally let us bound $\Phi_k$ for $k = 0, 1$. For $k = 0$, Eq.
{\eqref{Phikdef}} involves the sequence $(0, 0)$ and its extensions by zeros.
We then have by {\eqref{eq:Fextbd2}}:
\begin{equation}
  \Phi_0 = F_{\tmop{ext}} [(0, 0)] \leqslant 4 C_{} A \delta^2 .
  \label{Phi0est}
\end{equation}
For $k = 1$, Eq. {\eqref{Phikdef}} involves the sequences (1,0), (0,1) and
their extensions by zeros. By {\eqref{eq:Fextbd2}}:
\begin{equation}
  \Phi_1 = 2 F_{\tmop{ext}} [(1, 0)] \leqslant 8 C^2_{} A \delta^2 .
  \label{Phi1est}
\end{equation}
Finally we estimate \tmtextbf{$\Delta^{(2)}_k = \sum_{k' = k}^{\infty}
\Phi_{k'}$} summing {\eqref{Phikest}} in geometric progression, which is
possible since $2 C_{} \delta_{} \leqslant 1 / 2$, and adding
{\eqref{Phi0est}}, {\eqref{Phi1est}} when needed. We thus obtain:
\begin{eqnarray}
  \Delta^{(2)}_k & \leqslant & C (2 C \delta_{})_{}^k A \qquad (k \geqslant
  2),  \label{eq:Dk2est1}\\
  \Delta^{(2)}_1 & = & \Phi_1 + \Delta^{(2)}_2 \leqslant (8 C^2 + 16 C^3) A
  \delta^2, \nonumber\\
  \Delta^{(2)}_0 & = & \Phi_0 + \Phi_1 + \Delta^{(2)}_2 \leqslant (4 C + 8 C^2
  + 16 C^3) A \delta^2 . \nonumber
\end{eqnarray}
\section{One-loop coefficients \texorpdfstring{$I_1$}{I1} and \texorpdfstring{$I_2$}{I2}}\label{sec:I1I2}

Here we evaluate the coefficients $I_1$, $I_2$ in the fixed-point equations
from Section \ref{sec:beta}, and prove their needed properties. Thinking in
momentum space, computing the contributions to the effective $\nu$ and
$\lambda$ shown in {\eqref{I1I2def}} involves setting to zero the external
momenta in the corresponding Feynman diagrams (see footnote \ref{note:mom}).
To compute $I_1$ we consider the one-loop
diagram$\resizebox{30pt}{!}{\includegraphics{fig-1.pdf}}$, where the
vertex is the local quartic coupling and the field propagating in the loop is
the fluctuating component $\varphi$. This gives
\begin{equation}
  I_1 = 2 (N - 2) \int \frac{d^d k}{(2 \pi)^d} \frac{\rho (k)}{|k|^{d / 2 +
  \eps}}, \label{I1ans}
\end{equation}
where we denoted $\rho (k) = \chi (k) - \chi (\gamma k)$ . To work out the
combinatorial prefactor $2 (N - 2)$, rewrite the local quartic as $Q (\psi) =
\tfrac{1}{3} q_{abcd} \psi_a \psi_b \psi_c \psi_d$ (integration over $x$
understood) with $q_{abcd} = \Omega_{ab} \Omega_{cd} - \Omega_{ac} \Omega_{bd}
+ \Omega_{ad} \Omega_{bc}$ totally antisymmetric. Then $Q (\psi + \phi)$
contains the quadratic in $\phi$ term $2 q_{abcd} \psi_a \psi_b \phi_c \phi_d
\hspace{0.17em}$, from where we get the term $2 q_{abcd} \Omega_{cd} \psi_a
\psi_b = 2 (N - 2) \Omega_{a b} \psi_a \psi_b  \hspace{0.17em}$in the
effective action (times the $k$-integral). Identities $\Omega_{ac} \Omega_{bc}
= \delta_{ab}$, $\Omega_{ab} \Omega_{ab} = N$ are helpful.

Similarly, to compute $I_2$ we consider the diagrams
\raisebox{-3pt}{\includegraphics[width=1.06231142594779cm,height=0.503049980322708cm]{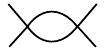}}(with
two local quartic vertices and two $\phi$-propagators) and
\raisebox{-4pt}{\resizebox{20pt}{!}{\includegraphics{fig-10.pdf}}}(where we start with the
diagram \raisebox{-3pt}{\resizebox{22pt}{!}{\includegraphics{fig-7.pdf}}} describing
the $H_{\text{6SL}}$ interaction, with the wavy line denoting
$\hat{\mathfrak{X}}_{\ast} (k)$, and contract two out of six external vertices
with the $\varphi$-propagator). Their contributions to $I_2$ are given by
\begin{eqnarray}
  \betadiag & = & - 4 (N - 8) \int \frac{d^d k}{(2 \pi)^d} \frac{\rho
  (k)^2}{|k|^{d + 2 \eps}} \hspace{0.17em}, \nonumber\\
  I^{\Xterm}_2 & = & - 8 (N - 8) \int \frac{d^d k}{(2 \pi)^d} \frac{\rho (k) R
  (k)}{|k|^{d + 2 \eps}} \hspace{0.17em}, 
\end{eqnarray}
where we wrote the Fourier transform of $\Xterm_{\ast} (x)$ as
\begin{equation}
  \widehat{\Xterm}_{\ast} (k) = - \, 8 \lambda^2  \hspace{0.17em} \frac{R
  (k)}{|k|^{\frac{d}{2} + \eps}} \hspace{0.17em}, \qquad R (k) = \sum_{n =
  1}^{\infty} \gamma^{4 \vareps n} \rho (k / \gamma^n) \hspace{0.17em} .
  \label{eq:XtermFP}
\end{equation}
To compute the prefactor in $\betadiag$, we use $Q (\psi + \phi) = 2 q_{abcd}
\psi_a \psi_b \phi_c \phi_d$ and the identity
\begin{equation}
  q_{abcd} q_{a' b' c' d'} \Omega_{cc'} \Omega_{dd'} = (N - 4) 
  \hspace{0.17em} \Omega_{ab} \Omega_{a' b'} + 2 \Omega_{aa'} \Omega_{bb'} + 2
  \Omega_{ab'} \Omega_{a' b} \hspace{0.17em} .
\end{equation}
The prefactor in $I^{\Xterm}_2$ can be computed similarly.\footnote{The
following argument explains why it is the double of $\betadiag$. Let $Q_a =
(\delta / \delta \psi_a) Q$, $Q_{ab} = (\delta^2 / \delta \psi_a \delta
\psi_b) Q$ be the first and second functional derivatives of $Q (\psi)$ at
$\psi = 0$. The \raisebox{-3pt}{\includegraphics{fig-0.pdf}} diagram is $-
\tfrac{1}{4} Q_{ab} Q_{cd} \Omega_{ac} \Omega_{bd}$ times the loop integral.
The $\Xterm$-term can be written as $- \tfrac{1}{2} \int d^d x d^d y \, Q_a
(x) Q_c (y) \Omega_{ac}  \Xterm  (x - y)$. It is then clear that the
$I^{\Xterm}_2$ diagram gives $- \tfrac{1}{2} Q_{ab} Q_{cd} \Omega_{ac}
\Omega_{bd}$ times the corresponding loop integral.} Thus
\begin{equation}
  I_2 = \betadiag + I^{\Xterm}_2 = - 4 (N - 8)  \int \frac{d^d k}{(2 \pi)^d}
  \frac{\rho (k)^2 + 2 \rho (k) R (k)}{|k|^{d + 2 \eps}} \hspace{0.17em} .
  \label{eq:I2}
\end{equation}
Now that we computed $I_1$ and $I_2$, let us discuss their properties. $I_1
\propto N - 2$ is explained by the fact that for $N = 2$ the quartic
interaction vanishes (see the Introduction). That $I_2 \ne 0$ for $N = 2$ is
not a contradiction, since for a vanishing quartic interaction the change in
$\lambda$ is anyway unphysical.

Note that perturbative beta-functions of symplectic fermion models can be
obtained from the beta-functions of bosonic $O (N_b)$ models by setting
formally $N_b = - N$. This is valid for local models
{\cite{LeClair:2007iy,Fei:2015kta}} and extends to nonlocal (long-range)
models considered here. The $(N - 8)$ factor in $I_2$ is thus related to the
well-known $(N_b + 8)$ factor in the bosonic $O (N_b)$ model one-loop
beta-function {\cite{Wilson:1971vs}}. The long-range bosonic $O (N_b)$ beta
function is known at three loops {\cite{Benedetti:2020rrq}}.\footnote{From the
two-loop level it differs from the local bosonic $O (N_b)$ beta function (e.g.
{\cite{ZinnJustin:2002ru}},(11.98)). Even the dependence on $N_b$ is
different, due to the absence of wavefunction renormalization diagrams in the
nonlocal model.} The two-loop term in {\cite{Benedetti:2020rrq}}, (3.16) is
proportional to $5 N_b + 22$ and does not vanish for $N_b \rightarrow - 8$.

Therefore, vanishing of $I_2$ for $N = 8$ is an accident unrelated to any
symmetry, which does not repeat in higher orders. With vanishing $\lambda^2$
and nonzero $\lambda^3$ term in $\lambda'$, there will be a perturbative fixed
point with $\lambda^{} = O \left( \sqrt{\varepsilon} \right)$ at $N = 8$, and
it should be possible to justify its existence non-perturbatively (see Section
\ref{extensions}). In this paper we stick to the generic case $N \ne 8$.

Because the integrals \ in {\eqref{I1ans}} and {\eqref{eq:I2}} are cutoff at
both UV and IR momenta, $I_1$ and $I_2$ depend analytically on $\varepsilon$.
At small $\varepsilon$ they behave as follows:

\begin{lemma}
  \label{lemI1I2}We have, for $| \varepsilon | \leqslant 1 / \log \gamma$,
  \begin{eqnarray}
    I_1 & = & 2 (N - 2) \left[ (1 - \gamma^{- d / 2}) \int \frac{d^d k}{(2
    \pi)^d} \frac{\chi (k)}{|k|^{d / 2}} + O (\varepsilon \log \gamma)
    \right],  \label{I1lem}\\
    I_2 & = & - 4 (N - 8)  \left[ \frac{S_d}{(2 \pi)^d} \log \gamma + O
    (\varepsilon (\log \gamma)^2) \right],  \label{I2lem}
  \end{eqnarray}
  where $S_d$ is the area of the unit sphere in $\mathbb{R}^d$ and the
  constants in $O$are $\gamma$- and $\varepsilon$-independent.
\end{lemma}

Note that $I_2$ is $\chi$-independent as $\varepsilon \rightarrow 0$, even
though $\betadiag$ and $I^{\Xterm}_2$ separately depend on $\chi$. This is not
accidental, see Remark \ref{chiindep}.

\tmtextit{Proof.} The $| k |^{- d / 2 - \varepsilon}$ factor in the integrand
in {\eqref{I1ans}} can be estimated as follows:
\begin{equation}
  \begin{array}{lll}
    \frac{1}{|k|^{d / 2 + \eps}} & = & \frac{1}{|k|^{d / 2}} + \frac{| k |^{-
    \varepsilon} - 1}{|k|^{d / 2}} = \frac{1}{|k|^{d / 2}} + \frac{O
    (\varepsilon \log \gamma)}{|k|^{d / 2}}  \qquad (1 / (2 \gamma) \leqslant
    | k | \leqslant 1) .
  \end{array} \label{denomin}
\end{equation}
The integral in {\eqref{I1ans}} is thus given, modulo $O (\varepsilon \log
\gamma)$ error, by
\begin{equation}
  \int \frac{d^d k}{(2 \pi)^d} \frac{\rho (k)}{|k|^{d / 2}} = (1 - \gamma^{- d
  / 2}) \int \frac{d^d k}{(2 \pi)^d} \frac{\chi (k)}{|k|^{d / 2}},
\end{equation}
thus proving {\eqref{I1lem}}. Let us next prove {\eqref{I2lem}}. Recall that
$\gamma \geqslant 2$. We have
\begin{eqnarray}
  &  & \tmop{supp} \rho \subset \{ 1 / (2 \gamma) \leqslant | k | \leqslant 1
  \}, \nonumber\\
  &  & \rho \equiv 1 \quad \tmop{on} \quad \{ 1 / \gamma \leqslant | k |
  \leqslant 1 / 2 \} . 
\end{eqnarray}
From here it follows that $R (k) = \gamma^{4 \varepsilon} \rho (k / \gamma) =
\gamma^{4 \varepsilon} (1 - \chi (k))$ on the support of $\rho$. We then
rewrite the numerator of the integrand in {\eqref{eq:I2}} as:
\begin{eqnarray}
  \rho (k)^2 + 2 \rho (k) R (k) & = & \rho (k)^2 + 2 \rho (k) (1 - \chi (k))
  [1 + (\gamma^{4 \varepsilon} - 1)] \nonumber\\
  & = & \rho (k)^2 + 2 \rho (k) (1 - \chi (k)) + O (\varepsilon \log \gamma),
  \nonumber\\
  & = & F (k) - F (\gamma k) + O (\varepsilon \log \gamma), \qquad F (k) = 2
  \chi (k) - \chi (k)^2 \hspace{0.17em}, 
\end{eqnarray}
and the $|k|^{- d - 2 \eps}$ factor similarly to {\eqref{denomin}}. Collecting
the error terms, the integral in {\eqref{eq:I2}} is
\begin{equation}
  \int \frac{d^d k}{(2 \pi)^d} \frac{F (k) - F (\gamma k)}{|k|^d} + O
  (\varepsilon (\log \gamma)^2) . \label{prom258}
\end{equation}
Like $\chi (k)$, the function $F (k) \equiv 1$ \ for $| k | \leqslant 1 / 2$
and vanishes for $| k | \geqslant 1$. The integral in {\eqref{prom258}} can
now be computed:
\begin{equation}
  \int \frac{d^d k}{(2 \pi)^d} \frac{F (k) - F (\gamma k)}{|k|^d} =
  \frac{S_d}{(2 \pi)^d} \log \gamma,
\end{equation}
e.g. by separating the integration region into the region close to the origin
plus the rest, and using the properties of $F (k)$ given above. In particular,
the answer is $\chi$-independent. Q.E.D.

\section{Finite volume and non-perturbative validity of \texorpdfstring{$H_{\tmop{eff}}$}{Heff}}\label{sec:non-pert}

In this appendix we will provide details mentioned in Remarks \ref{FinVol} and
\ref{R2}, regarding a definition of our model in finite volume, and a rigorous
derivation of Eq. {\eqref{eq:Hefflift}}. Our plan is as follows. First we
consider general aspects of gaussian and interacting Grassmann fields in
finite volume and with a UV cutoff, and explain that because the effective
number of Grassmann variables is finite, all path integrals are manifestly
well defined. Then we apply this to the effective action in finite volume, and
show that perturbation theory, if convergent, gives the correct answer.
Finally we pass to the infinite volume limit and show that it agrees with Eq.
{\eqref{eq:Hefflift}}.

\tmtextbf{General aspects.} Working in the finite volume $\mathcal{V}= [- V /
2, V / 2]^d$ with periodic boundary conditions, we Fourier-expand the fields
as
\begin{equation}
  \psi_a (x) = \frac{1}{V^d}  \sum_{k \in K_V} \psi_{a, k} e^{ikx},
  \label{psiak}
\end{equation}
where $K_V = (2 \pi / V) \mathbb{Z}^d \cap \tmop{supp} \chi$ is the finite set
of Fourier momenta which belong to $\tmop{supp} \chi$. We truncate away all
other momenta because they have zero propagator. The finite volume gaussian
measure $d \mu_{P, V}$ is a finite-dimensional measure over
Grassmann variables $\psi_{a, k}$:
\begin{eqnarray}
  d \mu_{P, V} (\psi) & = & \tmop{Pf}^{- 1} \prod_{\psi_{a, k}, k \in K_V} d
  \psi_{a, k} e^{S_{2, V} (\psi)},  \label{dmuPV2}\\
  S_{2, V} & = & \frac{1}{2 V^d}  \sum_{k \in K_V} \hat{P} (k)^{-
  1} \Omega_{ab} \psi_{a, k} \psi_{b, - k}, \nonumber
\end{eqnarray}
where the normalization factor $\tmop{Pf} > 0$ is the Pfaffian of $S_{2, V}$
which is an antisymmetric quadratic form in $\psi_{a, k}$'s. This is a
meaningful finite volume version of the formal Eq. {\eqref{S2}} in infinite
volume. The propagator $\langle \psi_a (x) \psi_b (y) \rangle =
\Omega_{ab} P_V  (x - y)$ where $P_V$ is the periodic version of
{\eqref{Pchi}}:
\begin{equation}
  P_V (x) = \frac{1}{V^d}  \sum_{k \in (2 \pi / V) \mathbb{Z}^d} \hat{P} (k)
  e^{ikx}, \quad \hat{P} (k) \text{ as in {\eqref{Pchi}}} .
\end{equation}
For fixed $x$ and $V \rightarrow \infty$, we have $P_V (x) \rightarrow P (x)$,
just as for the higher-order expectations. In this sense we can say that $d
\mu_{P, V} \rightarrow d \mu_{P}$, the infinite volume
measure defined in Section \ref{sec:ModDef}.

The interacting Grassmann measure is then defined as
\begin{equation}
  Z^{- 1}_V d \mu_{P, V} (\psi) e^{sH_V (\psi)}, \label{intFV}
\end{equation}
where $H_V$ is given by the finite volume analogue of {\eqref{Hpsi}}:
\begin{equation}
  H_V (\psi) = \sum_{\mathbf{A}} \int_{\mathcal{V}^l} d^d \bx \hspace{0.17em}
  H_V  (\bA, \bx) \Psi (\bA, \bx), \label{Hs}
\end{equation}
and $Z_V = \int d \mu_{P, V} (\psi) e^{sH_V (\psi)}$ is the
partition function. Factor $s$ multiplying $H_V (\psi)$ in {\eqref{intFV}} is
for further convenience, eventually we will set $s = 1$. Using Fourier
expansion {\eqref{psiak}}, we express $H_V (\psi)$ as a series in $\psi_{a,
k}$'s with finite coefficients (assuming that the kernels $H_V  (\bA, \bx)$
are in $L_1$). By Eq.{\eqref{dmuPV2}} and the usual rules of Grassmann
integration, $Z_V$ and $d \mu_{P, V} (\psi) e^{sH_V (\psi)}$ are
well defined and are polynomials in $s$, because there are only finitely many
$\psi_{a, k}$'s and only finitely many terms from the Taylor expansion of
$e^{sH_V (\psi)}$ will contribute. In particular $Z_V$ is finite.
The measure {\eqref{intFV}} will therefore be well defined as long as $Z_V
\neq 0$.

\tmtextbf{Effective action in finite volume.} Consider next Eq.
{\eqref{eq:Heff0}} in finite volume. Defining $d \mu_{g, V} (\phi)$ as $d
\mu_{P, V} (\psi)$, we consider
\begin{equation}
  I (s, \psi) = \int d \mu_{g, V} (\phi) e^{sH_V 
  (\psi + \phi)} .
\end{equation}
By the arguments as above, we have $I (s, \psi) = e^{sH_V
(\psi)} p (s, \psi)$ where $p (s, \psi)$ is a polynomial in $s$.

We would like to find $H^V_{\tmop{eff}}  (s, \psi)$ so that
\begin{equation}
  e^{H^V_{\tmop{eff}}  (s, \psi)} = I (s, \psi) . \label{HVeff}
\end{equation}
Let us define $H_{\tmop{eff}}^V  (s, \psi)$ by the perturbative
expansion (see Eqs. {\eqref{eq:Heff2}}, {\eqref{eq:Hefflift}}):
\begin{eqnarray}
  H^V_{\tmop{eff}}  (s, \psi) & = & \sum_{\bB}
  \int_{\mathcal{V}^{| \bB |}} d^d \bx \hspace{0.17em} H^V_{\tmop{eff}} 
  \left( s, \bB, \bx_{\bB} \right) \Psi (\bB, \bx_{\bB}), 
\end{eqnarray}
where $H^V_{\tmop{eff}}  \left( s, \bB, \bx_{\bB} \right)$ are given by the
finite volume analogue of Eq. {\eqref{eq:Hefflift}} replacing $H \rightarrow
sH_V$:
\begin{equation}
  H^V_{\eff}  (s, \bB, \bx_{\bB}) =\mathcal{A} \sum_{n = 1}^{\infty}
  \frac{s^n}{n!}  \sum_{\tmscript{\begin{array}{c}
    \sum \bB_i = \bB, \bA_i \supset \bB_i
  \end{array}}} (-)^{\#} \int_{\mathcal{V}^{| \barbB |}} d^d \bx_{\barbB}
  \hspace{0.17em} \hspace{0.17em} \mathcal{C}_V \left( \bx_{\barbB} \right) 
  \prod_{i = 1}^n H_V  (\bA_i, \bx_{\bA_i}), \label{eq:HeffliftV}
\end{equation}
where $\mathcal{C}_V$ is as in {\eqref{CxB}} only with finite-volume
propagators. From the arguments like in Key lemma, we will be able to show
that this series converges and defines $H^V_{\tmop{eff}}  \left( s, \bB,
\bx_{\bB} \right)$ as analytic $L_1$-valued functions in the disk $|s| < 2$
(Lemma \ref{appILemma} below, Part (b)). Since, by perturbation theory,
$e^{H^V_{\tmop{eff}}  (s, \psi)}$ and $I (s, \psi)$ have the same Taylor
series in $s$, we conclude Eq. {\eqref{HVeff}} is satisfied in the disk $|s| <
2$ where they are both analytic, in particular at $s = 1$. This proves that
{\eqref{eq:HeffliftV}} gives the correct effective action in finite volume.

\tmtextbf{Effective action in infinite volume.} For an infinite volume
interaction $H (\psi)$ given by {\eqref{Hpsi}}, we consider the corresponding
finite-volume interaction {\eqref{Hs}} with kernels given by periodization (we
are assuming translational invariance):
\begin{equation}
  H_V  \left( \bA, (0, x_2, \ldots, x_l) \right) = \sum_{r_i \in \mathbb{Z}^d,
  i = 2 \ldots l} H  \left( \bA, (0, x_2 + r_2 V, \ldots, x_l +
  r_l V) \right) . \label{HVdef}
\end{equation}
To prove that {\eqref{eq:Hefflift}} is the correct effective action in
infinite volume, we will show that it can be obtained as a $V \rightarrow
\infty$ limit of the kernels of $H^V_{\eff}$, in the precise sense of Part (c)
of the following lemma. (Part (b) was used above to justify the effective
action in finite volume.)

\begin{lemma}
  \label{appILemma}There exists $A > 0$ and $\delta >
  0$ such that, for any infinite volume interaction satisfying
  \begin{equation}
    \|H_l \|_w \leqslant A \delta^{\min (1, l / 2 - 1)}  \qquad (l
    \geqslant 2), \label{L1bounds}
  \end{equation}
  and defining the finite volume interactions by {\eqref{HVdef}} for any $V
  \geqslant 1$, we have
  
  (a) the kernels of $H_{\tmop{eff}}$ and of $H_{\tmop{eff}}^V$ given by Eqs.
  {\eqref{eq:Hefflift}} and by {\eqref{eq:HeffliftV}} with $s = 1$ are well
  defined (the series is convergent in $L_1$);
  
  (b) the kernels of $H_{\tmop{eff}}^V (s)$ defined by {\eqref{eq:HeffliftV}}
  are well defined and analytic $L_1$-valued functions in the disk $|s| < 2$;
  
  (c) for any $\bB$ we have $H_{\tmop{eff}}^V  \left( \bB, \bx
  \right) \rightarrow H_{\tmop{eff}}  \left( \bB, \bx \right)$ as
  $V \rightarrow \infty$ in the sense of $L_1$ norm on any fixed bounded
  subset of $(\mathbb{R}^d)^l$.
\end{lemma}

\tmtextit{Proof.} Claim (a) for $H_{\tmop{eff}}^V$ is a consequence of (b),
which we prove as follows: consider the $L_1$ norm (with, as usual, one of the
points fixed to the origin) of the $n$-th term of the series for
$H_{\tmop{eff}}^V$,
\begin{equation}
  \sum_{\tmscript{\begin{array}{c}
    \sum \bB_i = \bB, \bA_i \supset \bB_i
  \end{array}}} \frac{|s|^n}{n!}  \int_{\mathcal{V}^{| \bA |}, \hspace{0.17em}
  x_1 = 0} d^d \bx_{\bA} \left| \mathcal{C}_V \left( \bx_{\barbB} \right)
  \right| \hspace{-0.17em} \prod_{i = 1}^n \left| H_V (\bA_i,
  \bx_{\bA_i}) \right| \label{HVdef11}
\end{equation}
where $\bA = \bA_1 + \cdots + \bA_n$. Recall that $\mathcal{C}_V \left(
\bx_{\barbB} \right)$ is as in {\eqref{CxB}} with finite volume propagator
$g_V$ replacing $g$. Here:
\begin{equation}
  g_V (x) = \frac{1}{V^d}  \sum_{k \in (2 \pi / V) \mathbb{Z}^d} \hat{g} (k)
  e^{ikx} = \sum_{r \in \mathbb{Z}^d} g (x + rV) . \label{gVdef}
\end{equation}
From the Fourier representation of $g_V$ (first equality in {\eqref{gVdef}}),
we see that $g_V$ can be written in Gram form, as in {\eqref{E8}}, with $f_i$
and $h_i$ as in {\eqref{eq:fh}}, with the only difference that the finite
volume scalar product between $f_i$ and $h_j$ should be interpreted as $(f_i,
h_j) = \frac{1}{V^d}  \sum_{k \in (2 \pi / V) \mathbb{Z}^d} \lis{\hat{f}_i
(k)} \hat{h}_j (k)$. Therefore, the Gram-Hadamard bound {\eqref{eq:GHcond}}
holds, with $C_{\mathrm{GH}}$ replaced by $C_{\mathrm{GH}, V}$, which is
defined by the same expression as $C_{\mathrm{GH}}$, modulo the replacement of
$\int \frac{d^d k}{(2 \pi)^d}$ by the corresponding Riemann sum. Moreover,
from the real space representation of $g_V$ (second equality in
{\eqref{gVdef}}) and {\eqref{gbound0}}, we see that $g_V$ satisfies a bound
analogous to {\eqref{gbound0}} itself, with $|x / \gamma |$ replaced by $\|x\|
/ \gamma$ and $\|x\| = \min_{r \in \mathbb{Z}^d} |x + rV|$ the norm on the
torus, and with the constant $C_{\chi 1}$ replaced by a larger one, but still
independent of $V$ and $\gamma$. We denote by $M_V (x)$ the analogue of the
right side of {\eqref{gbound0}} with these two replacements. From these
considerations, we see that $\mathcal{C}_V \left( \bx_{\barbB} \right)$ is
bounded as in {\eqref{CxBbound}},
\begin{equation}
  \left| \mathcal{C}_V \left( \bx_{\barbB} \right) \right| \leqslant
  (C_{\mathrm{GH}, V})^{\frac{1}{2}  \sum_i l_i} 
  \sum_{\mathcal{T}} \prod_{(xx') \in \mathcal{T}} M_V
  (x - x') .
\end{equation}
Thanks to these considerations, proceeding as in Appendix \ref{sec:Snorm}, we
get the analogue of {\eqref{Kbound}}, namely
\begin{eqnarray}
  &  & \int_{\mathcal{V}^{| \bA |}, \hspace{0.17em} x_1 = 0} d^d \bx_{\bA}
  \left| \mathcal{C}_V \left( \bx_{\barbB} \right) \right| \hspace{-0.17em}
  \prod_{i = 1}^n \left| H_V (\bA_i, \bx_{\bA_i}) \right| 
  \label{eq:I.14}\\
  &  & \leqslant \hspace{1em} (C_{\mathrm{GH}, V})^{\frac{1}{2}  \sum_i l_i} 
  \sum_{\mathcal{T}} \int_{\mathcal{V}^{| \bA |}, \hspace{0.17em} x_1 = 0} d^d
  \bx_{\bA} \prod_{(xx') \in \mathcal{T}} M_V  (x -
  x')  \prod_{i = 1}^n \left| H_V (\bA_i, \bx_{\bA_i}) \right|,
  \nonumber
\end{eqnarray}
which, by computing the integrals, can be further bounded as
\begin{equation}
  \text{{\eqref{eq:I.14}}} \le (C_{\mathrm{GH}, V})^{\frac{1}{2}  \sum_i l_i}
  N_{\mathcal{T}} \|M_V \|_1^{n - 1}  \prod_{i = 1}^n \|H_{V, l_i} \|_1 .
  \label{Kbound.I}
\end{equation}
Now, by using {\eqref{HVdef}}, we see that $\|H_{V, l_i} \|_1 \le \|H_{l_i}
\|_w$, which is bounded by $A \delta^{\min (1, l / 2 - 1)}$,
thanks to {\eqref{L1bounds}}. Recalling also {\eqref{terms213}} and the fact
that $\mathcal{N}_{\mathcal{T}} \le n! 4^{\sum l_i}$, we find
\begin{equation}
  \text{{\eqref{HVdef11}}} \leqslant |s|^n A^n \|M_V \|_1^{n - 1} 
  \sum_{(l_i)_1^n} \prod_{i = 1}^n (C_V)^{\sum l_i} \delta^{\min (1, l_i / 2 -
  1)} \label{lat.vir}
\end{equation}
where $C_V = 16 N (d + 1) C_{\mathrm{GH}, V}^{1 / 2}$. Note that both $C_V$
and $\|M_V \|_1$ are uniformly bounded in $V$. Positive even integers $l_i$
satisfy $\sum_i l_i \geqslant l + 2 (n - 1)$, but here it will be enough to
extend the sum to arbitrary $l_i \geqslant 2$. Therefore, with a suitable
$V$-independent constant $C$, we get a bound
\begin{equation}
  \text{{\eqref{lat.vir}}} \le \left( \frac{C|s| \delta}{1 - C \delta}
  \right)^n, \label{lat.viru}
\end{equation}
from which summability in $n$ follows, for all $|s| < 2$, if $\delta$ is
sufficiently small. Of course, item (a) for $H_{\tmop{eff}}$ in infinite
volume follows from the same argument.\footnote{This case is also a
consequence of Key Lemma but we preferred to give an independent argument to
demonstrate how much simpler it is to show the convergence than, as in Key
Lemma, to get an optimal bound on the sum.}

\begin{remark}
  According to the discussion after Eq. {\eqref{Hs}}, we also need to make
  sure that the finite volume partition function is nonzero. The constant,
  $\psi$-independent term in the effective action can be estimated by the same
  argument as above, and it is given by the torus volume times a convergent
  series, in particular it is finite. Hence the partition function, which is
  its exponential, is nonzero. The effective action is thus well defined. Once
  we know that the $\psi$-independent term is finite, we may drop it as we did
  throughout.
\end{remark}

Let us now prove (c) (cf. {\cite{Graphene}}, App. D). We fix a bounded subset
of $(\mathbb{R}^d)^l$ that, without loss of generality, we assume to be
centered in the origin, and we call it $\mathcal{V}_0$. We want to prove that
the sum over $n$ of
\begin{equation}
\sum_{\scriptsize\begin{array}{c}
		\sum \bB_i = \bB, \\
		\bA_i \supset \bB_i
		\end{array}} \frac{1}{n!}  \int_{\mathcal{V}_0^{| \bB |}} d^d \bx_{\bB}
\left| \int_{\mathcal{V}^{| \barbB |}} d^d \bx_{\barbB} \hspace{0.17em}
\hspace{0.17em} \mathcal{C}_V \left( \bx_{\barbB} \right)  \prod_{i = 1}^n
H_V (\bA_i, \bx_{\bA_i})  - \int_{\mathbb{R}^{| \barbB |d}} d^d \bx_{\barbB}
\hspace{0.17em} \hspace{0.17em} \mathcal{C} \left( \bx_{\barbB} \right) 
\prod_{i = 1}^n H (\bA_i, \bx_{\bA_i}) \right| \label{eq:appI15}
\end{equation}
goes to zero as $V \to \infty$. We will in fact prove that the sum of
{\eqref{eq:appI15}} over $n$ goes to zero exponentially fast in $V$ as $V \to
\infty$. We rewrite the integral over $\mathcal{V}^{| \barbB |}$ by
multiplying the integrand by $1 = 1  (\tmop{St}_V (\bx_{\bA}) \le V / 4) + 1
(\tmop{St}_V (\bx_{\bA}) > V / 4)$ [here, if $\bx = (x_1, \ldots, x_l)$, the
finite volume Steiner diameter $\tmop{St}_V (\bx)$ is the length of the
shortest tree on the torus (possibly with extra vertices) which connects all
the points in $\bx$. Note that $\tmop{St}_V (\bx) \le \min_{\tmmathbf{r} \in
\mathbb{Z}^{dl}} \tmop{St} (\bx +\mathbf{r}V)$], and similarly for the
integral over $\mathbb{R}^{| \barbB | d}$, with the finite volume Steiner
diameter replaced by the standard, infinite volume, one. In view of this
manipulation, we bound {\eqref{eq:appI15}} from above by $|\mathcal{V}_0 | 
(R_{1, n} + R_{2, n} + R_{3, n})$, where, letting $x_1$ being the first
coordinate in the list $\bx_{\bB}$:
\begin{equation}
  R_{1, n} = \sum_{\tmscript{\begin{array}{c}
    \sum \bB_i = \bB\\
    \bA_i \supset \bB_i
  \end{array}}} \frac{1}{n!}  \int_{\mathcal{V}^{| \bA |}, \hspace{0.17em} x_1
  = 0} d^d \bx_{\bA} \left| \mathcal{C}_V \left( \bx_{\barbB} \right) \right|
  \prod_{i = 1}^n \left| H_V (\bA_i, \bx_{\bA_i}) \right| 
  (\tmop{St}_V (\bx_{\bA}) > V / 4), \nonumber
\end{equation}
\begin{equation}
  R_{2, n} = \sum_{\tmscript{\begin{array}{c}
    \sum \bB_i = \bB\\
    \bA_i \supset \bB_i
  \end{array}}} \frac{1}{n!}  \int_{\mathbb{R}^{| \bA | d}, \hspace{0.17em}
  x_1 = 0} d^d \bx_{\bA} \left| \mathcal{C}_V \left( \bx_{\barbB} \right)
  \right| \prod_{i = 1}^n \left| H_V (\bA_i, \bx_{\bA_i}) \right|
  1  (\tmop{St} (\bx_{\bA}) > V / 4),  \label{eq:appI.18}
\end{equation}
\begin{equation}
R_{3, n}=\hspace{-1em}  \sum_{\tmscript{\begin{array}{c}
		\sum \bB_i = \bB\\
		\bA_i \supset \bB_i
		\end{array}}} \hspace{-0.5em}\frac{1}{n!}  \int_{\mathbb{R}^{| \bA | d}, \hspace{0.17em}
	x_1 = 0} \hspace{-0.5em} d^d \bx_{\bA} \left| \mathcal{C}_V \left( \bx_{\barbB} \right) 
\prod_{i = 1}^n H_V (\bA_i, \bx_{\bA_i}) - \mathcal{C} \left( \bx_{\barbB} \right)  \prod_{i = 1}^n H
(\bA_i, \bx_{\bA_i}) \right| 1  (\tmop{St} (\bx_{\bA}) \le V / 4) \,,\nonumber
\end{equation}
where, in the definition of $R_{3, n}$, we used the fact that $1 (\tmop{St}_V
(\bx_{\bA}) \le V / 4)$ is the same as $1 (\tmop{St} (\bx_{\bA}) \le V / 4)$,
provided we identify the points of the torus $\mathcal{V}$ closer than $V / 4$
to the origin with the corresponding points in $\mathbb{R}^d$.

In order to bound $R_{1, n}$ we proceed as we did above for {\eqref{HVdef11}},
with only a few differences: consider the analogue of {\eqref{eq:I.14}} that,
compared with that equation, has the additional constraint $1  (\tmop{St}_V
(\bx_{\bA}) > V / 4)$ under the integral sign. In the second line, we multiply
and divide each factor $M_V  (x - x')$ by $w_V (x, x')$ and each factor $H_V 
(\bA_i, \bx_{\bA_i})$ by $w_V (\bx_{\bA_i})$, where $w_V (\bx)$ is the finite
volume analogue of $w (\bx)$, namely $w_V (\bx) = w (\bx) = e^{C_w 
(\tmop{St}_V (\bx) / \gamma)^{\sigma}}$. We collect together all the factors
$1 / w_V (x, x')$ and $1 / w_V (\bx_{\bA_i})$ and note that, on the support of
$1  (\tmop{St}_V (\bx_{\bA}) > V / 4)$,
\begin{equation}
  \left( \prod_{(xx') \in \mathcal{T}} \frac{1}{w_V (x, x')} \right)
  \hspace{0.17em} \left( \prod_{i = 1}^n \frac{1}{w_V (\bx_{\bA_i})} \right)
  \le e^{- C_w  (\tmop{St}_V (\bx_{\bA}) / \gamma)^{\sigma}} \le e^{- C_w  (V
  / (4 \gamma))^{\sigma}} .
\end{equation}
Therefore, we can bound the analogue of {\eqref{eq:I.14}} by the analogue of
the right side of {\eqref{Kbound.I}}, that is
\begin{equation}
  (C_{\mathrm{GH}, V})^{\frac{1}{2}  \sum_i l_i} N_{\mathcal{T}} e^{- C_w  (V
  / (4 \gamma))^{\sigma}} \|M_V \|_{w_V}^{n - 1}  \prod_{i = 1}^n \|H_{V, l_i}
  \|_{w_V} .
\end{equation}
Note also that, thanks to {\eqref{HVdef}} and {\eqref{L1bounds}}, $\|H_{V,
l_i} \|_{w_V} \le \|H_{l_i} \|_w \le A \delta^{\min (1, l_1 / 2 - 1)}$.
Putting things together, we get the analogue of {\eqref{lat.viru}}:
\begin{equation}
  R_{1, n} \le e^{- C_w  (V / (4 \gamma))^{\sigma}} \left( \frac{C \delta}{1 -
  C \delta} \right)^n,
\end{equation}
for a suitable $V$-independent constant $C$. Clearly, for $\delta$ small
enough, the sum over $n$ of $R_{1, n}$ converges and goes to zero
exponentially as $V \to \infty$. Analogous discussion and bounds are valid for
$R_{2, n}$.\footnote{This discussion also makes clear that finite-volume
convergence statements in Parts (a) and (b) can be easily generalized to
weighted $L_1$ norm with weight $w_V$.}

Let us now consider $R_{3, n}$. We rewrite the difference
\begin{equation}
  \mathcal{C}_V \left( \bx_{\barbB} \right)  \prod_{i = 1}^n H_V  (\bA_i,
  \bx_{\bA_i}) -\mathcal{C} \left( \bx_{\barbB} \right)  \prod_{i = 1}^n H
  (\bA_i, \bx_{\bA_i}) .
\end{equation}
in telescopic form as the sum of $n + 1$ terms, in each of which either a
difference $\mathcal{C}_V \left( \bx_{\barbB} \right) -\mathcal{C} \left(
\bx_{\barbB} \right)$ or $H_V  (\bA_i, \bx_{\bA_i}) - H (\bA_i,
\bx_{\bA_i})$ appears. The terms with $H_V - H$ can be bounded via
an analogue of {\eqref{Kbound.I}}, with the important difference that one of
the factors $\|H_{V, l_i} \|_1$ is replaced by (denoting $\bx_{\bA_i} = (x_1,
\ldots, x_{l_i})$ and $\tmmathbf{\mathbf{r}} = (r_1, \ldots, r_{l_i})$)
\begin{eqnarray}
  &  & \int_{\mathbb{R}^{| \bA_i | d}, \hspace{0.17em} x_1 = 0} d^d
  \bx_{\bA_i} \left| H_V (\bA_i, \bx_{\bA_i}) - H (\bA_i, \bx_{\bA_i}) \right|
  1 (\tmop{St} (\bx_{\bA_i}) \le V / 4)  \label{eq:appI.22}\\
  & \le & \sum_{\tmscript{\begin{array}{c}
    \mathbf{r} \in \mathbb{Z}^{dl_i}\\
    r_1 = 0, \hspace{0.17em} \mathbf{r} \neq \tmmathbf{0}
  \end{array}}} \quad \int_{\mathbb{R}^{dl_i}, \hspace{0.17em} x_1 = 0} d^d
  \bx_{\bA_i} \left| H (\bA, \bx_{\bA_i} + \tmmathbf{\mathbf{r}}
  V) \right| \frac{w (\bx_{\bA_i} +\mathbf{r}V)}{w (\bx_{\bA_i} +
  \tmmathbf{\mathbf{r}} V)} 1 (\tmop{St} (\bx_{\bA_i}) \le V / 4), \nonumber
\end{eqnarray}
where in passing from the first to the second line we used the definition
{\eqref{HVdef}} and we multiplied and divided by $w (\bx_{\bA_i}
+\mathbf{r}V)$. Now, note that, on the support of $1 (\tmop{St} (\bx_{\bA_i})
\le V / 4)$, $| \bx_{\bA_i} + \tmmathbf{\mathbf{r}} V| > V / 2$ for any
$\tmmathbf{\mathbf{r}} \neq \tmmathbf{0}$. Therefore, the second line of
{\eqref{eq:appI.22}} can be bounded from above by $\|H_{l_i} \|_w / w (V /
2)$, where $1 / w (V / 2) = e^{- C_w  (V / (2 \gamma))^{\sigma}}$ represents
the desired exponentially small gain as $V \to \infty$.

Consider now the contribution to $R_{3, n}$ associated with the difference
$\mathcal{C}_V \left( \bx_{\barbB} \right) -\mathcal{C} \left( \bx_{\barbB}
\right)$. Recall that both $\mathcal{C} (\bx_{\barbB})$ and $\mathcal{C}_V
(\bx_{\barbB})$ can be written in terms of the BBF formula that, see
{\eqref{eq:detNrep}}, can be written as
\begin{equation}
  \label{BBF.appI} \mathcal{C} (\bx_{\barbB}) = \sum_{\mathcal{T}}
  \prod_{(xx') \in \mathcal{T}} g (x - x')  \int d \mu_T (\mathbf{r}) \det
  \mathcal{N},
\end{equation}
where $\mathcal{N}=\mathcal{N} (\mathbf{r})$ is a Gram matrix, i.e., with
elements represented as a suitable scalar product. Of course, $\mathcal{C}_V
(\bx_{\barbB})$ admits a representation analogous to {\eqref{BBF.appI}}, with
$g$ replaced by $g_V$ and $\mathcal{N}$ replaced by $\mathcal{N}_V$. Using
{\eqref{BBF.appI}} and the analogous representation for $\mathcal{C}_V
(\bx_{\barbB})$, we write the difference $\mathcal{C}_V \left( \bx_{\barbB}
\right) -\mathcal{C} \left( \bx_{\barbB} \right)$ in telescopic form, as the
sum of terms in each of which either a difference $g_V  (x - x') - g (x - x')$
or $\det \mathcal{N}_V - \det \mathcal{N}$ appears. In the former terms,
recalling {\eqref{gVdef}}, we write $g_V (x) - g (x) = - \sum_{r \neq 0} g (x
+ rV)$ and, proceeding as we did for the bound of the terms with $H_V - H$, we
see that they are exponentially small in $V$, and their sum over $n$ too.

We are left with the term involving the difference $\det \calN_V - \det
\calN$, which we rewrite once again in telescopic form as
\begin{equation}
  \label{last} \det \calN_V - \det \calN = \sum_{i, j = 1}^s (\det
  \calN_V^{(i, j)} - \det \calN_V^{(i, j)'}),
\end{equation}
where $s$ is the linear size of the matrices, $\calN_V^{(i, j)}$ is the matrix
whose elements with label smaller or equal to (resp. larger than) $(i, j)$ in
the lexicographic order are equal to the elements of $\calN_V$ (resp.
$\calN$), and $(i, j)'$ is the label immediately preceding $(i, j)$ in the
lexicographic order (if $(i, j) = (1, 1)$, we interpret $\calN_V^{(1, 1)'}
\equiv \calN$). Since $\calN_V^{(i, j)}$ and $\calN_V^{(i, j)'}$ differ in
only one element, expanding in minors along row $i$ we have
\begin{equation}
  \det \calN_V^{(i, j)} - \det \calN_V^{(i, j)'} = (- 1)^{i + j}  \left(
  (\calN_V)_{i, j} - \calN_{i, j} \right) \det \widehat{\calN}_V^{(i, j)},
  \label{lastlast}
\end{equation}
where $\widehat{\calN}_V^{(i, j)}$ denotes the matrix $\calN_V^{(i, j)}$ with
both the $i$-th row and the $j$-th column removed. Recall that both $\calN_V$
and $\calN$ are Gram matrices; in particular, they can be written as
$(\calN_V)_{k, l} = (f_{V, k}, h_{V, l})$ and $\calN_{k, l} = (f_k, h_l)$ for
appropriate vectors $f_V, h_V, f, h$ in two apriori different Hilbert spaces
$\mathcal{H}_V$ and $\mathcal{H}$. Remarkably, also $\widehat{\calN}_V^{(i,
j)}$ is in Gram form, that is, for any $k \in \{1, \ldots, s\} \setminus
\{i\}$ and any $l \in \{1, \ldots, s\} \setminus \{j\}$, we can write
$(\widehat{\calN}_V^{(i, j)})_{k, l} = (F_k, H_l)$, where $(\cdummy, \cdummy)$
denotes the scalar product in $\mathcal{H}_V \oplus \mathcal{H}$, and $F_k,
H_l$ are the following vectors in $\mathcal{H}_V \oplus \mathcal{H}$:
\begin{equation}
  F_k = \left\{\begin{array}{ll}
    (f_{V, k}, 0) & \text{if } k < i\\
    (0, f_k) & \text{if } k > i
  \end{array}\right. \quad \text{and} \quad H_l = (h_{V, l}, h_l) .
\end{equation}
Therefore, $\det \widehat{\calN}_V^{(i, j)}$ can be bounded qualitatively in
the same way as $\det \calN_V$ or $\det \calN$, so that, using
{\eqref{lastlast}} into {\eqref{last}}, and recalling that $(\calN_V)_{i, j} -
\calN_{i, j}$ is proportional to $g_V - g$, we find that the term in $R_{3,
n}$ involving the difference $\det \calN_V - \det \calN$ is bounded
qualitatively as all the other terms, that is, they are exponentially small in
$V$, and their sum over $n$ too. This concludes the proof of Lemma
\ref{appILemma}. $\square$

Consider e.g. $H (\psi)$ corresponding to the fixed point whose existence we
proved. By Corollary \ref{FPaction}, this interaction satisfies bounds
{\eqref{FPactbnds}} which for sufficiently small $\varepsilon$ are stronger
than {\eqref{L1bounds}}. Therefore, the effective action is indeed given by
{\eqref{eq:Hefflift}} as we assumed all along.

\section{Fixed point in a formal power series expansion}\label{sec:formal}

In this appendix we will show that Eq. {\eqref{fy=0}} $f (y) = 0$ can be
solved in a formal power series expansion in $\varepsilon$. This is rather
easy, compared to the proof of the existence of an actual solution given in
Section \ref{sec:FP}. We introduce a positive grading function on the
couplings $y_i \in \{ \nu, \lambda, u_{\text{2R}}, u_{\text{4R}}, u_{\text{6R}},
(u_{\ell})_{\ell \geqslant 8} \}$:
\begin{equation}
  \tmop{gr} (\nu) = \tmop{gr} (\lambda) = 1, \quad \tmop{gr} (u_{\text{2R}}) =
  \tmop{gr} (u_{\text{4R}}) = 2, \quad \tmop{gr} (u_{\text{6R}}) = 3, \quad \tmop{gr}
  (u_{\ell}) = k (l) = \frac{l}{2} - 1 (l \geqslant 8) .
\end{equation}
\begin{table}[h]
  
  \[ \begin{array}{cccc}
       \hline
       \tmop{Function} & \tmop{Grading} & \tmop{Notable} \tmop{present}
       \tmop{terms} & \tmop{Notable} \tmop{absent} \tmop{terms}\\
       \hline
       e_{\nu} & \geqslant 2 & \lambda^2, \nu \lambda, u_{\text{4R}} & \nu, u_{\text{2R}},
       \lambda\\
       e_{\lambda} & \geqslant 3 & \nu \lambda^2, \lambda^3, \nu
       \mathfrak{X}_{\lambda}, \lambda \mathfrak{X}_{\lambda}, \lambda u_{4
       R}, u_{\text{6R}} & \lambda, u_{\text{4R}}, \lambda^2, \mathfrak{X}_{\lambda}, \nu
       \lambda, \nu u_{\text{4R}}\\
       e_{\text{2R}} & \geqslant 2 & u_{\text{2R}}, u_{\text{4R}}, \lambda^2 & \nu, \lambda\\
       e_{\text{4R}} & \geqslant 2 & u_{\text{4R}}, \lambda^2 & \lambda\\
       e_{\text{6R}} & \geqslant 3 & u_{\text{6R}}, \lambda^3, \nu \lambda^2, \nu
       \mathfrak{X}_{\lambda}, u_{\text{4R}} \lambda, u_8 & \mathfrak{X}_{\lambda},
       \lambda^2\\
       e_{\ell}, \hspace{1em} \ell \geqslant 8 & \geqslant k (l) & u_{\ell},
       \lambda^{k (l)}, \sum_{l_1 + \ldots + l_n \geqslant l + 2 (n - 1)}
       H_{\ell_1} \cdots H_{\ell_n} & \\
       \hline
     \end{array} \]
  \caption{\label{tab:grading}Grading of terms in functions $e_{y_i} .$ We
  only show the variables on which the terms depend. E.g. $u_{\text{2R}}$ and
  $\lambda^2$ in $e_{\text{2R}}$ stand for $R_{\text{2R}}^{\text{2R}} (u_{\text{2R}}) = D u_{\text{2R}}$ and
  $R_{\text{2R}}^{\text{4L}, \text{4L}} (\lambda, \lambda)$, respectively.}
\end{table}

We also define grading of a product as a sum of gradings. It is then easy to
check that each function $e_{y_i}$ is a sum of terms whose grading is
$\geqslant \tmop{gr} (y_i)$, with strict inequality for $\nu, \lambda$ (Table
\ref{tab:grading}). This motivates the following

\begin{theorem}
  \label{th:formal}Equation $f (y) = 0$ has a unique solution where couplings
  are formal power series in $\varepsilon$ starting from:
  \begin{equation}
    \nu = \frac{a}{b} \varepsilon + O (\varepsilon^2), \quad \lambda = -
    \frac{1}{b} \varepsilon + O (\varepsilon^2)^{}, \quad y_i = O
    (\varepsilon^{\tmop{gr} (y_i)}) \quad (y_i \in \{ u_{\rm{2R}}, u_{\rm{4R}}, u_{\rm{6R}}, (u_{\ell})_{\ell \geqslant 8} \}) .
  \end{equation}
\end{theorem}

\tmtextit{Proof.} Parameter $\varepsilon$ enters {\eqref{fy=0}} through the
explicit term $\varepsilon \lambda$. In addition, all the other coefficients
such as $a, b$ and the multilinear kernels from the r.h.s. of
{\eqref{betainiz}} also depend on $\varepsilon$. This dependence originates
from the fluctuation propagator $g (x)$ defined in {\eqref{eq:split}}, and it
is nonsingular as $\varepsilon \rightarrow 0$ in our setup involving the UV
and IR cutoffs. Below we will keep track only of the explicit dependence on
$\varepsilon$ from the $\varepsilon \lambda$ term, which we denote $\epsilon
.$ All other coefficients will be treated as constants. We will give an
algorithm to expand the solution as a formal power series in $\epsilon$. To
produce a power series in $\varepsilon$, one would have to set $\epsilon
\rightarrow \varepsilon$ and additionally expand all coefficients in
$\varepsilon$.

We start by rescaling the couplings $y_i \rightarrow \epsilon^{\tmop{gr}
(y_i)} y_i$. We will abuse notation denoting the rescaled couplings by the
same letters. We have to show that the rescaled couplings have unique power
series expansions starting at $O (1)$. The equations $f (y) = 0$ in terms of
the rescaled couplings can be written as (the explanations and the definition
of $e_{y_i, k}$ are given after {\eqref{eqyi}}) \
\begin{eqnarray}
  & \tmop{couplings} \tmop{of} \tmop{grading} 1 : & \left\{\begin{array}{l}
    - \nu - a \lambda = \sum_{k \geqslant 1} \epsilon^k e_{\nu, k}\\
    - \lambda - b \lambda^2 = \sum_{k \geqslant 1} \epsilon^k e_{\lambda, k +
    1}
  \end{array}\right.  \label{eqn}\\
  & \tmop{couplings} \tmop{of} \tmop{grading} 2 : & \left\{\begin{array}{l}
    (1 - D) u_{\text{2R}} - R_{\text{2R}}^{\text{4R}} u_{\text{4R}} = \sum_{k \geqslant 0} \epsilon^k
    e_{\text{2R}, k}\\
    (1 - D) u_{\text{4R}} = \sum_{k \geqslant 0} \epsilon^k e_{\text{4R}, k}
  \end{array}\right.  \label{eq2r}\\
  & \tmop{couplings} \tmop{of} \tmop{grading} 3 : & \left\{\begin{array}{l}
    (1 - D) u_{\text{6R}} - R_{\text{6R}}^8 u_8 = \sum_{k \geqslant 0} \epsilon^k e_{\text{6R},
    k}\\
    (1 - D) u_8 = \sum_{k \geqslant 0} \epsilon^k e_{8, k}
  \end{array}\right.  \label{eq6r}\\
  & \tmop{couplings} \tmop{of} \tmop{grading} \geqslant 4 : & (1 - D)
  u_{\ell} = \sum_{k \geqslant 0} \epsilon^k e_{\ell, k} \quad (\ell \geqslant
  10),  \label{eqyi}
\end{eqnarray}
For each coupling, $e_{y_i, k}$ denotes the part of $e_{y_i}$ which contains
the terms of grading exactly $\tmop{gr} (y_i) + k$. In addition we separated
the linear terms $D u_i$ as well as $R_{\text{2R}}^{\text{4R}} u_{\text{4R}}$ and $R_{\text{6R}}^8
u_8$ from $e_{u_i, 0}$ in {\eqref{eq2r}}-{\eqref{eqyi}}. With this definition
the remaining $e_{u_i, 0}$ are at least quadratic in its arguments. Note that
the r.h.s. of {\eqref{eqn}} are $O (\epsilon)$, while the other equations have
r.h.s. $O (1)$. Note also the shift $k \rightarrow k + 1$ in $e_{\lambda, k +
1}$ in \ {\eqref{eqn}}.

\tmtextbf{Step 1.} The $O (1)$ parts of $\nu$ and $\lambda$ known, $\nu =
\frac{a}{b} + O (\epsilon),$ $\lambda = - \frac{1}{b} + O (\epsilon)$, let us
solve for the $O (1)$ parts of the other couplings. Firstly, note that the $O
(1)$ parts of the r.h.s. of {\eqref{eq2r}}--{\eqref{eqyi}}, $e_{u_i, 0}$,
having grading exactly $\tmop{gr} (u_i)$ and being at least quadratic, are
computable in terms of the $O (1)$ parts of the couplings with smaller
grading. Secondly, the linear operators in the l.h.s. of
{\eqref{eq2r}}--{\eqref{eqyi}} are invertible. Indeed, the operator $(1 - D)$
is invertible on each irrelevant coupling subspace as is clear from definition
{\eqref{eq:Resclift}}.\footnote{\label{invert}We have $(1 - D)^{- 1} = 1 + D +
D^2 + \ldots$ and the series converges in $L_1$ if $\gamma^{- D_l - p} < 1$
which is the condition for irrelevance.} Eqs. {\eqref{eq2r}} and
{\eqref{eq6r}} involve a matrix-triangular operator with $(1 - D)$ on the
diagonal, hence also invertible. By these two observations, $O (1)$ parts of
all couplings are uniquely determined starting from $\nu$ and $\lambda$ and
going recursively up in grading.

\tmtextbf{Step 2a.} Now suppose we computed expansions of all couplings up to
and including $O (\epsilon^N)$ (call it ``inductive hypothesis 1''), and we
want to solve for the $\epsilon^{N + 1}$ terms. For any quantity $\alpha =
\sum^{} \epsilon^n \alpha_n$ we denote by $[\alpha]_n = \alpha_n$ the
$\epsilon^n$ coefficient. We start with {\eqref{eqn}} and take its
$\epsilon^{N + 1}$ part:
\begin{eqnarray}
  &  & - [\nu]_{N + 1} - a [\lambda]_{N + 1} = \sum_{k = 1}^{N + 1} [e_{\nu,
  k}]_{N + 1 - k}, \\
  &  & (- 1 - 2 b [\lambda]_0) [\lambda]_{N + 1} = b \sum_{k = 1}^N
  [\lambda]_k^{} [\lambda]_{N + 1 - k}^{} + \sum_{k = 1}^{N + 1} [e_{\lambda,
  k + 1}]_{N + 1 - k} . 
\end{eqnarray}
All the terms in the r.h.s. are computable by inductive hypothesis 1. Since $-
1 - 2 b [\lambda]_0 = 1$ we can compute first $[\lambda]_{N + 1}$ and then
$[\nu]_{N + 1}$.

\tmtextbf{Step 2b.} The remaining couplings are treated recursively going up
in grading as before. Suppose all couplings of grading lower than $u_i$ are
already known up to and including $O (\epsilon^{N + 1})$ (call it ``inductive
hypothesis 2''). Consider the equation for $u_i$ (if there are two couplings
having the same grading we should study their equations together as in Step 1)
and take its $\epsilon^{N + 1}$ part. In the l.h.s. we have an invertible
linear operator, same as in Step 1, acting on $[u_i]_{N + 1}$, while in the
r.h.s. we have
\begin{equation}
  \sum_{k = 0}^{N + 1} [e_{u_i, k}]_{N + 1 - k} .
\end{equation}
For $k \geqslant 1$ this is computable by inductive hypothesis 1, and for $k =
0$ by inductive hypothesis 2, since $e_{u_i, 0}$ has grading $\tmop{gr} (u_i)$
and is at least quadratic, which means it involves only lower-grading
couplings. Therefore, we can compute $[u_i]_{N + 1}$ and continue the
induction.\footnote{Steps 2a, 2b can be unified, at the price of rendering the
argument less explicit, by moving the nonlinear functions $e_{u_i, 0}$ to the
l.h.s. and noting that the Jacobian of the resulting nonlinear infinite matrix
function of $y$ in the l.h.s. is invertible at the point $(\nu, \lambda, u_i)
= \left( \frac{a}{b}, - \frac{1}{b}, u_i^{(0)} \right)$ where $u_i^{(0)}$ are
the $O (1)$ values of $u_i$ computed in Step 1.} This finishes the proof of
the theorem.

\begin{remark}
  At the level of formal series expansions bosonic and fermionic fixed point
  are quite analogous. Consider e.g. the bosonic model {\eqref{MFT}}. In
  perturbation theory, we could parametrize its fixed point by an interaction
  written in terms of kernels, like in {\eqref{Hpsi}}. We could derive a
  perturbative renormalization map acting on the sequence of kernels in the
  trimmed representation, similarly to Section \ref{detailsRG}. We could then
  find, exactly as in Theorem \ref{th:formal}, a fixed point in a formal power
  series in $\varepsilon$.
  
  This analogy breaks down beyond perturbation theory. As stated in Remark
  \ref{R2} and proved in Appendix \ref{sec:non-pert}, perturbative expansion
  captures the full fermionic effective action at small coupling. This does
  not hold for bosons due to large field effects (``instantons'') at
  arbitrarily small couplings. Furthermore, rigorous non-perturbative studies
  of bosonic models parametrize irrelevant interactions not by kernels as in
  {\eqref{Hpsi}}, but by a more complicated ``polymer expansion'' (see
  {\cite{Mitter:2005zx}} and Section \ref{bosons}).
  
  Another difference is that for bosons, the formal power series solution in
  $\varepsilon$ is expected to be only asymptotic, like the
  $\epsilon$-expansion series for the Wilson-Fisher fixed point in $d = 4 -
  \epsilon$ dimensions {\cite{Brezin:1976vw}}, necessitating Borel
  resummations for the critical exponents {\cite{Guida:1998bx}}. The same
  considerations should apply to the long-range bosonic model {\eqref{MFT}}.
  On the other hand, for fermions we have established in Section
  \ref{analyticity} that the fixed point depends analytically on
  $\varepsilon$. This implies that the formal power series solution will be
  convergent for small $\varepsilon$. For a direct proof of convergence of the
  fermionic power series expansion via tree expansion, see the next appendix.
\end{remark}

\section{Fixed point via the tree expansion}\label{App:trees}

In the main sections of this paper we provided an explicit rigorous
construction of a nontrivial RG fixed point, by finding the appropriate Banach
space, which the RG map acts on, and by proving its contractivity in an
appropriate neighborhood of this space. In the case of fermionic theories, as
in the case at hand, the fixed point can also be found by a different
strategy, which bypasses the construction of the Banach space and the
contractivity argument, and is based on an expansion in tree diagrams,
sometimes referred to as \ ``Gallavotti-Nicol{\`o}'' trees {\cite{GN85}}, see
also {\cite{Ga,BGbook}} and {\cite{Gentile:2001gb}} for a review in the
context of interacting fermionic theories. While the two constructions build
on the same general foundations from Sections
\ref{introduction}-\ref{detailsRG} (trimmed representation of the interaction,
weighted norm $\| \cdot \|_w$ for measuring the size of couplings, and the
norm bounds on $D$ and $R^{\ell_1, \ldots, \ell_n}_{\ell}$), the tree
expansion is closer in spirit to a direct combinatorial proof convergence of
the formal $\varepsilon$-expansion of Appendix \ref{sec:formal}. Let us
briefly describe here the construction of the fixed point via trees.

The starting point is the fixed point equation for the irrelevant couplings
$u_{\ell}$, with $\ell = \text{2R}, \text{4R}, \text{6R}, 8, 10, \ldots$, which we rewrite,
extracting the term $(n ; (\ell_1, \cdots, \ell_n)) = (1 ; \ell)$ explicitly:
\begin{equation}
  (1 - D) u_{\ell} = \sum_{n \geqslant 1} \sum^{\ast}_{(\ell_i)_1^n}
  R_{\ell}^{\ell_1, \ldots, \ell_n} (H_{\ell_1}, \ldots, H_{\ell_n}),
  \label{J.1}
\end{equation}
where the $\ast$ on the sum indicates the constraint that, if $n = 1$, then
$\ell_1 \neq \ell$. We look for a solution in power series in $\nu$ and
$\lambda$, $u_{\ell} = \sum_{k_1, k_2 \geqslant 0} u^{(k_1, k_2)}_{\ell}
\nu^{k_1} \lambda^{k_2}$, with $u^{(k_1, k_2)}_{\ell} = 0$ unless $k_1 + k_2
\geqslant | \ell | / 2 - 1$ and, moreover, $k_1 + k_2 \geqslant 2$ for $\ell =
\text{2R}, \text{4R}$ and $k_1 + k_2 \geqslant 3$ for $\ell = \text{6R}$. By plugging this
ansatz in {\eqref{J.1}}, we get\footnote{Note that since $u_{\ell}$ is
irrelevant, $(1 - D)$ is an invertable operator, see footnote \ref{invert}.}
\begin{equation}
  u^{(k_1, k_2)}_{\ell} = \sum_{n \geqslant 1} \sum^{\ast}_{(\ell_i)_1^n}
  \sum^{(k_1, k_2)}_{\tmscript{\begin{array}{c}
    \{ k_{_{1, i}}, k_{2, i} \}_{i = 1, 2}
  \end{array}}} (1 - D)^{- 1} R_{\ell}^{\ell_1, \ldots, \ell_n} (H^{(k_{1, 1},
  k_{2, 1})}_{\ell_1}, \ldots, H^{(k_{1, n}, k_{2, n})}_{\ell_n}), \label{J.2}
\end{equation}
where the third sum runs over integers $k_{1, i}, k_{2, i} \geqslant 0$, with
$i = 1, \ldots, n$, such that $k_{1, 1} + \cdots + k_{1, n} = k_1$ and $k_{2,
1} + \cdots + k_{2, n} = k_2$. Moreover, the arguments $H^{(k_{1, i}, k_{2,
i})}_{\ell_i}$ should be interpreted as being equal to $u^{(k_{1, i}, k_{2,
i})}_{\ell_i}$ if $\ell_i \neq \text{2L}, \text{4L}, \text{6SL}$, to $\nu$ if $\ell_i = 2
L$ (in which case we set $(k_{1, i}, k_{2, i}) = (1, 0)$), to $\lambda$ if
$\ell_i = \text{4L}$ (in which case we set $(k_{1, i}, k_{2, i}) = (0, 1)$), and to
$\mathfrak{X}_{\ast}$ if $\ell_i = \text{6SL}$ (in which case we set $(k_{1,
i}, k_{2, i}) = (0, 2)$). Eq.{\eqref{J.2}} can be graphically represented as
follows:\vspace{-1em}
\begin{center}
\raisebox{-0.00198258270119149\height}{\includegraphics[width=14.8909550045914cm,height=4.23470090515545cm]{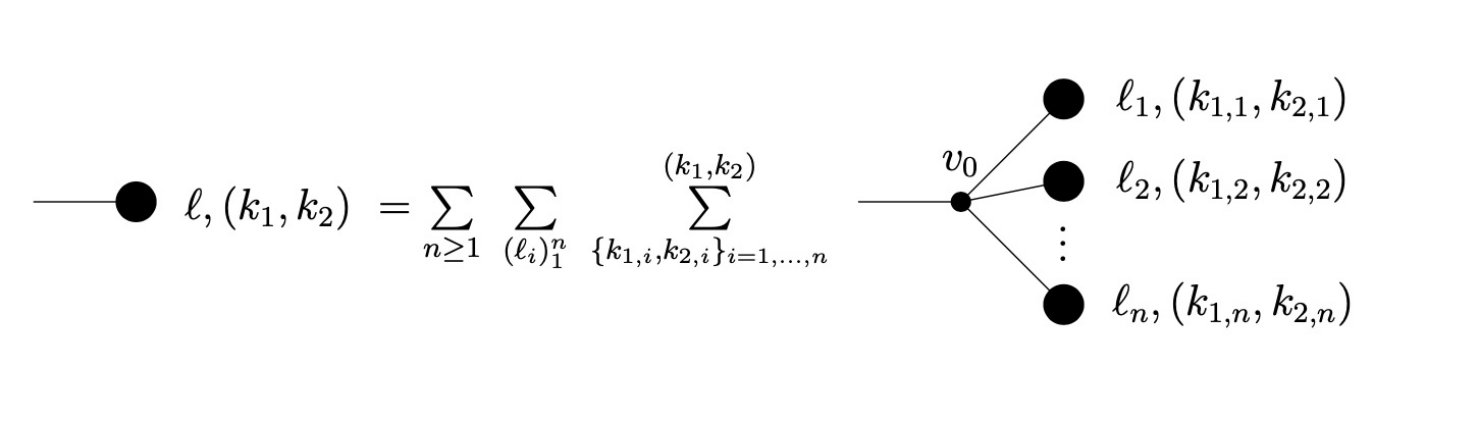}}
\end{center}\vspace{-1em}
where the vertex labelled $v_0$ in the right side represents the action of $(1
- D)^{- 1} R_{\ell}^{\ell_1, \ldots, \ell_n}$; we shall say that the $n$ lines
(or ``branches'') labelled $\ell_1, \ldots, \ell_n$ ``enter the vertex
$v_0$''; similarly, we'll say that the branch to the left of $v_0$ ``exits''
from $v_0$: it carries the label $\ell$ and its left endpoint is called
{\tmem{root}}. In the special cases ($\ell_i, (k_{1, i}, k_{2, i})
_{}) = (\text{2L}, (1, 0)), (\text{4L}, (0, 1)), (\text{6SL}, (0, 2))$, the big
dots in the right side will be reinterpreted as small dots with labels $\nu$,
$\lambda$, $\mathfrak{X}_{\ast}$, respectively:\vspace{-1em}
\begin{center}
\raisebox{-0.00503743641712335\height}{\includegraphics[width=14.8909550045914cm,height=1.666650268923cm]{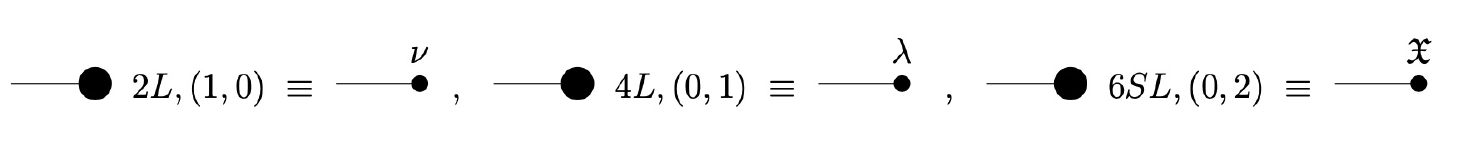}}
\end{center}\vspace{-1em}
Iterating the graphical equation above until the endpoints are all small dots
with labels $\nu$, $\lambda$, or $\mathfrak{X}_{\ast}$, we obtain an expansion
in tree diagrams of the following form:\vspace{-1em}
\begin{center}
\raisebox{-0.00136208974362385\height}{\includegraphics[width=14.8909550045914cm,height=6.16379706152433cm]{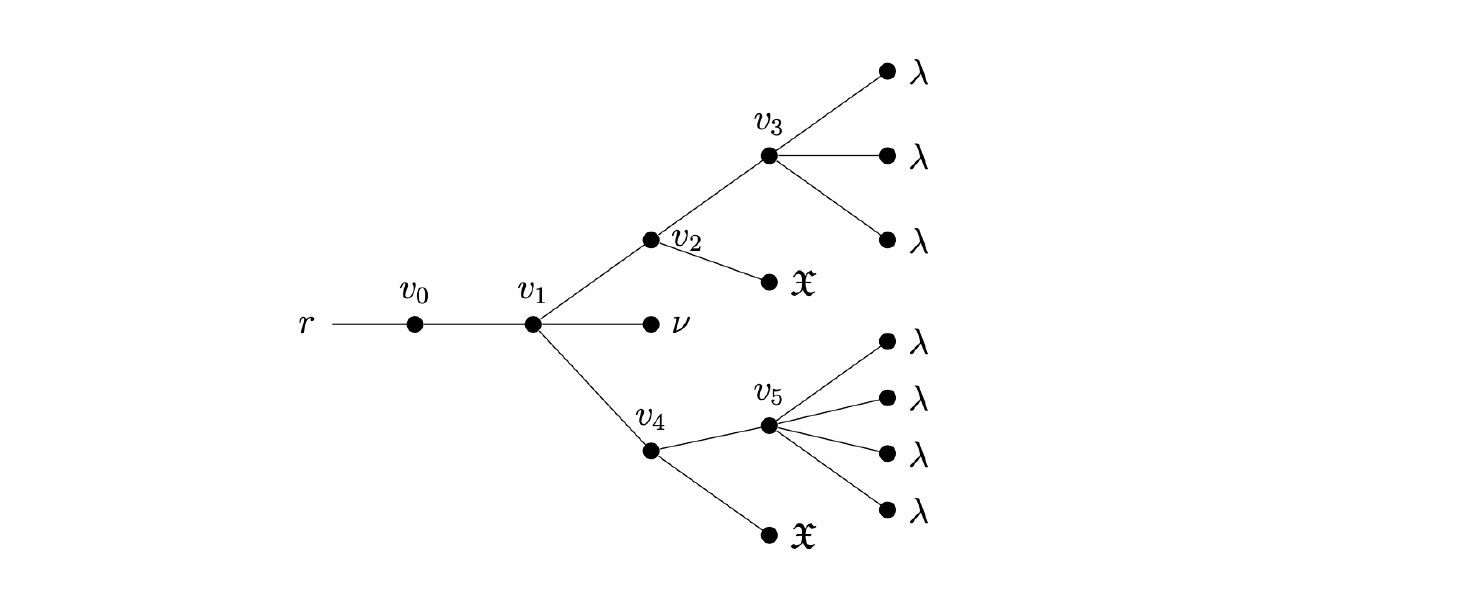}}
\end{center}\vspace{-1em}
In this example, the branch labels are left implicit; each vertex $v_r$, with
$r = 1, \ldots, 5$, is associated with the action of an operator $(1 - D)^{-
1} R^{\ell_1, \ldots, \ell_n}_{\ell}$, with $\ell$ the label of the line
exiting from $v_r$, $n$ the number of lines entering it, and $\ell_1, \ldots,
\ell_n$ their labels. If $m_{\nu}$, $m_{\lambda}$ and $m_{\mathfrak{X}}$ are
the numbers of endpoints of type $\nu, \lambda$ and $\mathfrak{X}_{\ast}$,
respectively (in the example, $m_{\nu} = 2$, $m_{\lambda} = 6$ and
$m_{\mathfrak{X}} = 2$), and $\ell_0$ is the label of the line exiting from
$v_0$, then the tree contributes to $u^{(m_{\nu}, m_{\lambda} + 2
m_{\mathfrak{X}})}_{\ell_0}$.

Let us now use the norm bounds on $D$ and $R^{\ell_1, \ldots,
\ell_n}_{\ell}$, see Sections \ref{sec:Dil} and \ref{sec:NormBounds}, to bound
the value of any such tree in the $\| \cdot \|_w$ norm, assuming $| \nu |, |
\lambda | \leqslant \delta$. The norm bounds on $D$ imply that the norm of $(1
- D)^{- 1}$ is bounded by some $d_{\gamma} > 1$ (uniformly in $\ell$). Using
this and Eq. {\eqref{Rest}}, the norm of the value of a tree is bounded by the
product of the norms of its endpoints, which is bounded by
\begin{equation}
  \delta^{m_{\nu} + m_{\lambda}} (B_{\gamma} \delta^2)^{m_{\mathfrak{X}}}
  \leqslant (B_{\gamma} \delta)^k, \qquad k = m_{\nu} + m_{\lambda} + 2
  m_{\mathfrak{X}}, \label{endpoint}
\end{equation}
times the product over vertices $v_r$ (i.e. all vertices $v$ which are not
endpoints):
\begin{equation}
  \prod_{v \text{ not e.p.}} d_{\gamma} C^{n_v - 1}_{\gamma} \gamma^{-
  D_{l_v}} C_0^{\sum^{n_v}_{i = 1} | \ell_i (v) |}, \label{prodJ.4}
\end{equation}
where $l_v = | \ell_v |$, with $\ell_v$ the label of the line exiting from
$v$, and $\ell_i (v)$ the label of the $i$-th line entering $v$. Note that
$\ell_i (v) = \ell_{v'}$ for some $v'$, which may be an endpoint. Therefore,
{\eqref{prodJ.4}} can be equivalently rewritten as
\begin{equation}
  C_0^{- | \ell_{v_0} |} \times \prod_{v \text{ not e.p.}} d_{\gamma} C^{n_v -
  1}_{\gamma} \gamma^{- D_{l_v}} C_0^{l_v} \times
  \prod_{v \text{ e.p.}} C_0^{l_v} .
\end{equation}
Replacing the first factor by 1, and including the factor {\eqref{endpoint}}
associated with the endpoints, the value of a single tree is bounded by
\begin{equation}
  (B_{\gamma} C_0^4 \delta)^k \prod_{v \text{ not e.p.}} d_{\gamma} C^{n_v -
  1}_{\gamma} \gamma^{- D_{l_v}} C_0^{l_v}, \label{singletree}
\end{equation}
where we also used that $\sum_{v \text{ e.p.}} l_v = 2 m_{\nu} + 4 m_{\lambda}
+ 6 m_{\mathfrak{X}} \leqslant 4 k$.

As the next step we have to sum over all possible values of $\ell_i$'s which
label tree branches. We take $\gamma$ sufficiently large so that $\gamma^{- (d
/ 4 - \varepsilon / 2)} C_0 < 1$, implying (recall $D_l = l (d / 4 -
\varepsilon / 2) - d$)
\begin{equation}
  \sum_{l = 2}^{\infty} \gamma^{- D_{l_{}}} (C_0^{})^l < \infty .
  \label{eq:J.7}
\end{equation}
Then the sum $\sum_{\{ \ell_v \}}$of {\eqref{singletree}} is bounded by
\begin{equation}
  (B_{\gamma} C_0^4 \delta)^k \prod_{v \text{ not e.p.}} \tilde{d}_{\gamma}
  C^{n_v - 1}_{\gamma} . \label{sumoverl}
\end{equation}
Furthermore, it is easy to see that
\begin{equation}
  \sum_{v \text{ not e.p.}} (n_v - 1) = m_{} - 1,
\end{equation}
where $m_. = m_{\nu} + m_{\lambda} + m_{\mathfrak{X}} \leqslant k$ is the
total number of the endpoints, and
\begin{equation}
  \sum_{v \text{ not e.p.}} 1 \leqslant \frac{1}{2} \sum_{v \text{ e.p.}} l_v,
  \label{vnotep}
\end{equation}
as implied by $| \ell | \leqslant \sum_i | \ell_i | - 2$ in each vertex $v$
(note the constraint {\eqref{rholbound}} for $n \geqslant 2$, and that for $n
= 1$ the sum in {\eqref{J.1}} starts from $l_1 = l + 2$). Using the last two
relations, we can estimate {\eqref{sumoverl}} by
\begin{equation}
  (B_{\gamma} C_0^4 C_{\gamma} \widetilde{d}^2_{\gamma}\delta)^k .
  \label{convest}
\end{equation}
Finally we have to sum this expression over all trees subject to the
constraint on the endpoints $m_{\nu} + m_{\lambda} + 2 m_{\mathfrak{X}} = k$.
The number of such trees is smaller than $4^k$ (see, e.g., Lemma A.1 of
{\cite{Gentile:2001gb}}). We conclude that the contribution to $u_{\ell}$ of
order $k$ is bounded in the $\| \cdot \|_w$ norm by $(A^{}_{\gamma}
\delta)^k$, for some $\gamma$-dependent constant $A_{\gamma}$. This implies
that the tree expansion for $u_{\ell}$ is convergent for $\delta \leqslant
\delta_0 (\gamma)$ small enough. This concludes the construction of the
irrelevant couplings $u_{\ell}$ in terms of the fixed point relevant couplings
$\nu$ and $\lambda$, which implies automatically that $u_{\ell}$ are analytic
in $\nu$ and $\lambda$ in the neighborhood $| \nu |, | \lambda | \leqslant
\delta_0 (\gamma)$ of the origin in $\mathbb{C}^2$. The same argument also
establishes analyticity of $u_{\ell}$ in $\varepsilon$ as long as it belongs
to the complex half-plane $\tmop{Re} \varepsilon < d / 6$ where couplings
$u_{\ell}$ are all irrelevant.

We are left with the beta function equations for the relevant couplings. Via
the same strategy, we find that they are given by the first two equations of
{\eqref{FPeqs}}, with $e_{\nu}^{(0)}$ and $e_{\lambda}^{(0)}$ expressed in
terms of two tree expansions, convergent for $\delta$ sufficiently small. In
conclusion, at the fixed point, $\nu$ and $\lambda$ satisfy
\begin{equation}
  \nu = \gamma^{\frac{d}{2} + \eps}  (\nu + I_1 \lambda) + e^{(0)}_{\nu} (\nu,
  \lambda, \varepsilon), \qquad \lambda = \gamma^{2 \eps}  (\lambda + I_2
  \lambda^2) + e^{(0)}_{\lambda} (\nu, \lambda, \varepsilon),
  \label{nulambdaTree}
\end{equation}
with $e_{\nu}^{(0)}$ and $e_{\lambda}^{(0)}$ two analytic functions of $\nu,
\lambda, \varepsilon$ for $| \nu |, | \lambda | \leqslant \delta_0 (\gamma)$,
$\tmop{Re} \varepsilon < d / 6$. Moreover, we have $e_{\nu}^{(0)}$ and
$e_{\lambda}^{(0)}$ of order $\delta^2$ and $\delta^3$, respectively where
$\delta = \max (| \nu |, | \lambda |)$ and for $\varepsilon$ in any compact
subset of $\tmop{Re} \varepsilon < d / 6$. $I_1$ and $I_2$ also depend
analytically on $\varepsilon$. By the analytic implicit function theorem,
these equations have a unique solution $\nu_{\ast} (\varepsilon)_{}$,
$\lambda_{\ast} (\varepsilon)$ which is defined in the disk $| \varepsilon |
\leqslant \varepsilon_0 (\gamma)$ and is analytically close to the lowest
order approximated solution $\lambda_0 = (1 - \gamma^{2 \varepsilon}) / I_2$,
$\nu_0 = I_1 \lambda_0 / \left( 1 - \gamma^{\frac{d}{2} + \varepsilon}
\right)$.

\begin{remark}
  The condition of $\gamma$ large is not truly required for the tree expansion
  to converge; in fact it converges for any $\gamma > 1$ (for $\lambda, \nu$
  sufficiently small). In the proof above, large $\gamma$ was needed due to
  the pessimistic way in which we bounded some combinatorial factors in the
  previous sections. Note that the origin of the factor $(C_0)^l$ in
  {\eqref{eq:J.7}} has to be traced back to the factors $C^{| \ell_i |}_0$ in
  {\eqref{rholbound}}. A critical rereading of the proof leading to those
  factors shows that the factor $C_0^{\sum^{n_v}_{i = 1} | \ell_i (v) |}$ in
  {\eqref{prodJ.4}} can be replaced by $C_{\text{R}} \binom{\sum^{n_v}_{i = 1} | \ell_i
  (v) |}{l_v} C_0^{\sum^{n_v}_{i = 1} | \ell_i (v) | - l_v}$.\footnote{See in
  particular the first equality in {\eqref{terms213}} and the second one in
  {\eqref{NTbound1}}. The factor $C_{\text{R}} > 1$ is the constant in
  {\eqref{TRfinal}}; it appears only if $l_v = 2, 4$. In the main text it was
  absorbed into $C_0$ (footnote \ref{CRabsorb}), but now we keep it explicit,
  since it blows up as $\gamma \rightarrow 1$, due to the blowup of constants
  $C_{1, 2, 3}$ from the end of Appendix \ref{sec:Trim}.} The product of the
  factors $C_0^{\sum^{n_v}_{i = 1} | \ell_i (v) | - l_v}$ over the vertices
  $v$ that are not endpoints gives $^{} C_0^{\sum^{}_{v \, e.p.} l_v -
  l_{v_0}}$, with $v_0$ the vertex attached to the root; this does not need
  any condition on $\gamma$. \ The product of $C_{\text{R}}$'s is bounded by $C_{\text{R}}^{2
  k}$ because of {\eqref{vnotep}}. Finally, the sum over the branch labels,
  given the type of endpoints, reduces to $\sum_{\{ \ell_v \}} \prod_{v
  \text{ not e.p.}} \gamma^{- D_{l_v}} \binom{\sum^{n_v}_{i = 1} | \ell_i (v)
  |}{l_v}$, which can be bounded as explained in Appendix A.6.1 of
  {\cite{Gentile:2001gb}}, leading to a factor smaller than $\left( \frac{1}{1
  - \gamma^{- \alpha}} \right)^{4 k}$, for some $\alpha > 0$ and any $\gamma >
  1$. 
\end{remark}

\begin{remark}
  In Wilsonian RG, the single RG step contains all the information about the
  fixed point, so that we should feel free to forget about what happened in
  the RG past. The fact that the tree expansion represents the fixed point
  kernels by tree diagrams with several levels may superficially seem to go
  against this standard idea. This is not the case: the tree expansion as
  presented in this appendix is just a way to solve the fixed point equation
  for a single RG step.\footnote{To make an analogy with something already
  seen, consider the construction of the fixed point sextic semilocal term in
  Section \ref{sec:intuitive}: $\Xterm_{\ast}$ is the solution to the single
  step equation {\eqref{eq:RGX}}, and its explicit expression in terms of
  $\lambda$ involves a sum over many ``levels'', i.e., the integers $n$ in
  {\eqref{fixedX}}.} In general, trees with several levels are exponentially
  suppressed, as compared to ``short'' ones: this is the so called
  {\tmem{short-memory property}} (see e.g. the Remark after (7.26) in
  {\cite{BGPS}}). 
\end{remark}

The tree expansion can be easily adapted to the construction of the full Wilsonian RG flow of the effective couplings from the ultraviolet to the infrared, see Section \ref{App:trees.sub} below for a brief discussion of this
fact, and to the
computation of correlation
functions and critical exponents. It has been used to construct them in
several 1D fermionic theories {\cite{BGPS,BMXYZ,BMWICMP,BFMTh}} and 2D
statistical mechanics models at criticality
{\cite{MaCMP,GMCMP,GMPRL,BFMCMP,BFMPRL,Dimers,GMTJSM,GMTCMP}}, whose
nontrivial fixed points are all in the Luttinger liquid universality class. In
these cases, the construction of the fixed point requires a proof that the
beta function for the quartic coupling is asympotically vanishing in the
infrared limit; historically, the proof of vanishing beta functions in these
models has first been proved via a comparison with the Luttinger model exact
solution {\cite{BGPS}}, and later via a combined use of Ward Identities and
Schwinger-Dyson equations {\cite{BMWICMP}}. We are not aware of a construction
of nontrivial fixed points in the Luttinger liquid universality class via
methods different from the tree expansion; it would be an interesting exercise
to reproduce their construction via a contraction argument in a suitable
Banach space of interactions, as done in this paper for long-range symplectic
fermions with quartic interaction.

The fermionic nature of models such as the one studied in this paper makes the
approach based on the tree expansion an extremely efficient tool for
constructing the RG fixed point, arguably simpler than the one based on the
contraction argument in a Banach space. However, we do not expect that the
tree expansion is as a general scheme as the other, which is, to date, the
only available technique for constructing nontrivial bosonic fixed points, see
Appendix \ref{literature}, and it looks the most promising (at least
conceptually) for approaching the non-perturbative problem of constructing
very non-Gaussian fixed points in the vicinity of approximate fixed points
(possibly computed via the truncation of some other alternative scheme, such
as numerical FRG). This explains the reason why in the main sections of this
paper we decided to follow the scheme based on the contraction argument in a
Banach space: it provides a benchmarch for other approaches, like the
Functional Renormalization Group (whose conceptual similarity allows for a
direct comparison of results), and it displays general features which do not
depend on the specific, fermionic, nature of the problem.

\subsection{On the flow of the effective couplings}\label{App:trees.sub}
A tree expansion analogous to the one described above for the construction of the fixed point can be used to compute the whole sequence of effective potentials associated with the Wilsonian RG flow 
from the ultraviolet to the infrared scales. 
Such a generalized tree expansion was described in several previous reviews on the subject, see in particular \cite{Gentile:2001gb}. 
Suppose that we interested in constructing 
the model formally defined by the interacting Grassmann measure $Z^{- 1} d \mu_P (\psi) e^{H^{(0)} (\psi)}$ at all distances (rather than being interested ``just" in the construction of its infrared fixed point, as done in this paper), 
where $d \mu_P (\psi)$ is the Grassmann Gaussian integration with the propagator $P(x)$ in \eqref{Pchi} (with fixed ultraviolet cutoff but without any infrared one) and $H^{(0)}$ is a local interaction 
like the one in \eqref{eq:V0}, with bare couplings $\nu_0,\lambda_0$. 
The partition function $\int d\mu_P (\psi) e^{H^{(0)} (\psi)}$ and the closely related generating function of correlations can be computed iteratively, by first integrating out momenta in the annulus of radii 
$\gamma^{-1}$ and $1$, then in the one of radii $\gamma^{-2}$ and $\gamma^{-1}$, and so on. In formulae, this means rewriting the propagator $P(x)$ as $\sum_{h\le 0}g^{(h)}(x)$, with 
$g^{(h)}(x)=\gamma^{h(d/2-\varepsilon)}g(\gamma^hx)$ and $g(x)$ the same as in \eqref{eq:split}; correspondingly, the fluctuation field $\psi$ is decomposed as $\psi=\sum_{h\le 0}\psi^{(h)}$ and the Grassmann 
Gaussian integration $d\mu_P(\psi)$ as $\prod_{h\le 0}d\mu_{g^{(h)}}(\psi^{(h)})$ (cf. with \eqref{eq:decompos}), in terms of which we define the sequence (cf. with \eqref{eq:Heff0}): 
$$e^{H_{\eff}^{(h-1)} (\psi)} = \int d \mu_{g^{(h)}} (\psi^{(h)}) e^{H_{\eff}^{(h)} (\psi+ \psi^{(h)})}, \quad h\le 0,$$
with $H^{(0)}_{\eff}\equiv H^{(0)}$. After appropriate rescaling, we obtain the effective potentials (cf. with \eqref{Hprime})
$$H^{(h)}(\psi)=H^{(h)}_{\eff}(\gamma^{h[\psi]}\psi(\cdot\ \gamma^h)),$$ 
which satisfy the RG equation 
$$R[H^{(h)}]=H^{(h-1)},$$
with $R=R(\varepsilon,\gamma)$ the same renormalization map introduced after \eqref{Hprime}. A mild generalization of the construction of this paper allows us to prove that,
for $\lambda_0$ {\it positive} and sufficiently small (we are taking here $\varepsilon$ positive and small, as well), there exists a choice of $\nu_0$ such that the whole sequence of effective potentials $\{H^{(h)}\}_{h\le 0}$ is well defined, they all belong to the same
Banach space (the same we used to construct the fixed point) and $\lim_{h\to-\infty}H^{(h)}=H_*$, where $H_*$ is the fixed point constructed above. Correspondingly, the local part of $H^{(h)}$ is parametrized by 
two running coupling constants $\nu_h,\lambda_h$, which interpolate between the bare values $\nu_0,\lambda_0$ and the fixed point values $\nu=\lim_{h\to-\infty}\nu_h$, $\lambda=\lim_{h\to-\infty}\lambda_h$.
In particular, the sequence $\{\nu_h,\lambda_h\}_{h\le0}$ is small, uniformly in the scale index $h$.

Remarkably, the effective potentials $H^{(h)}$ can be expressed in terms of a convergent tree expansion, analogous to the one described above, with the important difference that now the endpoints carry a scale label and are, therefore, associated with couplings $\nu_k$ or $\lambda_k$, computed at scales $h<k\le 0$. From the knowledge of the effective potentials, one can reconstruct all
the observables one is interested in, including the correlation functions computed at arbitrary finite distance (before any scaling limit is taken): these will be expressed as convergent expansions {\it in the 
whole sequence of running couplings} $\{\nu_h,\lambda_h\}_{h\le 0}$. See, e.g., Chapt.12-13-14 in \cite{Gentile:2001gb}. Note that the existence of such a convergent expansion does not imply
convergence of the naive perturbation theory in the bare couplings: schematically, the relation between $\lambda_h$ and $\lambda_0$ has the same features as the one between $\lambda(t)$
and $\lambda_0$ in \eqref{soltoyflow}; in particular, $\lambda_h$ is analytic in $\lambda_0$ non-uniformly in $h$, while it is Borel summable in $\lambda_0$ uniformly in $h$. Therefore, 
pre-scaling-limit observables are expected to be, at best, Borel summable in $\lambda_0$. On the contrary, observables at the fixed point, such as critical exponents, are expressed 
as convergent expansions in the fixed point couplings $\nu_*,\lambda_*$ only, and, therefore, recalling that $\nu_*$ and $\lambda_*$ are analytic in $\varepsilon$, they can be proved to be analytic functions of $\varepsilon$, as well. 

\section{Rigorously constructed bosonic fixed points}\label{literature}

{In this appendix we will mention some existing rigorous constructions of non-gaussian 
bosonic fixed points. Earlier works not directly focusing on such fixed points, but instrumental for acquiring rigorous RG control in bosonic theories, include \cite{BCGNOPS80,Gal79,gaw1980,Gawedzki1985,brydges1990}.}

In 1998, Brydges, Dimock and Hurd {\cite{Brydges1998}} gave the first
construction of a fixed point in a bosonic scalar field theory with a
long-range interactions. In analogy to {\eqref{aMFT}}, their bare action can
be written schematically as
\begin{equation}
  \tmop{MFT} (\varphi) + \nu \int d^d x\, \varphi^2 + \lambda \int d^d x\,
  \varphi^4, \label{MFT}
\end{equation}
i.e. a gaussian scale-invariant Mean Field Theory of a bosonic field $\varphi$
in $\mathbb{R}^d$ of dimension $[\varphi] = d / 4 - \varepsilon / 2$ with a
quadratic and quartic interactions. They considered the model in $d = 4$,
which necessitated adding to {\eqref{MFT}} one more relevant local interaction
$\int d^d x (\partial \varphi)^2$. Contrary to what the title of
{\cite{Brydges1998}} may suggest, it does not provide a rigorous definition of
the Wilson-Fisher fixed point in $d = 4 - \eps$. The two models differ already
in their perturbative critical exponents. E.g. the scaling dimension of
$\varphi$ gets corrections at $O (\varepsilon^2)$ in the Wilson-Fisher model,
while such corrections are absent in the model of {\cite{Brydges1998}} at any
order in $\eps$.

In 2000, Mitter and Scoppola {\cite{Mitter:1998vp}} studied a different model
perturbing MFT by a $\delta$-function interaction:
\begin{equation}
  \tmop{MFT} (\varphi) + g \int d x\, \delta^{(N)} (\varphi (x)),
\end{equation}
where $\varphi$ is an $N$-component field in $\mathbb{R}^{d = 1}$. The
$\delta$-function penalizes configurations when $\varphi (x)$ passes through
zero, which physically describes repulsion of a polymer from an impurity. The
scaling dimension of this interaction is $- N [\varphi]$. They constructed a
fixed point of this model in the case when $[\varphi]$ is negative and small
while $N$ is large so that $- N [\varphi] = 1 - \eps$ is close to marginality.

In 2003, Brydges, Mitter, and Scoppola {\cite{Brydges:2002wq}} constructed a
fixed point of exactly the model {\eqref{MFT}}. Following
{\cite{Mitter:1998vp}}, they used fluctuation covariance of finite support in
$x$-space, simplifying the proof compared to the construction in
{\cite{Brydges1998}}.\footnote{In this paper we used fluctuation covariance of
finite support in Fourier space, not in $x$-space, see {\eqref{eq:split}}.
Unlike in {\cite{Brydges:2002wq}}, our fluctuation covariance has zero
integral in $x$-space, which was useful at times although not essential.}
Although nominally $d = 3$ in {\cite{Brydges:2002wq}}, the proof should apply
also for $d = 1, 2$ {\cite{Mitter-talk,Mitter-comm}}. See also the nice review
in {\cite{Mitter:2005zx}}. Further work in this direction was done by
Abdesselam {\cite{Abdesselam:2006qg}} who constructed a full renormalization
group trajectory from MFT at short distances to the fixed point of
{\cite{Brydges:2002wq}} at long distances. More recently, Slade {\cite{Slade}}
considered an analogous fixed point for an $n$-component field $\varphi$. He
also considered the case $n = 0$, corresponding to the self-avoiding random
walk. This formal limit is analyzed rigorously by considering a theory of two
scalar bosons and two scalar fermions whose global symmetry is $\tmop{OSp} (2
| 2 )$. Unlike in our model, there is no quartic interactions for
fermions in {\cite{Slade}} because there are only two of them. Such a
``supersymmetric'' model was also studied earlier by Mitter and Scoppola
{\cite{Mitter:2007vy}}.

Physically, model {\eqref{MFT}} should describe the critical point of
the long-range Ising model {\cite{Fisher:1972zz}} (see also \cite{SUZUKI1972313}), and much is known or
conjectured about it. The critical point is expected to have conformal
invariance {\cite{Paulos:2015jfa}}. At $\varepsilon = \varepsilon_c (d)$ the
critical point should cross over to the local Wilson-Fisher fixed point plus a
decoupled Gaussian sector {\cite{Behan:2017emf}}. See also
{\cite{Benedetti:2020rrq}} for higher-loop perturbative computations of
critical exponents.

\small
\setlength{\parskip}{-2pt}

\providecommand{\href}[2]{#2}\begingroup\raggedright\endgroup

\end{document}